\documentclass[10pt,letterpaper]{article}
\usepackage[top=0.85in,left=1in,footskip=0.75in,marginparwidth=1in]{geometry}
\usepackage[utf8]{inputenc}
\usepackage{natbib}
\usepackage{cite}
\usepackage{nameref,hyperref}
\usepackage[right]{lineno}
\usepackage{microtype}
\DisableLigatures[f]{encoding = *, family = * }
\usepackage{changepage}
\usepackage[aboveskip=1pt,labelfont=bf,labelsep=period,singlelinecheck=off]{caption}
\makeatletter
\renewcommand{\@biblabel}[1]{\quad#1.}
\makeatother
\usepackage{lastpage,fancyhdr,graphicx}
\usepackage{epstopdf, epsfig}
\fancyheadoffset[L]{2.25in}
\fancyfootoffset[L]{2.25in}
\usepackage{color}
\definecolor{Gray}{gray}{.25}
\usepackage{sidecap}
\usepackage{wrapfig}
\usepackage[pscoord]{eso-pic}
\usepackage[fulladjust]{marginnote}
\reversemarginpar

\usepackage{mathabx}
\usepackage{morefloats}
\usepackage{upgreek}
\usepackage{bm}
\usepackage{ragged2e}

\input{definitions.sty}
\bmdefine{\bPhi}{\Phi}
\bmdefine{\bEqual}{=}
\bmdefine{\bcirc}{\circ}
\def\minf {\mathrm{M}_{\infty}}
\def\rec {\mathrm{Re}_c}

\begin{document}
\vspace*{0.35in}

\begin{flushleft}

{\Large
\textbf\newline{Bursting and reformation cycle of the laminar separation bubble over a NACA-0012 aerofoil: The underlying mechanism}
}
\newline\\
Eltayeb M. ElJack\textsuperscript{1,*}, and
Julio Soria\textsuperscript{2,3}\\
\bigskip
\bf{1} Mechanical Engineering Department, University of Khartoum, Khartoum, Sudan\\
\bf{2} Laboratory for Turbulence Research in Aerospace and Combustion, Department of Mechanical and Aerospace Engineering, Monash University, Melbourne, Australia\\
\bf{3} Aeronautical Engineering Department, King Abdulaziz University, Jeddah, Saudi Arabia\\
\bigskip
* emeljack@uofk.edu

\end{flushleft}

\section*{Abstract}
The present investigation shows that a triad of three vortices, two co-rotating vortices (TCV) and a secondary vortex lies beneath them and counter-rotating with them, is behind the quasi-periodic self-sustained bursting and reformation of the laminar separation bubble (LSB) and its associated low-frequency f\/low oscillation (LFO). The upstream vortex of the TCV (UV) is driven by the gradient of the oscillating-velocity across the laminar portion of the separated shear layer and is faithfully aligned with it. The UV front tip forms a half-saddle on the aerofoil surface just upstream the separation point of the shear layer, and its rear end forms a full-saddle. The downstream vortex of the TCV (DV) is submerged entirely inside the region of the LSB and aligned such that its front tip forms a full-saddle with the UV while its rear tip forms a half-saddle on the aerofoil surface upstream the reattachment point. The secondary vortex acts as a roller support that facilitates the rotation and orientation of the TCV. A global oscillation in the f\/low-f\/ield around the aerofoil is observed in all of the investigated angles of attack, including at zero angle of attack. The strength and dynamics of the TCV determine the time-scale of the global f\/low oscillation. The magnitude of oscillation increases as the angle of attack is increased. The f\/low switches between an attached-phase against an adverse pressure gradient (APG) and a separated-phase despite a favourable pressure gradient (FPG) in a periodic manner with some disturbed cycles. When the direction of the oscillating-f\/low is clockwise, it adds momentum to the boundary layer and helps it to remain attached against the APG and vice versa. The transition location along the separated shear layer moves upstream when the oscillating-f\/low rotates in the clockwise direction causing early-transition and vice versa. The best description of the mechanism is that of a whirligig. When the oscillating-f\/low rotates in the clockwise direction, the UV whirls in the clockwise direction and store energy until it is saturated, then the process is reversed. The strength and direction of rotation of the oscillating-f\/low; the transition location along the separated shear layer; the gradient of the oscillating-velocity across the laminar portion of the separated shear layer; the strength and dynamics of the triad of vortices; and the gradient of the oscillating-pressure along the aerofoil chord are interlinked, generate, and sustain the LFO. When these parameters are not synchronised, the LFO cycle is disturbed. The present investigation paves the way for formulating a time-accurate physics-based model of the LFO and stall prediction, and opens the door for control of its undesirable effects.

\section{Introduction}\label{sec:intorudction}
The laminar separation bubble (LSB), its stability, and its associated low-frequency oscillation (LFO) have been in the centre of aerodynamics research in the past $30$ years. The investigated conf\/igurations are two folds; an LSB on the suction surface of two-dimensional aerofoils and on f\/lat plates. The LSB is formed naturally in the former type of setup due to the curvature of the aerofoils. Whereas, in the latter the LSB is induced by an imposed adverse pressure gradient (APG). Most of the studies are compressible; however, several incompressible investigations are carried out recently as will be discussed later. The overwhelming majority of previous studies are around clean aerofoils; however, considerable research work investigated the LSB and the LFO around iced aerofoils. A landmark work on LSBs was carried out by~\citet{gaster1967structure}. Gaster wanted to eliminate the effect of the aerofoil geometry and generate data of different LSBs. He invented a brilliant model that allowed him to vary both the Reynolds number and the pressure distribution. The model was a f\/lat plate with an adjustable pressure distribution. He used the model to carry out a series of experiments that provided enough data sets for him to characterise the LSB and its bursting process. He found that the structure of the LSB depends on two parameters. The f\/irst parameter is the Reynolds number of the separated boundary layer, and the second parameter is a function of the pressure rise over the region occupied by the bubble. Then he determined conditions for the bursting of short bubbles by a unique relationship between these two parameters.\newline
\citet{zaman1987effect} conducted wind--tunnel measurements at NASA Langley Research Center. They measured the lift, the drag, and wake velocity under acoustic excitation at chord Reynolds number ranging from $4\times10^4$ to $1.4\times10^5$. The authors observed a sharp spike in the velocity spectra at a considerably low frequency, lower than the bluff-body shedding frequency. They noted that the low frequency varies with the free-stream velocity and the Strouhal number was in the order of $0.02$. Zaman and co-workers continued the work of~\citet{zaman1987effect} at NASA Lewis Research Center by conducting a series of experiments on the characteristics, stability and control of the LSB.~\citet{zaman1988natural} and~\citet{zaman1989natural} investigated the LFO phenomenon experimentally for the f\/low-f\/ield around two-dimensional aerofoil model at Reynolds number in the range of $1.5\times10^4$--$3\times10^5$. The frequency of oscillation is considered to be low if the Strouhal number is less than $0.2$. The authors reported that the phenomenon could not be reproduced initially in a relatively cleaner wind tunnel at NASA Lewis research centre. However, the phenomenon was produced by either raising the turbulence level of the free-stream or exciting the f\/low by using acoustic waves. The authors studied the f\/luctuation of the lift coeff\/icient where it has a quasi-periodic trend between stalled and non-stalled conditions. In the numerical study, they used a thin layer approximation to solve two-dimensional Navier--Stokes equations with Baldwin--Lomax turbulence model. The authors studied the sensitivity of the lift coeff\/icient to the grid size and verif\/ied it with experimental data. They concluded that the LFO phenomenon occurs when there is a trailing-edge stall or a thin aerofoil stall.~\citet{zaman1989low} extended the work on the LSB and investigated the LFO over a NACA-0012 with an ``iced'' leading-edge experimentally and numerically. They observed the LFO in their measurements, however, capturing the phenomenon numerically seemed to depend on the turbulence model and the Reynolds number. Acoustic excitation at an appropriate frequency can not only diminish the laminar separation, but also it can remove the LSB entirely.~\citet{zaman1991control} aimed at controlling the laminar separation using acoustic excitation at Reynolds number ranging from $2.5\times10^4$ to $10^5$.~\citet{zaman1992effect} explored the effect of acoustic excitation on separated f\/lows over aerofoils experimentally. He found that the acoustic excitation has improved the aerodynamic performance of the aerofoil. Furthermore, the effect of excitation became more pronounced when the amplitude of excitation was increased. However, the Strouhal number at which the optimum increase occurred was shifted towards a much lower Strouhal number than that expected from linear inviscid instability of the separated shear layer.\newline
A leading-edge separation bubble induced by a simulated ice was investigated experimentally by~\citet{bragg1992measurements}. A two-dimensional NACA-0012 aerofoil equipped with an interchangeable leading-edge was used in the experiment. The leading-edge portion was replaced by an aerofoil with ice contour attached to it to simulate the iced leading-edge. The authors reported that they measured bubbles with a length longer than $35\%$ of the chord. The LFO was observed at a Strouhal number of about $0.0185$ which is comparable to the Strouhal number of $0.02$ observed in previous research.~\citet{bragg1993low} used a conf\/iguration similar to that employed by~\citet{zaman1989natural} and measured the frequencies of the LFO with hot-wire measurements in the wake of the aerofoil. The measurements were conducted at Reynolds numbers of up to $1.3\times10^6$. The LFO took place at a Strouhal number of about $0.02$ and increased slightly with the Reynolds number.~\citet{bragg1996flow} studied an experiment of the f\/low-f\/ield around an LRN(1)-1007 aerofoil at Reynolds number ranging from $3\times10^5$ to $1.25\times10^6$ and angles of attack in the range of $0^{\circ}$--$28^{\circ}$. The LFO phenomenon was captured using hot-wire spectra measurements in the aerofoil wake where the LFO phenomenon is observed in the range of angles of attack of $14.4^{\circ} \leq \alpha \leq 16.6^{\circ}$. The authors reported that the Strouhal number increased slightly with the Reynolds number and signif\/icantly with the angle of attack. The Strouhal number of the oscillation is an order of magnitude lower than that of bluff-body shedding. The surface oil f\/low and laser sheet visualisation experiments showed the growth and bursting of the leading-edge separation bubble and its role in the f\/low oscillation.~\citet{broeren1998low} investigated the LFO and the intensity of f\/low oscillation in near stall conditions. The authors tested $12$ aerofoils having different stalling characteristics over a range of angles of attack at a Reynolds number of $3\times10^5$. They concluded that there is a distinct relationship between stall type and low-frequency/large-scale unsteady f\/low. The authors reported that a combination of thin aerofoil and trailing-edge stall types results in an LFO of magnitudes which are nearly double that for pure thin aerofoil stall type.~\citet{broeren1999flowfield} utilized the aerofoil used earlier by~\citet{bragg1996flow} at a chord Reynolds number of $3\times 10^5$ and the angle of attack of $15^{\circ}$. The authors used conditionally averaged Laser Doppler velocimeter (LDV) to resolve a fully separated boundary layer. The boundary layer is fully separated due to the merging of the leading-edge and the trailing-edge separation when they grow in time. Their data showed the periodic switching between stalling and non-stalling behaviour, which involved a leading-edge separation bubble, and how the bubble interacted with the trailing-edge separation.~\citet{broeren2001spanwise} studied f\/ive different aerofoil conf\/igurations (NACA-2414, NACA-64A010, LRN-1007, E374 and Ultra-Sport) by measuring the wake velocity across the spanwise direction, and using mini-tufts for f\/low visualisation. They found that all the stall types are dependent on the type of the aerofoil. They concluded that the LFO phenomenon always occur in the aerofoils that exhibits a thin-aerofoil stall or the combination of both thin-aerofoil and trailing-edge stall.\newline
\citet{ansell2013characterization} and~\citet{ansell2015characterization} conducted an experiment to investigate the LFO phenomenon in the f\/low-f\/ield around a NACA-0012 aerofoil with a horn ice shape at the leading-edge and compared it to a clean aerofoil. They obtained the lift coeff\/icient by duly integrating the pressure coeff\/icient around the aerofoil at mid-span location. The authors correlated the lift coeff\/icient and the location of the shear layer reattachment. They found that the elongation of the LSB takes place at the maximum lift and the shrinkage of LSB presents when the lift coeff\/icient decreases.~\citet{ansell2015unsteady} simultaneously measured the unsteady pressure around a NACA-0012 aerofoil at mid-span locations, the unsteady velocity at the wake of the aerofoil, and surface--oil f\/low visualisation. They identif\/ied three distinct modes of unsteadiness at various locations throughout the f\/low f\/ield:
\begin{enumerate}
  \item A regular mode that features unsteadiness due to vortical motion along the separated shear layer and vortex shedding from the laminar separation bubble.
  \item A shear-layer f\/lapping mode dominating the upstream portion of the separation bubble at a Strouhal number of $St = 0.0185$.
  \item A low-frequency mode characterised by a low-frequency oscillation of the aerofoil circulation.
\end{enumerate}
The average Strouhal number of the regular mode corresponds to $St = 0.60$, and the average vortex convection velocity was found to be $0.45 U_\infty$. The obtained values of the Strouhal number of the low-frequency mode varies with the aerofoil angle of attack. The work initiated at NASA Langley and Lewis showed that the LFO indeed exists on clean and iced aerofoils, explored the range of Strouhal number of its existence, examined the effect of acoustic excitation on the LFO, and suggested means to control its undesired effects on aerofoils performance. However, the underlying mechanism behind the LFO and the primary cause that triggers the LSB instability are not revealed.\newline
A research group that has contributed with copious amounts of research on the topic of the LSB and its associated LFO is located at the university of Stuttgart, Germany. The vast majority of the research done in this group is on a f\/lat plate and an adjustable pressure distribution model.~\citet{maucher1994direct} performed DNS to investigate the LSB on a f\/lat plate. The LSB was induced by increasing the pressure gradient in the streamwise direction. An undisturbed LSB exhibited low-frequency large oscillation, whereas, a disturbed LSB showed a higher frequency oscillation corresponding to the Tollmien-Schlichting waves (TS waves).~\citet{maucher2000dns} reported that the unsteady results of a Direct simulation exhibited temporal amplif\/ications of the three-dimensional small amplitude disturbances. The authors explained the growth of the perturbations by the roll-up of the separated boundary layer that entrains the three-dimensional f\/luctuations. They observed that the f\/low eventually undergoes transition to turbulence.\newline
\citet{rist2002instabilities} and~\citet{rist2003instability} investigated the instability and transition in the LSB that was induced on a f\/lat plate. He reported several experimental and numerical works in which different underlying mechanisms are discussed. The author observed that the size of the bubble is very sensitive to small amplitude disturbances. The author reported that a new type of secondary instability was found which derive the growth of three-dimensional disturbances. He concluded that the nature of the primary instability of most LSB that induced by invoking adverse pressure gradient is convective.~\citet{rist2002investigations} investigated temporally growing unstable disturbance using linear stability theory and compared the observations with that of two dimensional DNS. The authors used analytical base mean f\/low and found that the LSB become absolutely unstable when the maximum reversed f\/low is $20\%$ of the free-stream. They also examined the effect of Reynolds number and the wall-distance of the separated shear layer and found two unstable modes at large wall-distances. A time-growing instability was identif\/ied in the two dimensional DNS of an LSB.\newline
\citet{marxen2003combined} studied an LSB induced on a f\/lat plate by means of an APG that was imposed by a displacement body at a downstream location. Measurements in the water tunnel were carried out using LDA and PIV. Additionally, DNS was performed to study the phenomenon numerically. The authors reported that the transition in the LSB is driven by convective primary amplif\/ication of two-dimensional TS waves. They observed that variations in the f\/low-f\/ield in the spanwise direction causes break down of the two-dimensional rolls into three-dimensional structures. They also observed that the amplitude of the disturbances upstream the transition location is negligible because of their absolute instability.~\citet{marxen2007numericala} studied an LSB formed on a f\/lat plate using DNS and an LSB on the suction surface of an aerofoil using an LES. The authors reported that a short LSB is formed when a low-amplitude forcing is applied to both cases. When they switched the forcing off the DNS case, the bubble burst from a short to a long one.~\citet{marxen2007numericalb},~\citet{marxen2008direct} and~\citet{marxen2008laminar} investigated bubble bursting on a f\/lat plate by means of numerical simulation of forced and unforced f\/low. They simulated three cases: 1) a f\/low forced to undergo transition to turbulence using volume forcing; 2) transition to turbulence was triggered by introducing small disturbance into a laminar f\/low; and 3) unforced f\/low. The authors reported that the bubble burst when saturated disturbances were not able to attach the f\/low. When disturbance forcing was switched off, the transition process was no longer able to immediately reattach the f\/low. It was concluded that the disturbance input is essential to maintain a short bubble. The authors reported that they observed that in the forced simulations the vortical structures are large spanwise vortices, whereas, in the unforced simulation the structures are small streamwise vortices. It was concluded that the large-scale spanwise vortices are responsible for the quick reattachment of the f\/low. The short bubble size was found to be inversely proportional to the amplitude of the forcing.\newline
\citet{marxen2010mean} studied the mutual interaction between the mean f\/low and transition in an LSB that was induced by an APG on a f\/lat plate. Transition to turbulence was forced by introducing small amplitude disturbances upstream the LSB. They reported that the transition process altered the mean f\/low and the pressure distribution. The authors observed that the f\/low was stabilised with respect to small linear perturbations and the separation region was reduced due to the deformation of the mean f\/low.~\citet{marxen2011effect} used small disturbance forcing with various amplitudes to investigate stability and transition in an LSB formed on a f\/lat plate. They observed that the resulting bubbles were different with respect to their mean f\/low, linear stability characteristics, and the distance between transition and reattachment locations in the mean sense. The authors observed that when the disturbance amplitude was reduced below a certain value, the bubble started to grow and became a long bubble. They concluded that the failure of the transition process to reattach the f\/low was attributed to the orientation and size of the transitional vortical structures. The authors observed that the distance between the transition and reattachment locations, in the mean sense, is inversely proportional to the amplitude of forcing. They also observed that bursting of the bubble occurred in the three-dimensional simulation case and not in the two-dimensional simulation even in the absence of forcing. The authors demonstrated that short bubbles could be both convectively and absolutely unstable. They concluded that the bubble in their conf\/iguration was absolutely unstable.~\citet{marxen2013vortex} characterised vortex formation and vortex breakup process in an LSB formed on a f\/lat plate. They observed that the transition process is dominated by the vortex formation and breakup process. The authors showed that multiple secondary instability mechanisms are active during the vortex formation and breakup process. Despite the considerable amount of research done in this group, the fundamental question of why the bubble burst remains unanswered.\newline
Disturbances developing in a f\/low-f\/ield over a f\/lat plate was investigated by~\citet{haggmark2001numerical}. They induced an LSB in the f\/low-f\/ield and used controlled forcing of low-amplitude two-dimensional instability waves. They modelled the experiment numerically using two-dimensional DNS and a good agreement is found between experiments, simulations and linear stability theory. A region with exponential disturbance growth is observed in the separated shear layer associated with a highly two-dimensional f\/low. A local maximum in the disturbance amplitude develops at the inf\/lection point in the mean velocity prof\/ile, indicating an inviscid type of instability.\newline
\citet{diwan2006bursting} conducted velocity and pressure measurements and carried out f\/low visualisation in a wind tunnel to investigate the bursting of an LSB induced on a f\/lat plate. They suggested a ref\/ined bursting criterion which considers not only the length of the bubble but also the maximum height of the bubble. The authors revised the bursting criterion suggested by~\citet{gaster1967structure} and proposed a ref\/ined bursting criterion involving the maximum height of the bubble. They formulated a single parameter criterion and argued that their criterion is more universal than the existing criteria.~\citet{diwan2007laminar} used hot-wire anemometry to trace a wave packet as it is advected downstream in a f\/low conf\/iguration similar to that of~\citet{diwan2006bursting}. The wave packet is initiated by small amplitude disturbance that was introduced upstream the separation location. The authors reported that the effect of forcing varies from quenching the bubble to modifying its characteristics and control it depending on the amplitude of the disturbance.~\citet{diwan2009origin} performed a detailed experimental and theoretical investigation of the linear instability mechanisms associated with an LSB that was induced on a f\/lat plate. The authors observed that small disturbance amplitudes originate at the front portion of the LSB and grow in space with maximum amplitude close to the reattachment location. They introduced an inf\/initesimal disturbance into the boundary layer upstream the separation location and tracked the wave packet while it was advected downstream, in the mean sense. The authors concluded that the primary instability mechanism in an LSB is inf\/lectional in nature and it originates upstream the separation location. The authors argued that only when the separated shear layer move signif\/icantly away from the wall, a description by the Kelvin--Helmholtz instability is then relevant.\newline
\citet{tanaka2004flow} investigated the f\/low-f\/ield around a NACA-0012 aerofoil near stall conditions. The author performed experimental measurements by means of Particle Image Velocimetry (PIV). He was aiming to understand the self-sustained LFO. The f\/low Reynolds number was $1.3\times10^5$, and the measurements were carried out at various angles of attack near stall conditions. In the range of investigated angles of attack, he found that the frequency of the LFO increased when the angle of attack was increased. The f\/low oscillation had a maximum amplitude at the angle of attack of $11.5^{\circ}$. One of the major conclusions of his work that he observed large vortices that play an essential role in the reattachment of the f\/low. However, his observations were overlooked in the literature despite their importance. The author observed that when the f\/low is massively separated, a large vortex that bends the shear layer towards the aerofoil surface is formed near the leading-edge. This vortex introduces the free-stream into the separated f\/low region. He concluded that this vortex might be a part of the mechanism that forces the f\/low to reattach and suggested further investigations on the nature of this vortex.~\citet{rinoie2004oscillating} conducted an experiment very similar to that carried out by~\citet{tanaka2004flow}. They observed that the f\/low was oscillating between a small bubble formed near the leading-edge of about a $10\%$ of the chord length and a massively separated f\/low. However, they did not look into the large vortices observed by~\citet{tanaka2004flow}.\newline
This review cannot be considered complete without noting the signif\/icant contribution from Queen Mary, University of London, that has been documented in the PhD theses of~\citet{mcgregor1954regions},~\citet{gaster1963stability},~\citet{horton1968laminar}, and~\citet{woodward1970investigation}. The state of our understanding of the LSB and its associated LFO is due to the exceptional contribution from the research group located at the university of Southampton and has been led by professor Sandham. The work carried out in this group is numerical, and a custom code was developed and optimised to simulate the LSB and the LFO phenomena. The code used in the present work is a modif\/ied version of the code developed by this group. The existence of such a code made the current study possible.\newline
\citet{alam2000direct} investigated an LSB on a f\/lat plate induced by the action of a suction prof\/ile applied as the upper boundary condition. They showed that the separation bubble is considered absolutely unstable if the separation bubble sustains a maximum reverse velocity percentage in the range of $15\%$--$20\%$ of the local free-stream velocity. The three-dimensional bubbles with turbulent reattachment have maximum reverse f\/low of less than $8\%$ and it is concluded that for these bubbles the primary instability is convective in nature.~\citet{sandham2008transitional} modelled the laminar-turbulent transition by an absolute instability. The results are qualitatively comparable to existing experimental data; however, the method needs further development. Sandham used a time-accurate viscous-inviscid interaction method (VII) for the coupled potential f\/low and integral boundary-layer equations to study several aerofoil models near stall conditions. The author modelled the laminar-turbulent transition by an absolute instability which sustains the transition of the laminar bubble without using upstream perturbations. The Strouhal number of the LFO was pointed out to be dependent on the shape of the aerofoil. The author reported that the LFO phenomenon could be captured by a simple model of boundary layer transition and turbulence because the bubble bursting occurs as a consequence of potential f\/low/boundary layer interactions.\newline
\citet{jones2010stability} performed a linear stability analysis of the f\/low around a NACA-0012 aerofoil at an angle of attack of $5^{\circ}$. The authors observed no evidence of local absolute instability. They conf\/irmed this by performing two-dimensional linear stability analysis via forced Navier--Stokes simulations. Forced Navier--Stokes simulations did, however, determine that the time-averaged f\/low-f\/ield is unstable because of an acoustic-feedback instability, in which instability waves convecting over the trailing-edge of the aerofoil generate acoustic waves that propagate upstream to some location of receptivity and generate further instability waves within the boundary layer.~\citet{almutairi2010intermittent} used LES to study a low-Reynolds number f\/low, $\rec=5\times10^4$ around a NACA-0012 aerofoil at the angle of attack of $5^{\circ}$. The primary objective was to validate their LES code using the DNS data of~\citet{jones2008direct}.\newline
\citet{almutairi2013large} investigated the phenomenon of natural LFO using the LES code developed by~\citet{almutairi2010intermittent}. They studied the f\/low-f\/ield around a NACA-0012 aerofoil at a Reynolds number of $5\times10^4$, Mach number of $0.4$, and a range of angles of attack; $\alpha = 9.25^{\circ}$, $9.29^{\circ}$, $9.4^{\circ}$ and $9.6^{\circ}$. The authors observed a self-sustained natural LFO in all of the investigated angles of attack.~\citet{almutairi2015large} performed an LES of the f\/low around a NACA-0012 at a Reynolds number of $1.3\times10^5$ and $\alpha=11.5^{\circ}$. They described the role of transition and how the large three-dimensional scales break down to small scales, which are followed by a turbulent region. They found that the dominant f\/low mode is of Strouhal number of $0.00826$ which is in very good agreement with that of the experiment of~\citet{rinoie2004oscillating}. Dynamic Mode Decomposition (DMD) analysis identif\/ied another dominant frequency that features the trailing-edge shedding. The authors concluded that the LFO is linked with the trailing-edge vortex shedding. When the f\/low is separated, the strong vortex shedding at the trailing-edge generates acoustic waves that travel upstream and excite the separated shear layer and force it to undergo transition to turbulence and reattach the f\/low. However, when the f\/low is attached the vortex shedding at the trailing-edge dies down, and the feedback is cut off.~\citet{eljack2018influence} took up the work of~\citet{almutairi2015large} and used a combination of trailing-edge shedding frequencies to eliminate the low-frequency f\/low oscillation induced by bubble bursting and improve the aerofoil performance. The numerical simulations carried out by this group shed light on different aspects of the LSB and the LFO phenomena. A model that predicts the phenomenon was even developed. Nevertheless, fundamental questions such as: why the bubble burst and the f\/low separate? And why the f\/low reattach after it was separated? Were not answered.\newline
\citet{eljack2017high} carried out a careful grid sensitivity study that is based on the characteristics of the bubble rather than being based on the static aerodynamic coeff\/icients. He found that the characteristics of the bubble, the angle of attack at which the bubble starts to become unstable, and the angle of attack at which the LSB becomes an open bubble are highly sensitive to the grid distribution. The author performed LESs over sixteen angles of attack around the inception of stall and compared the aerodynamic coeff\/icients to recent and previous experimental data. The author showed that, qualitatively, at relatively low angles of attack, the LSB is present on the upper surface of an aerofoil and remains intact. At moderate angles of attack, the LSB abruptly and intermittently bursts, which causes an oscillation in the f\/low f\/ield and consequently the aerodynamic forces. At higher angles of attack, the LSB exhibits a quasi-periodic switching between long and short bubbles, which results in a global low-frequency f\/low oscillation. As the aerofoil approaches the full stall angle, the f\/low remains separated with intermittent and abrupt random reattachments. The aerofoil eventually undergoes a full stall as the angle of attack is further increased.\newline
\citet{eljack2018bursting} used the data sets generated by~\citet{eljack2017high} to characterised the f\/low-f\/ield. The authors verif\/ied and validated the LES data, then used a conditional time-averaging to describe the f\/low-f\/ield in detail and investigate the effects of the angle of attack on the characteristics of the f\/low-f\/ield. They reported the existence of three distinct angle-of-attack regimes. At relatively low angles of attack, f\/low and aerodynamic characteristics are not much affected by the LFO. At moderate angles of attack, the f\/low-f\/ield undergoes a transition regime in which the LFO develops until it reaches a quasi-periodic switching between separated and attached f\/low. At high angles of attack, the f\/low-f\/ield and the aerodynamic characteristics are overwhelmed by a quasi-periodic and self-sustained LFO until the aerofoil approaches the angle of full stall. The authors concluded that most of the observations reported in the literature about the LSB and its associated LFO are neither thresholds nor indicators for the inception of the instability, but rather are consequences of it. However, the mechanism that generates and sustains the LFO is not revealed.\newline
In the present work, the conditional time-averaging def\/ined and used by~\citet{eljack2018bursting} and a conditional phase-averaging will be used to shed light on the underlying mechanism that generates and sustains the LFO phenomenon and answer fundamental questions like why does the f\/low separate despite the strongly favourable pressure gradient? Why does the f\/low reattach after it separates? And why does the LSB reform after it bursts?

\section{Results and Discussion}
LESs were carried out for the f\/low around the NACA-0012 aerofoil at sixteen angles of attack ($\alpha = 9.0^{\circ}$--$10.1^{\circ}$ at increments of $0.1^{\circ}$ as well as $\alpha = 8.5^{\circ}$, $8.8^{\circ}$, $9.25^{\circ}$, and $10.5^{\circ}$). The simulation Reynolds number and Mach number were $\rec=5\times10^4$ and $\minf = 0.4$, respectively. The free-stream f\/low direction was set parallel to the horizontal axis for all simulations ($u=1$, $v=0$, and $w=0$). The entire domain was initialised using the free-stream conditions. The simulations were performed with a time-step of $10^{-4}$ non-dimensional time units. The samples for statistics are collected once transition of the simulations has decayed and the f\/low became stationary in time after $50$ f\/low through times which is equivalent to $50$ non-dimensional time units. Aerodynamic coeff\/icients (lift coeff\/icient $(\mathrm{C}_{\!_\mathrm{L}})$, drag coeff\/icient $(\mathrm{C}_{\!_\mathrm{D}})$, skin-friction coeff\/icient $(\mathrm{C}_{\!_\mathrm{f}})$, and moment coeff\/icient $(\mathrm{C}_{\!_\mathrm{m}})$) were sampled for each angle of attack at a frequency of $10,000$ to generate $2.5$ million samples over a time period of $250$ non-dimensional time units. The locally-time-averaged and spanwise ensemble-averaged pressure, velocity components, and Reynolds stresses were sampled every $50$ time-steps on the $x$--$y$ plane. A data set of $20,000$ $x$--$y$ planes was recorded at a frequency of $204$ for each angle of attack. The reader is referred to~\citet{eljack2018bursting} for more details.

\subsection{Conditional time-average of the oscillating-f\/low-f\/ield} \label{section_conditional_time-average}
\begin{figure}
\begin{center}
\begin{minipage}{220pt}
\centering
\includegraphics[width=220pt, trim={0mm 0mm 0mm 0mm}, clip]{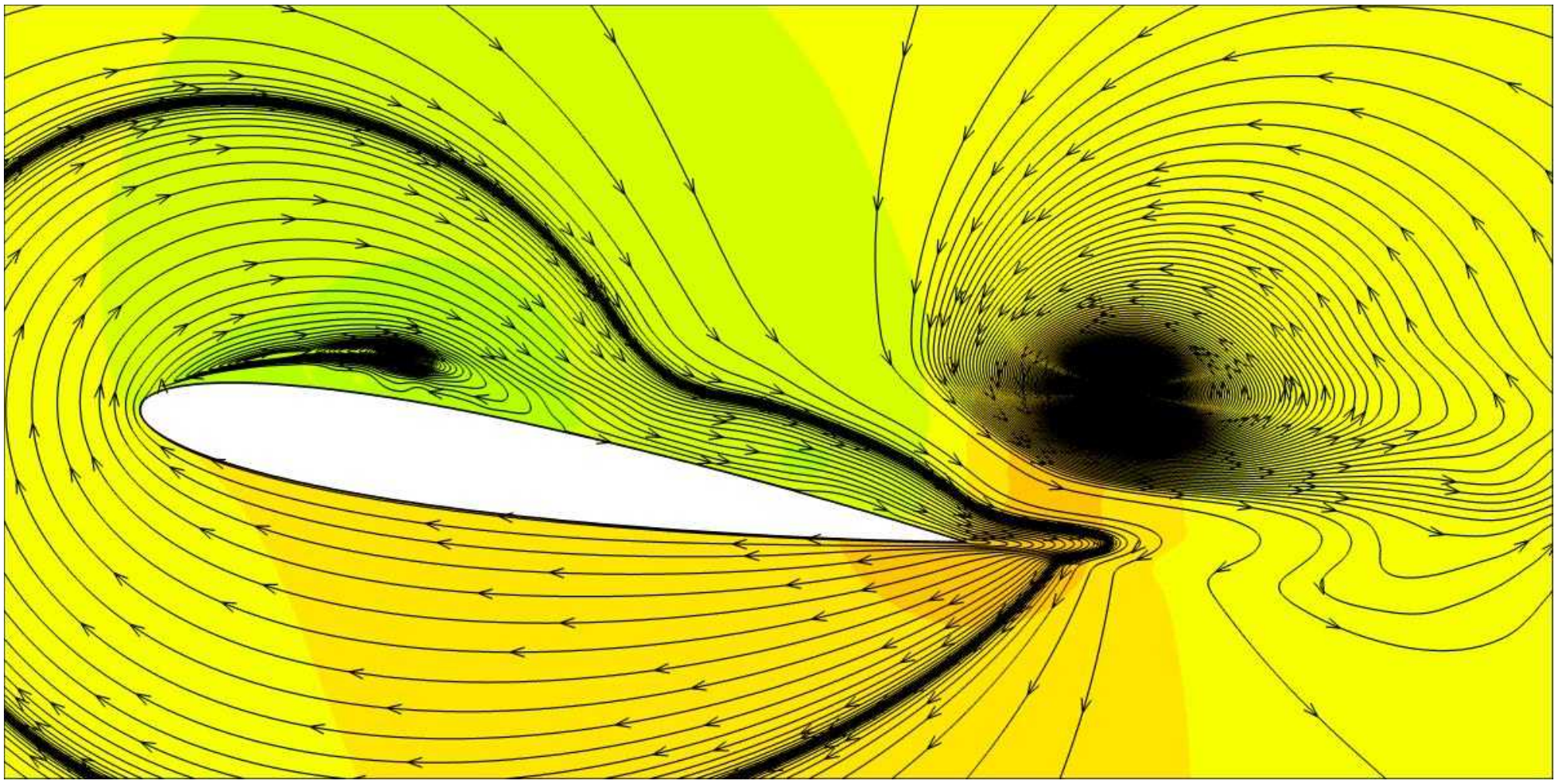}
\textit{$\alpha = 9.25^{\circ}$}
\end{minipage}
\begin{minipage}{220pt}
\centering
\includegraphics[width=220pt, trim={0mm 0mm 0mm 0mm}, clip]{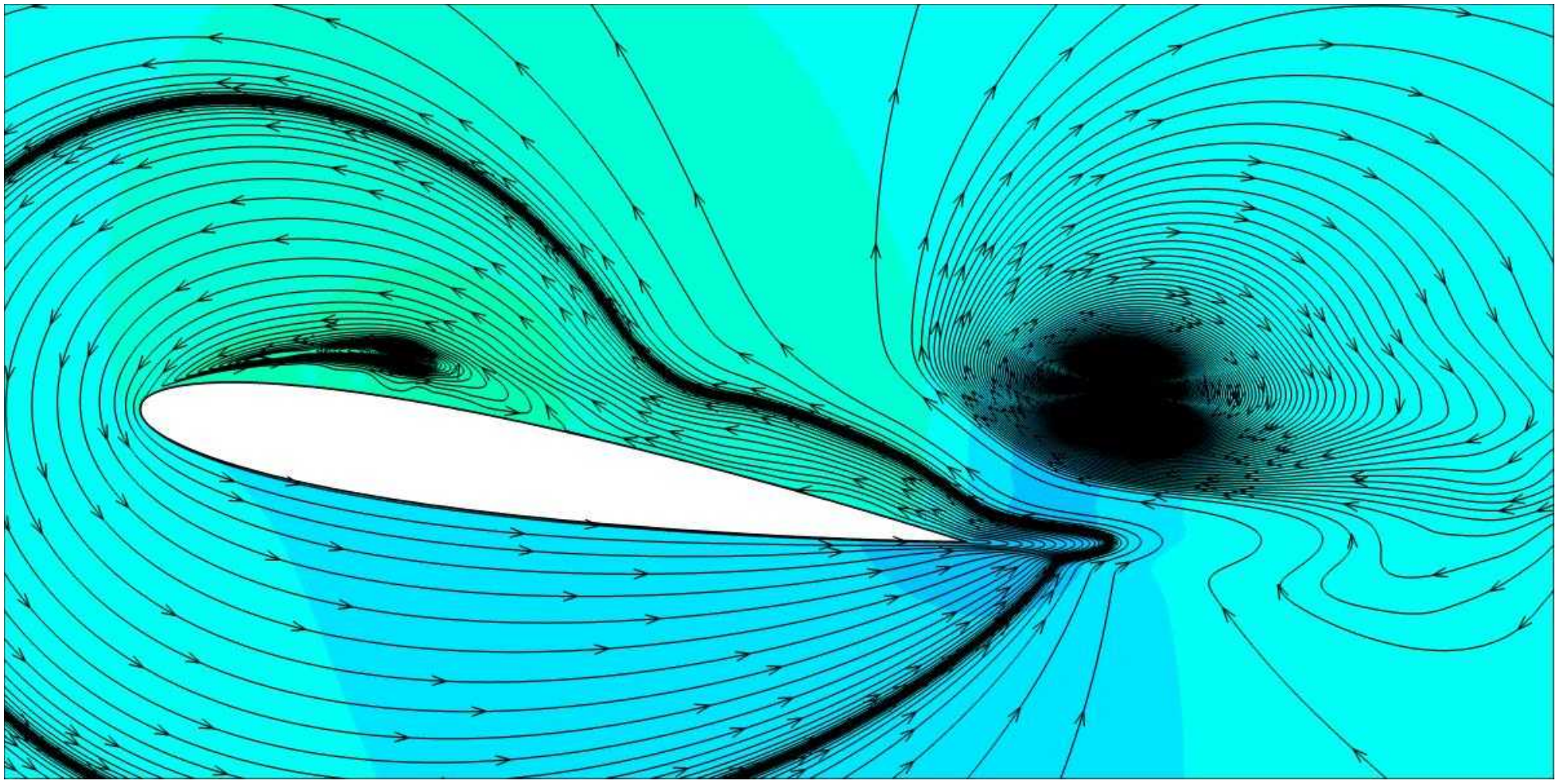}
\textit{$\alpha = 9.25^{\circ}$}
\end{minipage}
\begin{minipage}{220pt}
\centering
\includegraphics[width=220pt, trim={0mm 0mm 0mm 0mm}, clip]{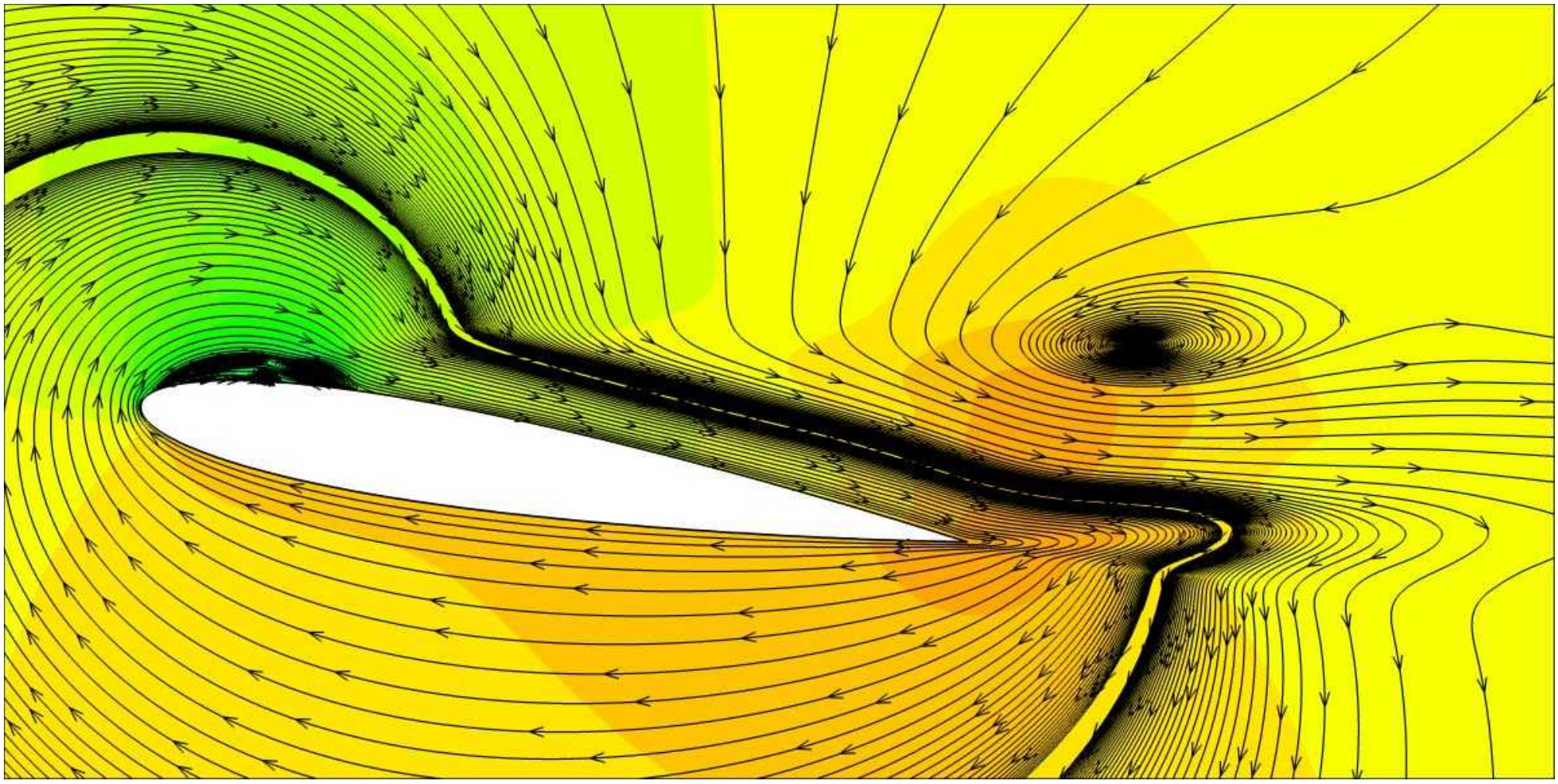}
\textit{$\alpha = 9.4^{\circ}$}
\end{minipage}
\begin{minipage}{220pt}
\centering
\includegraphics[width=220pt, trim={0mm 0mm 0mm 0mm}, clip]{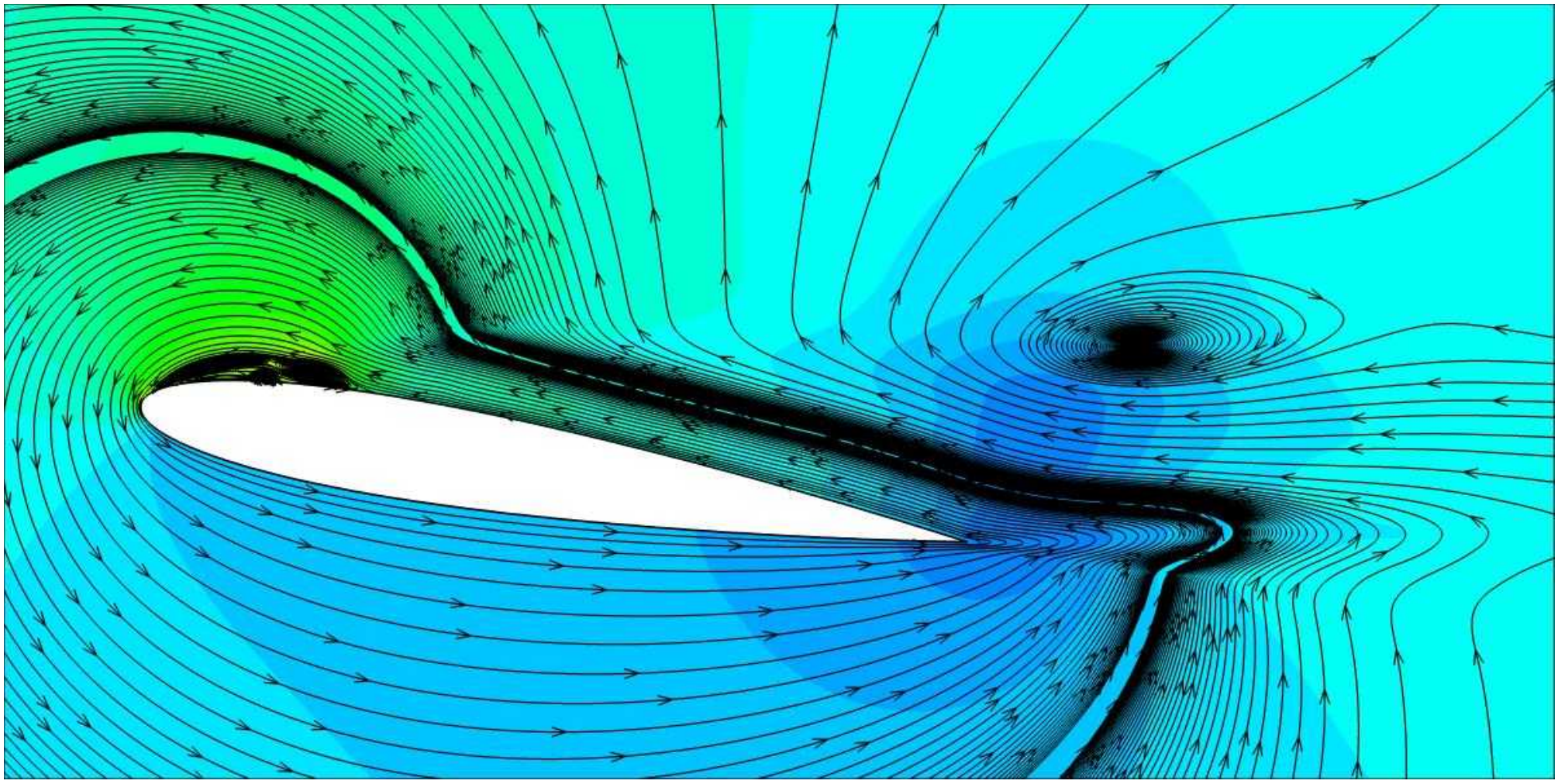}
\textit{$\alpha = 9.4^{\circ}$}
\end{minipage}
\begin{minipage}{220pt}
\centering
\includegraphics[width=220pt, trim={0mm 0mm 0mm 0mm}, clip]{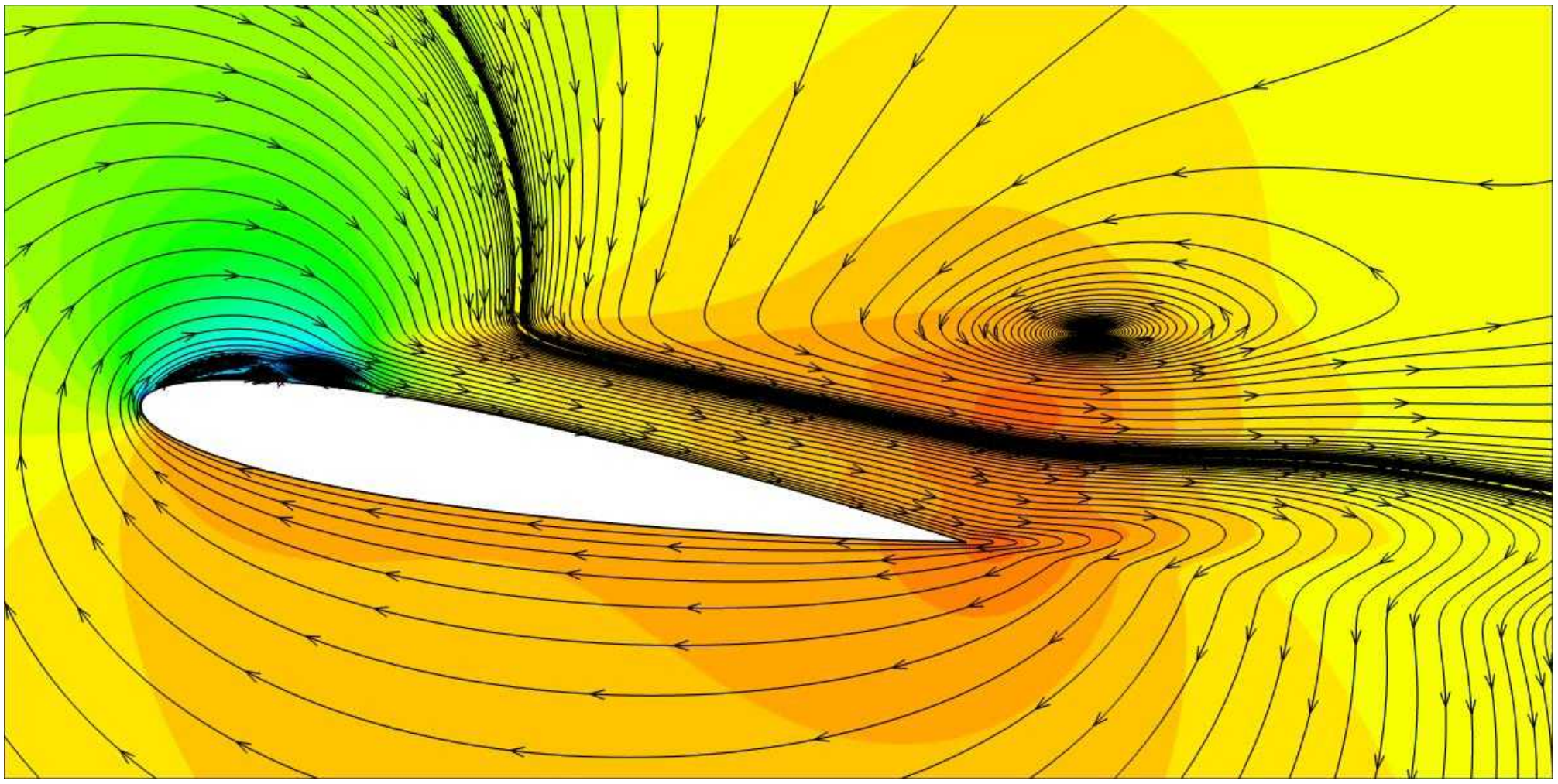}
\textit{$\alpha = 9.5^{\circ}$}
\end{minipage}
\begin{minipage}{220pt}
\centering
\includegraphics[width=220pt, trim={0mm 0mm 0mm 0mm}, clip]{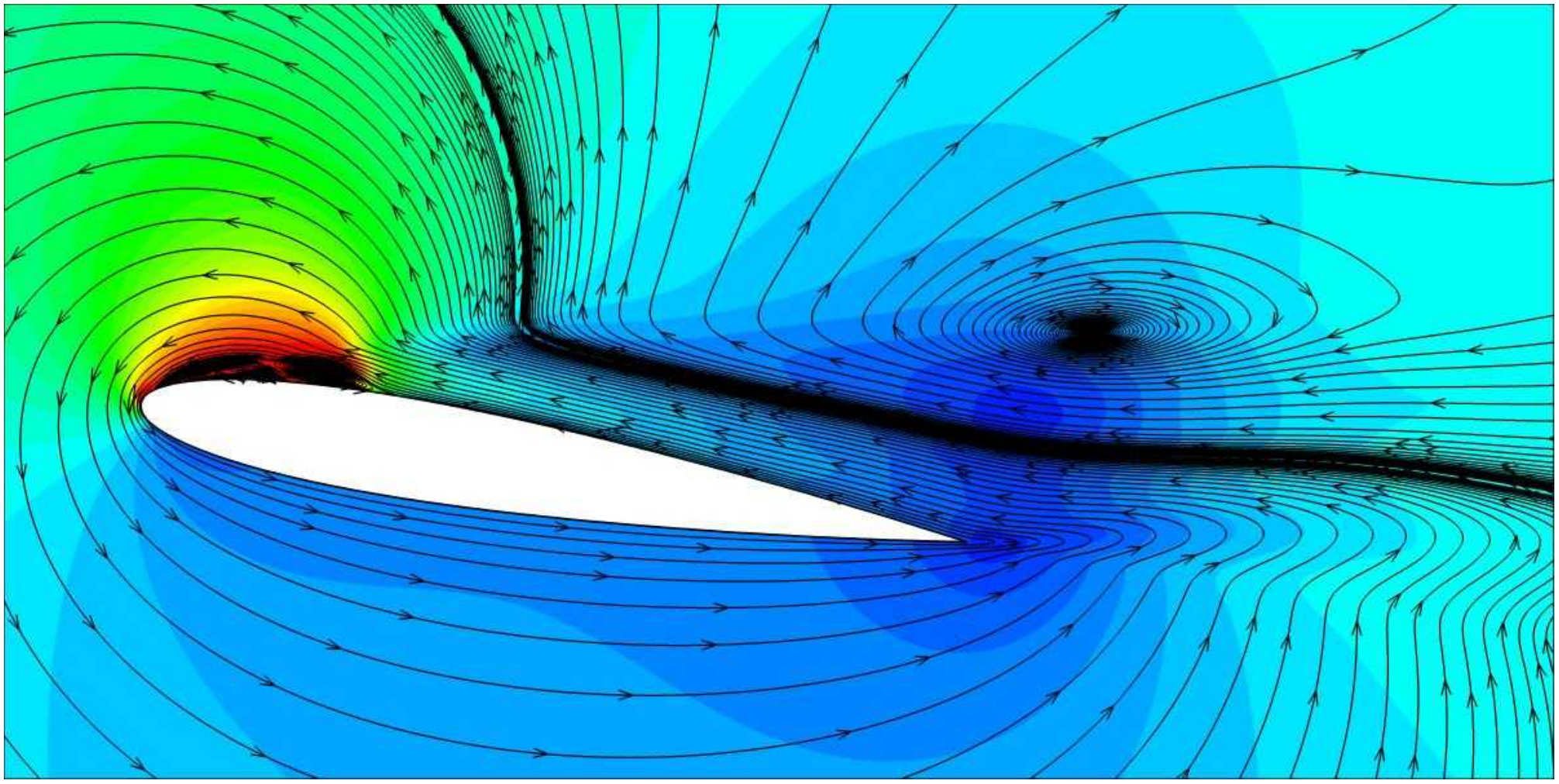}
\textit{$\alpha = 9.5^{\circ}$}
\end{minipage}
\begin{minipage}{220pt}
\centering
\includegraphics[width=220pt, trim={0mm 0mm 0mm 0mm}, clip]{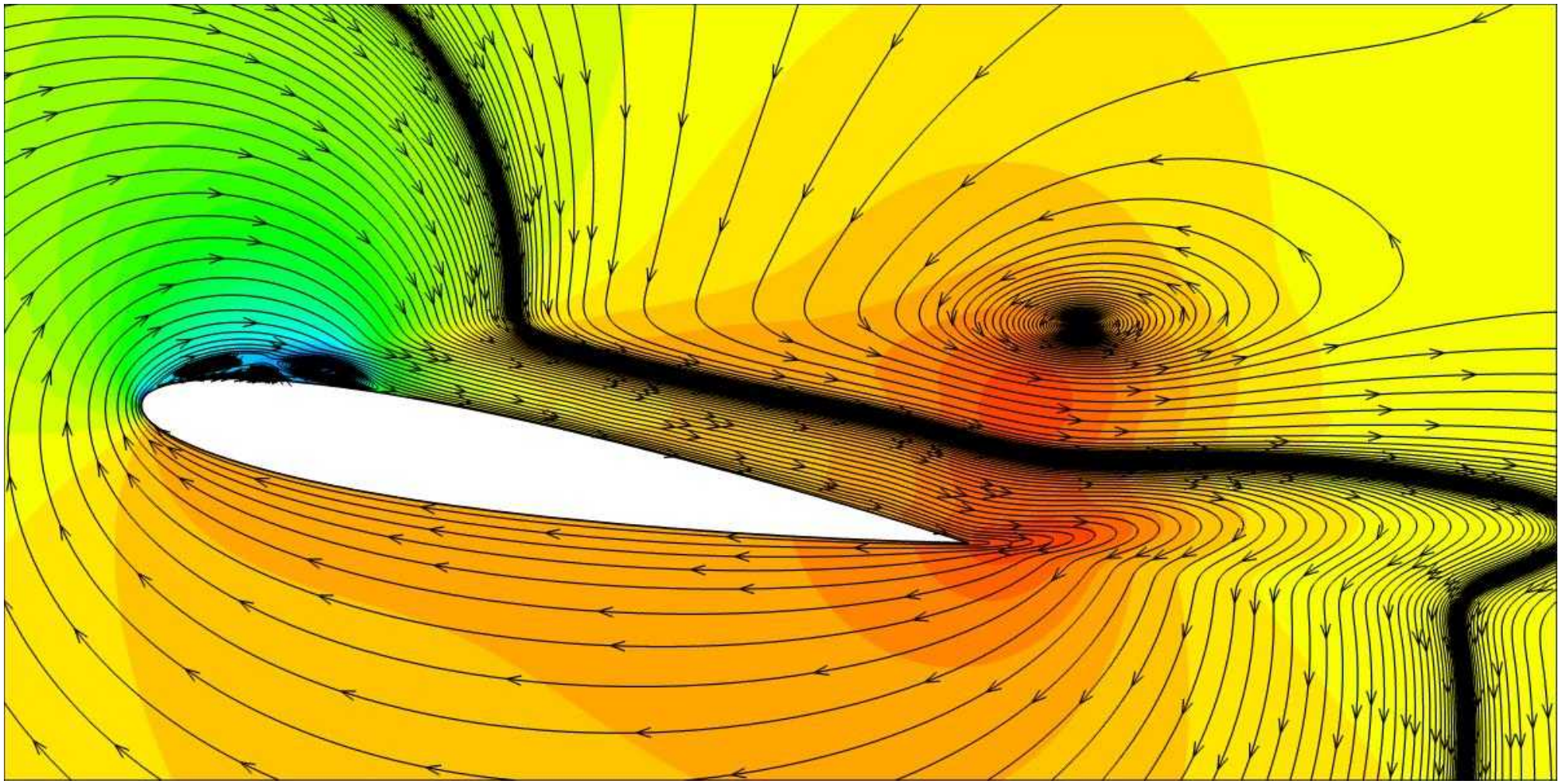}
\textit{$\alpha = 9.6^{\circ}$}
\end{minipage}
\begin{minipage}{220pt}
\centering
\includegraphics[width=220pt, trim={0mm 0mm 0mm 0mm}, clip]{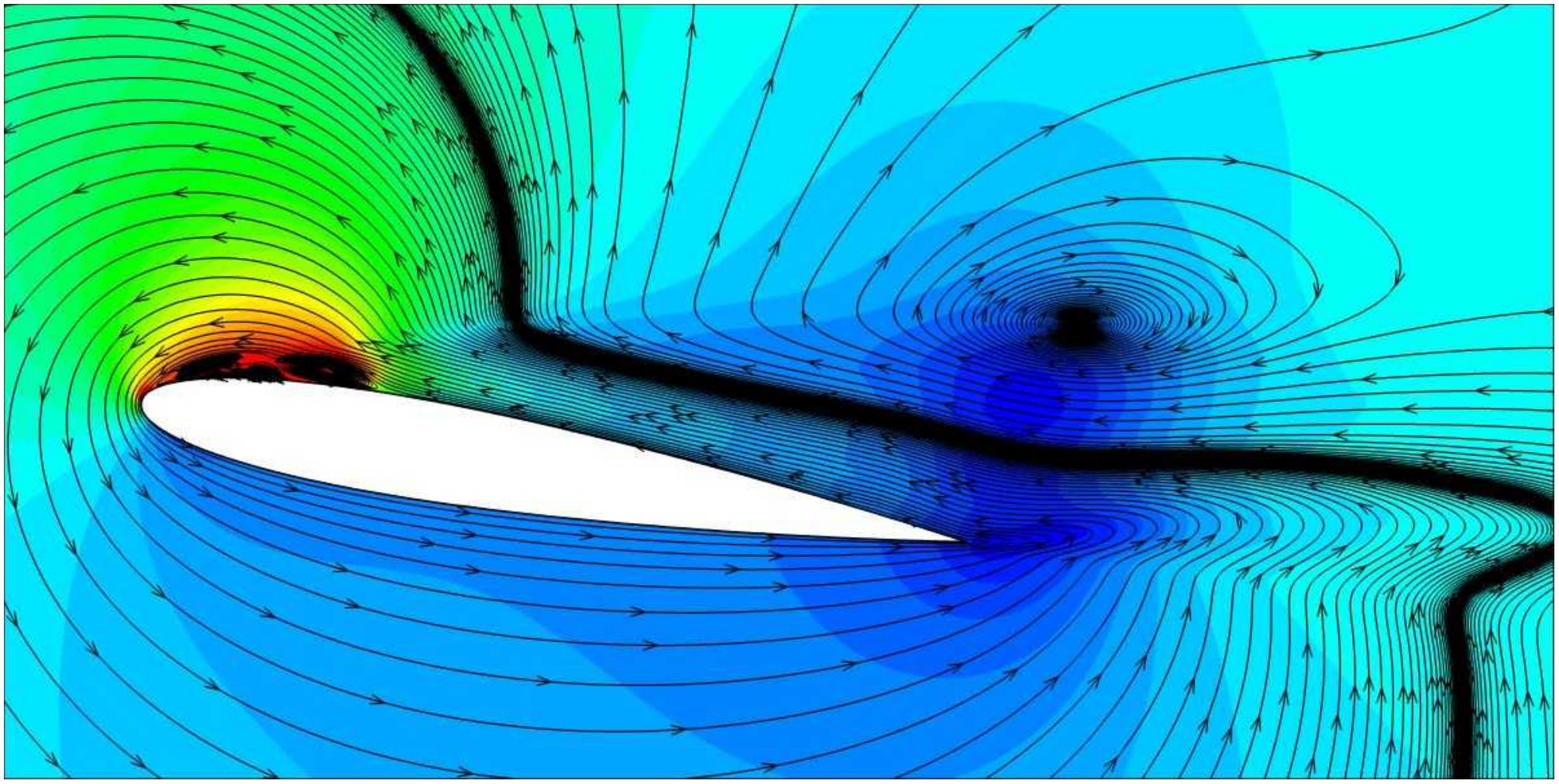}
\textit{$\alpha = 9.6^{\circ}$}
\end{minipage}
\begin{minipage}{220pt}
\centering
\includegraphics[width=220pt, trim={0mm 0mm 0mm 0mm}, clip]{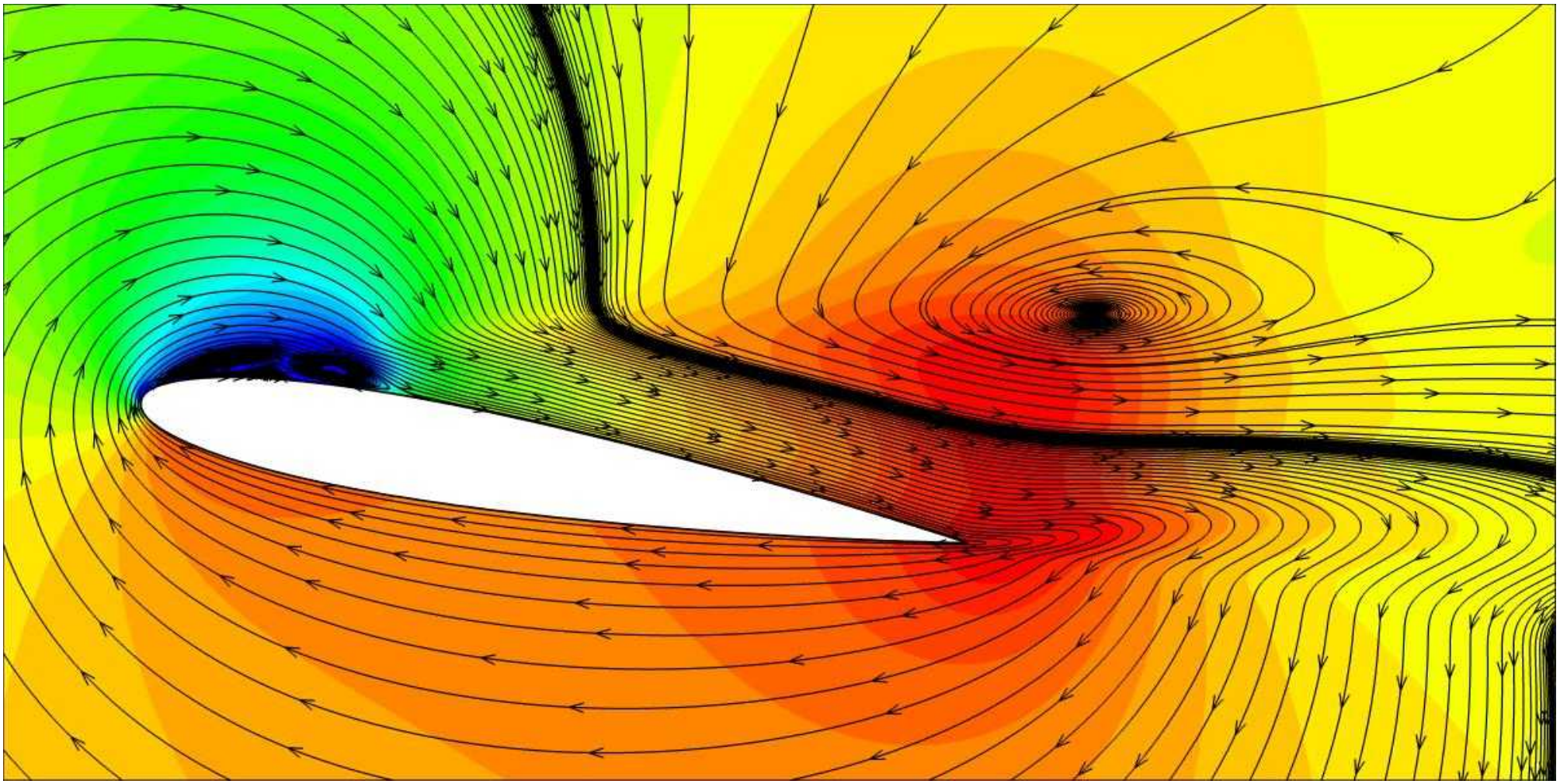}
\textit{$\alpha = 9.7^{\circ}$}
\end{minipage}
\begin{minipage}{220pt}
\centering
\includegraphics[width=220pt, trim={0mm 0mm 0mm 0mm}, clip]{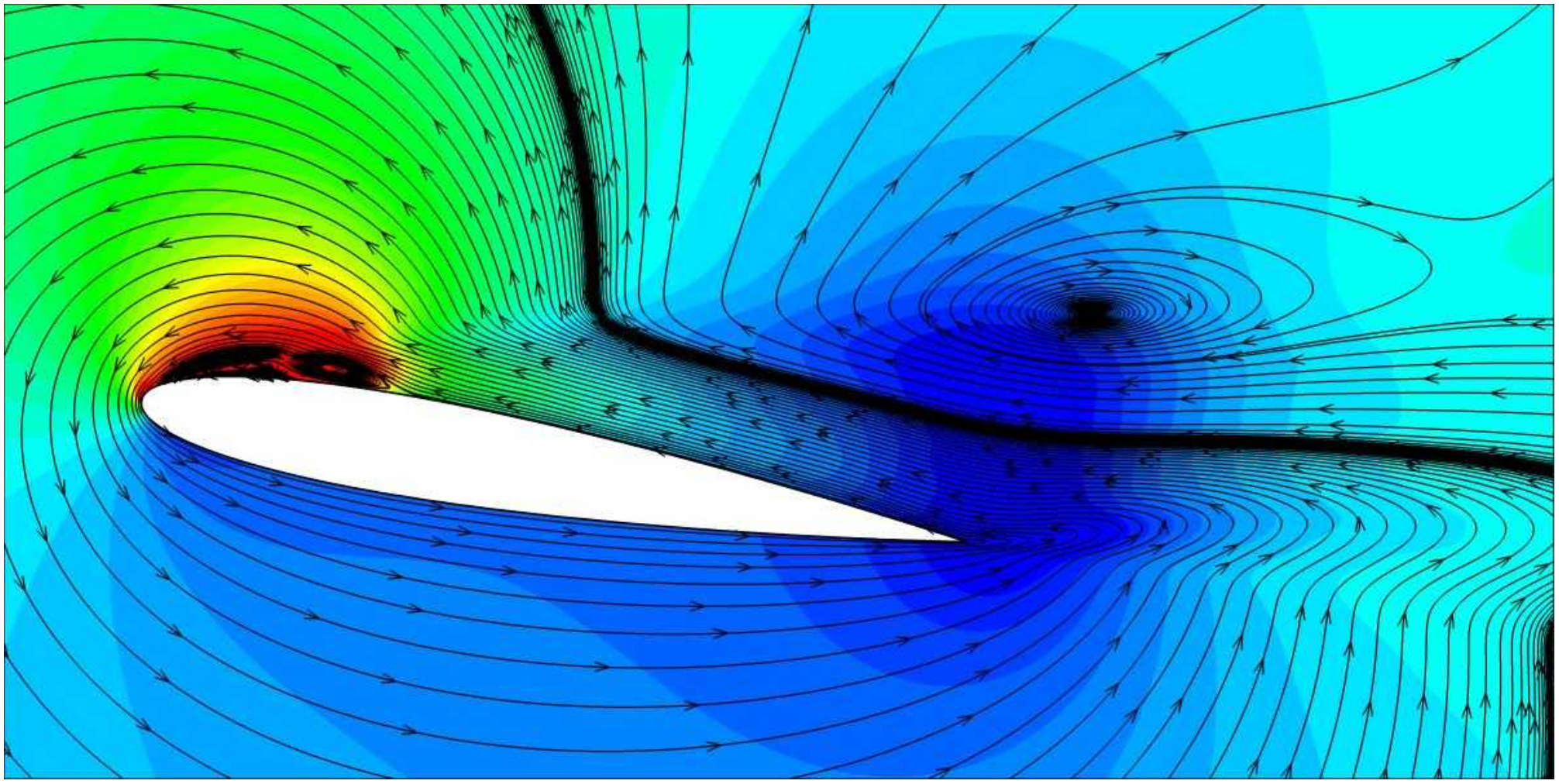}
\textit{$\alpha = 9.7^{\circ}$}
\end{minipage}
\caption{Streamlines patterns of the time-averaged attached and separated phases of the oscillating-f\/low-f\/ield superimposed on colour maps of the time-averaged attached and separated phases of the oscillating-pressure-f\/ield for the angles of attack $\alpha = 9.25^{\circ}$--$9.7^{\circ}$. Left: time-averaged attached-phase of the oscillating-f\/low-f\/ield estimated from the high-lift f\/low-f\/ield minus the mean-lift f\/low-f\/ield, $\Delta{\overline{U}}^\mathrm{+}$, $\Delta{\overline{V}}^\mathrm{+}$, and $\Delta{\overline{P}}^\mathrm{+}$. Right: time-averaged separated-phase of the oscillating-f\/low-f\/ield estimated from the low-lift f\/low-f\/ield minus the mean-lift f\/low-f\/ield, $\Delta{\overline{U}}^\mathrm{-}$, $\Delta{\overline{V}}^\mathrm{-}$, and $\Delta{\overline{P}}^\mathrm{-}$.}
\label{time-average_oscillating-flow1}
\end{center}
\end{figure}

\begin{figure}
\begin{center}
\begin{minipage}{220pt}
\centering
\includegraphics[width=220pt, trim={0mm 0mm 0mm 0mm}, clip]{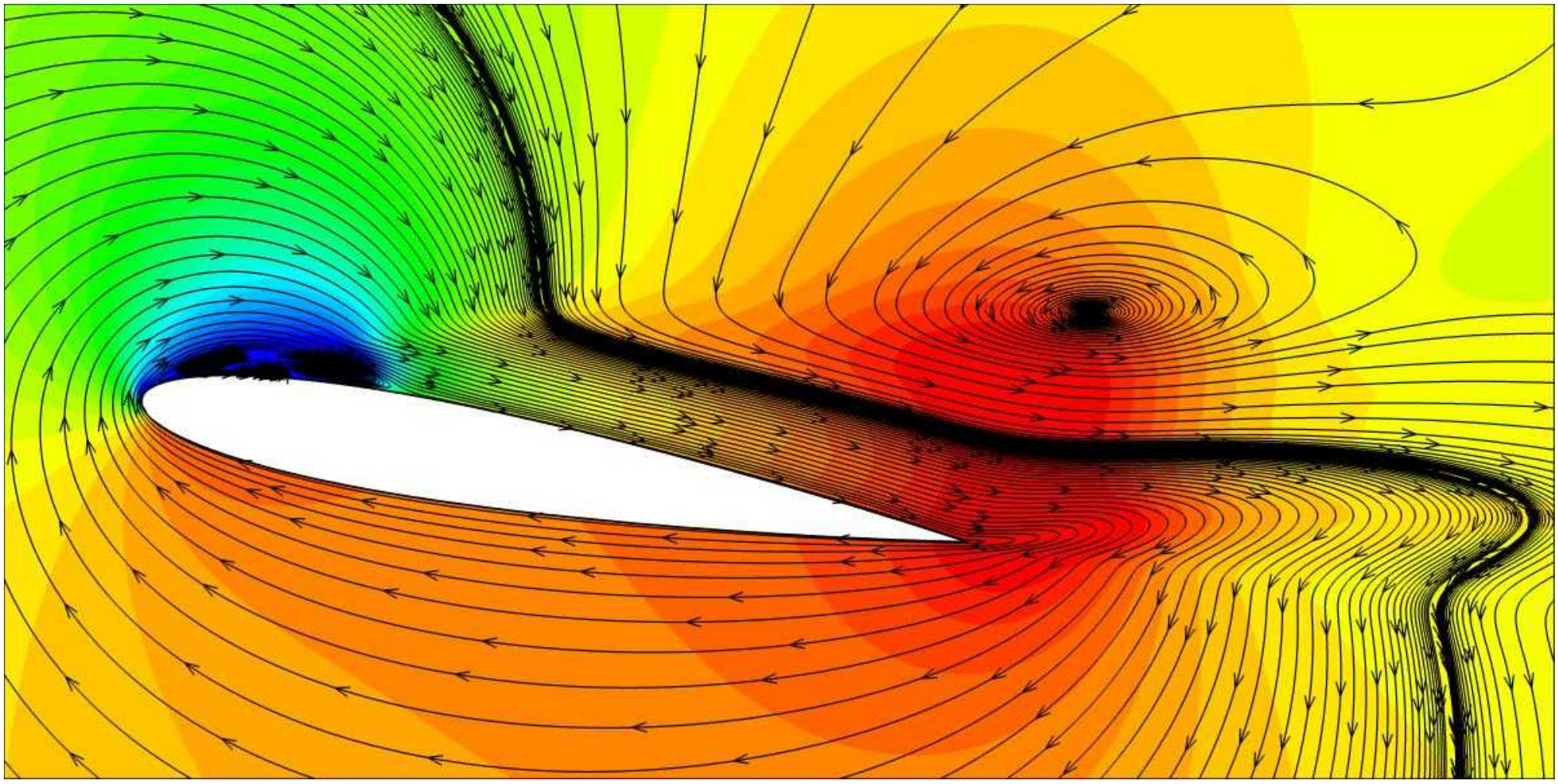}
\textit{$\alpha = 9.8^{\circ}$}
\end{minipage}
\begin{minipage}{220pt}
\centering
\includegraphics[width=220pt, trim={0mm 0mm 0mm 0mm}, clip]{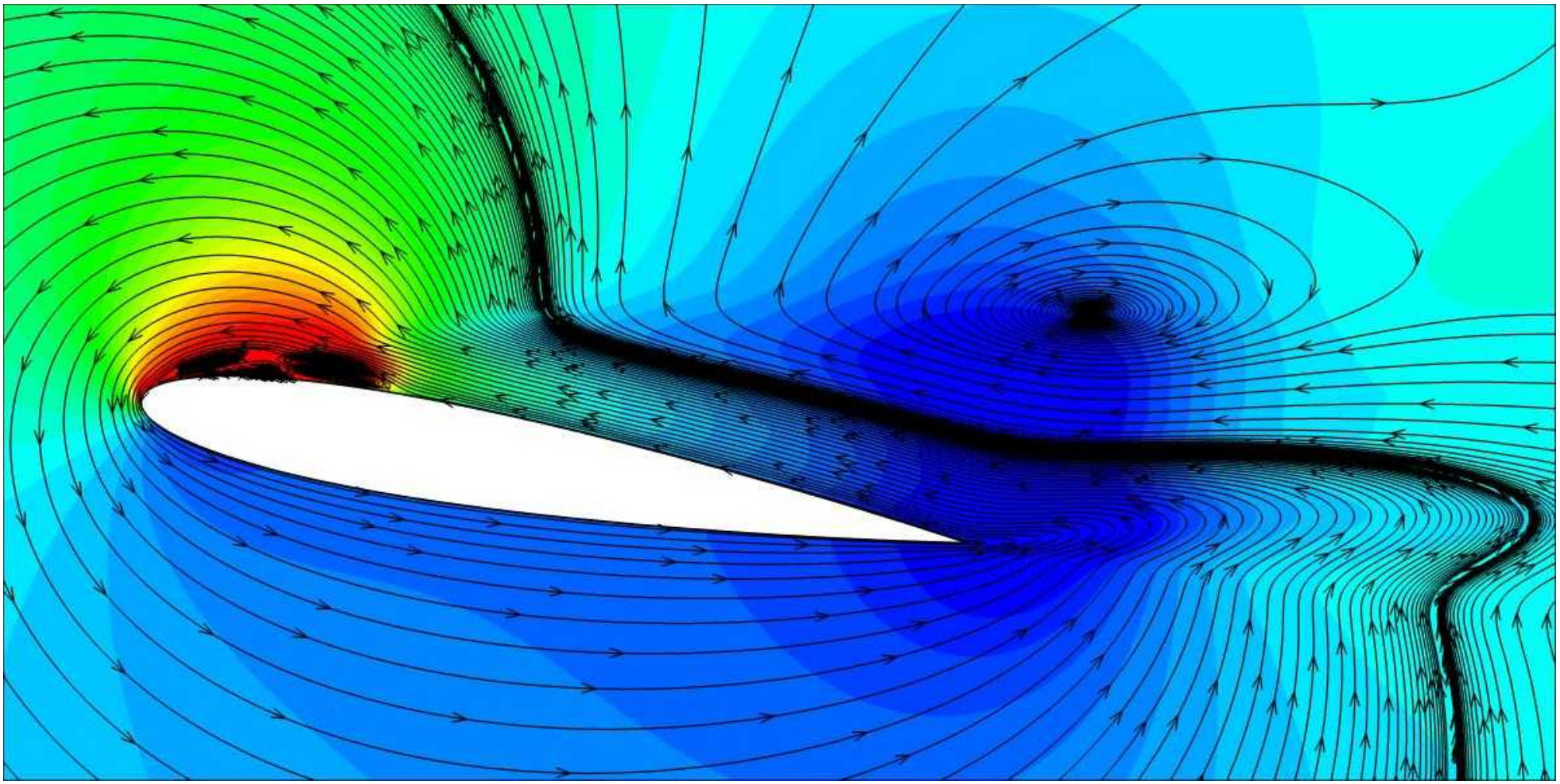}
\textit{$\alpha = 9.8^{\circ}$}
\end{minipage}
\begin{minipage}{220pt}
\centering
\includegraphics[width=220pt, trim={0mm 0mm 0mm 0mm}, clip]{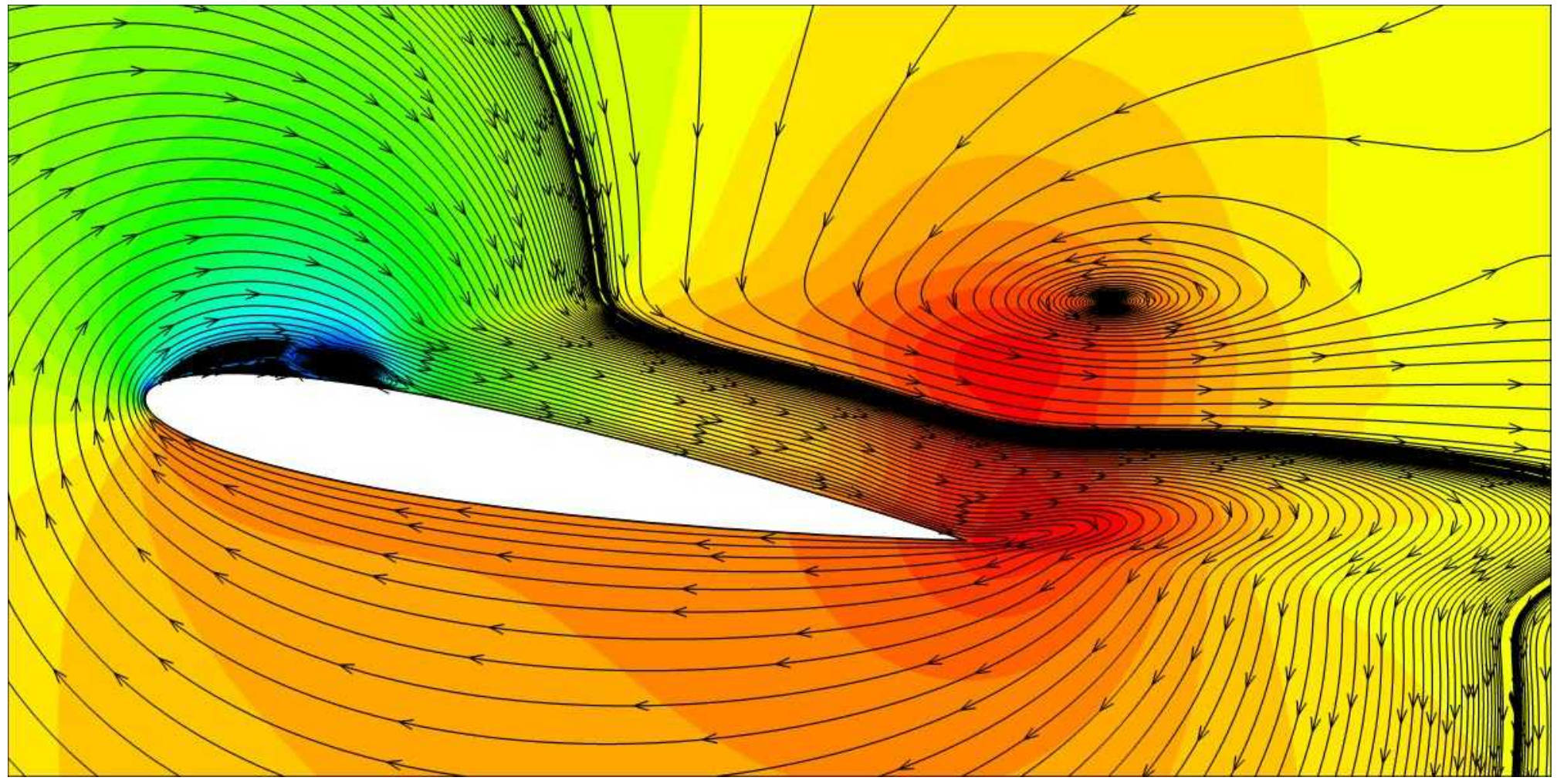}
\textit{$\alpha = 9.9^{\circ}$}
\end{minipage}
\begin{minipage}{220pt}
\centering
\includegraphics[width=220pt, trim={0mm 0mm 0mm 0mm}, clip]{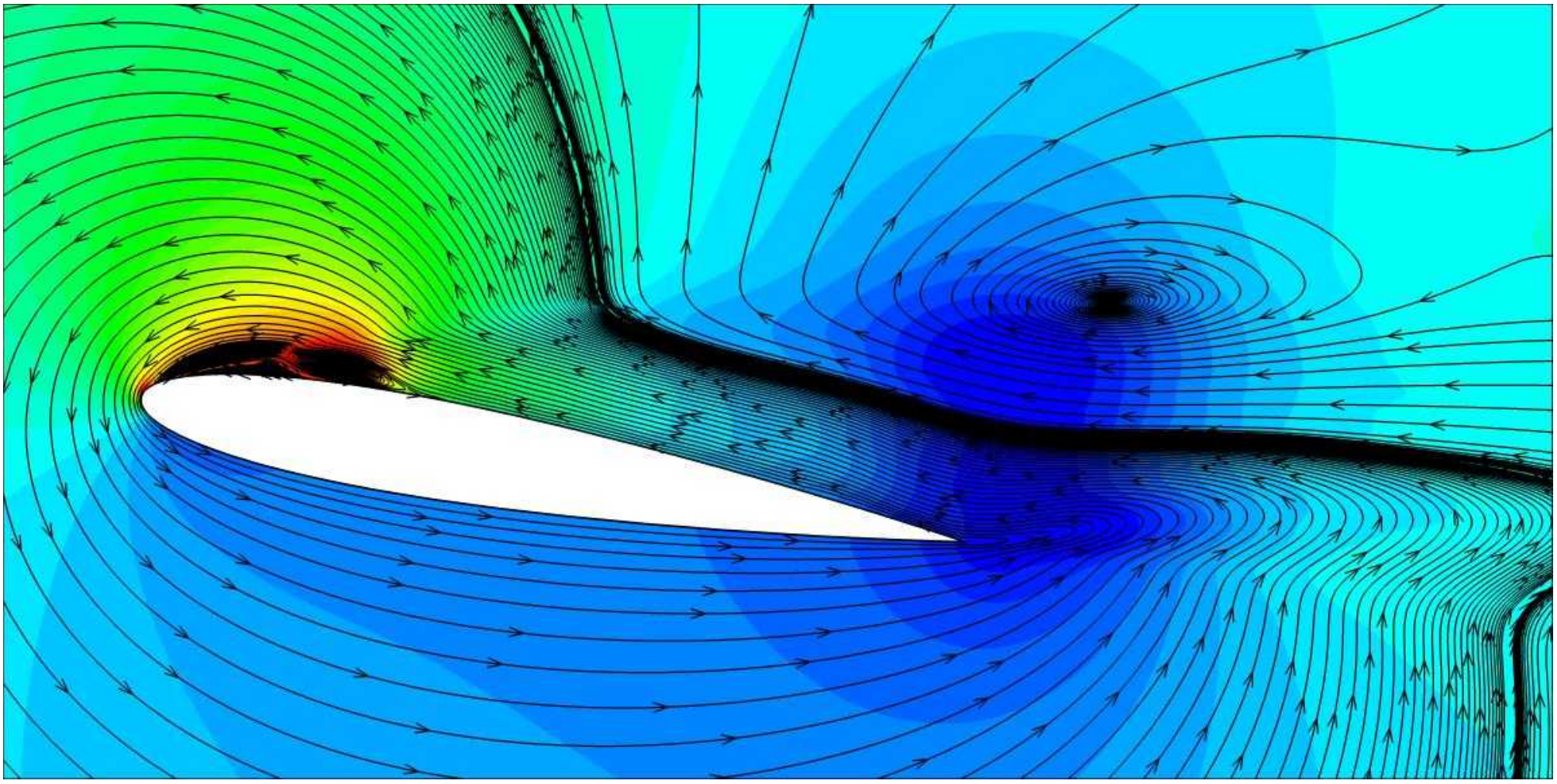}
\textit{$\alpha = 9.9^{\circ}$}
\end{minipage}
\begin{minipage}{220pt}
\centering
\includegraphics[width=220pt, trim={0mm 0mm 0mm 0mm}, clip]{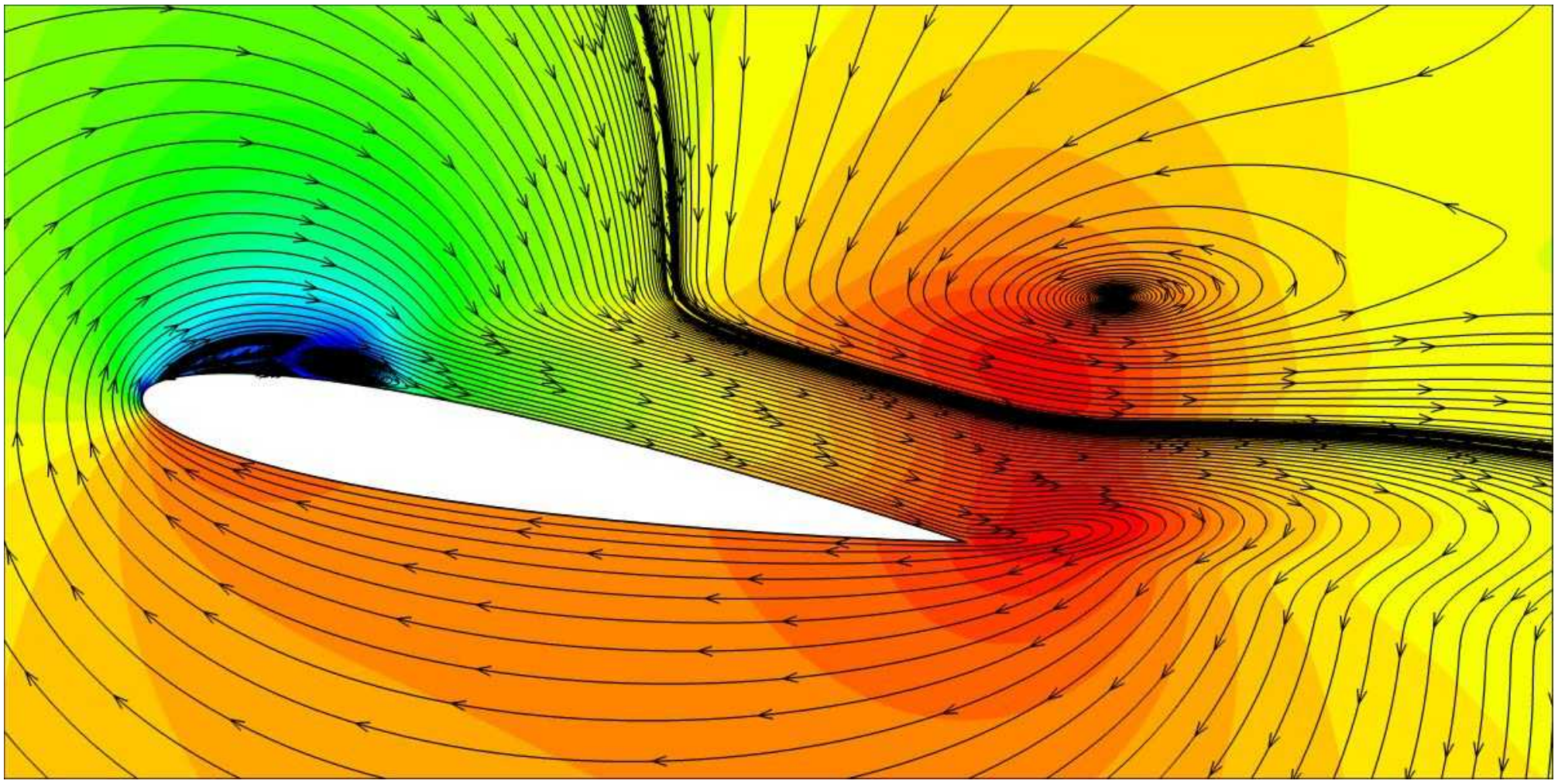}
\textit{$\alpha = 10.0^{\circ}$}
\end{minipage}
\begin{minipage}{220pt}
\centering
\includegraphics[width=220pt, trim={0mm 0mm 0mm 0mm}, clip]{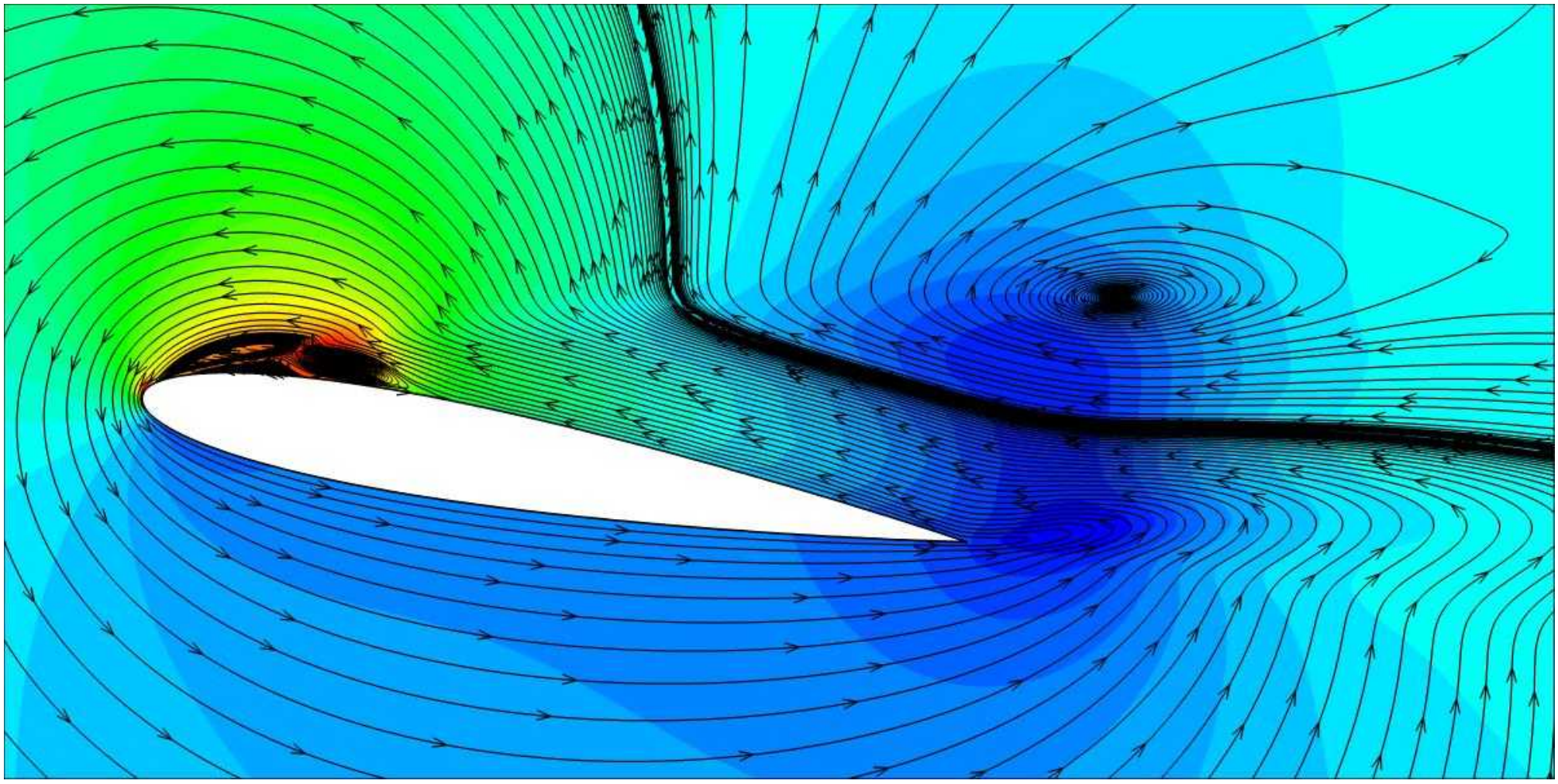}
\textit{$\alpha = 10.0^{\circ}$}
\end{minipage}
\begin{minipage}{220pt}
\centering
\includegraphics[width=220pt, trim={0mm 0mm 0mm 0mm}, clip]{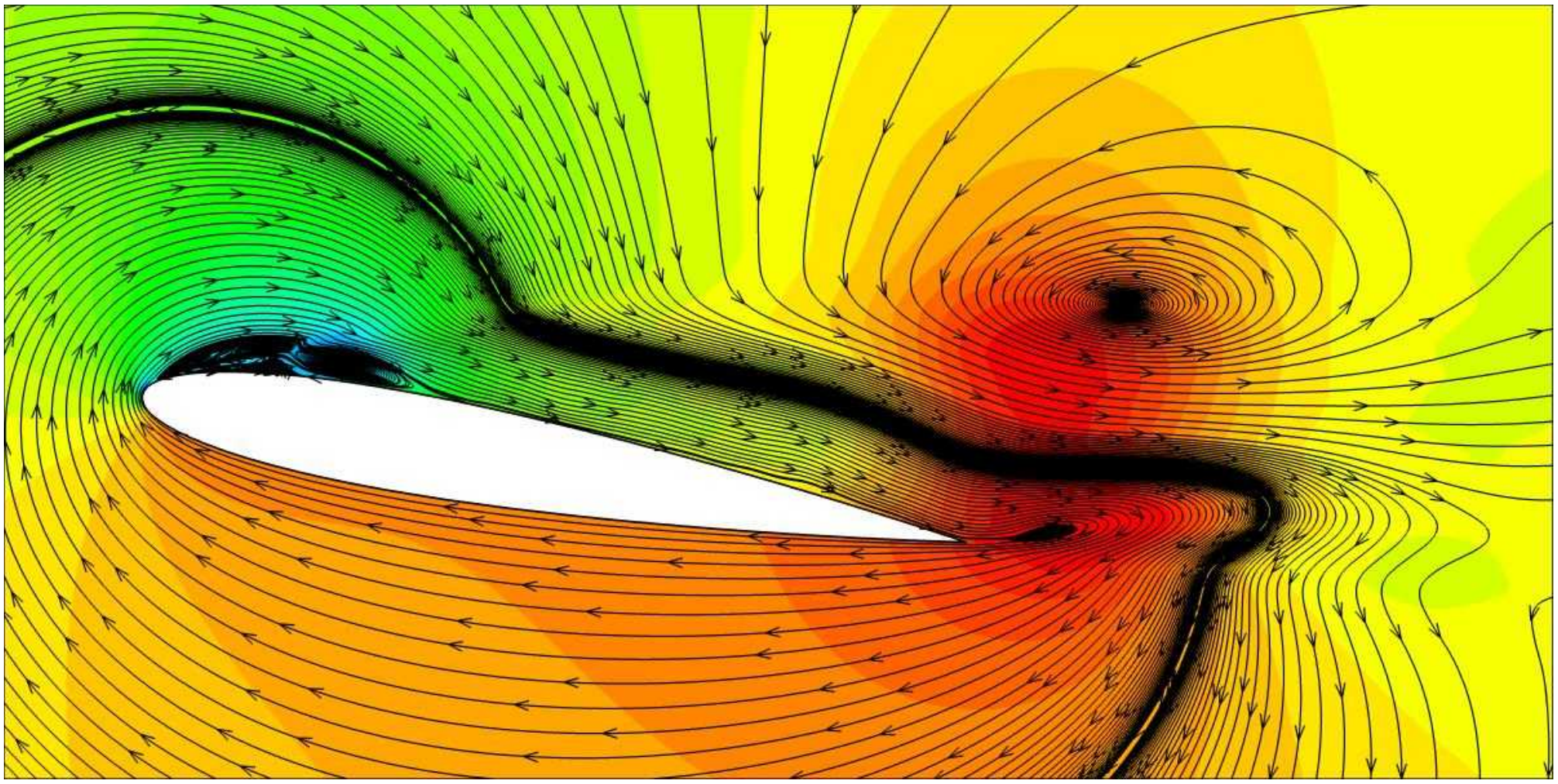}
\textit{$\alpha = 10.1^{\circ}$}
\end{minipage}
\begin{minipage}{220pt}
\centering
\includegraphics[width=220pt, trim={0mm 0mm 0mm 0mm}, clip]{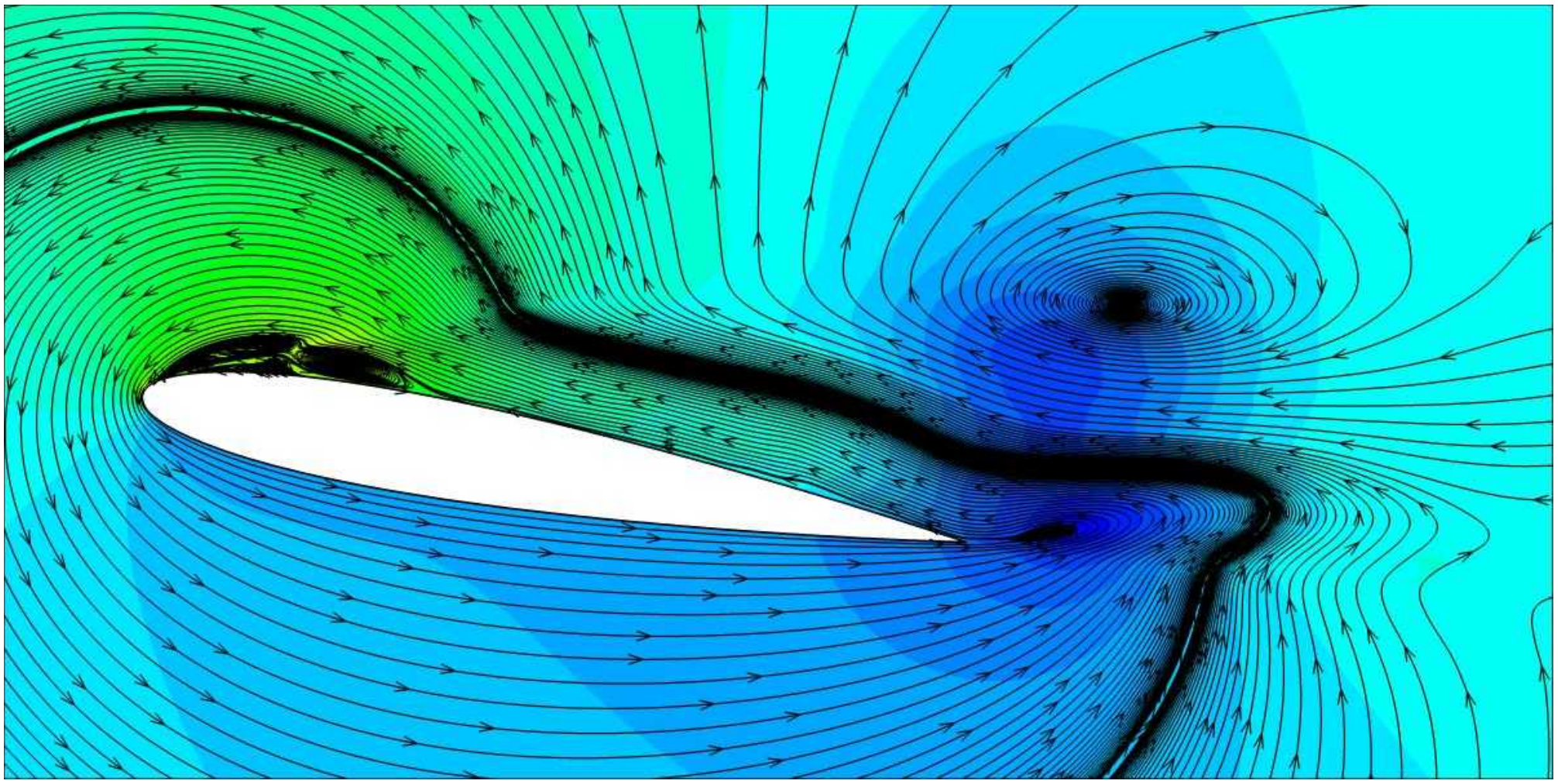}
\textit{$\alpha = 10.1^{\circ}$}
\end{minipage}
\begin{minipage}{220pt}
\centering
\includegraphics[width=220pt, trim={0mm 0mm 0mm 0mm}, clip]{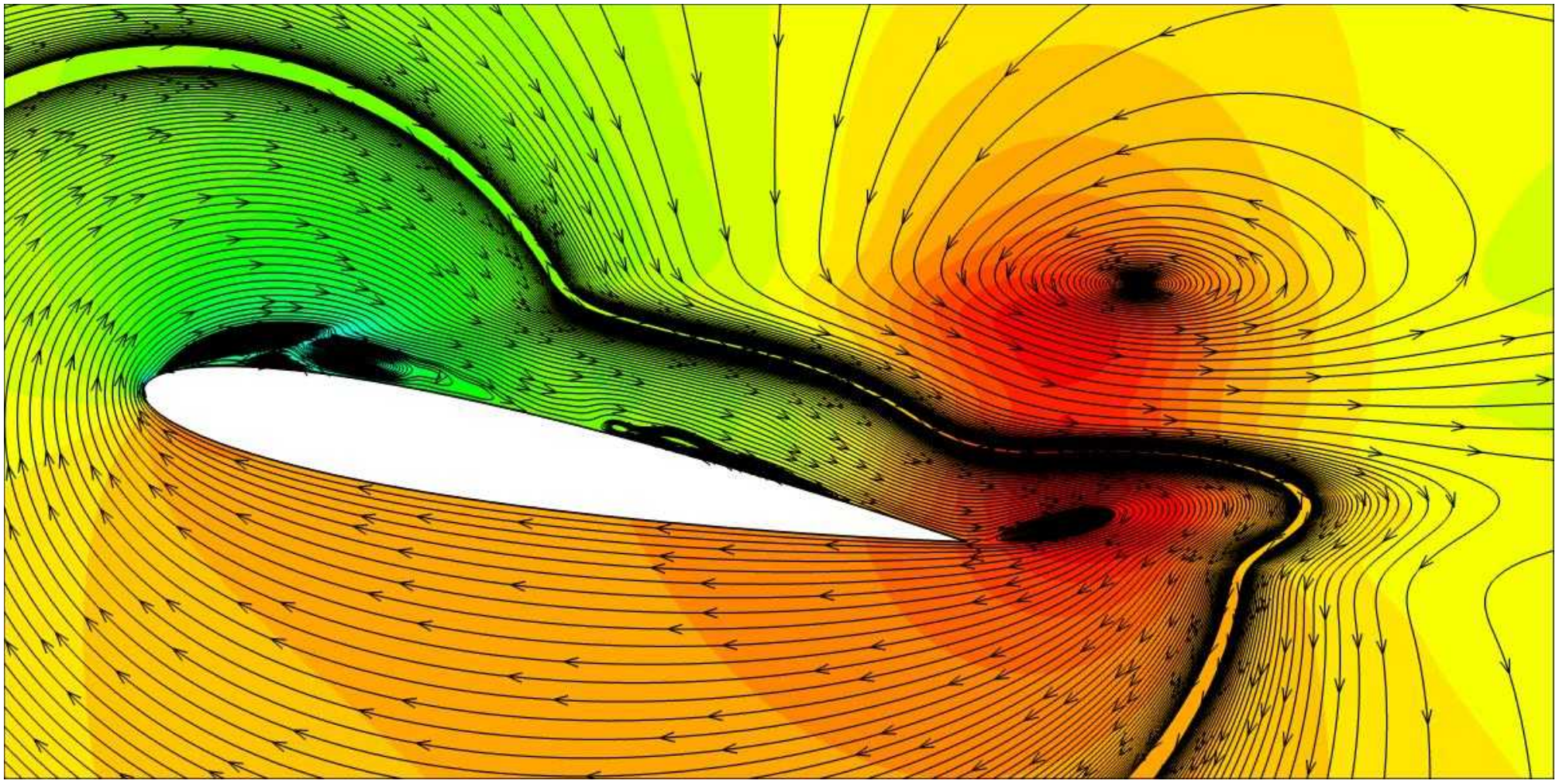}
\textit{$\alpha = 10.5^{\circ}$}
\end{minipage}
\begin{minipage}{220pt}
\centering
\includegraphics[width=220pt, trim={0mm 0mm 0mm 0mm}, clip]{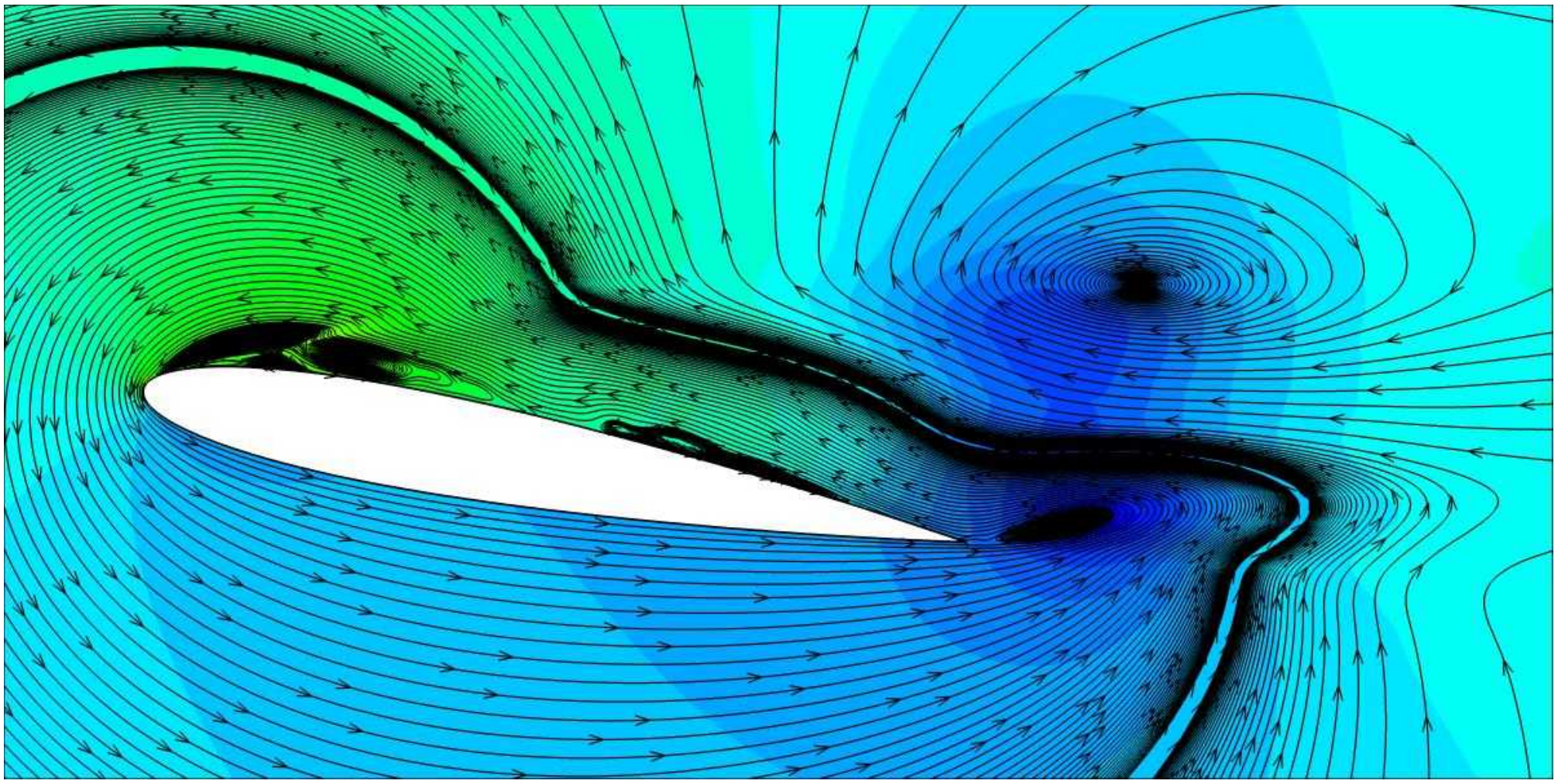}
\textit{$\alpha = 10.5^{\circ}$}
\end{minipage}
\caption{Streamlines patterns of the time-averaged attached and separated phases of the oscillating-f\/low-f\/ield superimposed on colour maps of the time-averaged attached and separated phases of the oscillating-pressure-f\/ield for the angles of attack $\alpha = 9.8^{\circ}$--$10.5^{\circ}$. Left: time-averaged attached-phase of the oscillating-f\/low-f\/ield estimated from the high-lift f\/low-f\/ield minus the mean-lift f\/low-f\/ield, $\Delta{\overline{U}}^\mathrm{+}$, $\Delta{\overline{V}}^\mathrm{+}$, and $\Delta{\overline{P}}^\mathrm{+}$. Right: time-averaged separated-phase of the oscillating-f\/low-f\/ield estimated from the low-lift f\/low-f\/ield minus the mean-lift f\/low-f\/ield, $\Delta{\overline{U}}^\mathrm{-}$, $\Delta{\overline{V}}^\mathrm{-}$, and $\Delta{\overline{P}}^\mathrm{-}$.}
\label{time-average_oscillating-flow2}
\end{center}
\end{figure}

\begin{figure}
\begin{center}
\begin{minipage}{220pt}
\centering
\includegraphics[width=220pt, trim={0mm 0mm 0mm 0mm}, clip]{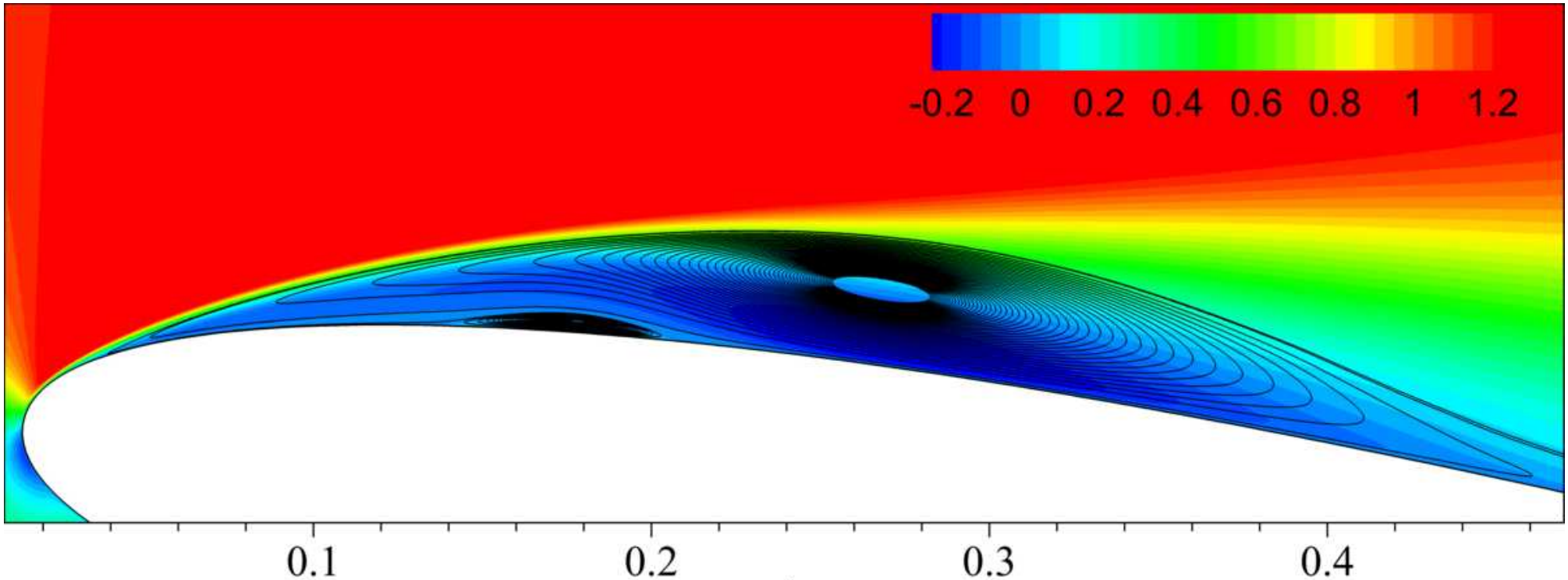}
\textit{Time-averaged attached-phase of the f\/low-f\/ield.\\ $ \qquad $}
\end{minipage}
\begin{minipage}{220pt}
\centering
\includegraphics[width=220pt, trim={0mm 0mm 0mm 0mm}, clip]{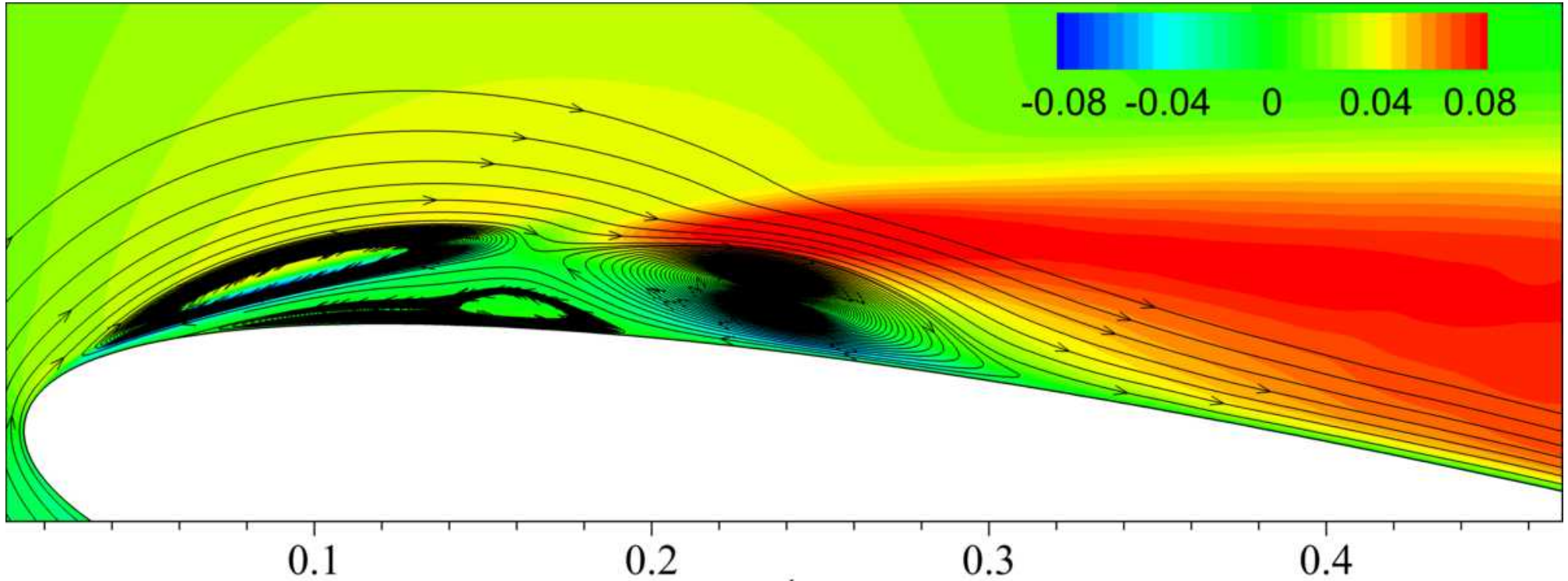}
\textit{Time-averaged attached-phase of the oscillating-f\/low.}
\end{minipage}
\begin{minipage}{220pt}
\centering
\includegraphics[width=220pt, trim={0mm 0mm 0mm 0mm}, clip]{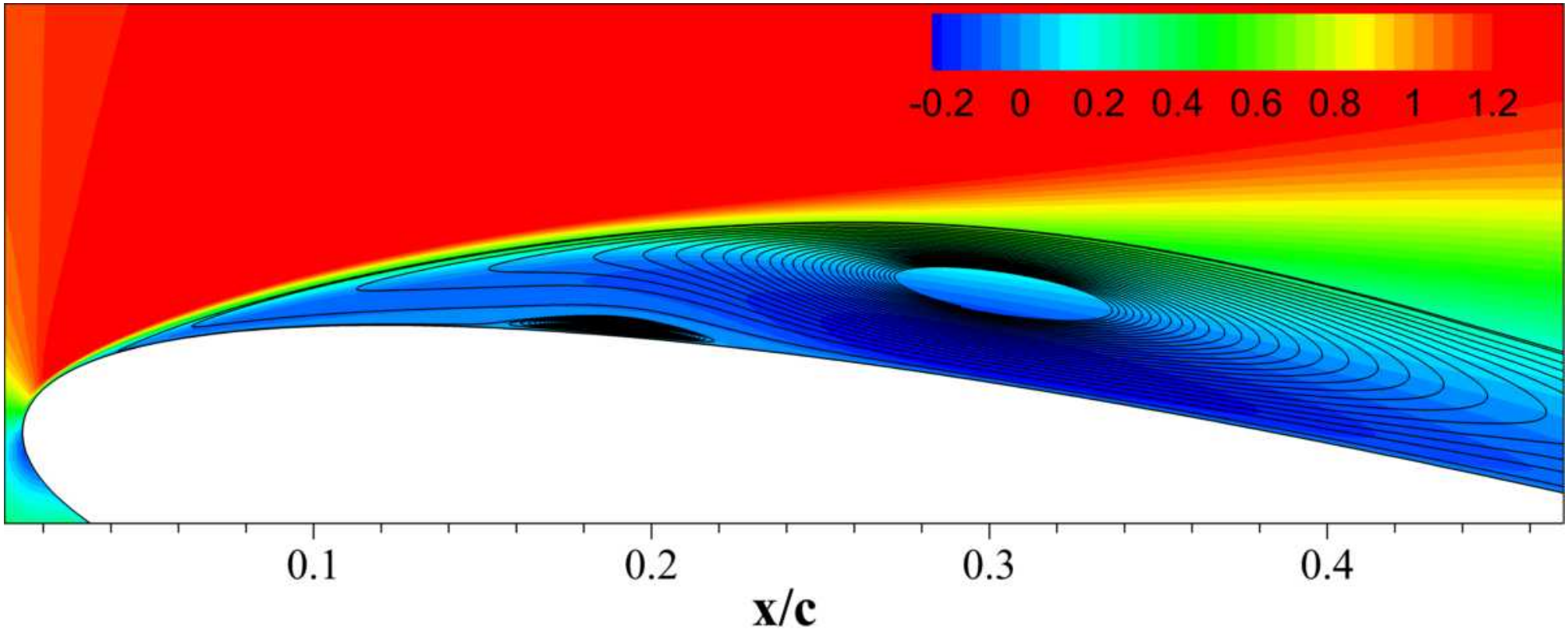}
\textit{Time-averaged separated-phase of the f\/low-f\/ield.}
\end{minipage}
\begin{minipage}{220pt}
\centering
\includegraphics[width=220pt, trim={0mm 0mm 0mm 0mm}, clip]{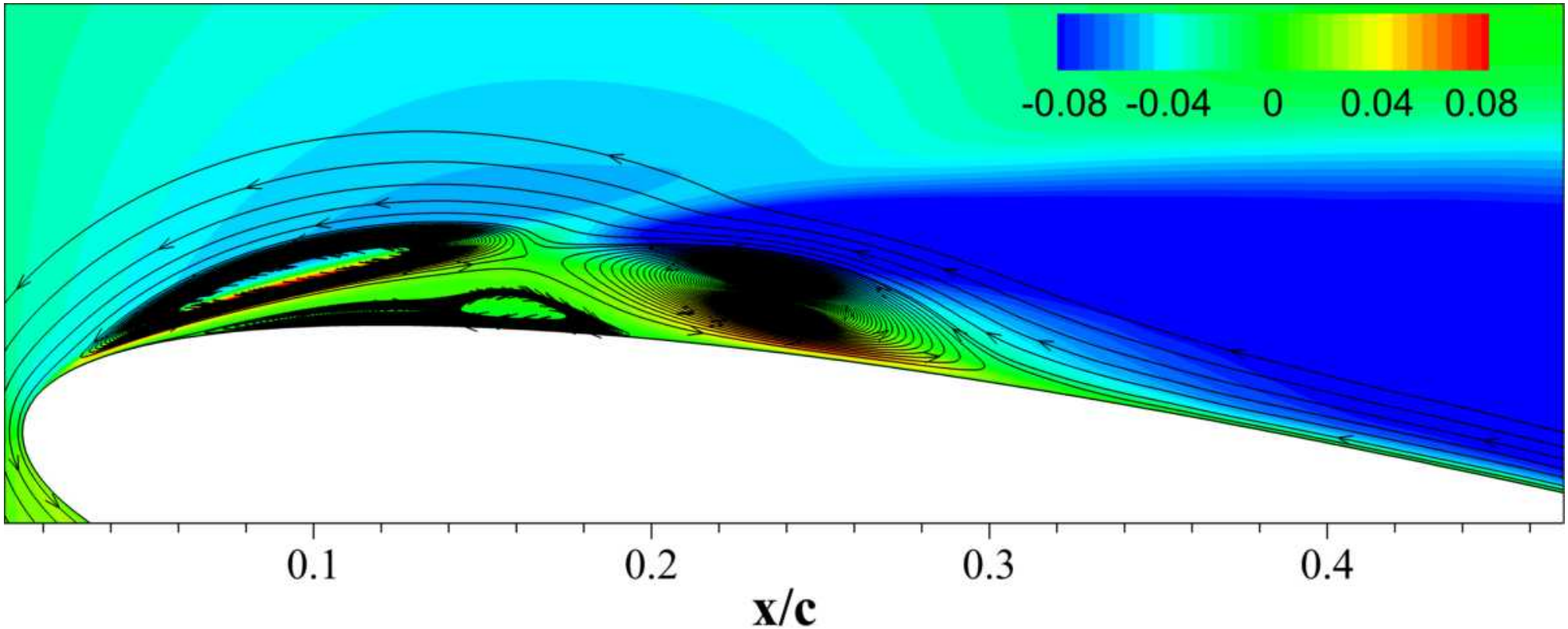}
\textit{Time-averaged separated-phase of the oscillating-f\/low.}
\end{minipage}
\caption{Streamlines patterns of the f\/low-f\/ield and the oscillating-f\/low-f\/ield superimposed on colour maps of the streamwise velocity component and the oscillating-streamwise-velocity, respectively, for the angle of attack of $9.6^{\circ}$. Left: time-averaged attached-phase (top) and separated-phase (bottom) f\/low-f\/ields, $\widehat{U}$ \& $\widehat{V}$ and $\widecheck{U}$ \& $\widecheck{V}$, respectively. Right: time-averaged attached-phase (top) and separated-phase (bottom) of the oscillating-f\/low-f\/ield, $\Delta{\overline{U}}^\mathrm{+}$ \& $\Delta{\overline{V}}^\mathrm{+}$ and $\Delta{\overline{U}}^\mathrm{-}$ \& $\Delta{\overline{V}}^\mathrm{-}$, respectively.}
\label{time-average_oscillating-flow_960}
\end{center}
\end{figure}

The conditional time-averaging, described and used in~\citet{eljack2018bursting}, def\/ines the time-average of an instantaneous variable of the f\/low-f\/ield, $\psi$, on three different levels a high-lift average, a low-lift average, and a mean-lift average. The mean-lift average of the variable ($\overline{\psi}$) is simply the time-average of all data samples of the variable. The high-lift average of the variable ($\widehat{\psi}$) is the time-average of all data samples that have values higher than the mean values. The low-lift average of the variable ($\widecheck{\psi}$) is the average of all data samples that have values less than the mean values. When the lift coeff\/icient signal is Gaussian, the high-lift and low-lift averages of the variable represent the attached and separated f\/low-f\/ield in the statistical sense, respectively. In this sense, one can def\/ine a time-averaged surplus ($\Delta{\overline{\psi}}^\mathrm{+}$) and def\/icit ($\Delta{\overline{\psi}}^\mathrm{-}$) of the f\/low variable above and below that of the mean-f\/low, respectively. Thus, the time-averaged surplus and def\/icit of the variable are given by subtracting the mean-lift average of the variable from the high-lift average of the variable and the low-lift average of the variable, respectively. Hence, the time-averaged surplus of the f\/low-f\/ield is estimated from the high-lift f\/low-f\/ield minus the mean-lift f\/low-f\/ield, $\Delta{\overline{\psi}}^\mathrm{+} = \widehat{\psi} - \overline{\psi}$. The time-averaged def\/icit of the f\/low-f\/ield is estimated from the low-lift f\/low-f\/ield minus the mean-lift f\/low-f\/ield, $\Delta{\overline{\psi}}^\mathrm{-} = \widecheck{\psi} - \overline{\psi}$. For a Gaussian lift coeff\/icient signal, the time-averaged surplus and def\/icit of the variable above and below the mean feature the time-averaged attached-phase and separated-phase of the oscillating-f\/low-f\/ield, respectively.\newline
F\/igures~\ref{time-average_oscillating-flow1} and~\ref{time-average_oscillating-flow2} show streamlines patterns of the time-averaged attached-phase of the oscillating-f\/low-f\/ield, $\Delta{\overline{U}}^\mathrm{+}$ and $\Delta{\overline{V}}^\mathrm{+}$, and the time-averaged separated-phase of the oscillating-f\/low-f\/ield, $\Delta{\overline{U}}^\mathrm{-}$ and $\Delta{\overline{V}}^\mathrm{-}$, superimposed on colour maps of the time-averaged attached-phase of the oscillating-pressure-f\/ield, $\Delta{\overline{P}}^\mathrm{+}$, and the time-averaged separated-phase of the oscillating-pressure-f\/ield, $\Delta{\overline{P}}^\mathrm{-}$, respectively, at the angles of attack $\alpha = 9.25^{\circ}$--$10.5^{\circ}$. As seen in the f\/igures, the time-averaged attached and separated phases of the oscillating-f\/low-f\/ield are rotating in a closed loop around the aerofoil in the clockwise and the anti-clockwise direction, respectively. A triad of three vortices, two co-rotating vortices (TCV) and a secondary vortex lies beneath them, and counter-rotating with them is formed in the vicinity of the leading-edge on the suction surface of the aerofoil in a region almost coincide with that of the LSB. A large vortex is located above the aerofoil trailing-edge and counter-rotates with the oscillating-f\/low-f\/ield adjacent to the aerofoil surface. This vortex acts as a wheel that pushes and guides the oscillating-f\/low-f\/ield towards the aerofoil surface and forces it to be directed tangentially to the aerofoil surface. The origin of this vortex is not known yet, i.e., is it created locally by the oscillating-f\/low-f\/ield or does it originate somewhere upstream the trailing-edge and advects downstream by the f\/low. Hereafter, this vortex will be referred to as the ``guide-vortex''. The shape and strength of the guide-vortex vary with the angle of attack as seen in the f\/igures. The time-averaged attached-phase of the f\/low-f\/ield sets the conditions for the LSB to be reformed after it bursts. Consequently, the pressure drops in the vicinity of the LSB and an adverse pressure gradient (APG) is formed along the aerofoil chord. When the LSB burst, the f\/low separates creating a high-pressure region in the vicinity of the leading-edge and a low-pressure region in the vicinity of the trailing-edge. Hence, a favourable pressure gradient (FPG) along the aerofoil chord is formed. When the direction of the time-averaged oscillating-f\/low-f\/ield is clock-wise, it adds momentum to the boundary layer, and the f\/low remains attached against the APG. When the oscillating-f\/low-f\/ield rotates in the anti-clockwise direction, it subtracts momentum from the boundary layer and drives the f\/low to separate despite the FPG. It seems that this triad of vortices, the gradient of the oscillating-pressure, and changes in the turbulent kinetic energy along the separated shear layer are interlinked, govern, and sustain the LFO. However, the details of the rotation of this triad and how it adjusts the frequency of the f\/low oscillation are lost in the averaging process. Most importantly, how this triad controls the bursting and stability of the LSB.\newline
F\/igure~\ref{time-average_oscillating-flow_960} shows zoom-in plots of streamlines patterns of the f\/low-f\/ield and streamlines patterns of the oscillating-f\/low-f\/ield superimposed on colour maps of the streamwise velocity component of the f\/low-f\/ield and the oscillating-streamwise-velocity, respectively, for the angle of attack of $9.6^{\circ}$. The left-hand side of the f\/igure shows the time-averaged attached (top) and the time-averaged separated (bottom) f\/low-f\/ields, and the right-hand side of the f\/igure displays the time-averaged attached (top) and the time-averaged separated (bottom) oscillating-f\/low-f\/ields. As seen in the f\/igure, the upstream vortex of the TCV (UV) is driven by the gradient of the oscillating-velocity across the laminar portion of the separated shear layer and is faithfully aligned with it. The UV front tip forms a half-saddle on the aerofoil surface just upstream the separation point of the shear layer, and its rear end forms a full-saddle. The downstream vortex of the TCV (DV) is submerged entirely inside the region of the LSB and aligned such that its front tip forms a full-saddle with the UV while its rear tip forms a half-saddle on the aerofoil surface upstream the reattachment point. The secondary vortex act as a roller support that facilitates the rotation and orientation of the TCV. The focus of the DV of the TCV is located just upstream and below that of the LSB, in the mean sense. However, the orientation of the triad of vortices, in the mean sense, changes with the angle of attack. Most importantly, instantaneous as well as spatial orientation of the triad of vortices seems to play a profound role on the stability of the LSB and the sustainability of the LFO.\newline
F\/igures~\ref{time-average_oscillating-flow3} and~\ref{time-average_oscillating-flow4} show streamlines patterns of the time-averaged attached and separated phases of the oscillating-f\/low-f\/ield superimposed on colour maps of the time-averaged attached and separated f\/low phases of the oscillating-spanwise-vorticity for the angles of attack $\alpha = 9.25^{\circ}$--$10.5^{\circ}$. The left-hand side of the f\/igures shows the time-averaged attached-phase of the oscillating-f\/low-f\/ield estimated from the high-lift f\/low minus the mean-lift f\/low, $\Delta{\overline{U}}^\mathrm{+}$ and $\Delta{\overline{V}}^\mathrm{+}$. The right-hand side of the f\/igures shows the time-averaged separated-phase of the oscillating-f\/low-f\/ield estimated from the low-lift f\/low minus the mean-lift f\/low, $\Delta{\overline{U}}^\mathrm{-}$ and $\Delta{\overline{V}}^\mathrm{-}$. The f\/igures are zoomed in such a way that the orientation of the triad of vortices and how it varies with the angle of attack can be examined. The red dashed line displays the streamline that bounds the LSB in the mean sense. The black dash-dot line denotes the contour line at which the time-averaged oscillating-streamwise-velocity equals zero. It is evident that the oscillating-f\/low-f\/ield rotates in the clockwise direction during the attached-phase and rotates in the anti-clockwise direction during the separated-phase of the f\/low-f\/ield. It is noted that each vortex of the triad of vortices is aligned along a part of the dash-dot line.\newline
The conditional time-averaging employed in the present study to obtain the attached and separated phases of the oscillating-f\/low-f\/ield takes into account contributions from all vortical structures, in the mean sense, regardless of their frequency and size. Thus, the time-averaged oscillating-f\/low-f\/ield captures the spatial evolution of all f\/low modes in a single snapshot. However, some oscillating-f\/low features could be obscured considering their relatively low amplitude. Therefore, the appearance of the time-averaged oscillating-f\/low-f\/ield ref\/lects the shape of f\/low features that have the highest magnitude.\newline
At the angle of attack of $9.0^{\circ}$, not shown here, the time-averaged attached and separated phases of the oscillating-f\/low-f\/ield form a single vortex above the streamline that bound the LSB. The vortex is stretched and break into two co-rotating vortices as the angle of attack is increased to $9.25^{\circ}$. These two co-rotating vortices are similar to the previously discussed TCV. However, it seems that these two co-rotating vortices gain momentum as the angle of attack increases. Thus, they become more pronounced in shape and magnitude as the angle of attack is increased. The TCV at the angle of attack of $9.25^{\circ}$ does not have enough energy to change the f\/low-f\/ield globally. The TCV could only extract energy from the oscillating-f\/low-f\/ield and redistribute this energy to remain in equilibrium. While doing this, the TCV elongate the time-scale of the oscillating-f\/low-f\/ield cycle and consequently reduce the frequency of the f\/low-oscillation below that of the bluff-body shedding frequency. Furthermore, the presence of the TCV in the f\/low-f\/ield changes the magnitude of the oscillating-f\/low-f\/ield, decreases the mean lift coeff\/icient, increases the mean drag coeff\/icient, and deteriorates the performance of the aerofoil.\newline
At the angle of attack of $9.4^{\circ}$ a drastic change in the time-averaged attached and separated phases of the oscillating-f\/low-f\/ield takes place. The triad of vortices has enough energy to dominate the time-averaged oscillating-f\/low-f\/ield. The UV of the TCV elongates and moves upstream until its front tip forms a half-saddle that rests on the aerofoil surface just upstream the separation point of the shear layer. The UV is anchored to the laminar portion of the separated shear layer; thus, it preserves its shape and position, and remains erected above the separated zone. The secondary vortex, which lies beneath the TCV, is combined into one coherent vortex at this angle of attack. Whereas, at the angles of attack of $9.0^{\circ}$ and $9.25^{\circ}$ it was less coherent and splattered into two or three small vortices, in the mean sense. The most dramatic and important change that takes place in the orientation of the TCV is that of the DV. The downstream vortex is elongated just like the UV and oriented such that its rear tip forms a half-saddle on the aerofoil surface and entirely resides in the region of the LSB. However, the DV could preserve neither its shape nor its position. Its front tip forms a full-saddle with the UV, and its orientation with respect to the UV is dependent on the angle of attack, in the mean sense. What is important about this orientation is that the TCV are aligned in such a way that the UV could slide over the DV. While the UV is extracting energy from the mean f\/low, it gains more and more momentum until it is saturated. Then, the UV slides over the DV, expands, and advects downstream. When the UV expands, it changes the direction of rotation of the oscillating-f\/low-f\/ield. Hence, if the f\/low is attached, it will separate as a consequence of the expansion of the UV and vice versa. Thus, the conditions for the LSB bursting is that the UV of the TCV has enough energy and the TCV are oriented in such a way that the UV could slide above the DV. The ability of the UV to extract energy from the mean f\/low depends primarily on the gradient of the oscillating-velocity across the laminar portion of the separated shear layer. Therefore, the primary cause that triggers the instability of the LSB and initiates the LFO is the gradient of the oscillating-velocity across the laminar portion of the separated shear layer. This does not only reveal the bursting condition for the LSB, but it also reveals the mechanism that governs and sustains the LFO.\newline
As the angle of attack is further increased to $9.5^{\circ}$, the UV expands intermittently. Hence, the LSB is intact with intermittent occasions on which it bursts. At this angle of attack, the mean f\/low is attached. Therefore, the occasional expansion of the UV and the subsequent bursting of the LSB separate the f\/low intermittently. The transition of the strength and orientation of the triad of vortices continues in the same manner until the LFO process becomes regular and more pronounced in the magnitude of oscillation at the angle of attack of $9.8^{\circ}$. As the angle of attack increases, the mean f\/low becomes separated, and the LSB does not form. The UV of the TCV extracts less energy from the mean f\/low as the gradient of the oscillating-velocity across the laminar portion of the separated shear layer decreases. However, the strength of the UV reaches the threshold required for it to expand and reattach the f\/low-f\/ield intermittently. As the f\/low reattaches, the conditions for the formation of the LSB are met, and the LSB forms in the vicinity of the leading-edge. Therefore, the f\/low-f\/ield at this angle of attack remains separated with frequent reattachments. As the angle increases, the UV expands less frequently, and the f\/low-f\/ield remains separated for longer time intervals. The aerofoil eventually undergoes a full stall when the angle of attack is further increased.

\subsection{The origin of the oscillating-f\/low-f\/ield}
The oscillating-f\/low and the velocity gradient across the laminar portion of the separated shear layer are observed in all of the investigated angles of attack including at zero angle of attack. However, the magnitude and frequency of the oscillation vary with the angle of attack. The origin of the oscillation is the instability at the trailing-edge that generates and sustains the von Karman alternating vortices. In harmony with the alternating vortices at the trailing-edge, the stagnation point at the vicinity of the leading-edge pitches up and down. The Strouhal number of the oscillating-f\/low at low to moderate angles of attack f\/luctuates around that of the bluff-body shedding of around $0.2$. However, the interaction between the tailing-edge instability, the triad of vortices, and the gradient of the oscillating-pressure shifts the frequency of the resultant oscillating-f\/low to lower values. The more energy the UV of the TCV extracts from the mean-f\/low, the more it will be able to elongate the time-scale of the cycle of the oscillating-f\/low and consequently lowers its frequency. The ability of the UV to extract energy from the mean-f\/low depends primarily on the gradient of the oscillating-velocity across the laminar portion of the separated shear layer. As the gradient of the oscillating-velocity is proportional to the angle of attack, the UV gains more momentum as the angle of attack increases and can further lower the frequency of the LFO and increases its magnitude of oscillation.

\subsection{Conditional phase-average of the oscillating-f\/low-f\/ield}
\begin{figure}
\begin{center}
\includegraphics[width=420pt, trim={0mm 0mm 0mm 0mm}, clip]{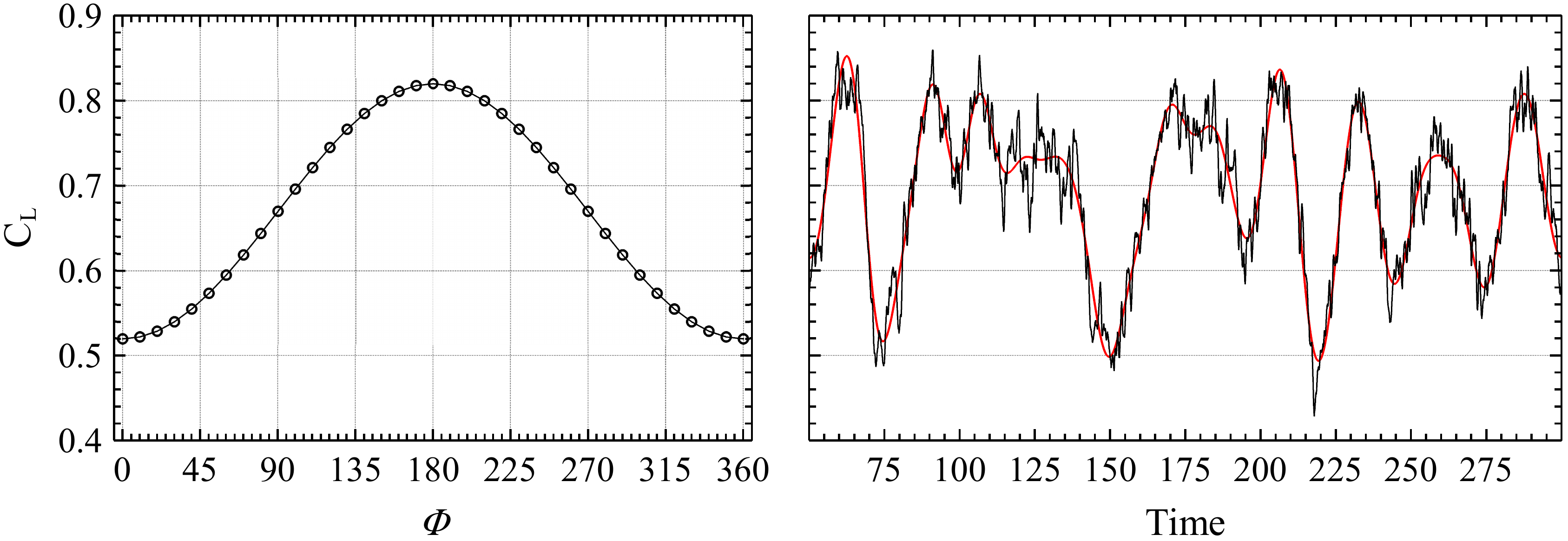}
\caption{The conditional phase-averaging process. Left: One sinusoidal cycle with a magnitude range of that of the low-pass f\/iltered lift coeff\/icient divided into $36$ equally distributed phases. Right: The black line displays the time history of the lift coeff\/icient, and the red line denotes the low-pass f\/iltered lift coeff\/icient at the angle of attack of $9.80^{\circ}$.}
\label{conditional_phase_average}
\end{center}
\end{figure}

\begin{figure}
\begin{center}
\begin{minipage}{220pt}
\centering
\includegraphics[width=220pt, trim={0mm 0mm 0mm 0mm}, clip]{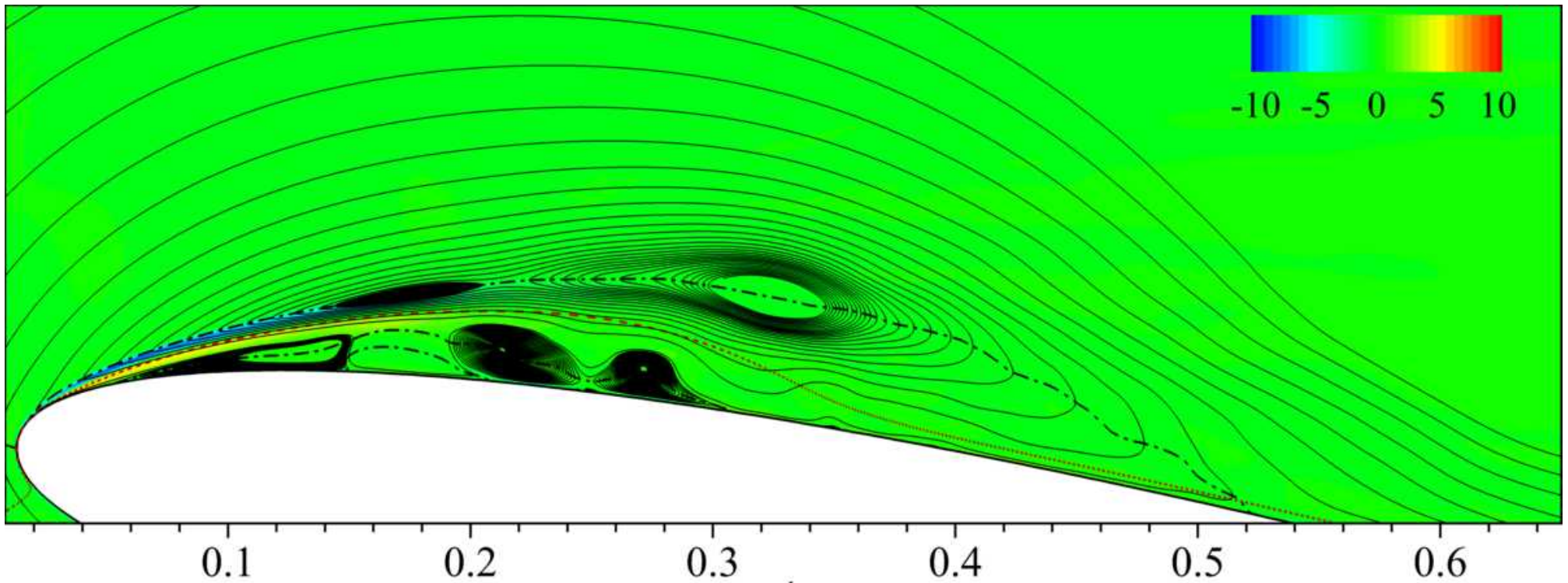}
\textit{$\alpha = 9.25^{\circ}$.}
\end{minipage}
\medskip
\begin{minipage}{220pt}
\centering
\includegraphics[width=220pt, trim={0mm 0mm 0mm 0mm}, clip]{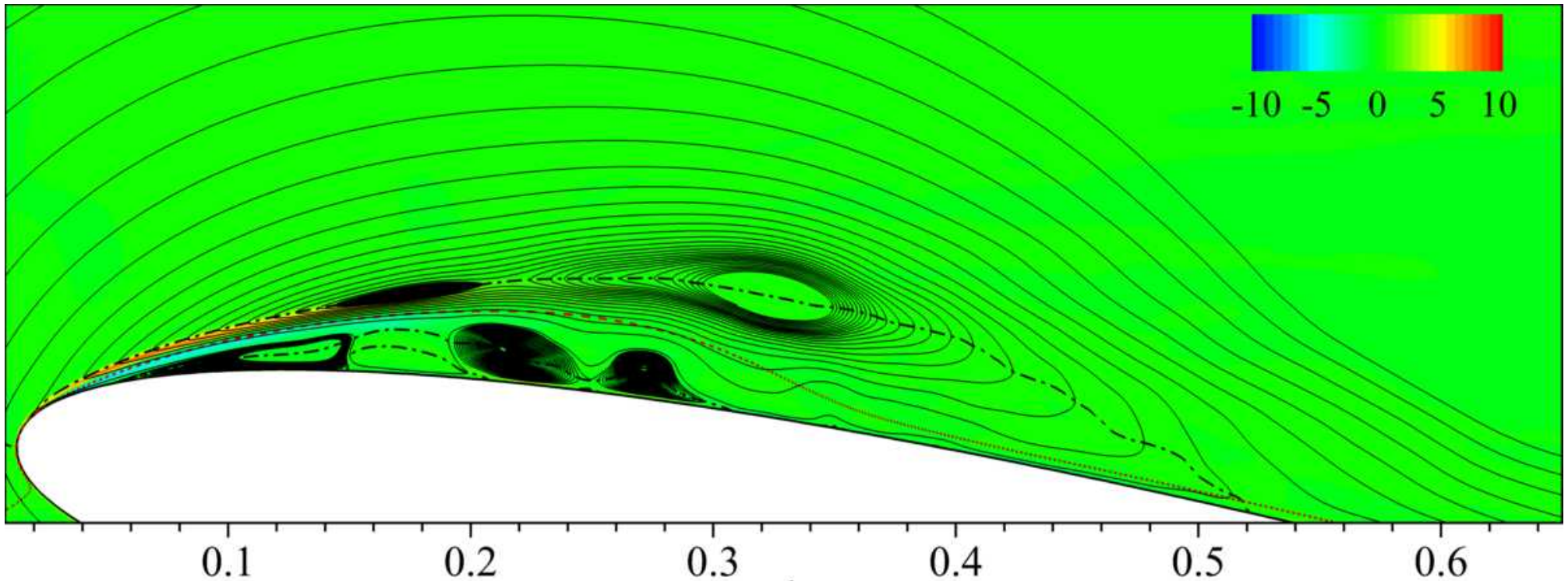}
\textit{$\alpha = 9.25^{\circ}$.}
\end{minipage}
\medskip
\begin{minipage}{220pt}
\centering
\includegraphics[width=220pt, trim={0mm 0mm 0mm 0mm}, clip]{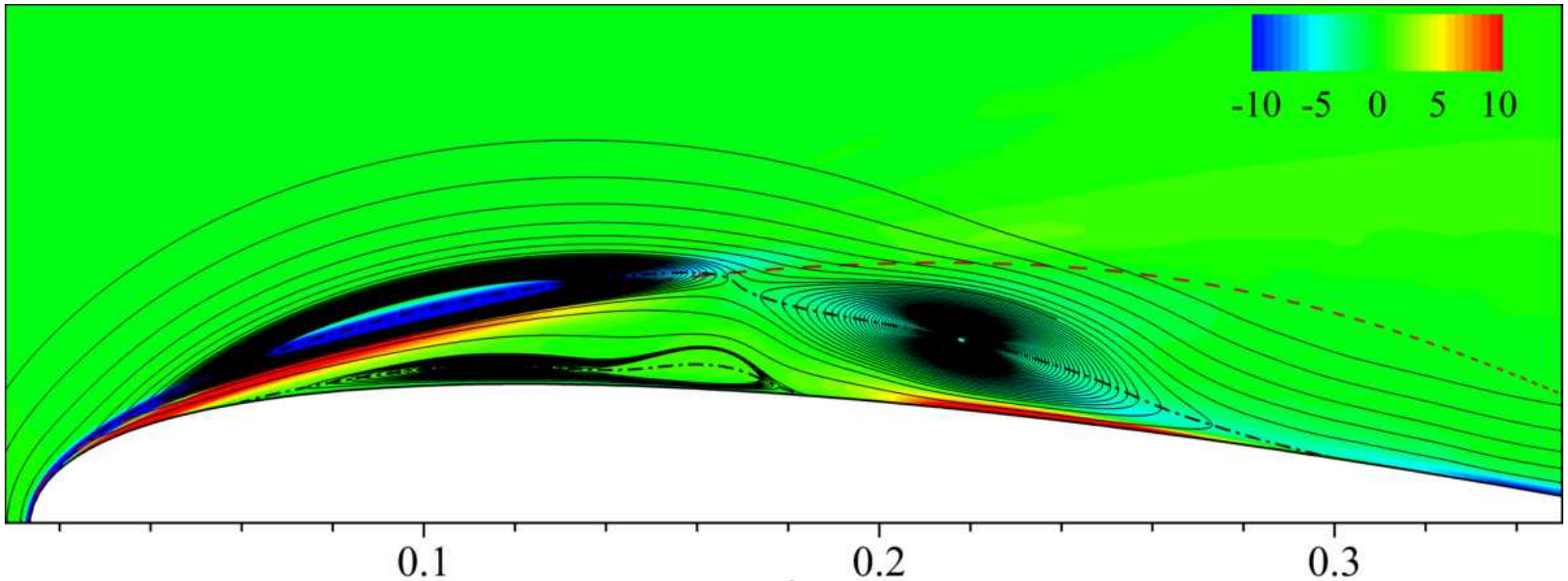}
\textit{$\alpha = 9.4^{\circ}$.}
\end{minipage}
\medskip
\begin{minipage}{220pt}
\centering
\includegraphics[width=220pt, trim={0mm 0mm 0mm 0mm}, clip]{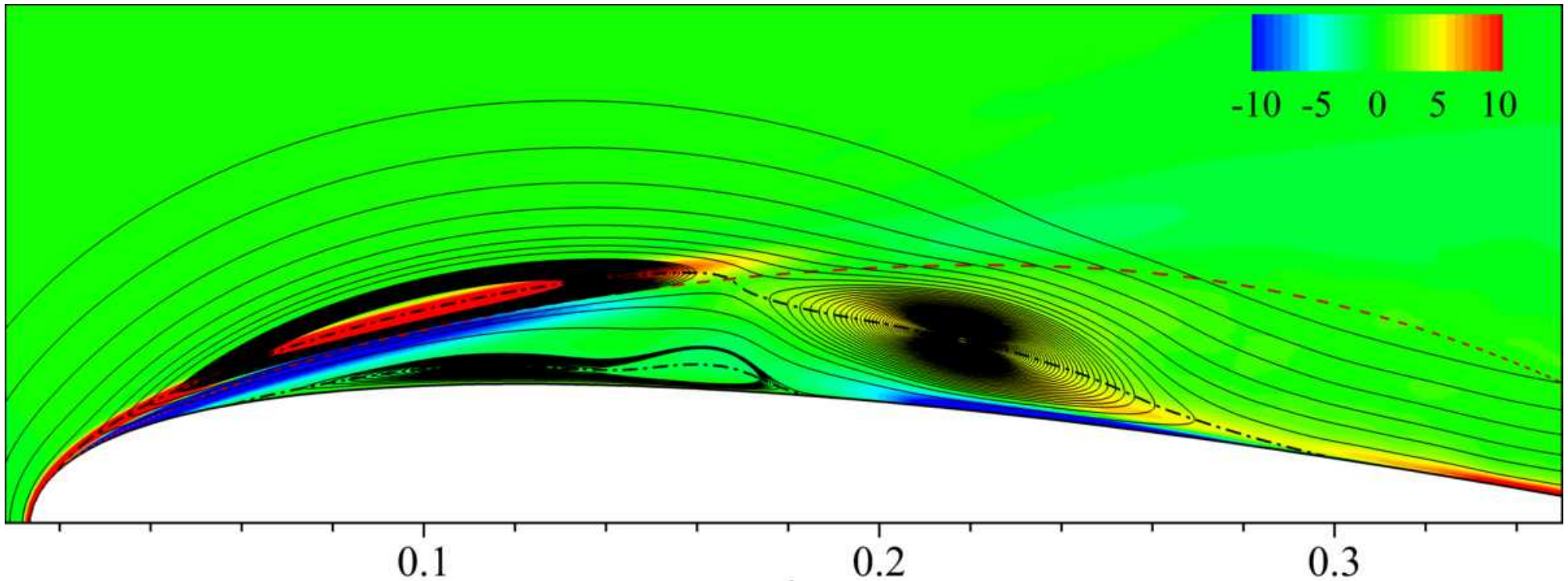}
\textit{$\alpha = 9.4^{\circ}$.}
\end{minipage}
\medskip
\begin{minipage}{220pt}
\centering
\includegraphics[width=220pt, trim={0mm 0mm 0mm 0mm}, clip]{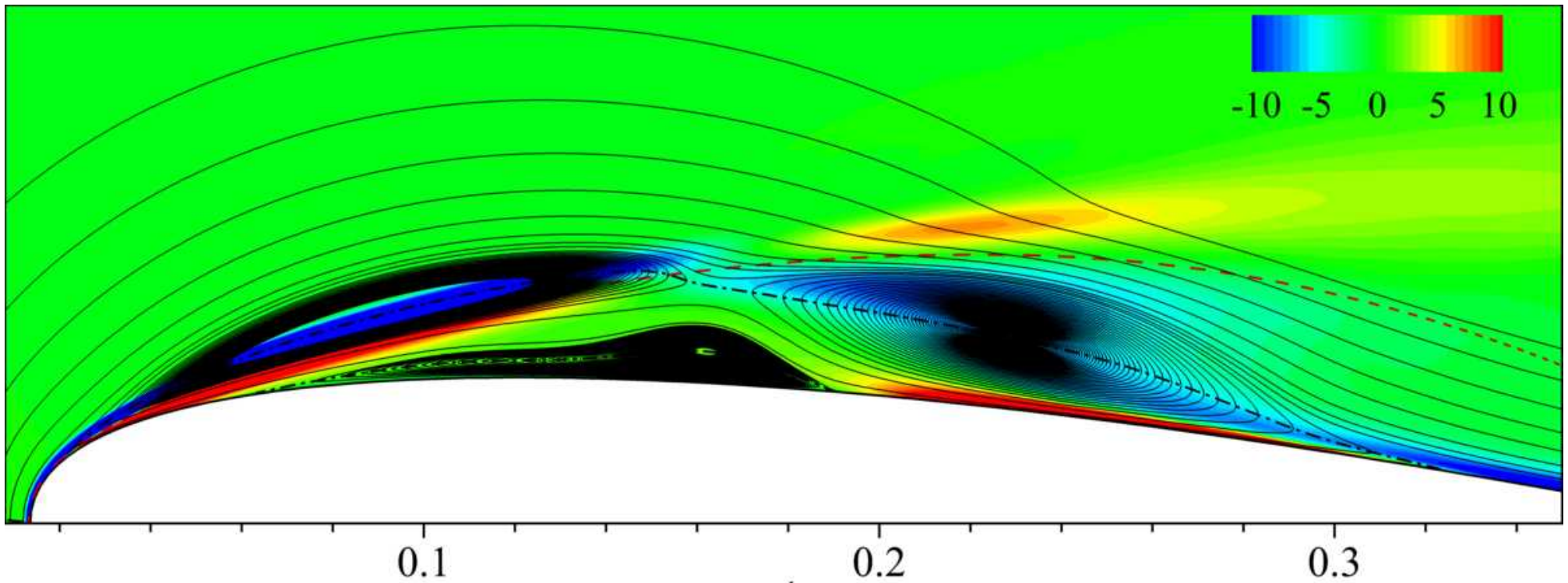}
\textit{$\alpha = 9.5^{\circ}$.}
\end{minipage}
\medskip
\begin{minipage}{220pt}
\centering
\includegraphics[width=220pt, trim={0mm 0mm 0mm 0mm}, clip]{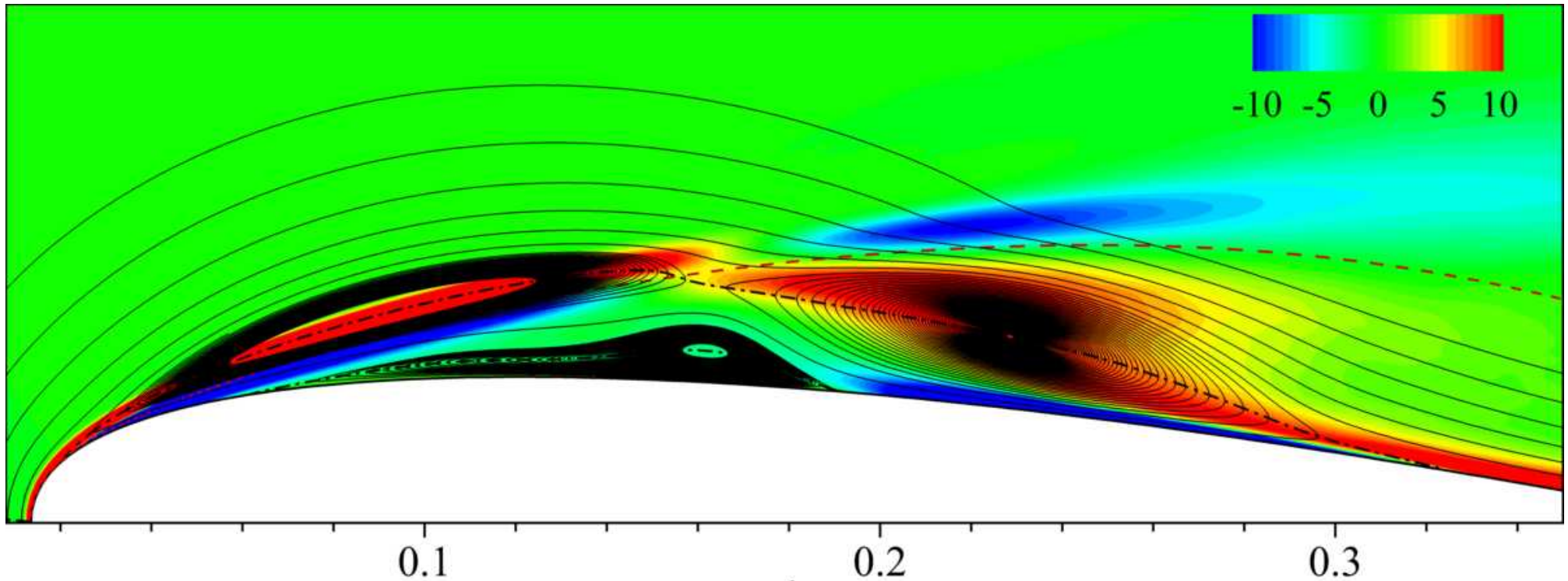}
\textit{$\alpha = 9.5^{\circ}$.}
\end{minipage}
\medskip
\begin{minipage}{220pt}
\centering
\includegraphics[width=220pt, trim={0mm 0mm 0mm 0mm}, clip]{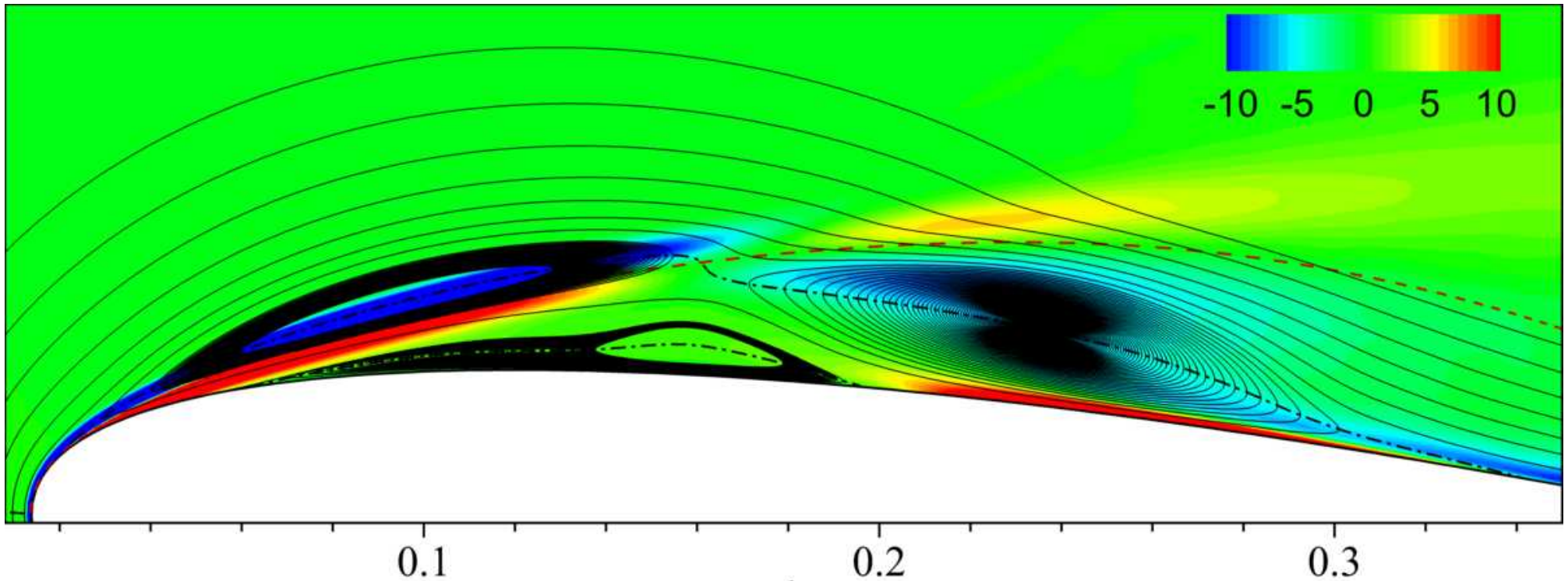}
\textit{$\alpha = 9.6^{\circ}$.}
\end{minipage}
\medskip
\begin{minipage}{220pt}
\centering
\includegraphics[width=220pt, trim={0mm 0mm 0mm 0mm}, clip]{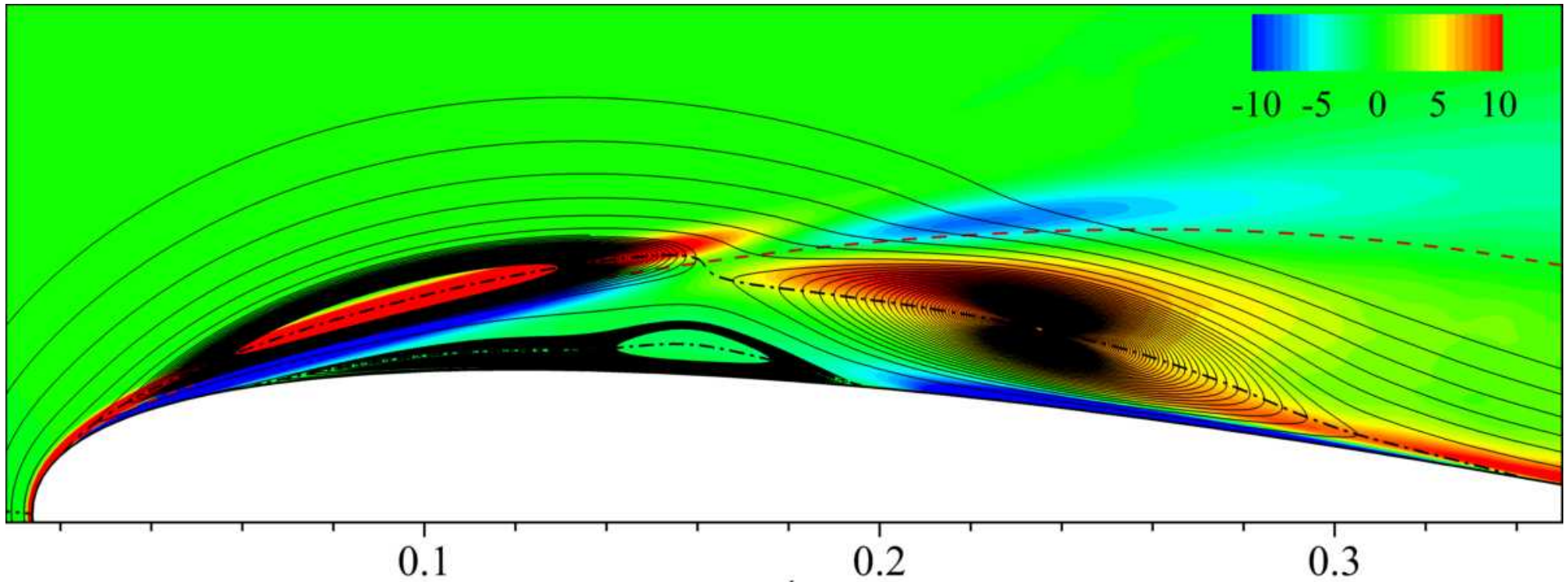}
\textit{$\alpha = 9.6^{\circ}$.}
\end{minipage}
\medskip
\begin{minipage}{220pt}
\centering
\includegraphics[width=220pt, trim={0mm 0mm 0mm 0mm}, clip]{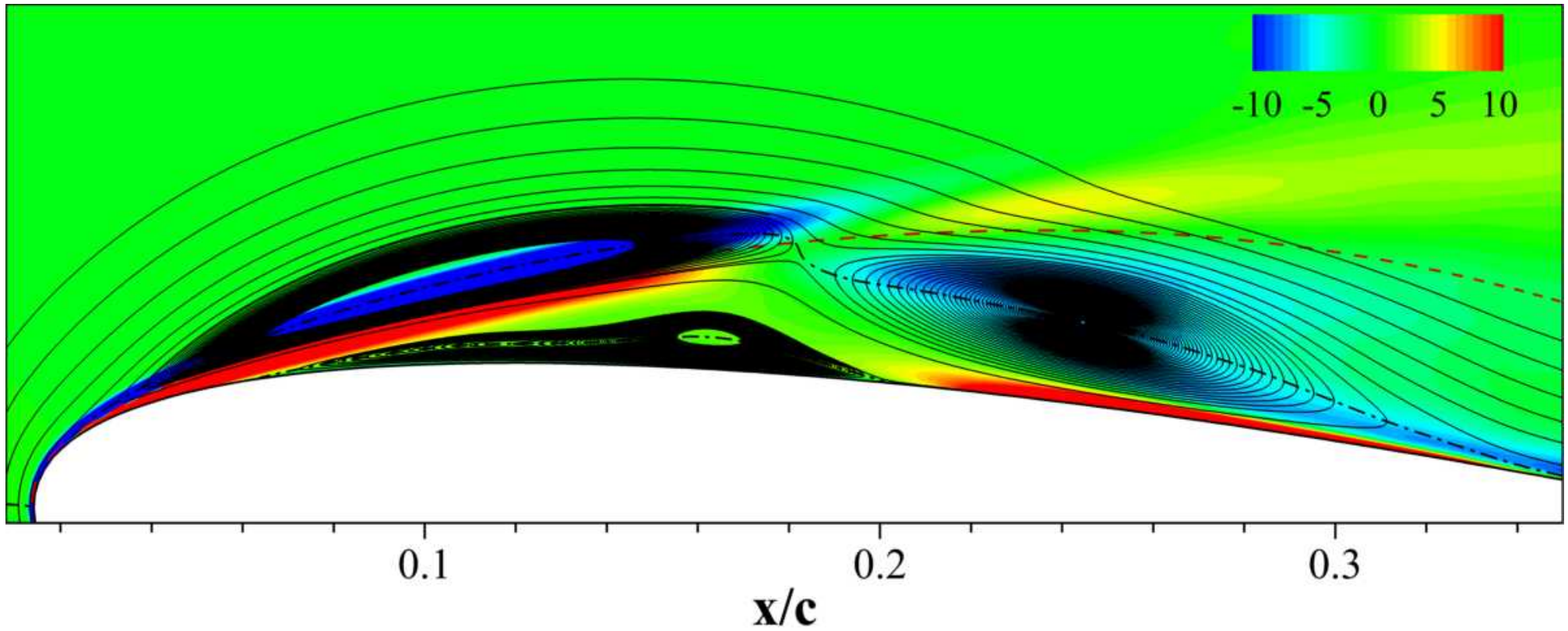}
\textit{$\alpha = 9.7^{\circ}$.}
\end{minipage}
\begin{minipage}{220pt}
\centering
\includegraphics[width=220pt, trim={0mm 0mm 0mm 0mm}, clip]{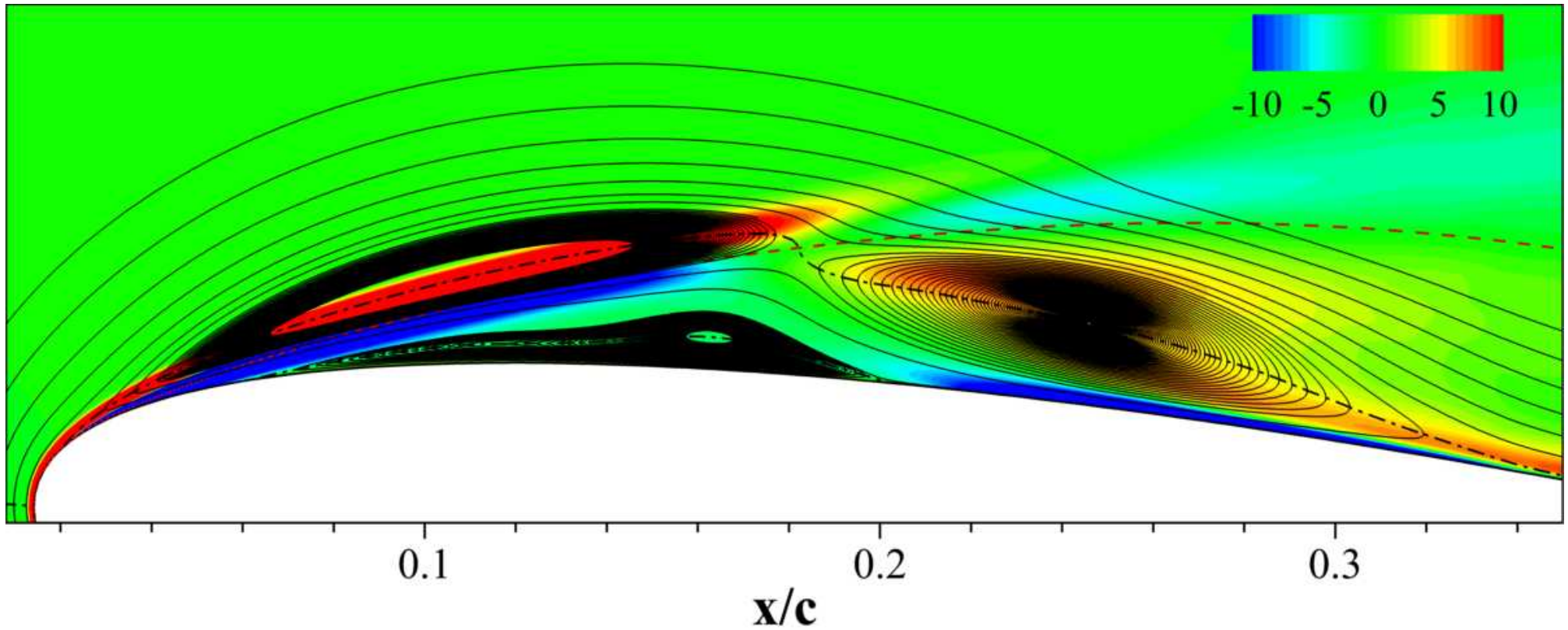}
\textit{$\alpha = 9.7^{\circ}$.}
\end{minipage}
\caption{Streamlines patterns of the time-averaged attached and separated phases of the oscillating-f\/low-f\/ield superimposed on colour maps of the time-averaged attached and separated phases of the oscillating-spanwise-vorticity for the angles of attack $\alpha = 9.25^{\circ}$--$9.7^{\circ}$. The red dashed line displays the streamline that bounds the LSB in the mean sense. The black dash-dot line denotes the contour line at which the time-average oscillating-streamwise-velocity equals zero. Left: time-average attached-phase of the oscillating-f\/low-f\/ield estimated from the high-lift f\/low minus the mean-lift f\/low-f\/ield, $\Delta{\overline{U}}^\mathrm{+}$ and $\Delta{\overline{V}}^\mathrm{+}$. Right: time-average separated-phase of the oscillating-f\/low estimated from the low-lift f\/low minus the mean-lift f\/low-f\/ield, $\Delta{\overline{U}}^\mathrm{-}$ and $\Delta{\overline{V}}^\mathrm{-}$.}
\label{time-average_oscillating-flow3}
\end{center}
\end{figure}

\begin{figure}
\begin{center}
\begin{minipage}{220pt}
\centering
\includegraphics[width=220pt, trim={0mm 0mm 0mm 0mm}, clip]{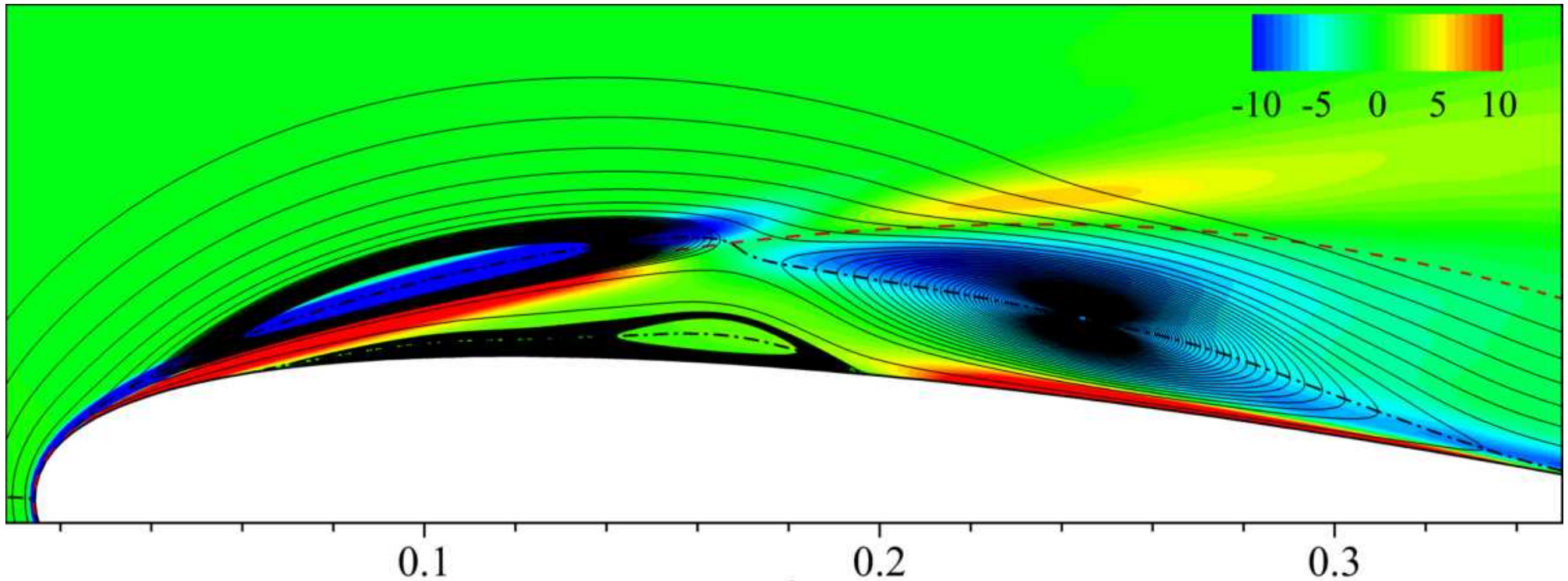}
\textit{$\alpha = 9.8^{\circ}$.}
\end{minipage}
\medskip
\begin{minipage}{220pt}
\centering
\includegraphics[width=220pt, trim={0mm 0mm 0mm 0mm}, clip]{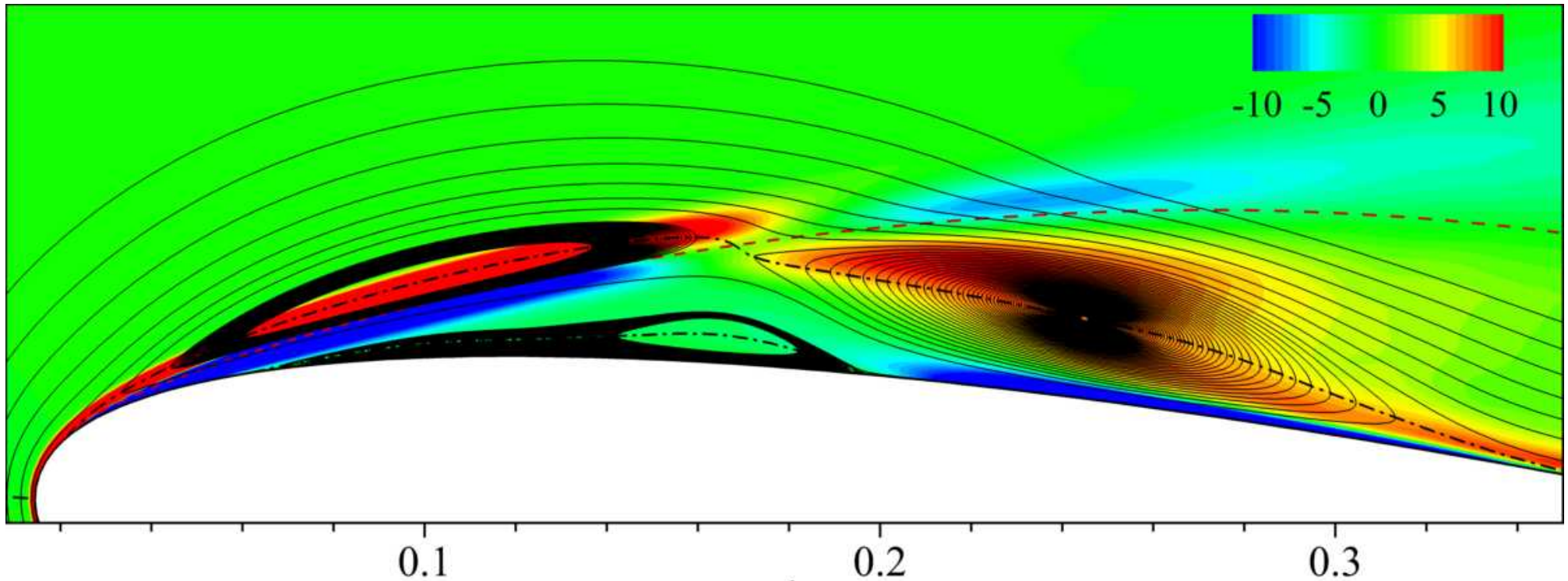}
\textit{$\alpha = 9.8^{\circ}$.}
\end{minipage}
\medskip
\begin{minipage}{220pt}
\centering
\includegraphics[width=220pt, trim={0mm 0mm 0mm 0mm}, clip]{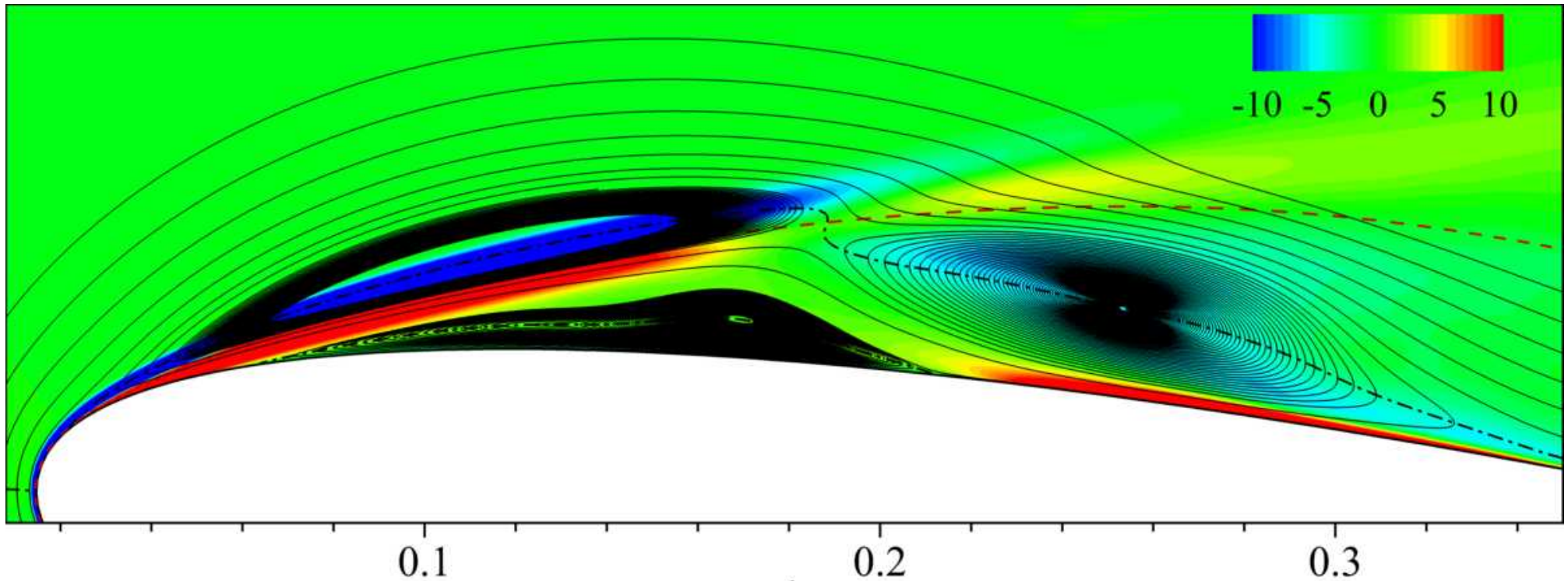}
\textit{$\alpha = 9.9^{\circ}$.}
\end{minipage}
\medskip
\begin{minipage}{220pt}
\centering
\includegraphics[width=220pt, trim={0mm 0mm 0mm 0mm}, clip]{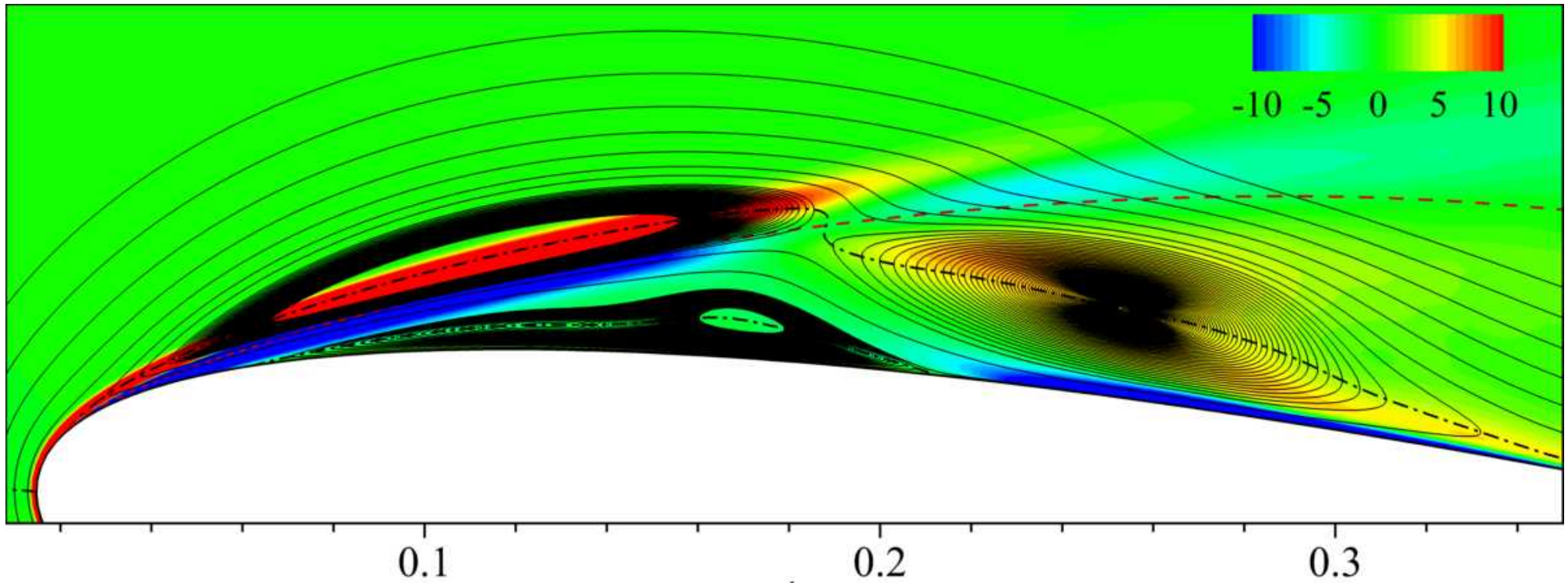}
\textit{$\alpha = 9.9^{\circ}$.}
\end{minipage}
\medskip
\begin{minipage}{220pt}
\centering
\includegraphics[width=220pt, trim={0mm 0mm 0mm 0mm}, clip]{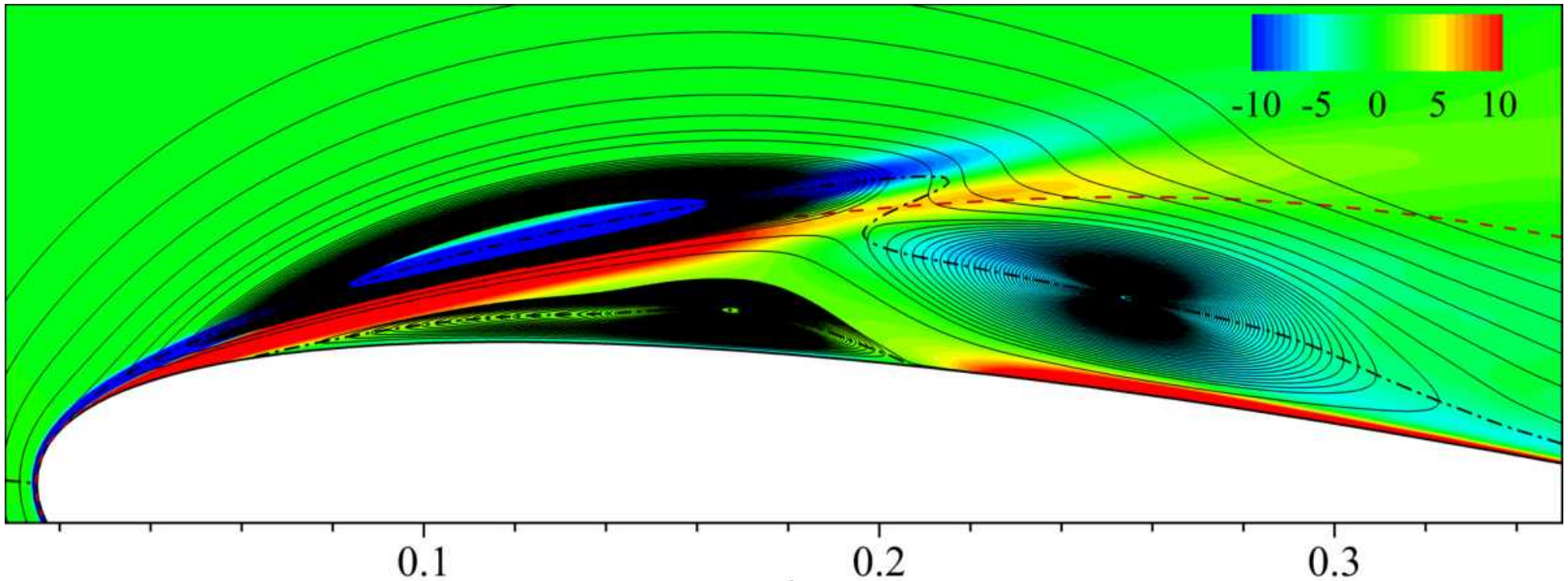}
\textit{$\alpha = 10.0^{\circ}$.}
\end{minipage}
\medskip
\begin{minipage}{220pt}
\centering
\includegraphics[width=220pt, trim={0mm 0mm 0mm 0mm}, clip]{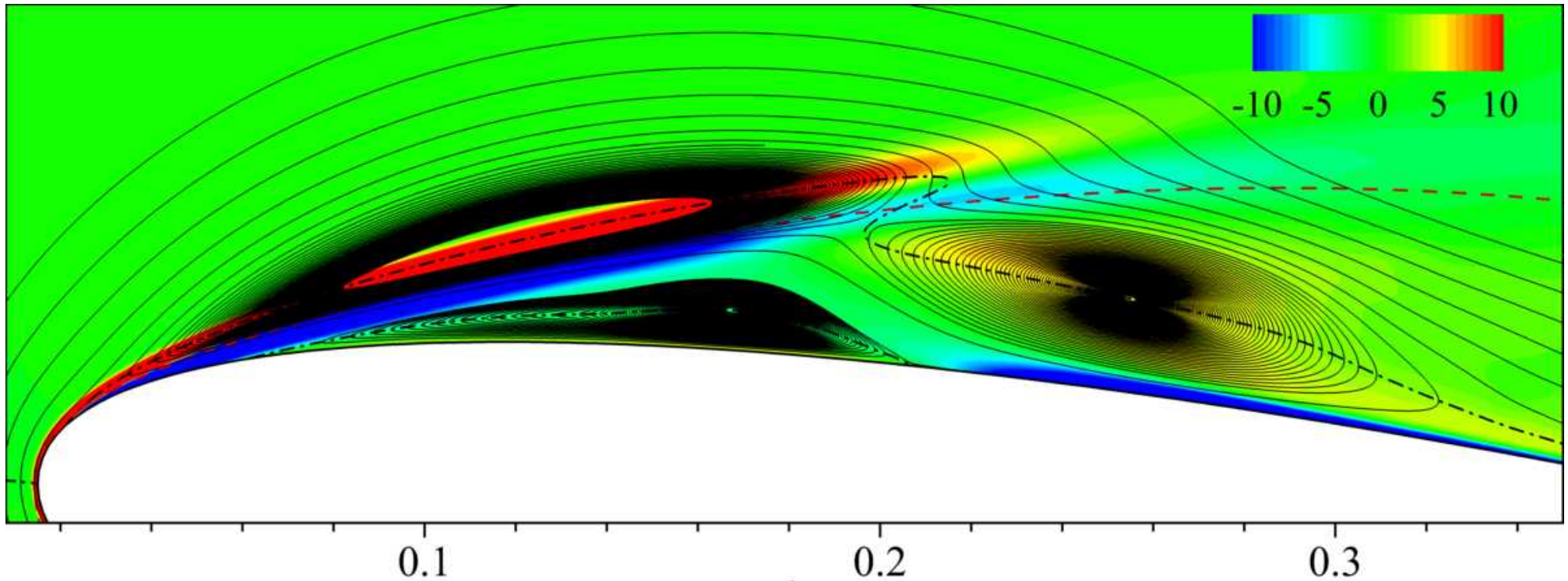}
\textit{$\alpha = 10.0^{\circ}$.}
\end{minipage}
\medskip
\begin{minipage}{220pt}
\centering
\includegraphics[width=220pt, trim={0mm 0mm 0mm 0mm}, clip]{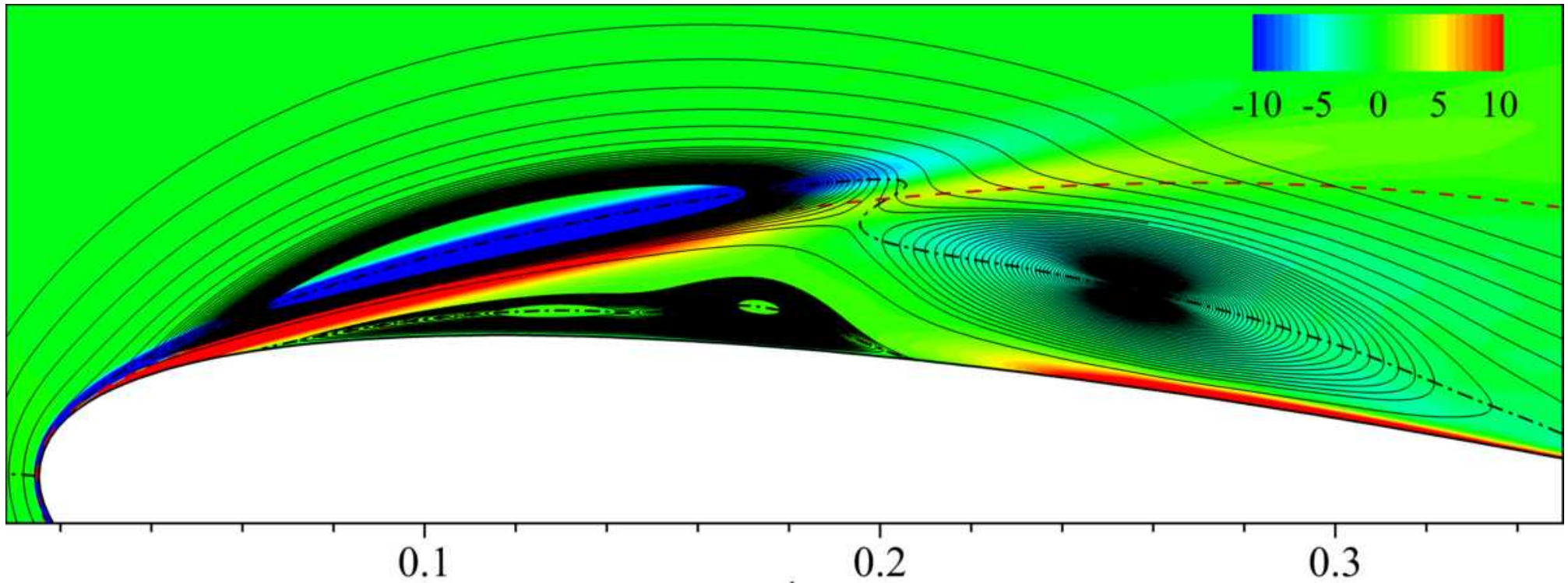}
\textit{$\alpha = 10.1^{\circ}$.}
\end{minipage}
\medskip
\begin{minipage}{220pt}
\centering
\includegraphics[width=220pt, trim={0mm 0mm 0mm 0mm}, clip]{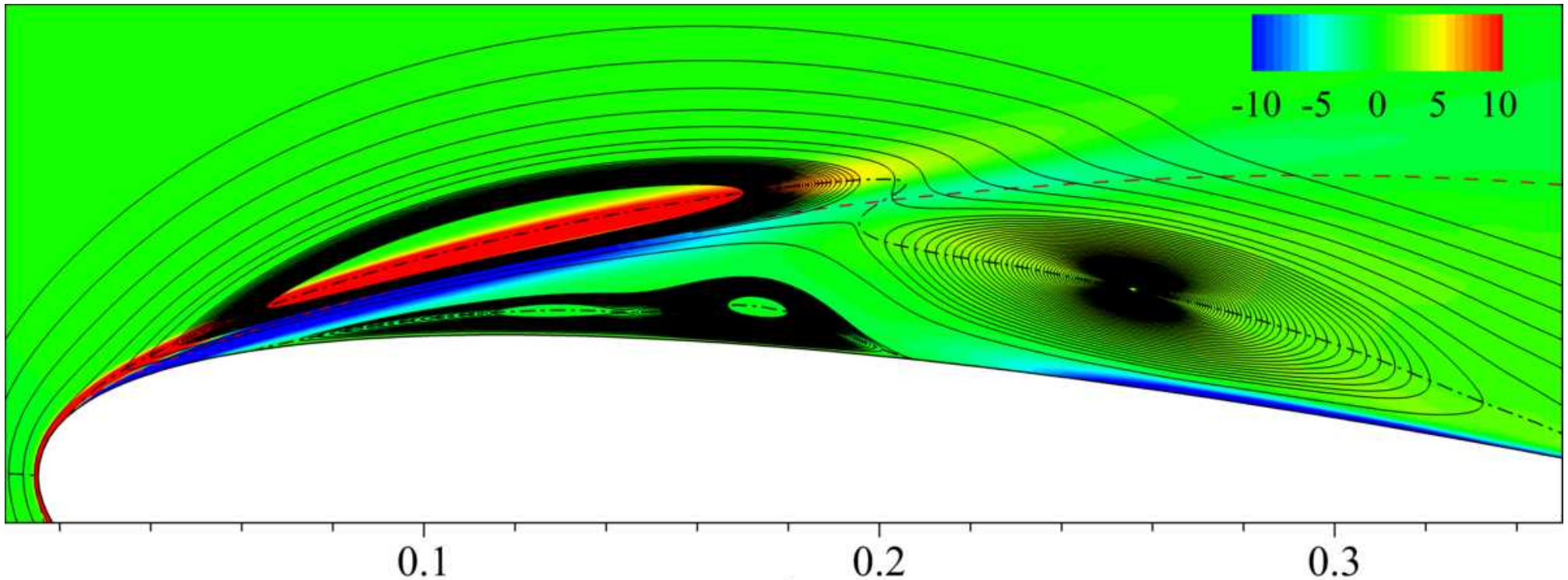}
\textit{$\alpha = 10.1^{\circ}$.}
\end{minipage}
\medskip
\begin{minipage}{220pt}
\centering
\includegraphics[width=220pt, trim={0mm 0mm 0mm 0mm}, clip]{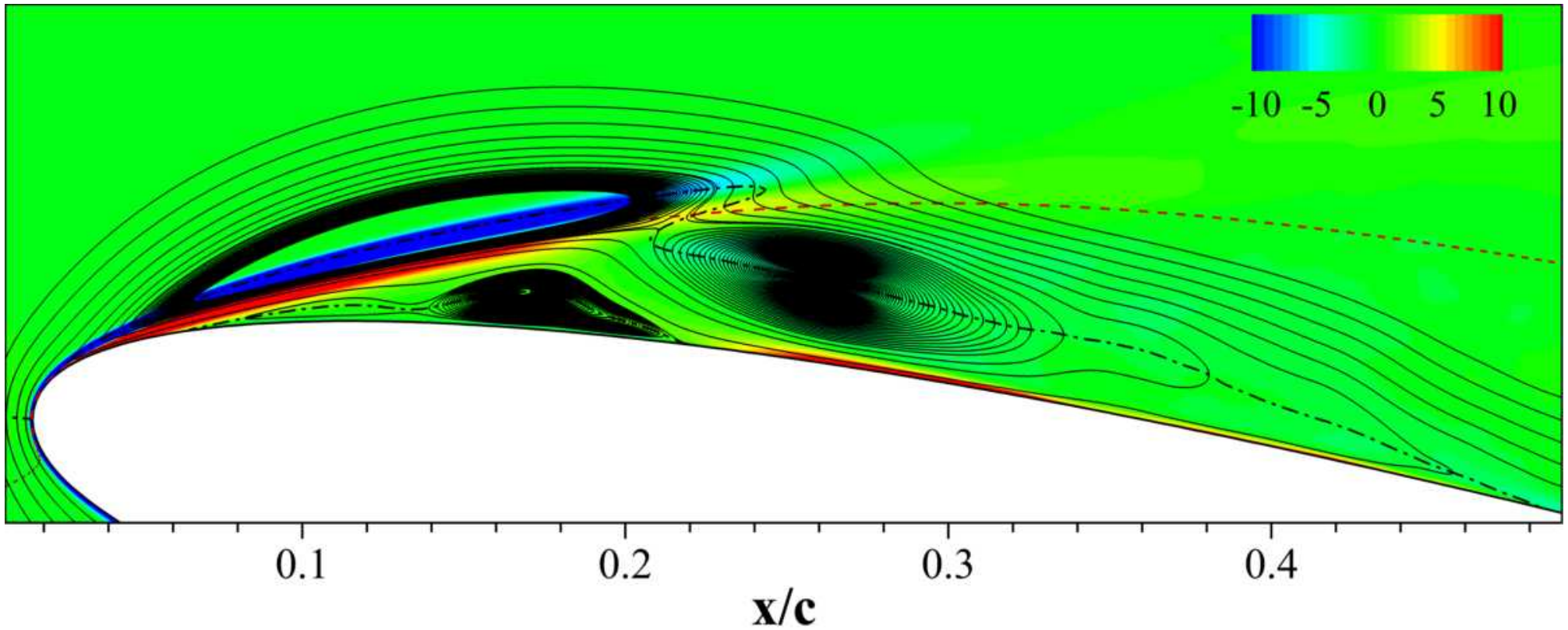}
\textit{$\alpha = 10.5^{\circ}$.}
\end{minipage}
\begin{minipage}{220pt}
\centering
\includegraphics[width=220pt, trim={0mm 0mm 0mm 0mm}, clip]{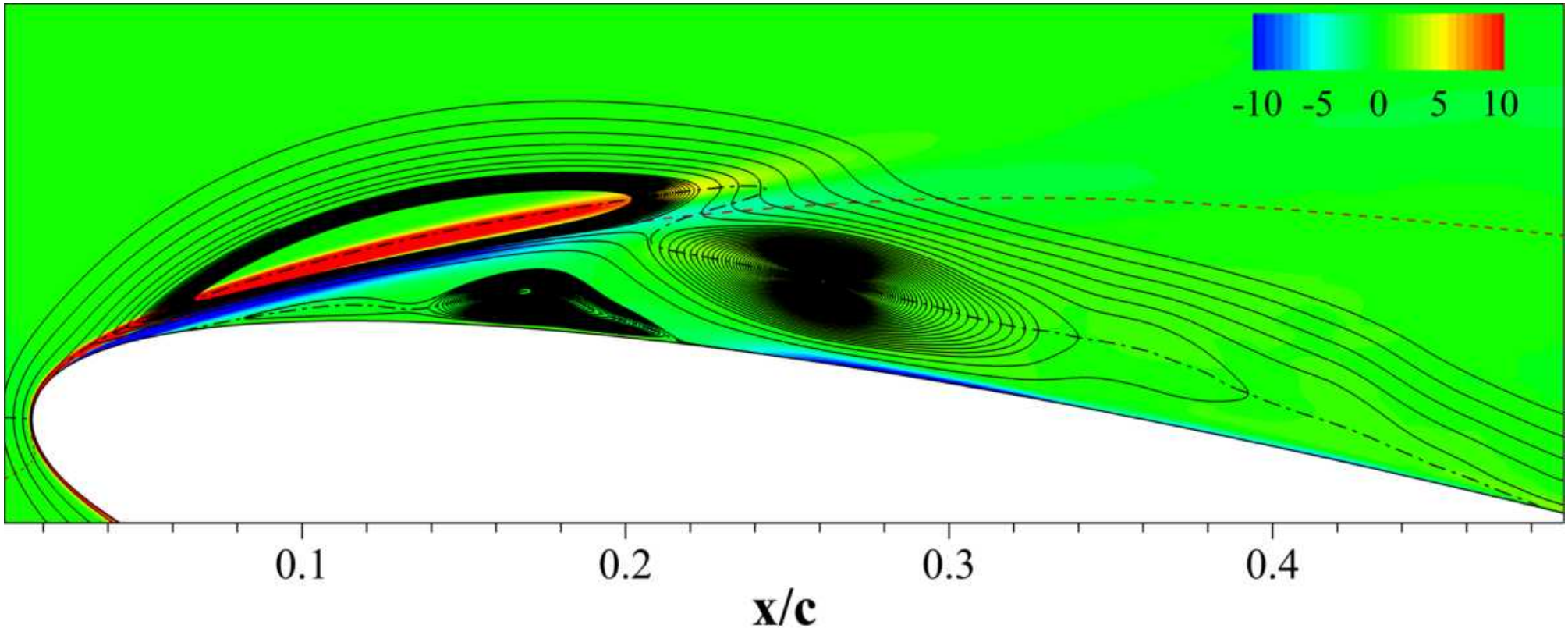}
\textit{$\alpha = 10.5^{\circ}$.}
\end{minipage}
\caption{Streamlines patterns of the time-averaged attached and separated phases of the oscillating-f\/low-f\/ield superimposed on colour maps of the time-averaged attached and separated phases of the oscillating-spanwise-vorticity for the angles of attack $\alpha = 9.8^{\circ}$--$10.5^{\circ}$. The red dashed line displays the streamline that bounds the LSB in the mean sense. The black dash-dot line denotes the contour line at which the time-average oscillating-streamwise-velocity equals zero. Left: time-average attached-phase of the oscillating-f\/low-f\/ield estimated from the high-lift f\/low minus the mean-lift f\/low-f\/ield, $\Delta{\overline{U}}^\mathrm{+}$ and $\Delta{\overline{V}}^\mathrm{+}$. Right: time-average separated-phase of the oscillating-f\/low estimated from the low-lift f\/low minus the mean-lift f\/low-f\/ield, $\Delta{\overline{U}}^\mathrm{-}$ and $\Delta{\overline{V}}^\mathrm{-}$.}
\label{time-average_oscillating-flow4}
\end{center}
\end{figure}

\begin{figure}
\begin{center}
\begin{minipage}{220pt}
\centering
\includegraphics[width=220pt, trim={0mm 0mm 0mm 0mm}, clip]{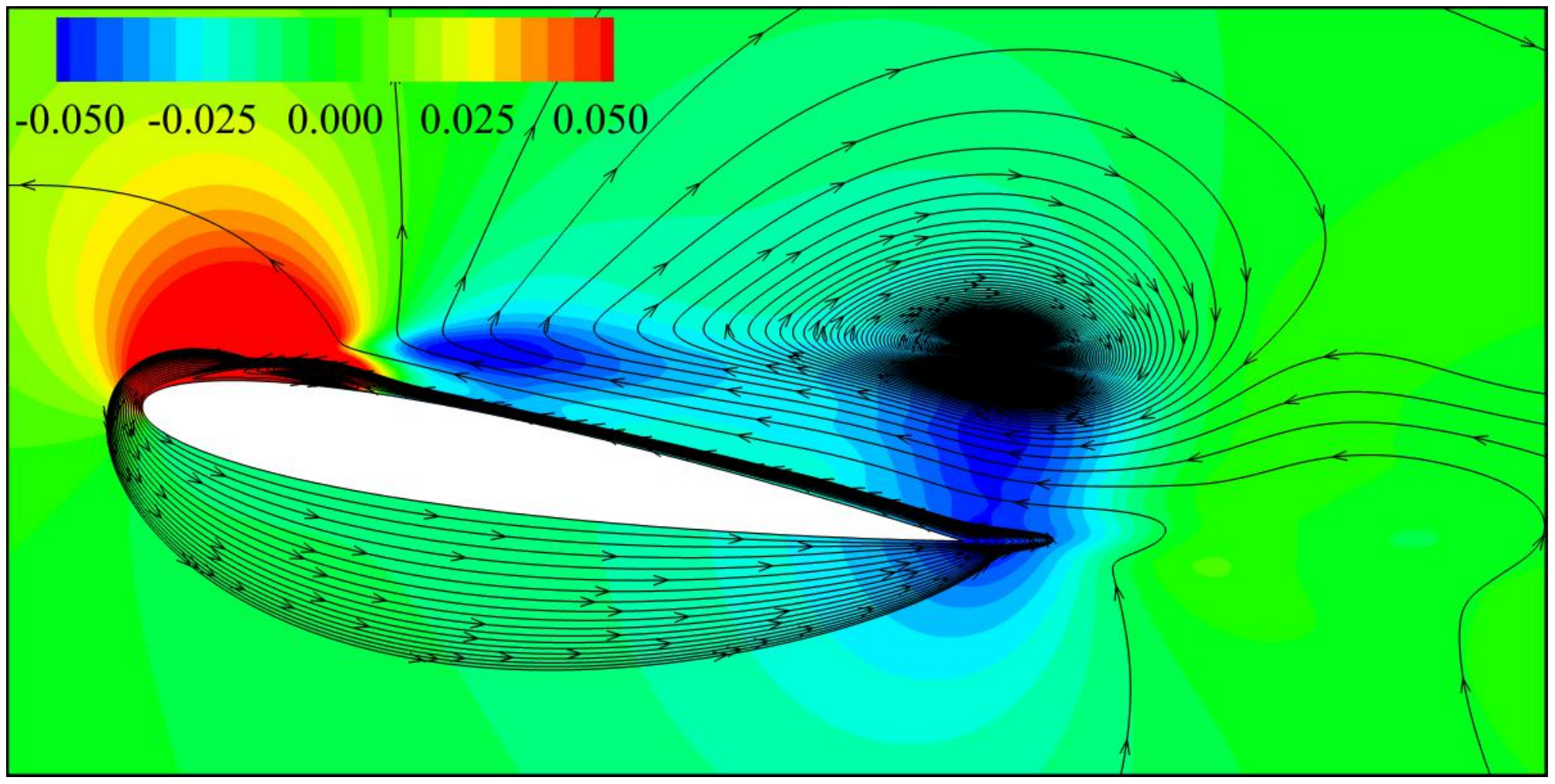}
\textit{$\Phi = 0^{\circ}$}
\end{minipage}
\medskip
\begin{minipage}{220pt}
\centering
\includegraphics[width=220pt, trim={0mm 0mm 0mm 0mm}, clip]{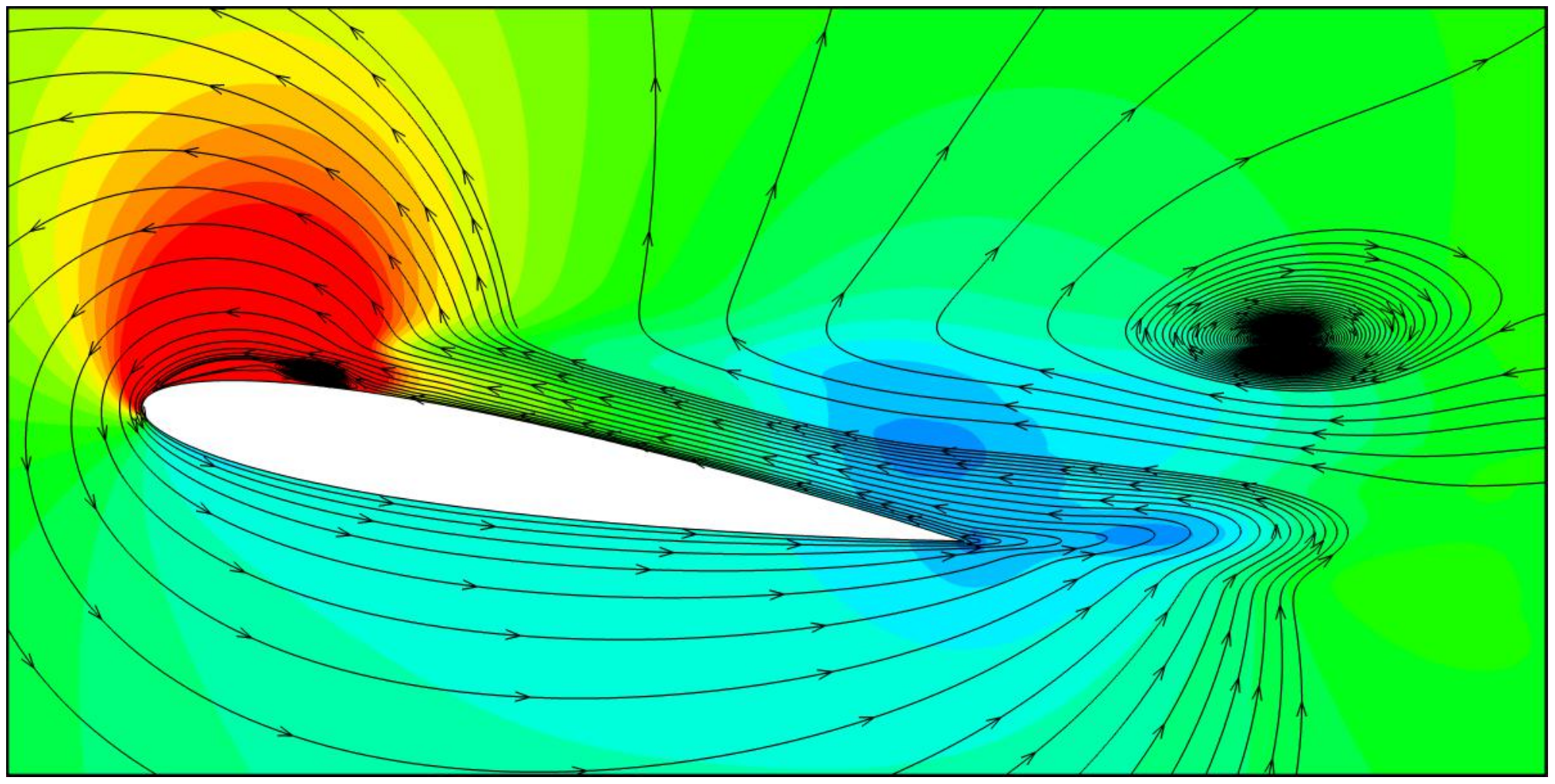}
\textit{$\Phi = 40^{\circ}$}
\end{minipage}
\medskip
\begin{minipage}{220pt}
\centering
\includegraphics[width=220pt, trim={0mm 0mm 0mm 0mm}, clip]{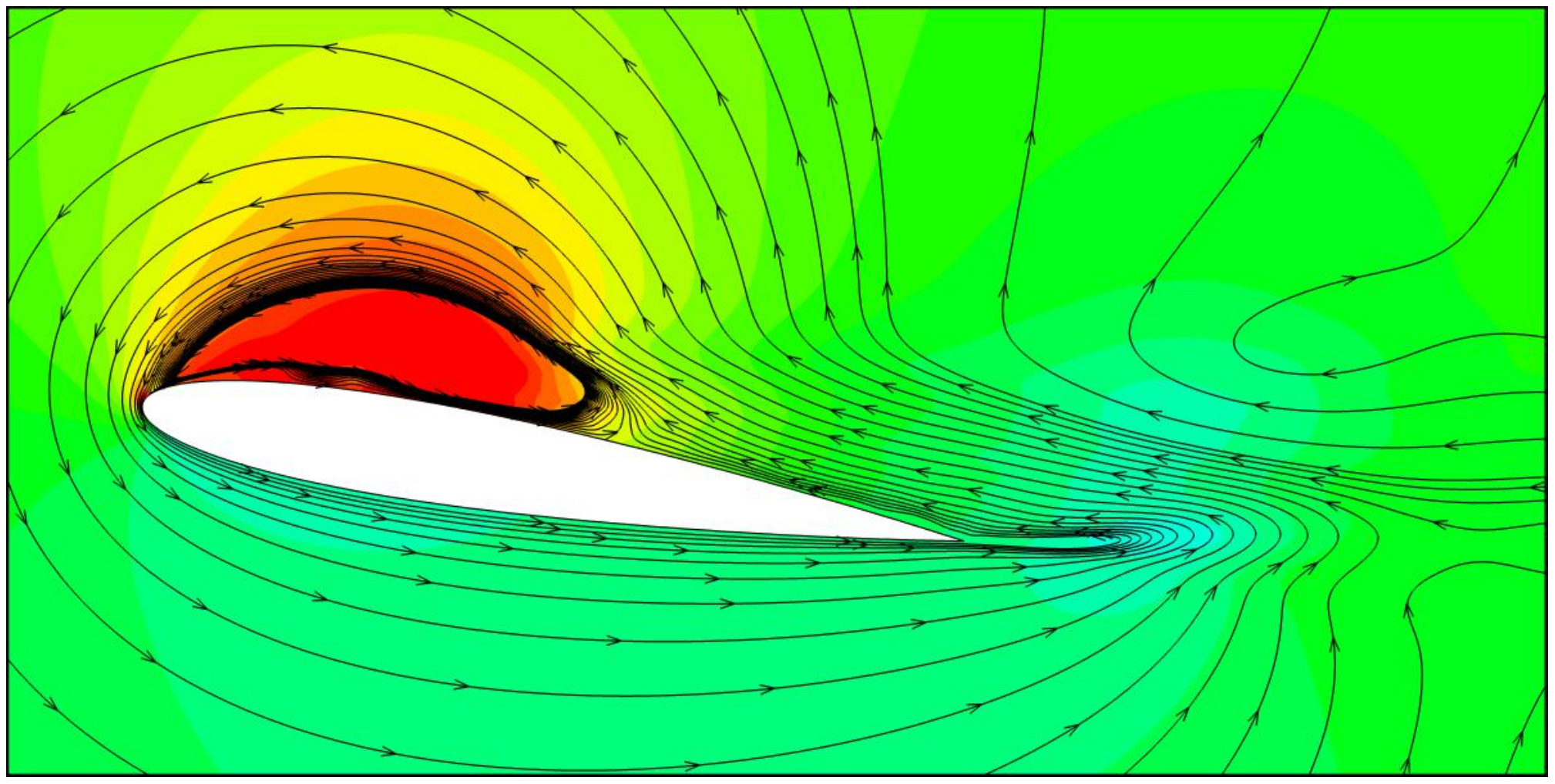}
\textit{$\Phi = 90^{\circ}$}
\end{minipage}
\medskip
\begin{minipage}{220pt}
\centering
\includegraphics[width=220pt, trim={0mm 0mm 0mm 0mm}, clip]{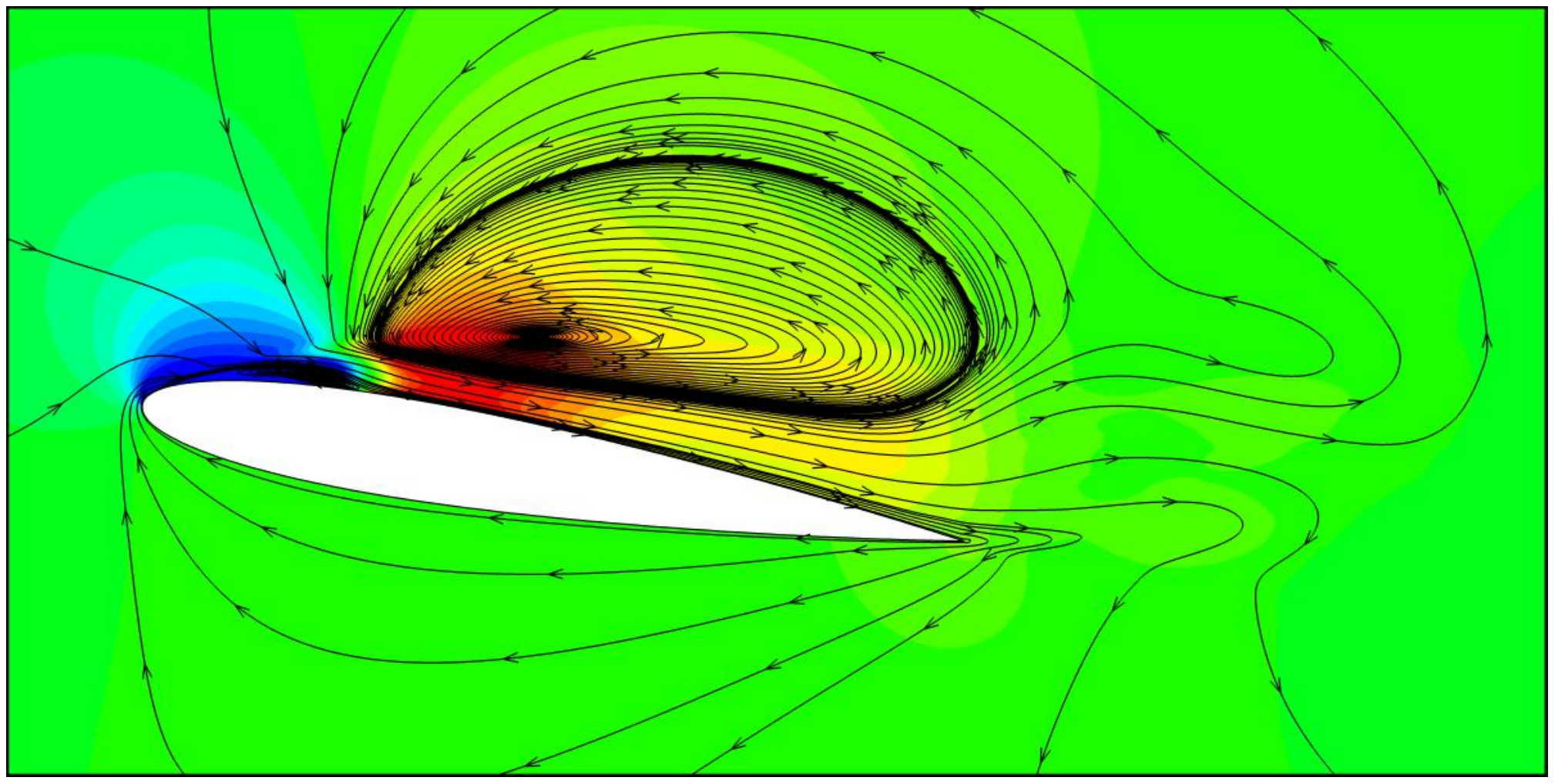}
\textit{$\Phi = 130^{\circ}$}
\end{minipage}
\medskip
\begin{minipage}{220pt}
\centering
\includegraphics[width=220pt, trim={0mm 0mm 0mm 0mm}, clip]{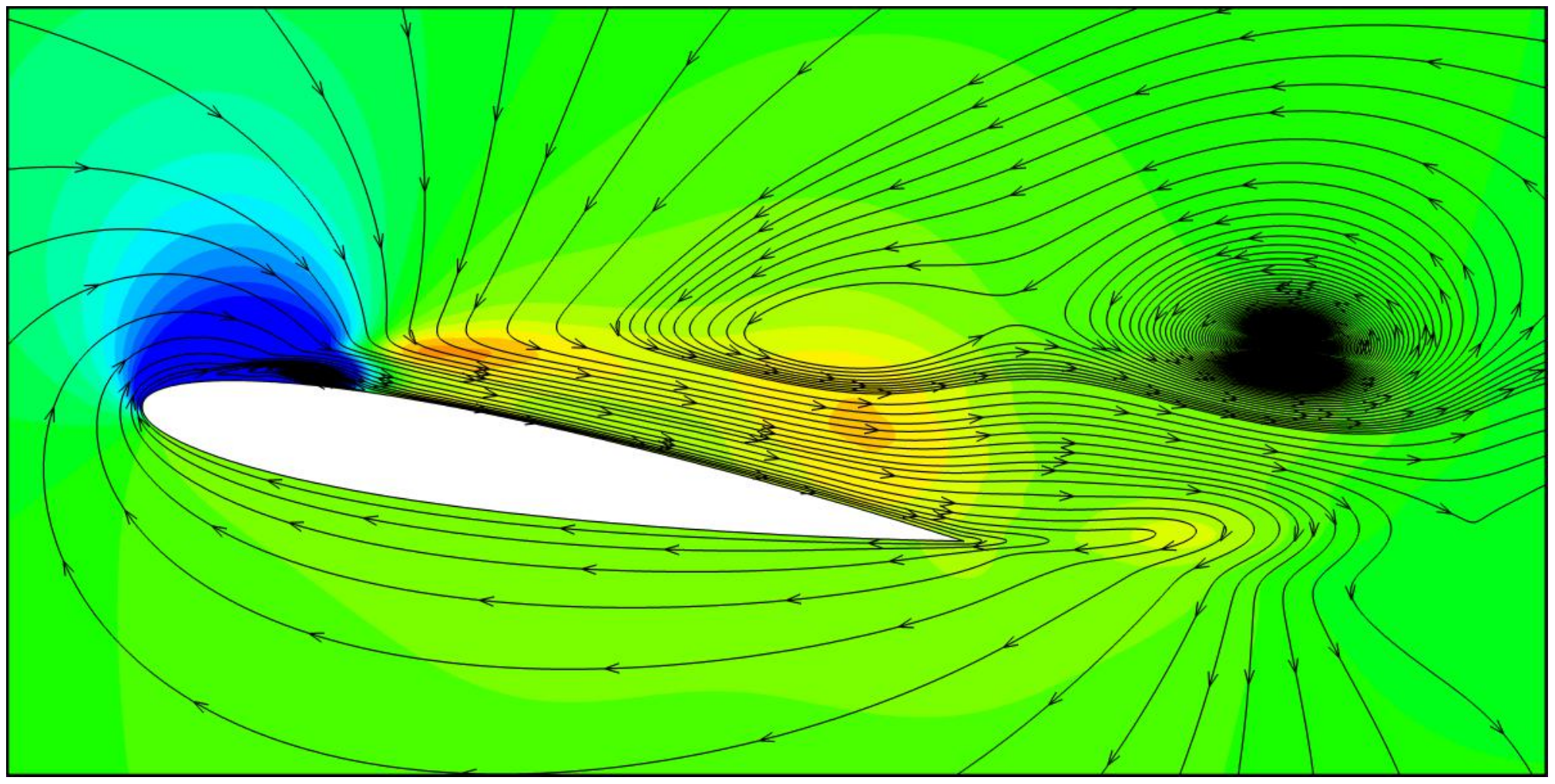}
\textit{$\Phi = 180^{\circ}$}
\end{minipage}
\medskip
\begin{minipage}{220pt}
\centering
\includegraphics[width=220pt, trim={0mm 0mm 0mm 0mm}, clip]{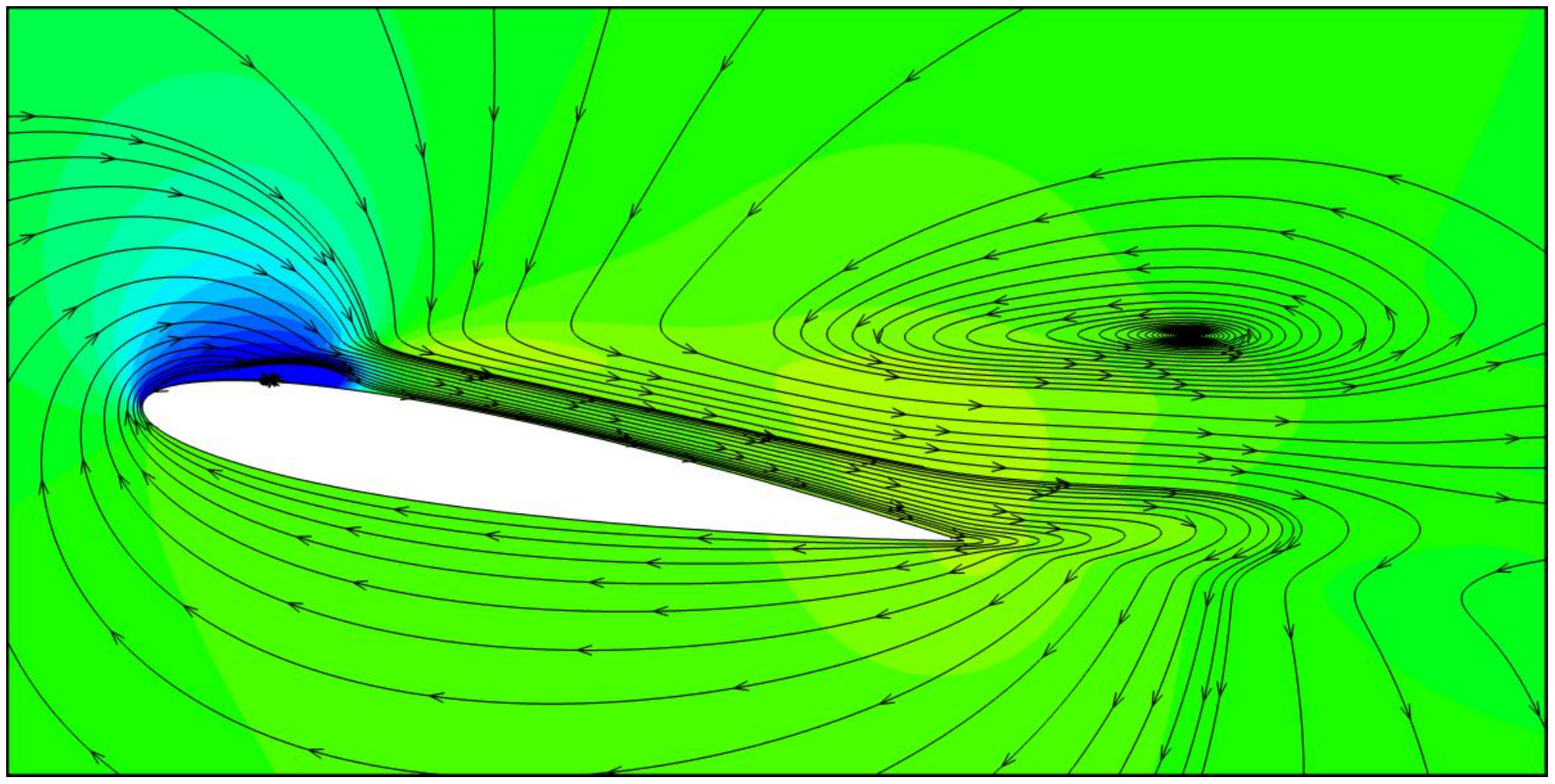}
\textit{$\Phi = 220^{\circ}$}
\end{minipage}
\medskip
\begin{minipage}{220pt}
\centering
\includegraphics[width=220pt, trim={0mm 0mm 0mm 0mm}, clip]{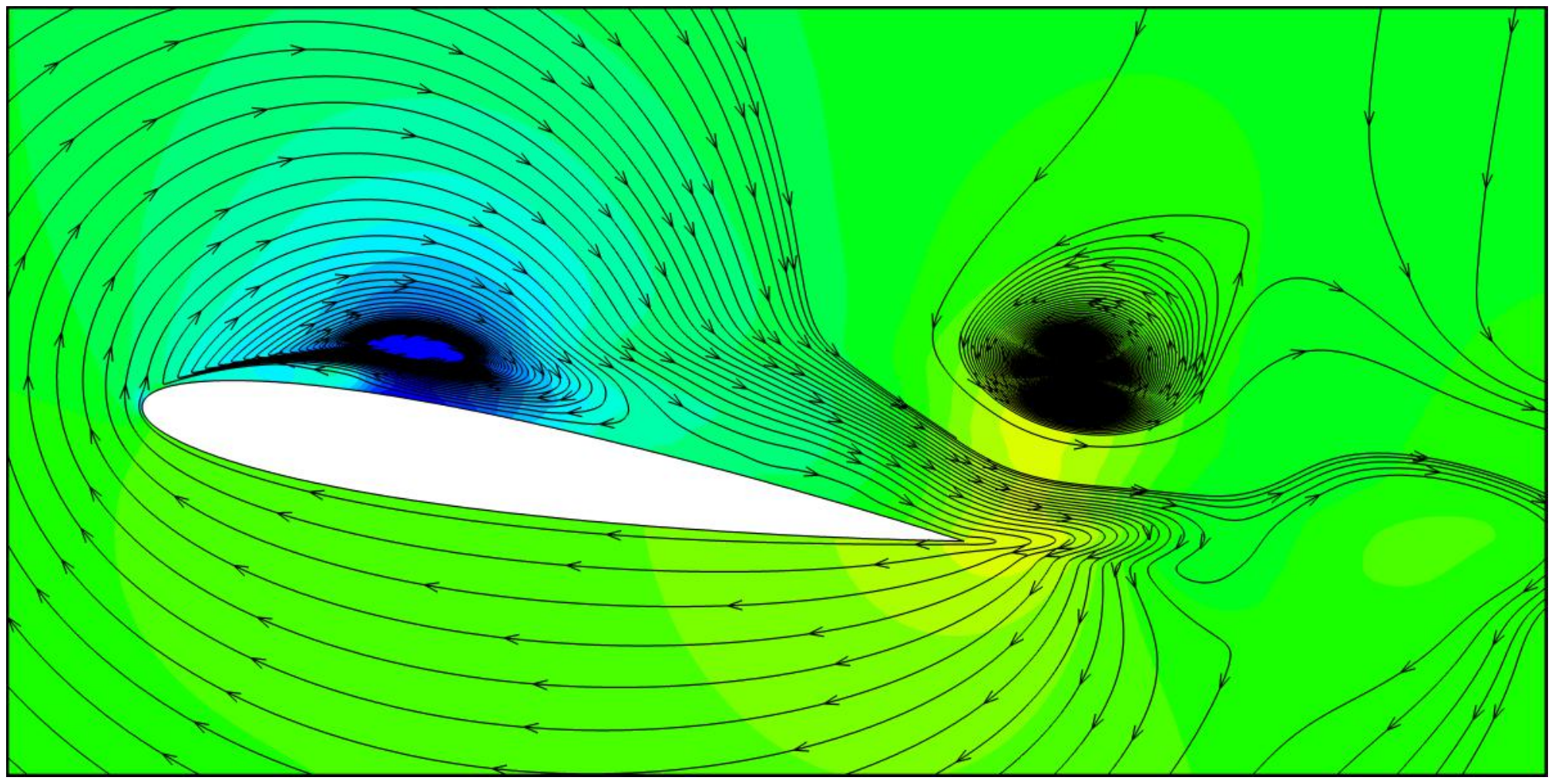}
\textit{$\Phi = 270^{\circ}$}
\end{minipage}
\begin{minipage}{220pt}
\centering
\includegraphics[width=220pt, trim={0mm 0mm 0mm 0mm}, clip]{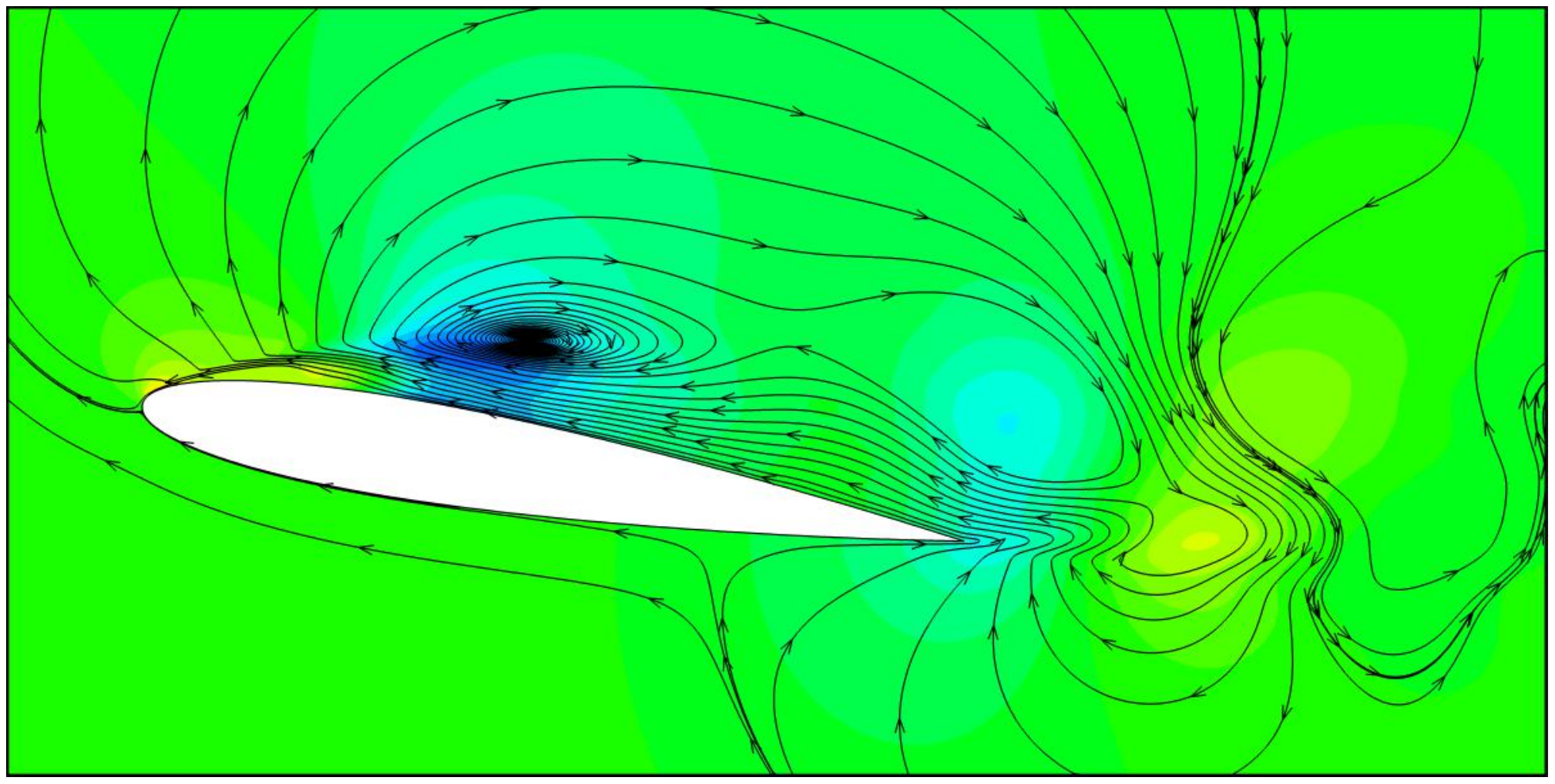}
\textit{$\Phi = 310^{\circ}$}
\end{minipage}
\caption{Streamlines patterns of the phase-averaged oscillating-f\/low-f\/ield superimposed on colour maps of the phase-averaged oscillating-pressure for the angle of attack of $9.4^{\circ}$.}
\label{phase-average_oscillating-flow_940}
\end{center}
\end{figure}

\begin{figure}
\begin{center}
\begin{minipage}{220pt}
\centering
\includegraphics[width=220pt, trim={0mm 0mm 0mm 0mm}, clip]{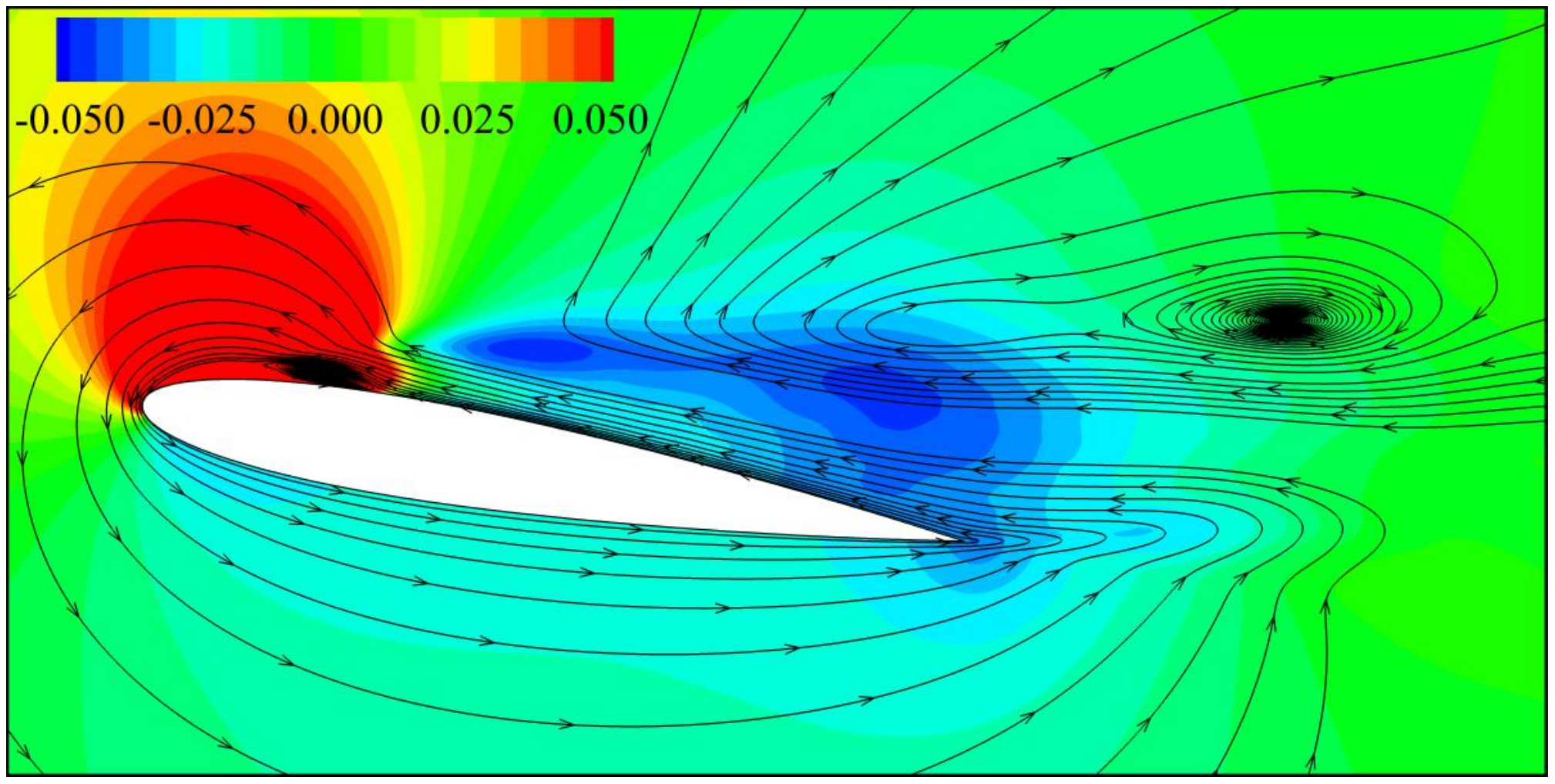}
\textit{$\Phi = 0^{\circ}$}
\end{minipage}
\medskip
\begin{minipage}{220pt}
\centering
\includegraphics[width=220pt, trim={0mm 0mm 0mm 0mm}, clip]{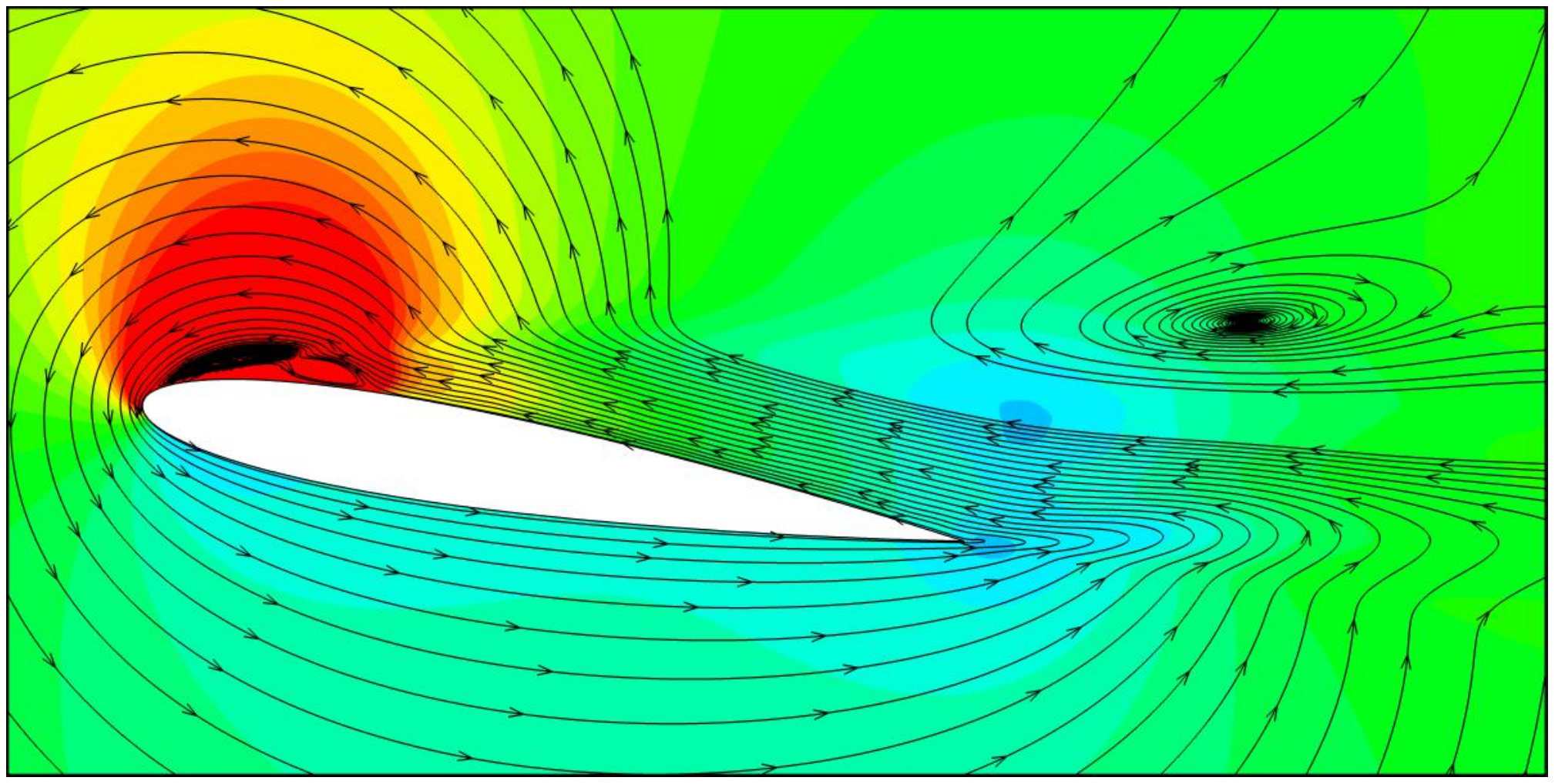}
\textit{$\Phi = 40^{\circ}$}
\end{minipage}
\medskip
\begin{minipage}{220pt}
\centering
\includegraphics[width=220pt, trim={0mm 0mm 0mm 0mm}, clip]{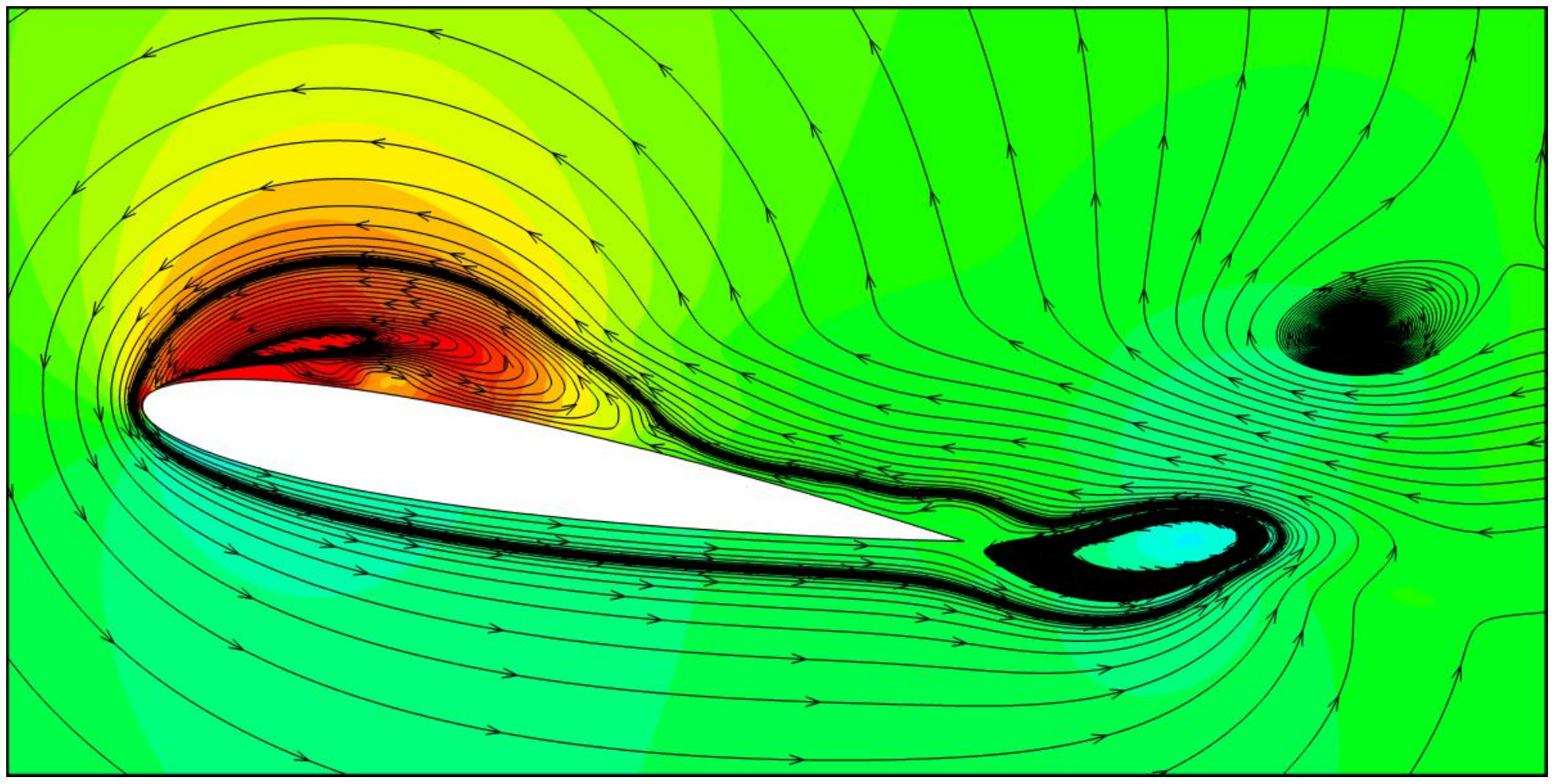}
\textit{$\Phi = 90^{\circ}$}
\end{minipage}
\medskip
\begin{minipage}{220pt}
\centering
\includegraphics[width=220pt, trim={0mm 0mm 0mm 0mm}, clip]{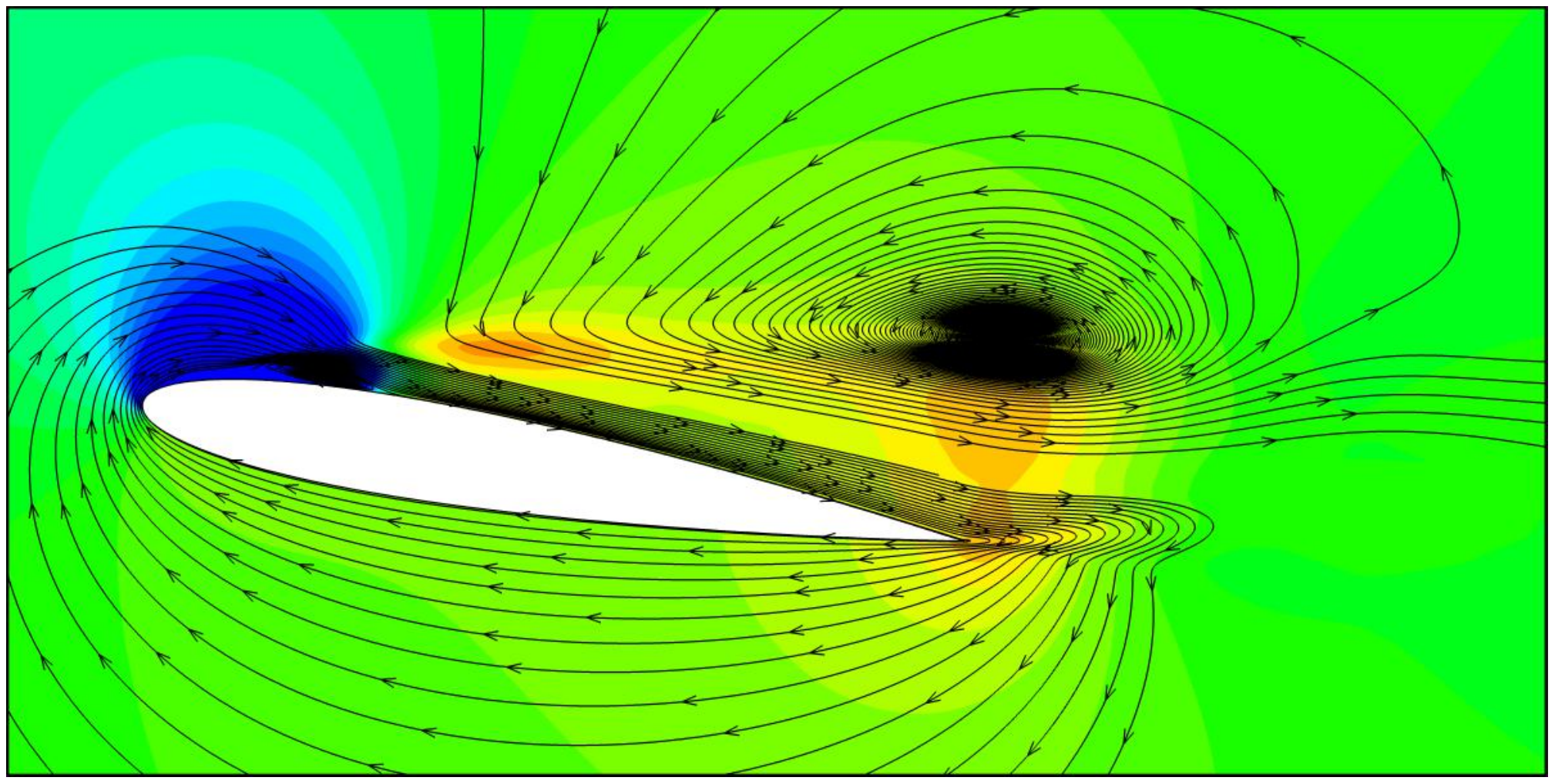}
\textit{$\Phi = 130^{\circ}$}
\end{minipage}
\medskip
\begin{minipage}{220pt}
\centering
\includegraphics[width=220pt, trim={0mm 0mm 0mm 0mm}, clip]{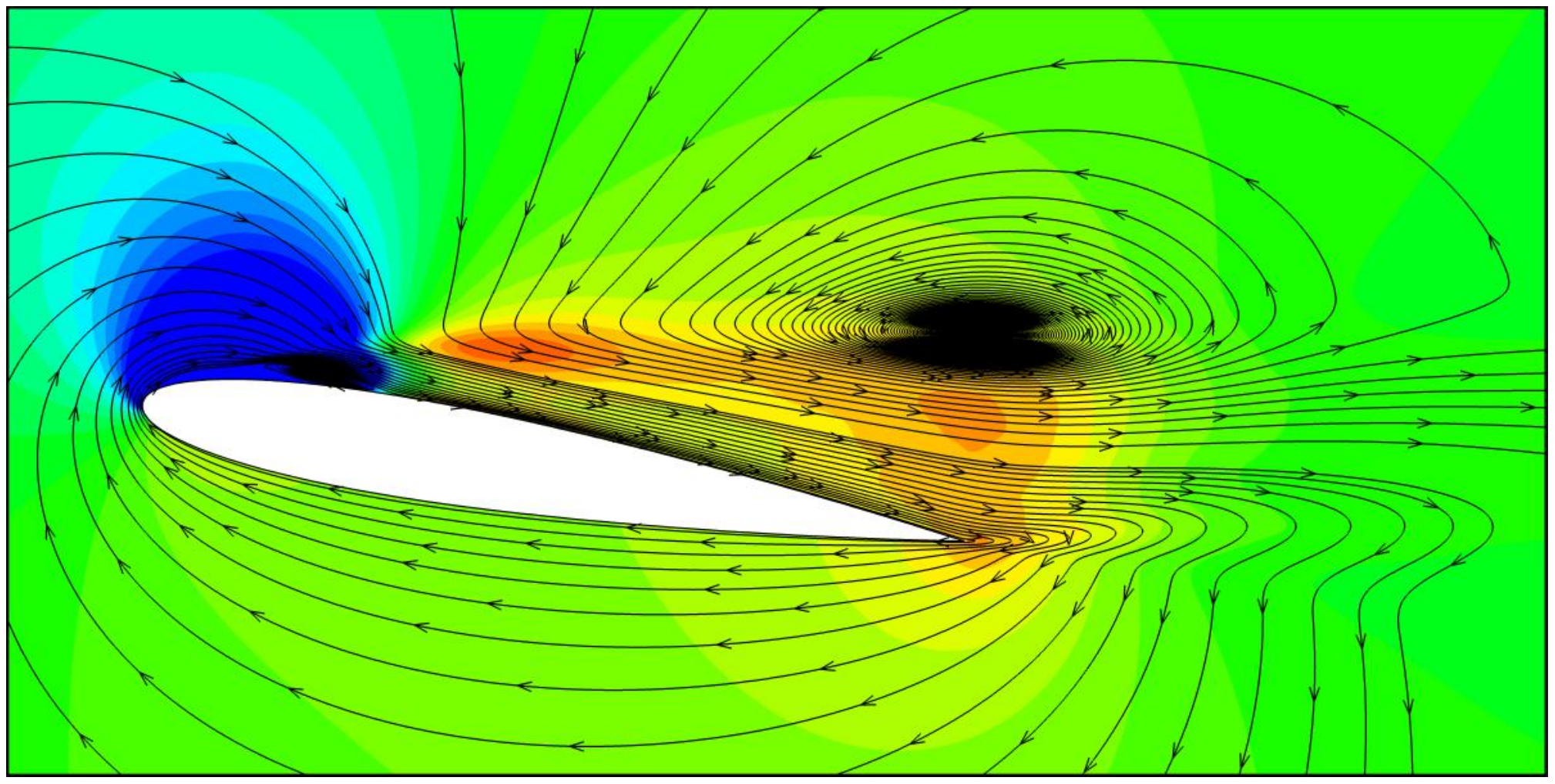}
\textit{$\Phi = 180^{\circ}$}
\end{minipage}
\medskip
\begin{minipage}{220pt}
\centering
\includegraphics[width=220pt, trim={0mm 0mm 0mm 0mm}, clip]{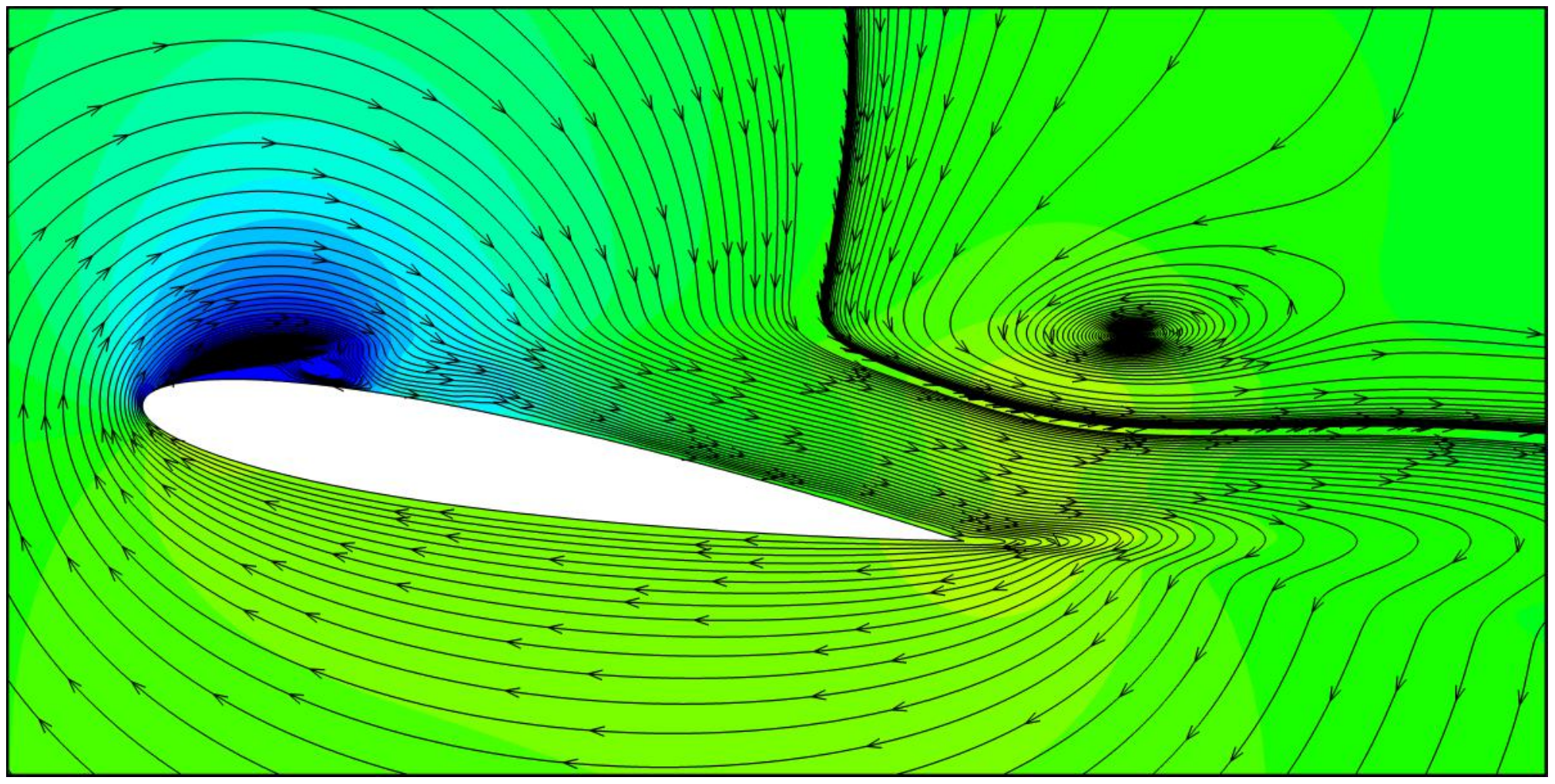}
\textit{$\Phi = 220^{\circ}$}
\end{minipage}
\medskip
\begin{minipage}{220pt}
\centering
\includegraphics[width=220pt, trim={0mm 0mm 0mm 0mm}, clip]{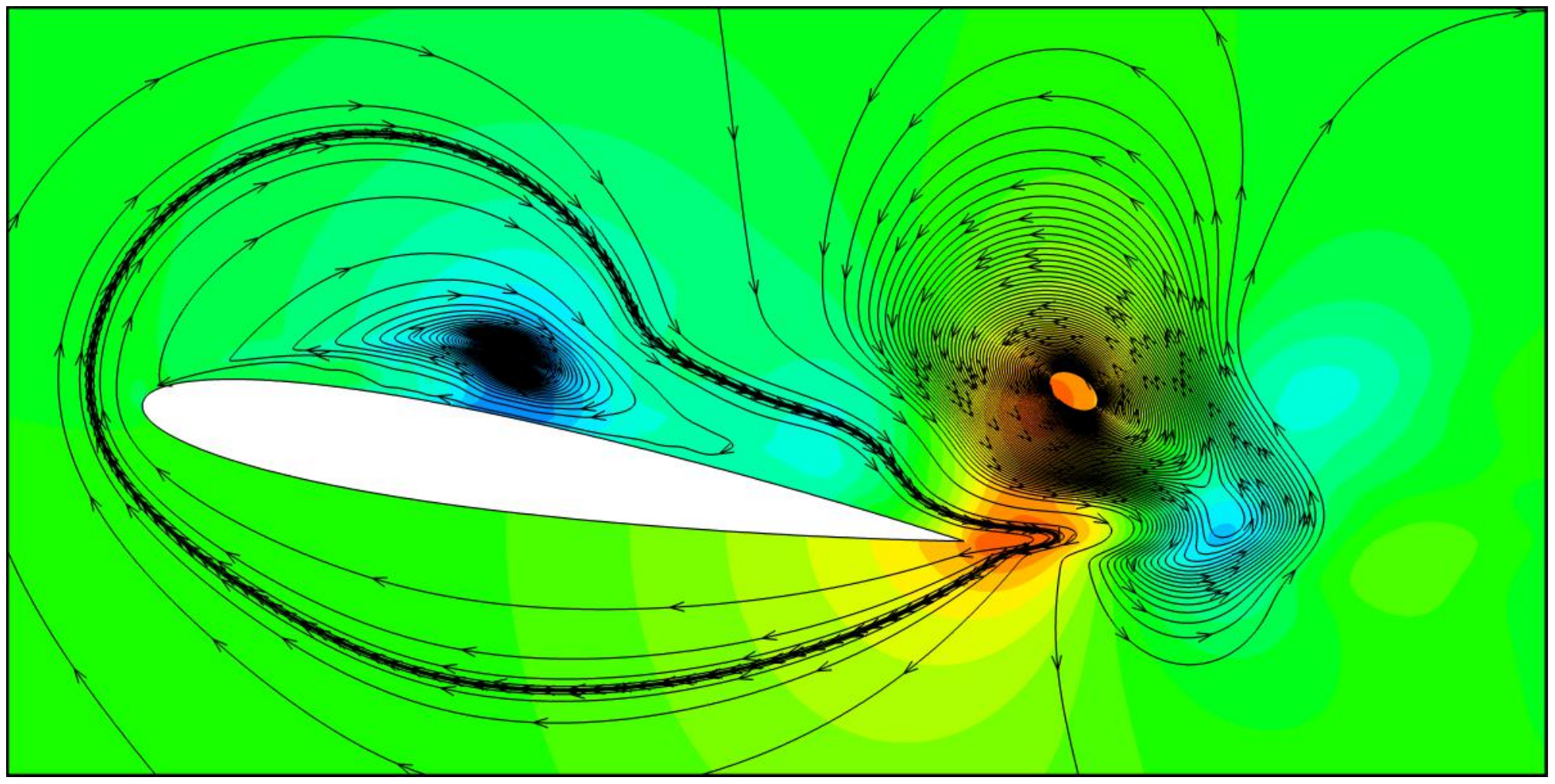}
\textit{$\Phi = 270^{\circ}$}
\end{minipage}
\begin{minipage}{220pt}
\centering
\includegraphics[width=220pt, trim={0mm 0mm 0mm 0mm}, clip]{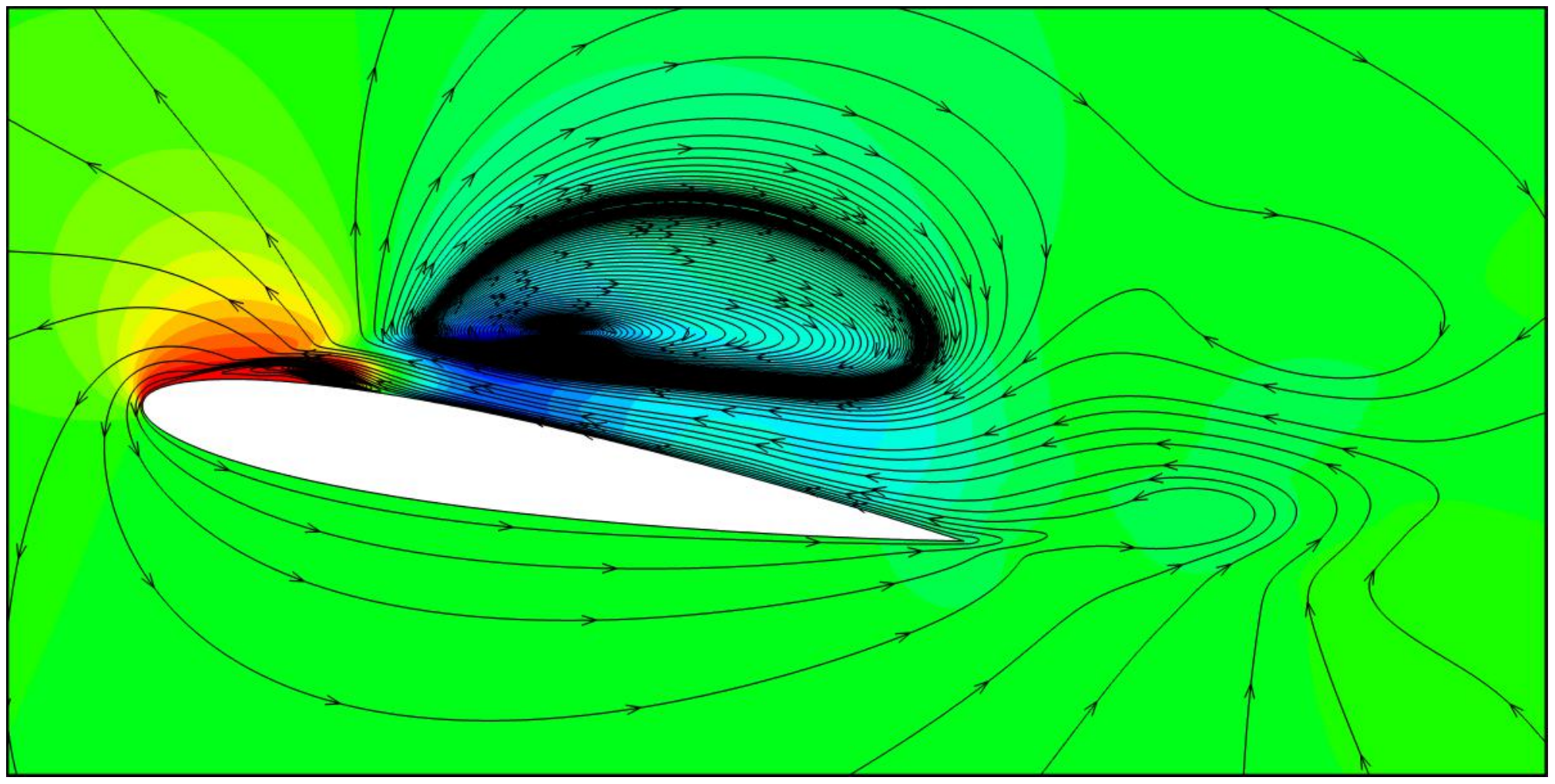}
\textit{$\Phi = 310^{\circ}$}
\end{minipage}
\caption{Streamlines patterns of the phase-averaged oscillating-f\/low-f\/ield superimposed on colour maps of the phase-averaged oscillating-pressure for the angle of attack of $9.5^{\circ}$.}
\label{phase-average_oscillating-flow_950}
\end{center}
\end{figure}

\begin{figure}
\begin{center}
\begin{minipage}{220pt}
\centering
\includegraphics[width=220pt, trim={0mm 0mm 0mm 0mm}, clip]{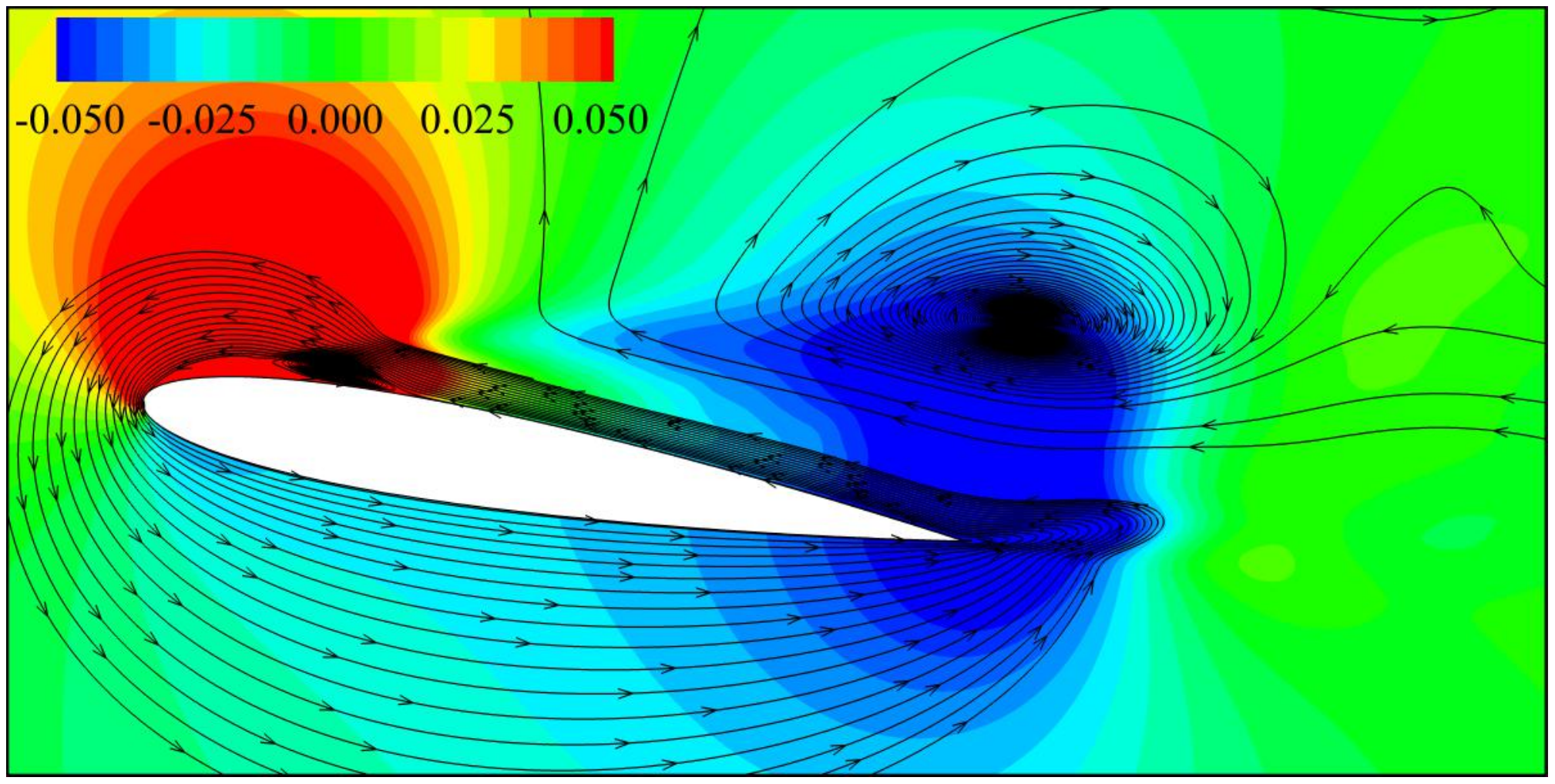}
\textit{$\Phi = 0^{\circ}$}
\end{minipage}
\medskip
\begin{minipage}{220pt}
\centering
\includegraphics[width=220pt, trim={0mm 0mm 0mm 0mm}, clip]{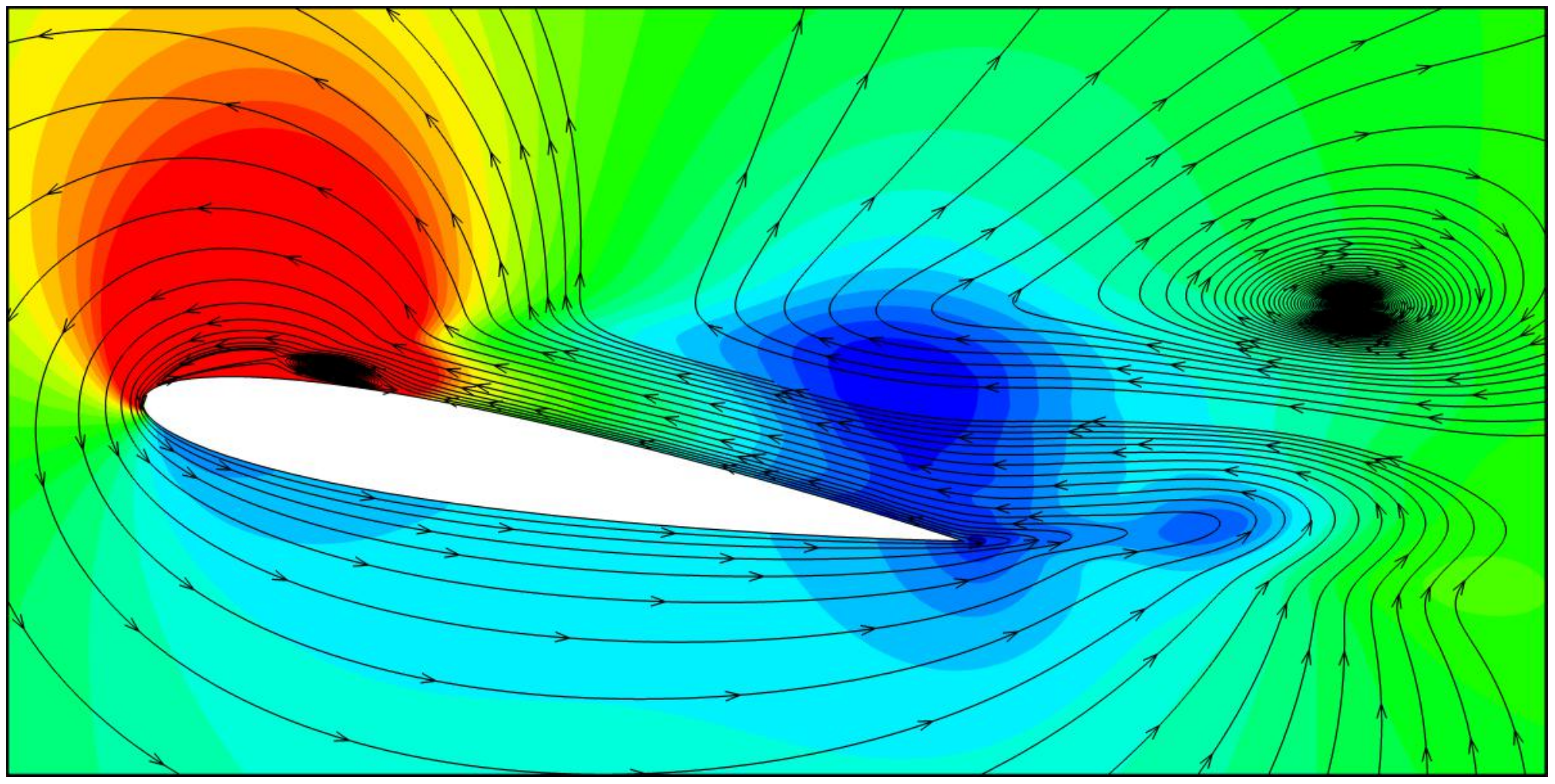}
\textit{$\Phi = 40^{\circ}$}
\end{minipage}
\medskip
\begin{minipage}{220pt}
\centering
\includegraphics[width=220pt, trim={0mm 0mm 0mm 0mm}, clip]{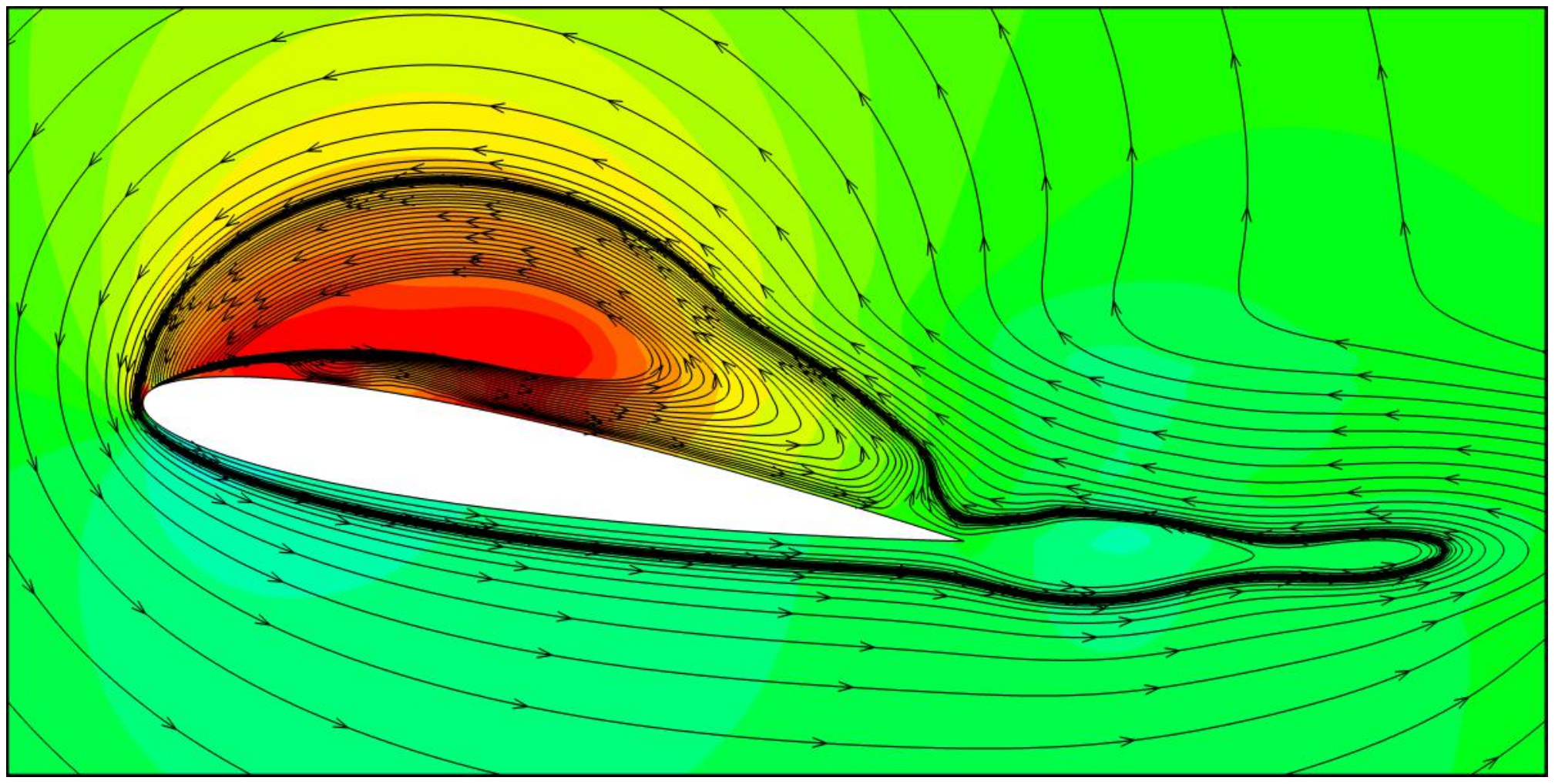}
\textit{$\Phi = 90^{\circ}$}
\end{minipage}
\medskip
\begin{minipage}{220pt}
\centering
\includegraphics[width=220pt, trim={0mm 0mm 0mm 0mm}, clip]{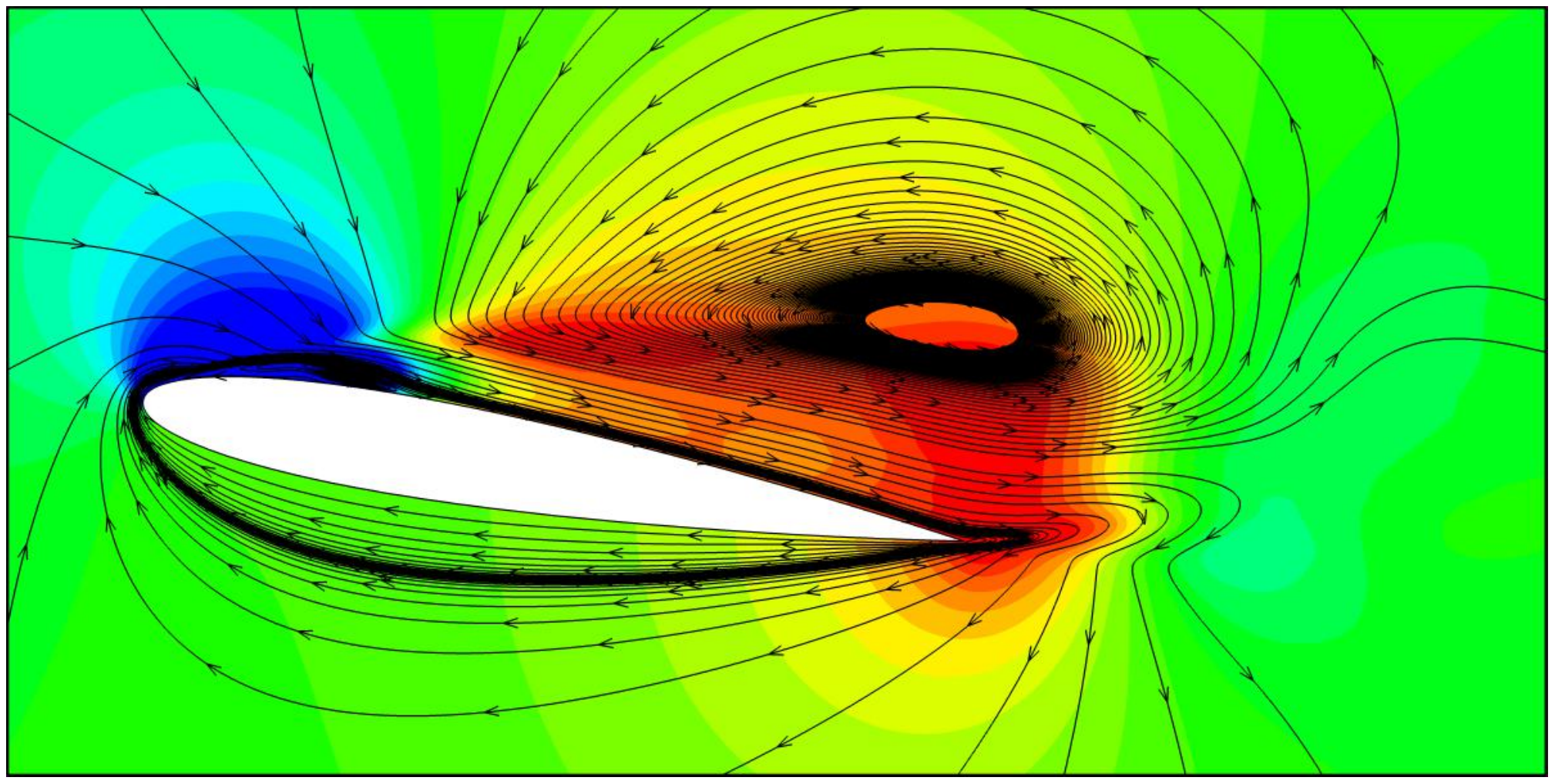}
\textit{$\Phi = 130^{\circ}$}
\end{minipage}
\medskip
\begin{minipage}{220pt}
\centering
\includegraphics[width=220pt, trim={0mm 0mm 0mm 0mm}, clip]{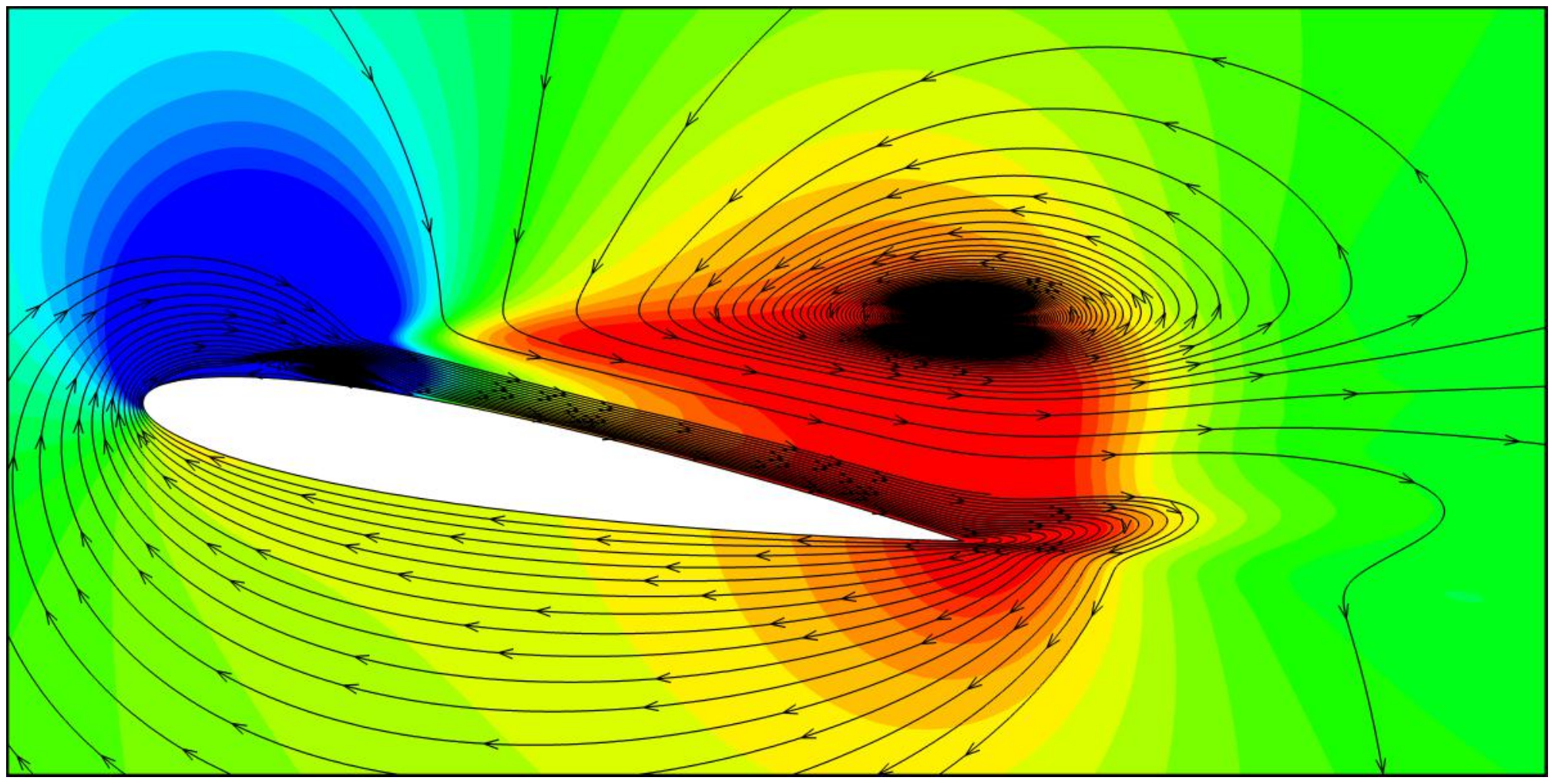}
\textit{$\Phi = 180^{\circ}$}
\end{minipage}
\medskip
\begin{minipage}{220pt}
\centering
\includegraphics[width=220pt, trim={0mm 0mm 0mm 0mm}, clip]{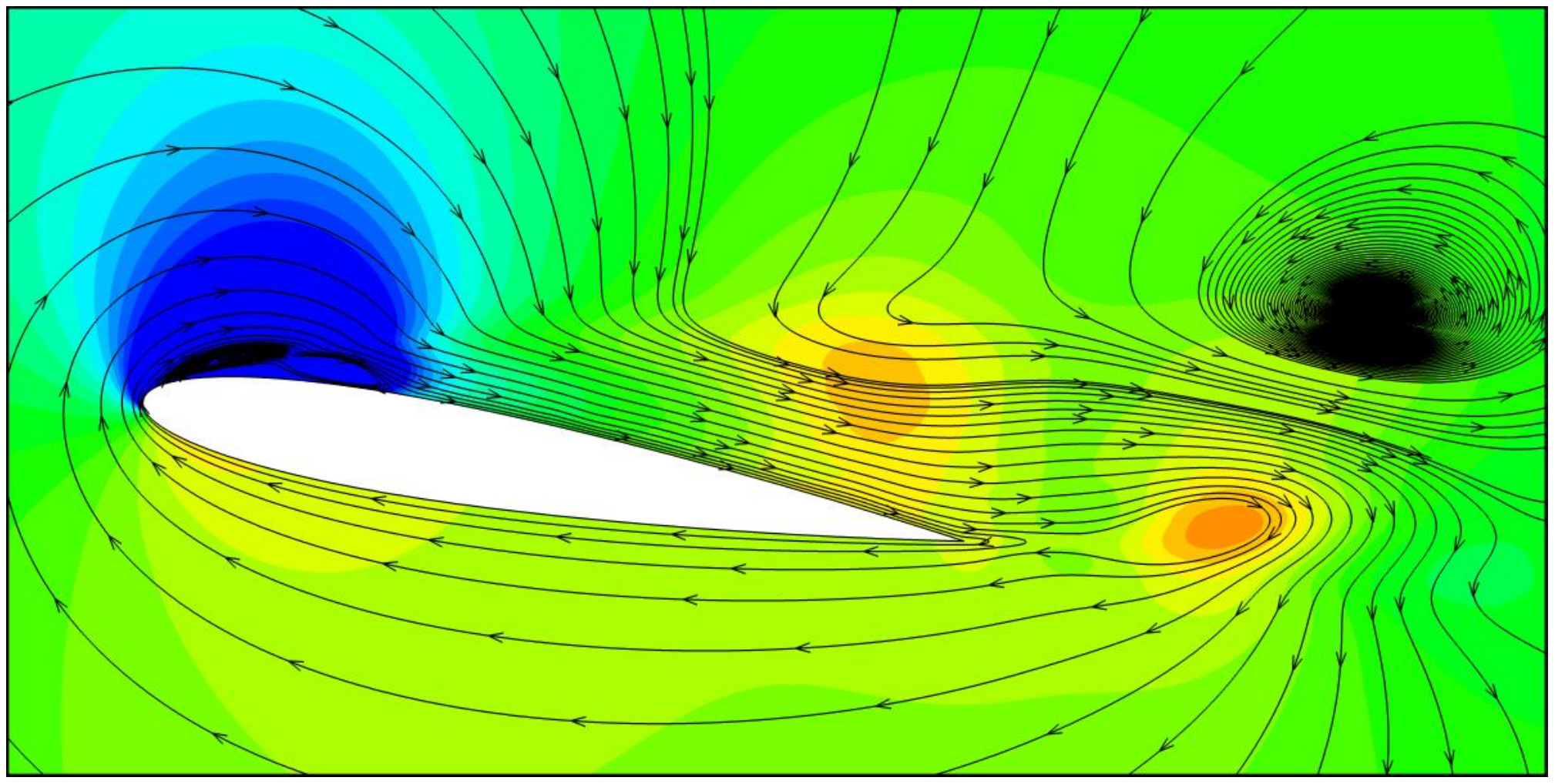}
\textit{$\Phi = 220^{\circ}$}
\end{minipage}
\medskip
\begin{minipage}{220pt}
\centering
\includegraphics[width=220pt, trim={0mm 0mm 0mm 0mm}, clip]{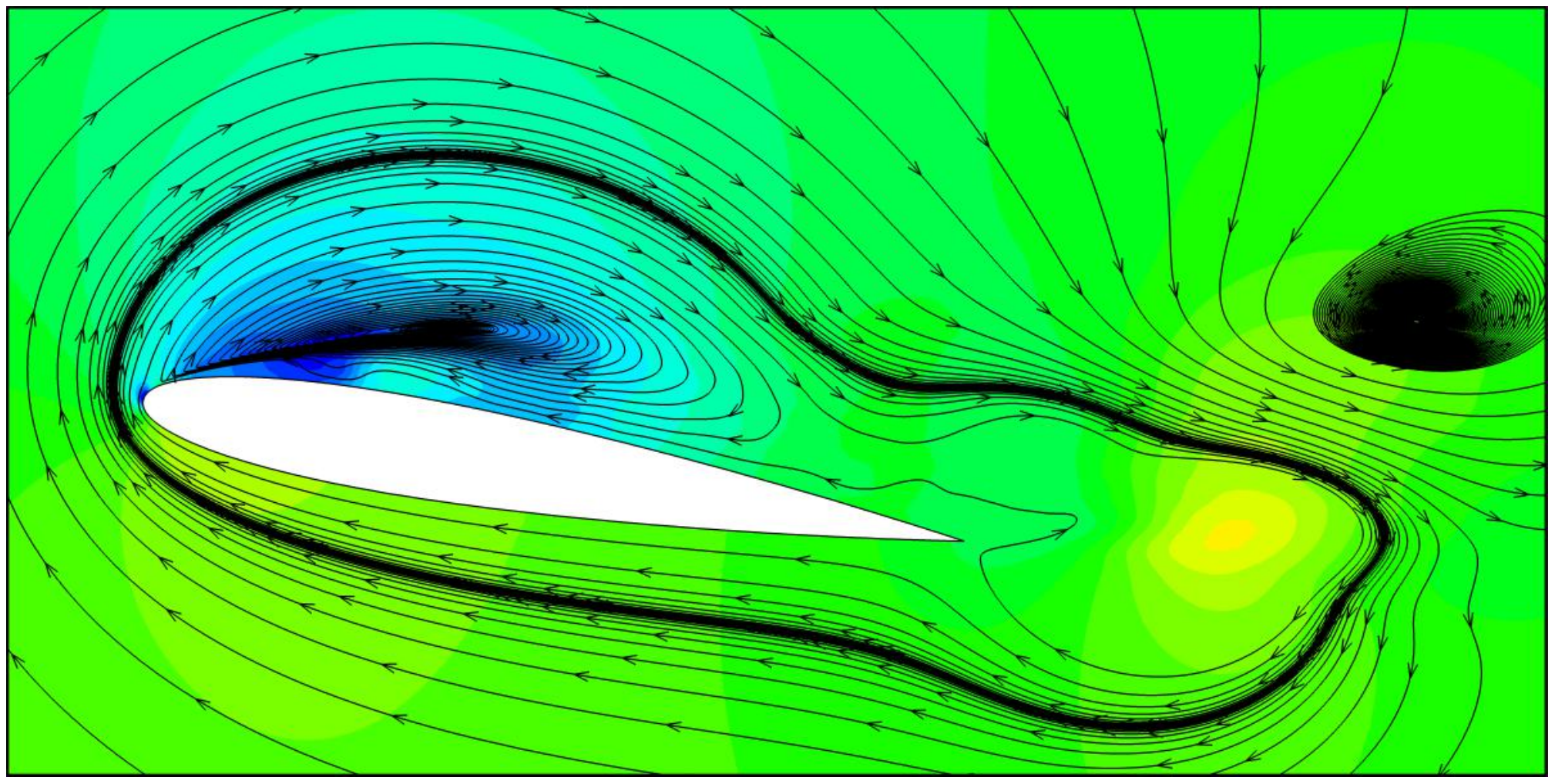}
\textit{$\Phi = 270^{\circ}$}
\end{minipage}
\begin{minipage}{220pt}
\centering
\includegraphics[width=220pt, trim={0mm 0mm 0mm 0mm}, clip]{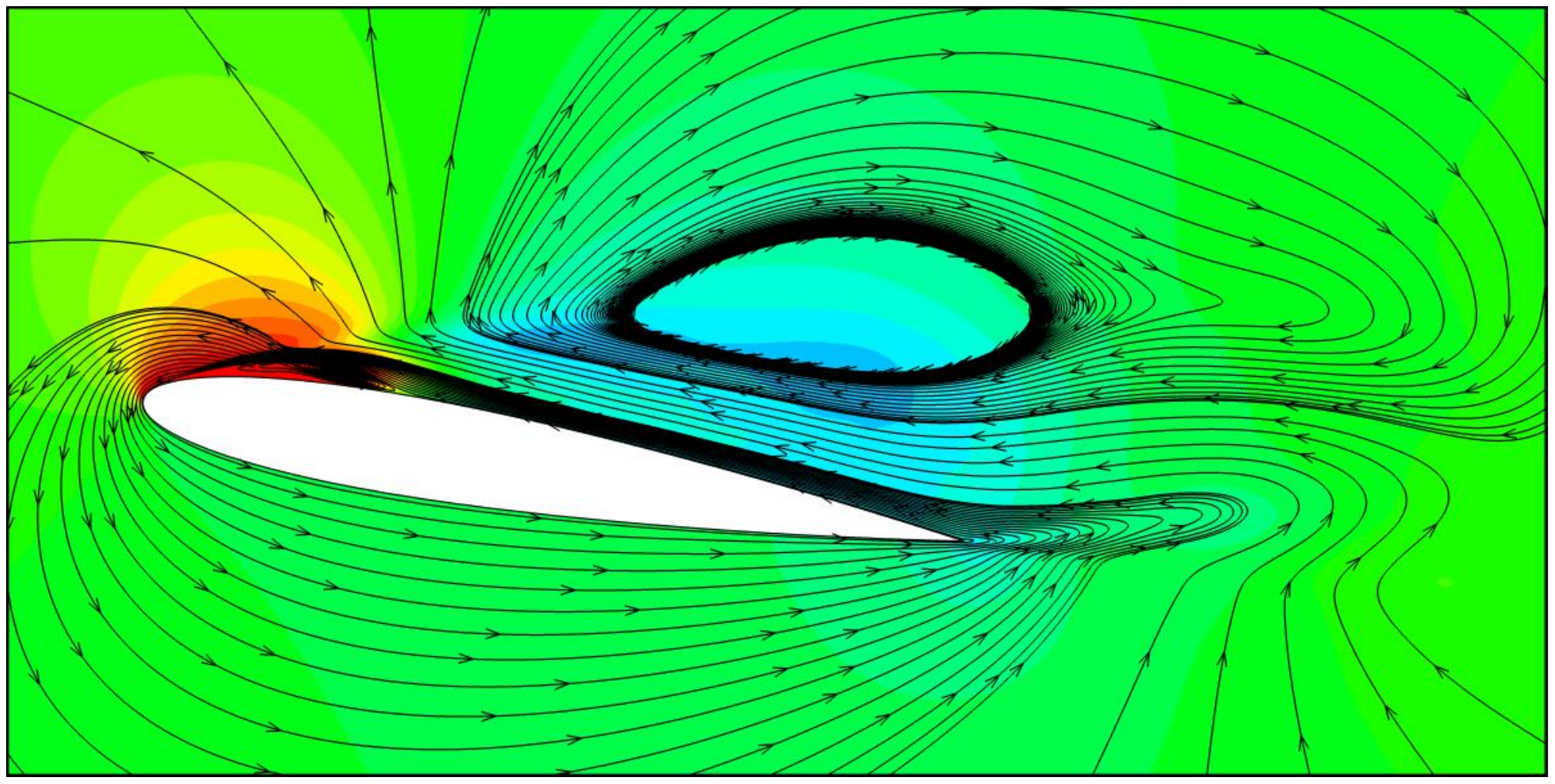}
\textit{$\Phi = 310^{\circ}$}
\end{minipage}
\caption{Streamlines patterns of the phase-averaged oscillating-f\/low-f\/ield superimposed on colour maps of the phase-averaged oscillating-pressure for the angle of attack of $9.7^{\circ}$.}
\label{phase-average_oscillating-flow_970}
\end{center}
\end{figure}

\begin{figure}
\begin{center}
\begin{minipage}{220pt}
\centering
\includegraphics[width=220pt, trim={0mm 0mm 0mm 0mm}, clip]{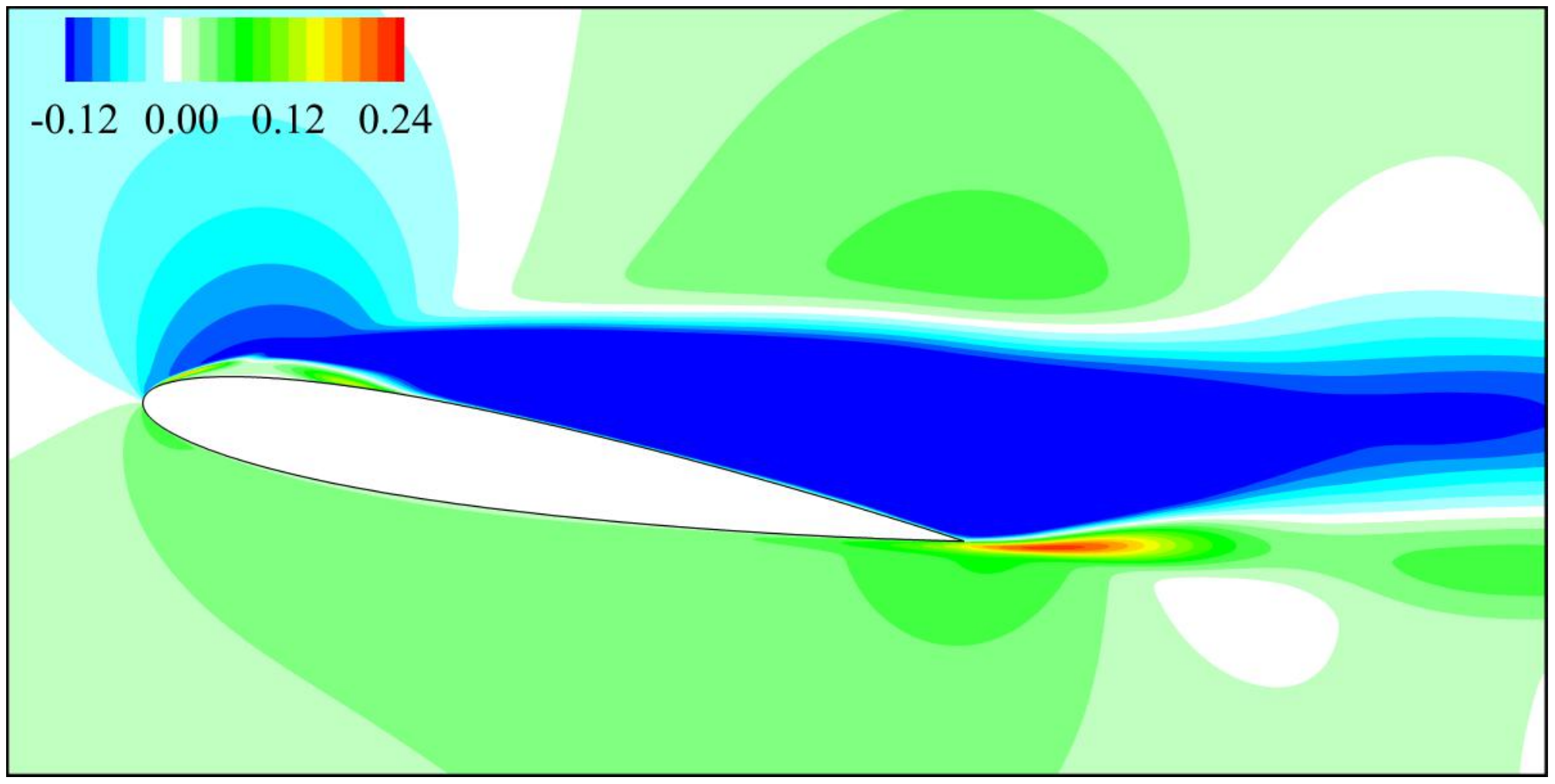}
\textit{$\Phi = 0^{\circ}$}
\end{minipage}
\medskip
\begin{minipage}{220pt}
\centering
\includegraphics[width=220pt, trim={0mm 0mm 0mm 0mm}, clip]{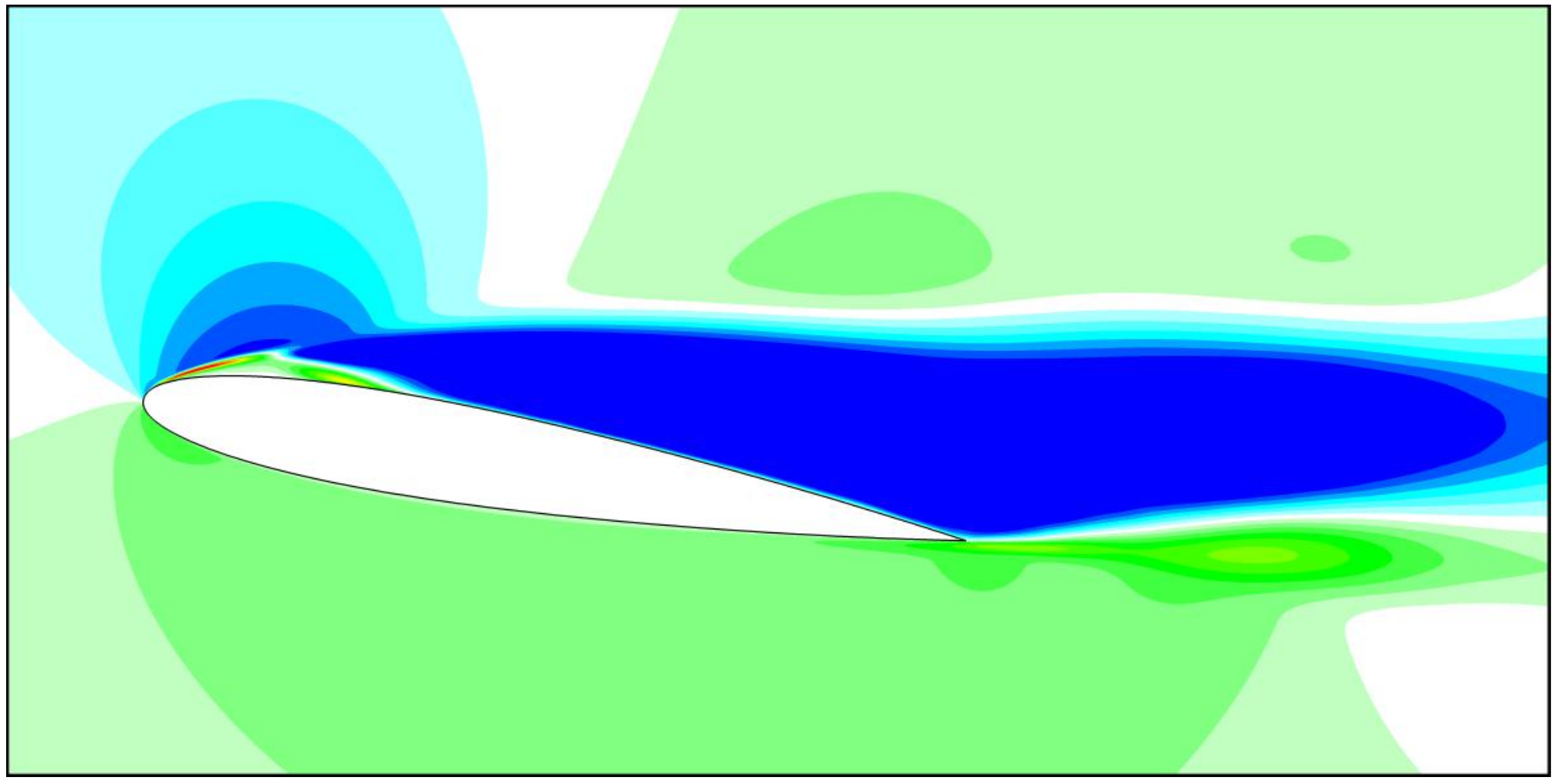}
\textit{$\Phi = 40^{\circ}$}
\end{minipage}
\medskip
\begin{minipage}{220pt}
\centering
\includegraphics[width=220pt, trim={0mm 0mm 0mm 0mm}, clip]{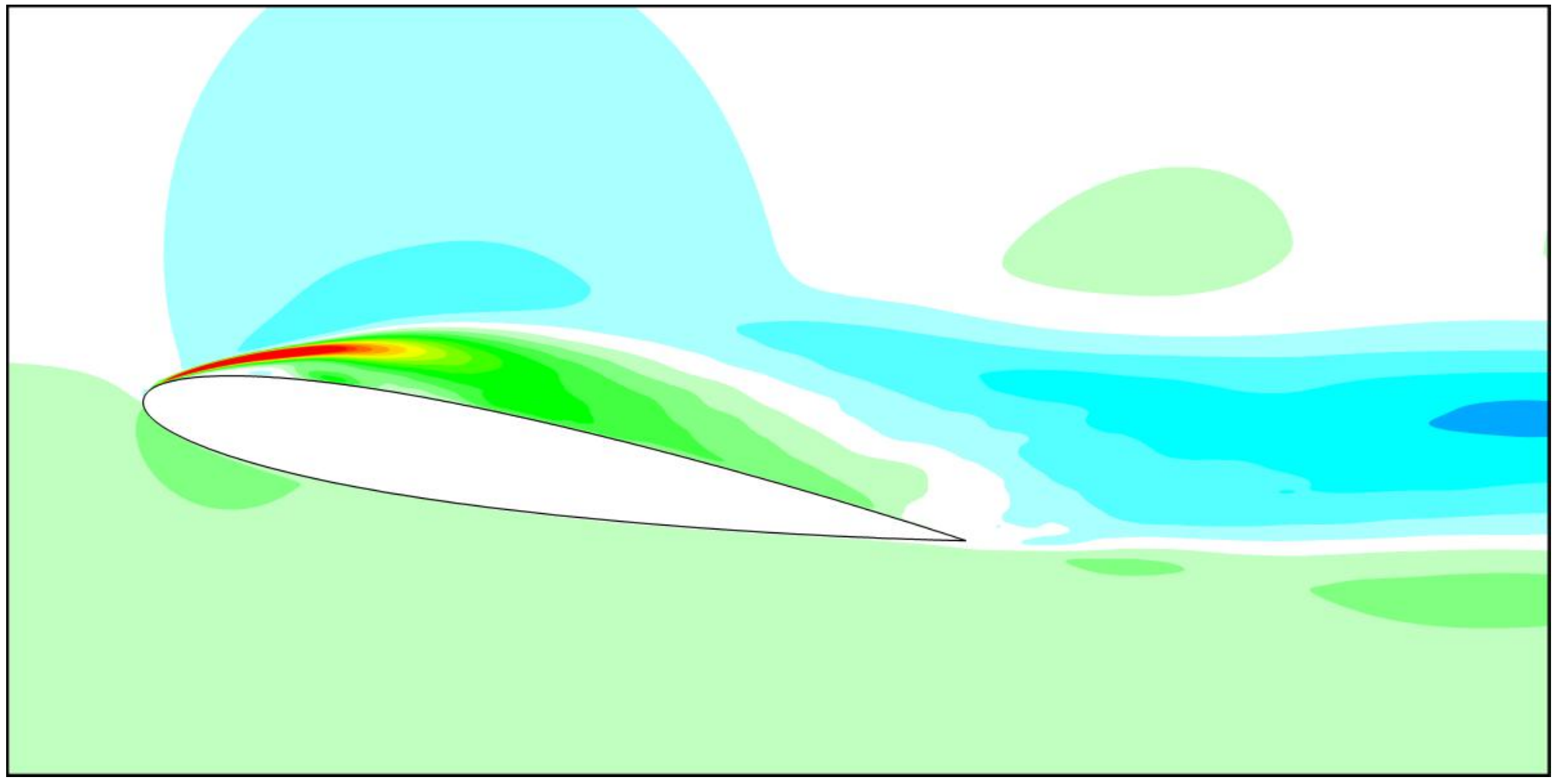}
\textit{$\Phi = 90^{\circ}$}
\end{minipage}
\medskip
\begin{minipage}{220pt}
\centering
\includegraphics[width=220pt, trim={0mm 0mm 0mm 0mm}, clip]{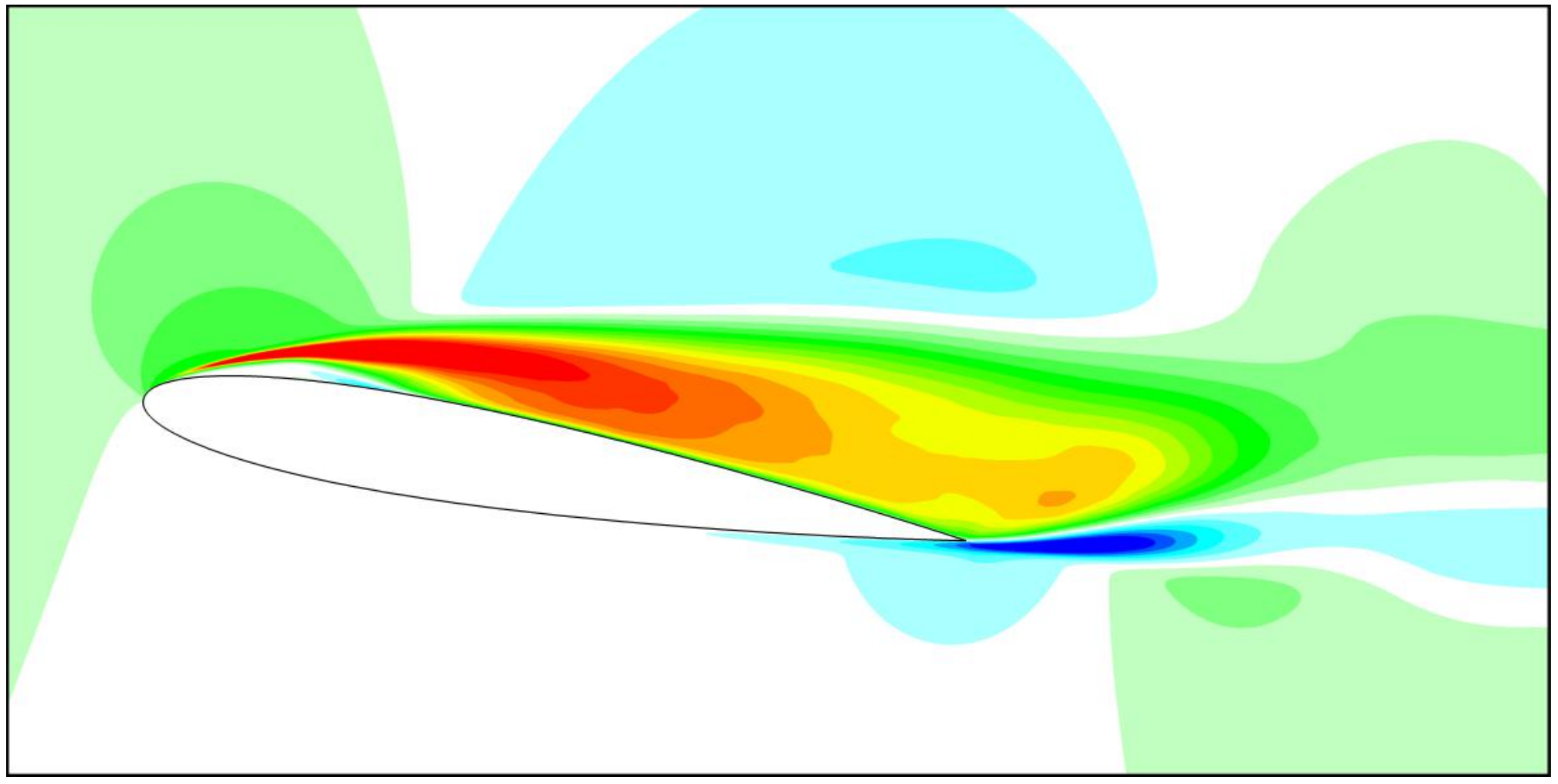}
\textit{$\Phi = 130^{\circ}$}
\end{minipage}
\medskip
\begin{minipage}{220pt}
\centering
\includegraphics[width=220pt, trim={0mm 0mm 0mm 0mm}, clip]{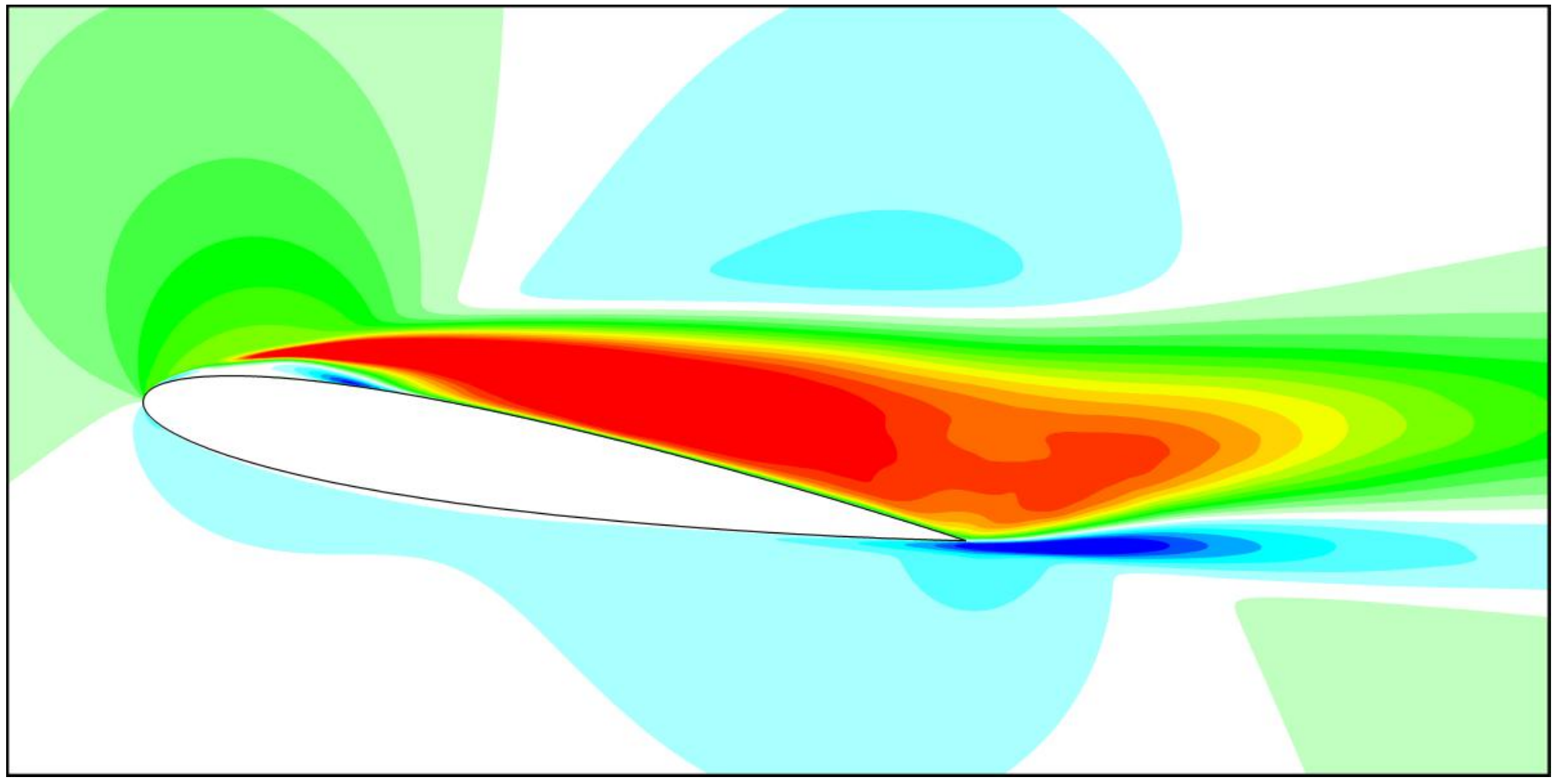}
\textit{$\Phi = 180^{\circ}$}
\end{minipage}
\medskip
\begin{minipage}{220pt}
\centering
\includegraphics[width=220pt, trim={0mm 0mm 0mm 0mm}, clip]{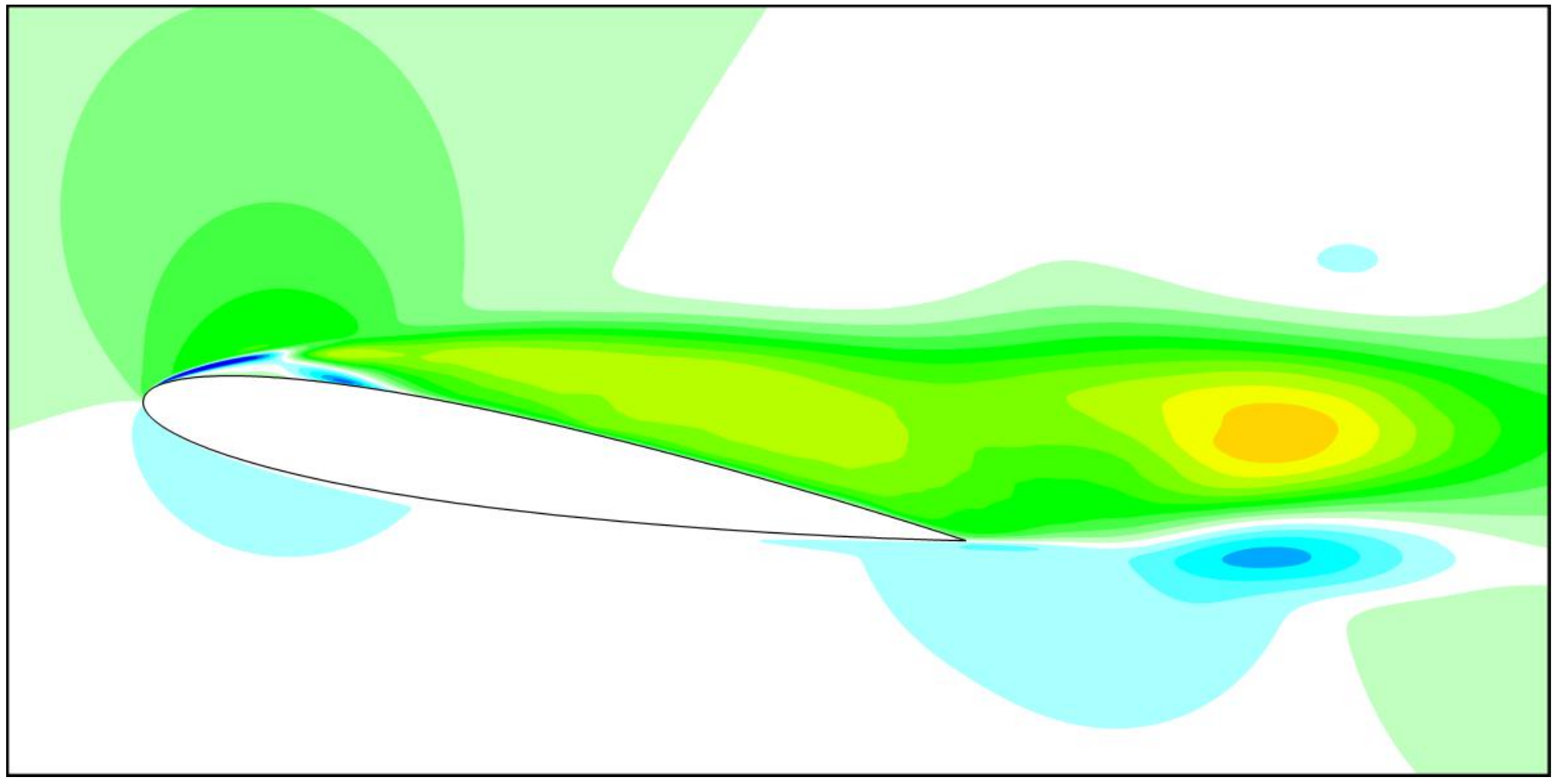}
\textit{$\Phi = 220^{\circ}$}
\end{minipage}
\medskip
\begin{minipage}{220pt}
\centering
\includegraphics[width=220pt, trim={0mm 0mm 0mm 0mm}, clip]{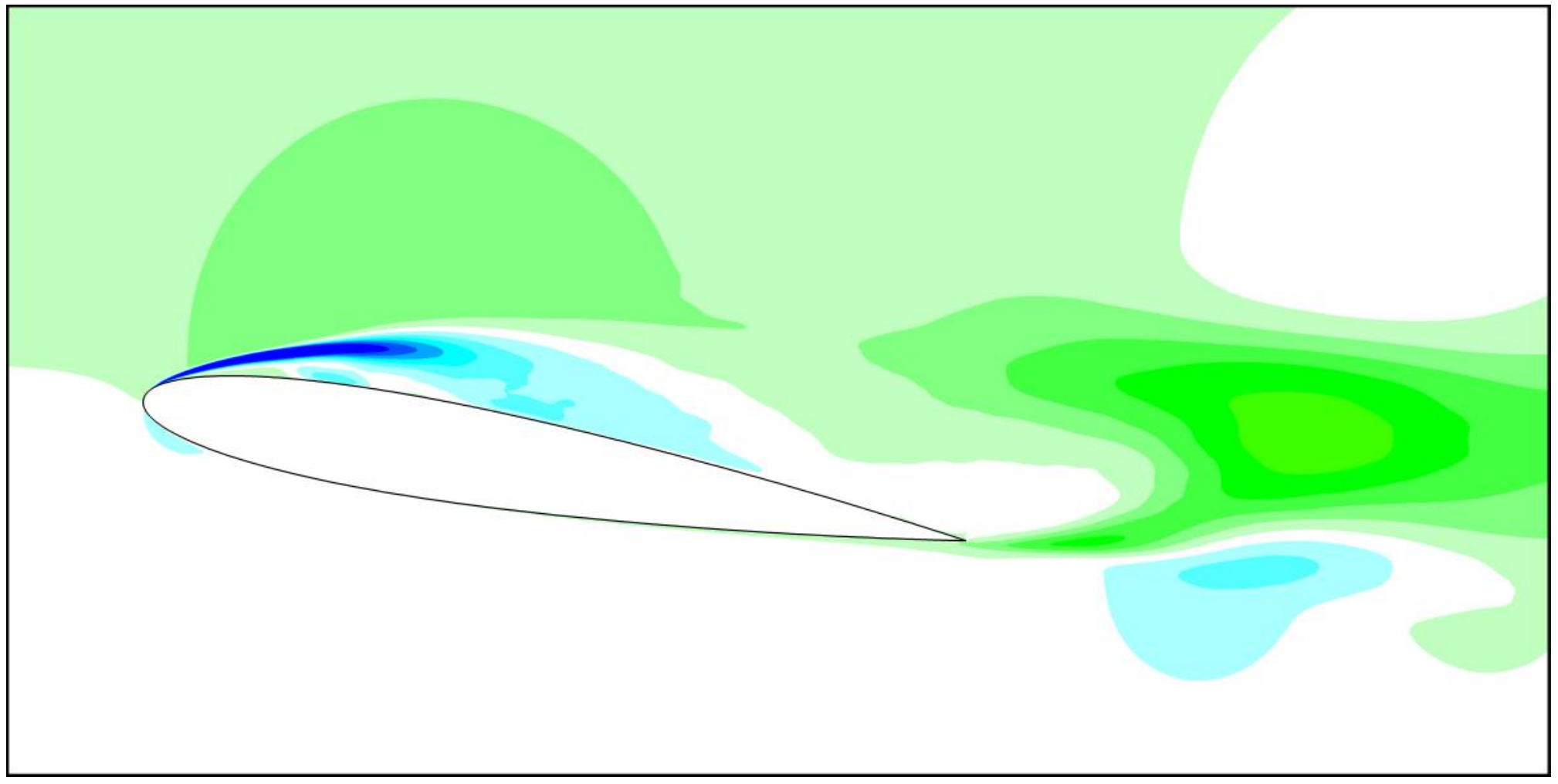}
\textit{$\Phi = 270^{\circ}$}
\end{minipage}
\begin{minipage}{220pt}
\centering
\includegraphics[width=220pt, trim={0mm 0mm 0mm 0mm}, clip]{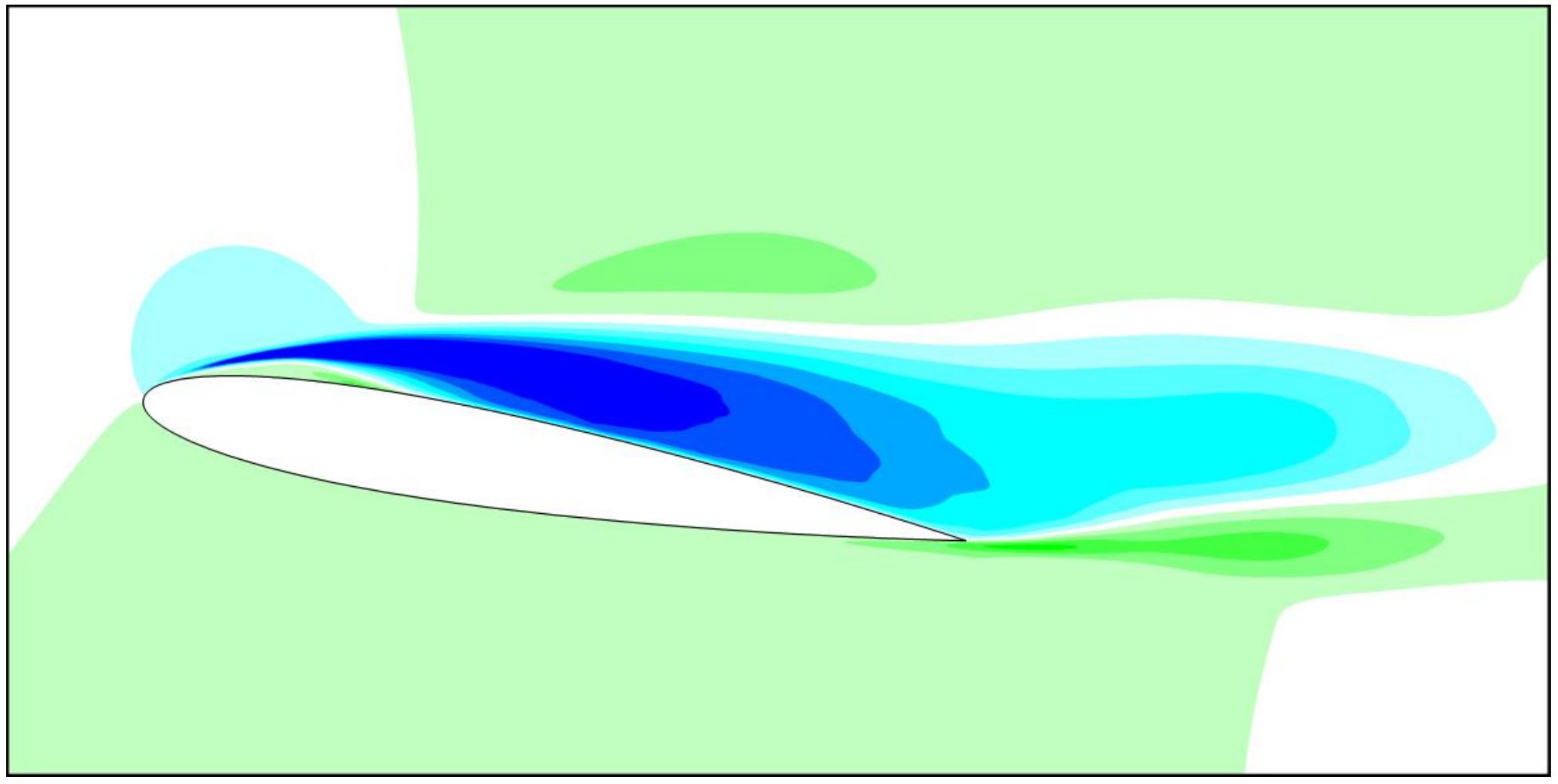}
\textit{$\Phi = 310^{\circ}$}
\end{minipage}
\caption{Colour maps of the phase-averaged oscillating-streamwise-velocity for the angle of attack of $9.7^{\circ}$.}
\label{phase-average_oscillating-streamwise_velocity_970}
\end{center}
\end{figure}

The conditional time-average of the oscillating-f\/low-f\/ield not only characterised the behaviour of the f\/low in the mean sense, but it also revealed a profound triad of vortices that sets the bursting conditions of the LSB and sustains the LFO. However, temporal and spatial information are lost in the averaging process. Therefore, the conditional-averaging implemented in $\S$~\ref{section_conditional_time-average} was extended to cover a broader range of levels; hence, help to keep some phase information. The previously def\/ined conditional time-average on three levels (high-lift, low-lift, and man lift) was expanded to a series of $37$ intervals that give continues phase information of one complete cycle. F\/igure~\ref{conditional_phase_average} illustrates the conditional phase-averaging process. The right-hand side of the f\/igure shows time history of the lift coeff\/icient at the angle of attack of $9.80^{\circ}$. The black line denotes the lift coeff\/icient signal, and the red line displays the low-pass f\/iltered lift coeff\/icient. The left-hand side of the f\/igure shows one sinusoidal cycle with a magnitude range of that of the low-pass f\/iltered lift coeff\/icient divided into $36$ equally distributed phases. The phase-averaging process is implemented in the same manner described and used in~\citet{eljack2018bursting} for the conditional time-averaging. Low-pass f\/iltering the data before conditionally phase-averaging it improves the phase-averaging process signif\/icantly. The primary objective is to examine the evolution of the low-frequency f\/low modes; thus, f\/iltering out the high frequencies does not affect the resulting phase-averaged f\/low-f\/ield. The low-pass f\/iltered lift coeff\/icient signal was split into an ascending part in which the lift is increasing with time and a descending part. The phase-averaging process was applied to the ascending low-pass f\/iltered lift coeff\/icient and its corresponding low-pass f\/iltered data points for the phases intervals from $\Phi = 0^{\circ}$ to $\Phi = 180^{\circ}$. The descending low-pass f\/iltered lift coeff\/icient and its corresponding low-pass f\/iltered data points are used in the phase-averaging process in the phases intervals from $180^{\circ} < \Phi < 360^{\circ}$. The f\/low-f\/ield studied in the present work involves exceptionally low frequency and one cycle span over $25$ non-dimensional time units. Collecting data that covers one hundred cycles is not feasible in numerical simulation. However, the conditional phase-averaging described above enhances the statistics and give much better estimate. Hereafter, the conditionally phase-average and phase-average terms will be used interchangeably. For a f\/low variable $\psi$ the decomposition is as follows:
\begin{equation}\label{decomposition}
  \psi = \overline{\psi} + \underbrace{\widetilde{\psi}+{\psi}\myprime}_{{\psi}\mydprime}
\end{equation}
Where, $\overline{\psi}$ is the mean of the variable, $\widetilde{\psi}$ is the conditionally phase-averaged, and ${\psi}\myprime$ is the f\/luctuations of the variable ``turbulent-f\/luctuations''.
\newline
F\/igure~\ref{conditional_phase_average} shows that the f\/low is fully separated at the phase angle $\Phi = 0^{\circ}$ and the lift coeff\/icient is at its minimum value. F\/igure~\ref{phase-average_oscillating-flow_940} shows streamline patterns of the phase-averaged oscillating-f\/low-f\/ield superimposed on colour maps of the phase-averaged oscillating-pressure for the angle of attack of $9.4^{\circ}$. As shown in the f\/igure, at the phase angle $\Phi = 0^{\circ}$ the oscillating-f\/low is rotating in the anti-clockwise direction. Thus, the oscillating-f\/low adjacent to the aerofoil surface on the suction surface subtracts momentum from the boundary layer and separates the f\/low. The colour maps of the oscillating-pressure indicate that the f\/low is separated and an FPG is building up along the aerofoil chord. The oscillating-f\/low rotates in the anti-clockwise direction and reduces the gradient of the streamwise velocity component across the shear layer. Thus, the transition location is expected to be at its furthest downstream position. The TCV rotate in the anti-clockwise direction, the UV of the TCV is storing energy, and the secondary vortex rotates in the clock-wise direction.\newline
At the phase angle of $\Phi = 40^{\circ}$ the UV of the TCV a little bit pop-up above the DV. The magnitude of the oscillating-pressure decreases, and the FPG along the aerofoil chord weakens. The oscillating-f\/low-f\/ield losses momentum, and the gradient of the streamwise velocity across the shear layer increases. Hence, the transition point is expected to move a little bit upstream. At the phase angle of $\Phi = 90^{\circ}$ the UV is saturated with energy, expands, and advects downstream. The UV of the TCV is rotating in the anti-clockwise direction. Thus, as the UV expands, the direction of the oscillating-f\/low f\/lips globally and the oscillating-f\/low rotates in the clockwise direction. Consequently, the oscillating f\/low adds momentum to the boundary layer and the f\/low-f\/ield attaches as seen at $\Phi = 130^{\circ}$. An APG starts to build up along the aerofoil chord. There are $40$ or $50$ degrees between each of the snapshots, and the most important ones are these between $40$ and $130$ because they contain the information on how the UV expands and advects.\newline
As the oscillating-f\/low rotates in the clockwise direction so are the TCV. The UV after it expands, advects downstream, and goes through a series of modif\/ications, it eventually forms the guide-vortex mentioned earlier and its origin was not clear. The f\/low-f\/ield is attached during the phases $\Phi = 180^{\circ}$ and $\Phi = 220^{\circ}$. At the phase angle of $\Phi = 180^{\circ}$ the f\/low is fully attached, the APG at its maximum magnitude and the lift coeff\/icient is at its maximum value as seen in f\/igure~\ref{conditional_phase_average}. After that, the UV raises a little bit above the DV, again, at the phase angle $\Phi = 220^{\circ}$. The UV, rotating in the clockwise direction now, continues to rise above the DV and gain more momentum until it saturates with energy, expands and the f\/low separates again. The guide-vortex continues to deform, and the f\/low continues to separate until the lift coeff\/icient returns to its minimum at the phase angle of $360^{\circ}$. Meanwhile, the point of maximum turbulent kinetic energy, transition location, moves upstream and downstream in accordance with the direction of rotation of the oscillating-f\/low.\newline
It is noted that at the angle of attack of $9.4^{\circ}$ the gradient of the oscillating-velocity across the laminar portion of the separated shear layer occasionally reaches the threshold required for the UV of the TCV to expand. Accordingly, the UV expands and the LSB bursts intermittently. As the mean-f\/low is attached at this angle of attack, the f\/low remains attached and separates intermittently. Therefore, the lift coeff\/icient is at its maximum value with intermittent occasions at which the lift drops below its maximum value. However, the gradient of the oscillating-velocity is relatively small, and the UV extracts relatively small fraction of energy from the mean f\/low. Thus, the amplitude of the intermittent oscillations in the f\/low-f\/ield and the lift coeff\/icient is relatively small. The gradient of the oscillating-velocity and the capacity of the UV to store energy before expanding are proportional to the angle of attack. Consequently, the UV causes more pronounced oscillations in the oscillating-f\/low-f\/ield and the aerodynamic coeff\/icients as the angle of attack increases. Furthermore, the time required for the oscillating-f\/low-f\/ield to change its direction of rotation is proportional to the initial energy input by the UV of the TCV. Hence, the UV capability in reducing the frequency of the LFO and increasing the amplitude of oscillation increases as the angle of attack is increased. As shown in f\/igures~\ref{time-average_oscillating-flow3} and~\ref{time-average_oscillating-flow4} the magnitude of the oscillating-spanwise-vorticity increases as the angle of attack is increased. This is indicative that the gradient of the oscillating-velocity across the laminar portion of the separated shear layer increases as the angle of attack is increased. Thus, the strength of the UV is proportional to the angle of attack. Consequently, the magnitude of the oscillating-f\/low increases as shown in f\/igure~\ref{phase-average_oscillating-flow_950}. The expansion, advection, and reformation of the UV become more regular and pronounce in magnitude in the angle of attack of $9.5^{\circ}$ as seen in the f\/igure.\newline
F\/igure~\ref{phase-average_oscillating-flow_970} shows streamline patterns of the phase-averaged oscillating-f\/low-f\/ield superimposed on colour maps of the phase-averaged oscillating-pressure for the angle of attack of $9.7^{\circ}$. F\/igure~\ref{phase-average_oscillating-streamwise_velocity_970} shows colour maps of the phase-averaged oscillating-streamwise-velocity for the angle of attack of $9.7^{\circ}$. At the phase angle $\Phi = 0^{\circ}$, the f\/low is separated, the oscillating-streamwise-velocity is negative on the suction surface of the aerofoil, and the FPG along the aerofoil chord is at its maximum magnitude in the cycle. The magnitude of the oscillating-streamwise-velocity increases and the amplitude of the FPG decreases at the phase angle of $\Phi = 40^{\circ}$. The magnitude of the oscillating-streamwise-velocity and the amplitude of the FPG f\/luctuate around zero at the phase angle of $90^{\circ}$. As the oscillating-f\/low decays, the f\/low reaches an equilibrium state, and the oscillation in the f\/low vanishes. However, the UV of the TCV is saturated with energy. Thus, the UV expands, the f\/low-f\/ield attaches, and a new imbalance is created at the phase angle of $90^{\circ}$. As the f\/low attaches, the oscillating-streamwise-velocity becomes positive, and an APG builds up along the aerofoil chord as seen in the phase angles of $130^{\circ}$, $180^{\circ}$, and $220^{\circ}$. The f\/low is fully attached at $\Phi = 180^{\circ}$, the oscillating-streamwise-velocity at its maximum magnitude above the mean streamwise velocity, and the APG is at its maximum amplitude. After that, the oscillating-streamwise-velocity and the APG decrease until they become zero at $\Phi = 270^{\circ}$. The oscillating-f\/low-f\/ield has no enough momentum to keep the boundary layer attached. Consequently, the f\/low reaches a new equilibrium, and the UV of the TCV expands and repels the attached-f\/low and forces it to separate abruptly at $\Phi = 310^{\circ}$. As the process reverses its direction, the oscillating-streamwise-velocity becomes negative again, and an FPG builds up along the aerofoil chord as seen in $\Phi = 310^{\circ}$ and $0^{\circ}$. The transition process starts at the angle of attack of $9.25^{\circ}$ and continues until the LFO becomes self-sustained at the angle of attack of $9.7^{\circ}$

\subsection{The underlying mechanism} \label{mechanism}
The conditional phase-averaging of the data revealed the details of how the triad of vortices controls the time-scale of the LFO. However, the process is not perfectly periodic especially at this low Reynolds number. Therefore, the above discussion on the underlying mechanism is summarised in f\/igure~\ref{model}. The LFO cycle is assumed to be sinusoidal and divided into eight equally distributed phases. The sequence of phase angles and events matches that of the above data analysis. The self-sustained separation and reattachment of the f\/low repeat periodically in the following sequence of events:\newline
$\bPhi \bEqual \bm{0}^{\bcirc}$\textbf{:} The f\/low is fully separated. The LSB is not present. The oscillating-streamwise-velocity is negative and at its minimum amplitude. The oscillating-f\/low and the TCV rotate in the anti-clockwise direction. The gradient of the oscillating-velocity across the laminar portion of the separated shear layer at its maximum magnitude, and the energy extraction by the UV of the TCV at its maximum rate. The velocity gradient across the separated shear layer at its minimum magnitude. Thus, the transition process via the Kelvin--Helmholtz instability is delayed, and the transition location is at its furthest downstream position, ``late-transition''. The pressure gradient along the aerofoil chord is favourable and at its maximum magnitude. The f\/luctuating lift coeff\/icient is negative and at its minimum magnitude.\newline
$\bPhi \bEqual \bm{45}^{\bcirc}$\textbf{:} The oscillating-streamwise-velocity, the velocity gradient across the separated shear layer, and the f\/luctuating lift coeff\/icient increase. Thus, the transition location moves a little pit upstream. The gradient of the oscillating-velocity across the laminar portion of the separated shear layer, the rate of energy extraction by the UV of the TCV, and the gradient of the pressure along the aerofoil chord decrease.\newline
$\bPhi \bEqual \bm{90}^{\bcirc}$\textbf{:} The amplitude of the oscillating-streamwise-velocity, the velocity gradient across the separated shear layer, the gradient of the oscillating-velocity across the laminar portion of the separated shear layer, the rate of energy extraction by the UV of the TCV, the gradient of the pressure along the aerofoil chord, and the f\/luctuating lift coeff\/icient become almost zero. Hence, the transition location is determined naturally by the Kelvin--Helmholtz instability. Therefore, the f\/low-f\/ield reaches an equilibrium state. The UV of the TCV saturates with energy, expands, advects downstream, and starts a new imbalance. The direction of the oscillating-f\/low f\/lips globally and the TCV rotate in the clockwise direction. The oscillating-f\/low adds momentum to the separated boundary layer, and the f\/low attaches.\newline
$\bPhi \bEqual \bm{135}^{\bcirc}$\textbf{:} The f\/low is attached. The LSB is present and intact. The oscillating-streamwise-velocity becomes positive. The oscillating-f\/low rotates in the clockwise direction. The velocity gradient across the separated shear layer increases. Thus, the transition process via the Kelvin--Helmholtz instability is accelerated, and the transition location moves a little pit upstream. An adverse pressure gradient forms along the aerofoil chord. The f\/luctuating lift coeff\/icient is positive.\newline
$\bPhi \bEqual \bm{180}^{\bcirc}$\textbf{:} The f\/low is fully attached. The oscillating-streamwise-velocity is positive and at its maximum amplitude. The oscillating-f\/low and the TCV rotate in the clockwise direction. The gradient of the oscillating-velocity across the laminar portion of the separated shear layer at its maximum magnitude, and the energy extraction by the UV of the TCV at its maximum rate. The velocity gradient across the separated shear layer at its maximum value. Consequently, the transition process via the Kelvin--Helmholtz instability is further accelerated, and the transition location is at its furthest upstream position, ``early-transition''. The adverse pressure gradient along the aerofoil chord at its maximum magnitude. The f\/luctuating lift coeff\/icient is positive and at its maximum amplitude.\newline
$\bPhi \bEqual \bm{225}^{\bcirc}$\textbf{:} The oscillating-streamwise-velocity, the velocity gradient across the separated shear layer, the gradient of the oscillating-velocity across the laminar portion of the separated shear layer, the rate of energy extraction by the UV of the TCV, the gradient of the pressure along the aerofoil chord, and the f\/luctuating lift coeff\/icient decrease. The transition location moves a little pit downstream.\newline
$\bPhi \bEqual \bm{270}^{\bcirc}$\textbf{:} The f\/low-f\/ield reaches a new equilibrium state. The UV of the TCV saturates with energy, expands, advects downstream, and starts a new disequilibrium. The direction of the oscillating-f\/low f\/lips globally and the TCV rotate in the anti-clockwise direction. The oscillating-f\/low subtracts momentum from the boundary layer, and the f\/low separates again.\newline
$\bPhi \bEqual \bm{315}^{\bcirc}$\textbf{:} The oscillating-streamwise-velocity becomes negative again. The oscillating-f\/low rotates in the anti-clockwise direction. The velocity gradient across the separated shear layer decreases. Thus, the transition process via the Kelvin--Helmholtz instability is further delayed, and the transition location moves a little pit downstream. A favourable pressure gradient forms along the aerofoil chord. The f\/luctuating lift coeff\/icient is negative and decreasing.\newline
Thus, the stability of the LSB and the sustainability of the LFO phenomenon, in a naturally evolving f\/low, are controlled by the following parameters:
\begin{enumerate}
  \item The strength and the direction of rotation of the oscillating-f\/low.
  \item The velocity gradient across the separated shear layer.
  \item The transition location along the separated shear layer.
  \item The gradient of the oscillating-velocity across the laminar portion of the separated shear layer.
  \item The strength and dynamics of the triad of vortices.
  \item The ability of the UV of the TCV to extract energy from the mean-f\/low.
  \item The gradient of the oscillating-pressure along the aerofoil chord.
\end{enumerate}
These parameters, hereafter the LFO-control-parameters, are interlinked. The process is self-sustained and quasi-periodic in a manner that once the oscillating-f\/low reaches an equilibrium, the UV expands and creates a new disequilibrium. However, sometimes the TCV reverse the process before the cycle is completed and the periodic cycle is disturbed. The present model conf\/irms the observations of~\citet{tanaka2004flow} about the guide-vortex that play an important role in the phenomenon. Most importantly, the model presented in the current work shows that the elegant modelling of the phenomenon carried out by~\citet{sandham2008transitional} is accurate in many aspects despite the fact that the underlying mechanism was not revealed at the time.

\subsection{The disturbed cycles and the route to a full stall}
\begin{figure}
\begin{center}
\begin{minipage}{220pt}
\centering
\includegraphics[width=220pt, trim={0mm 0mm 0mm 0mm}, clip]{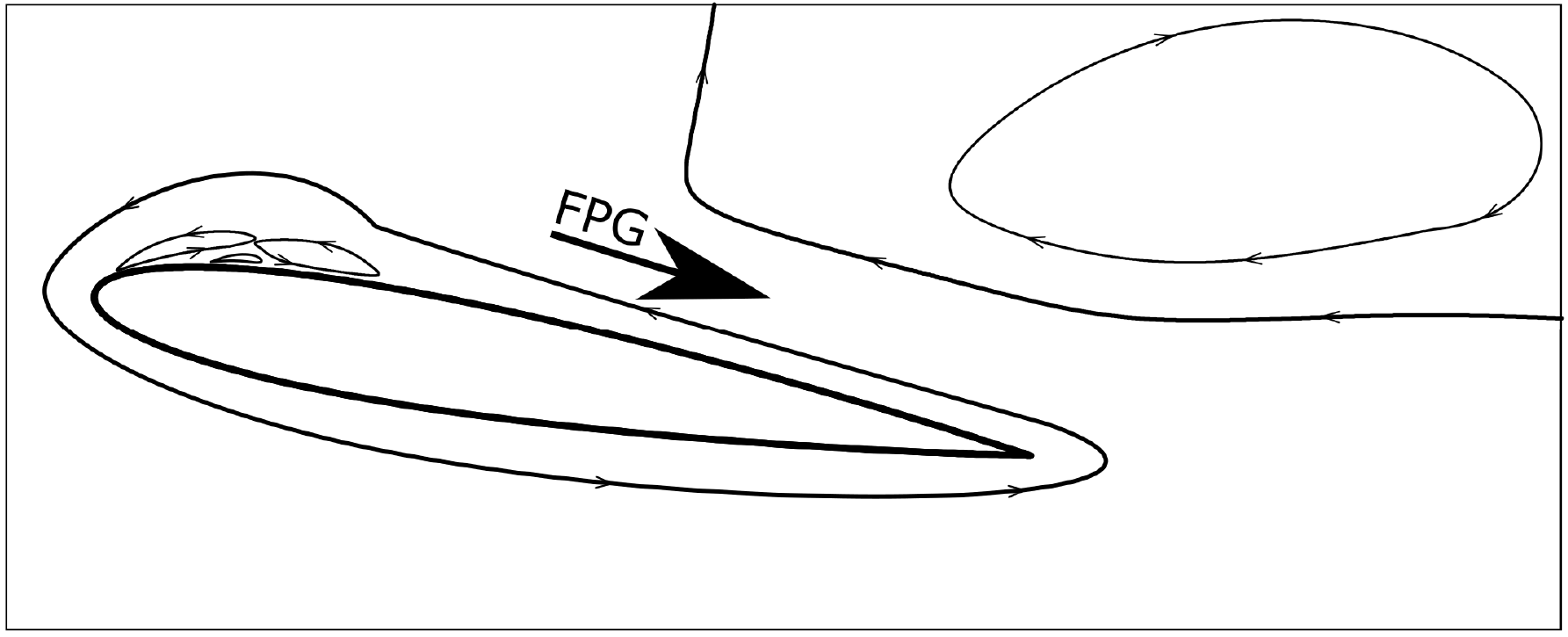}
\textit{$\Phi = 0^{\circ}$}
\end{minipage}
\medskip
\begin{minipage}{220pt}
\centering
\includegraphics[width=220pt, trim={0mm 0mm 0mm 0mm}, clip]{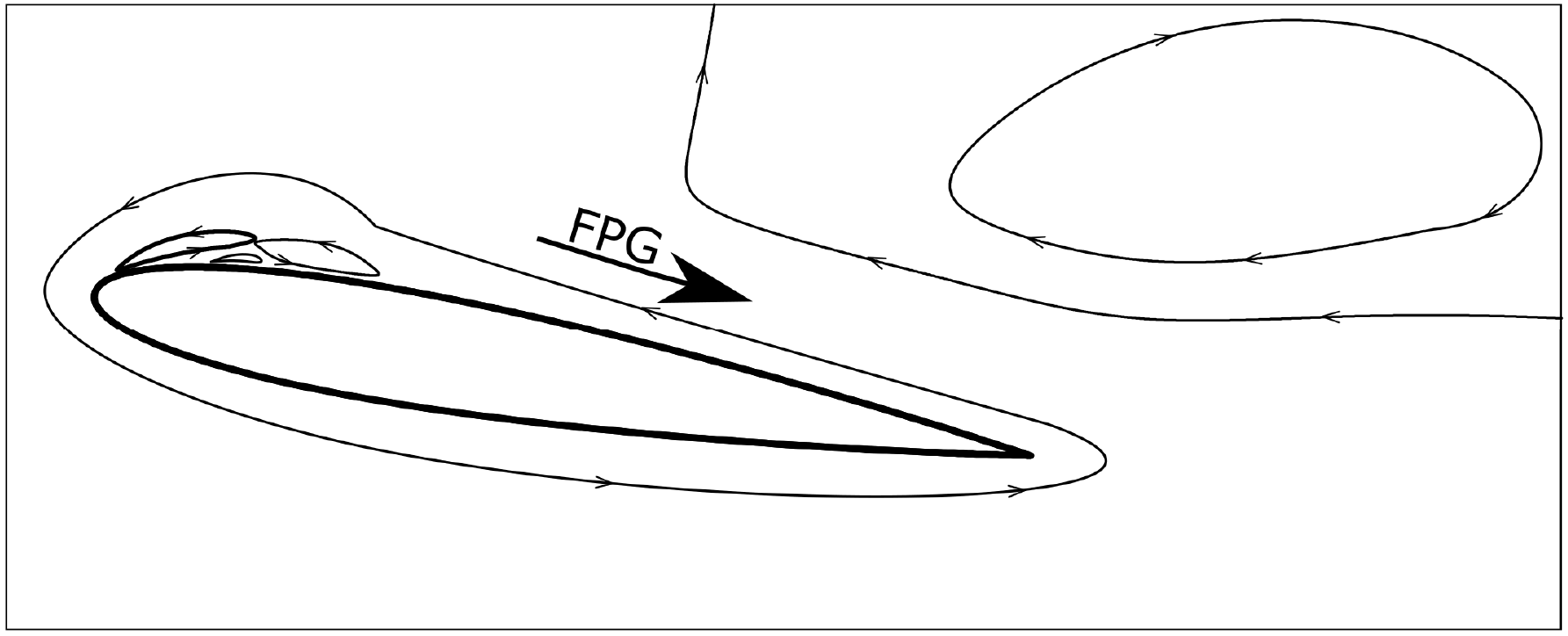}
\textit{$\Phi = 45^{\circ}$}
\end{minipage}
\medskip
\begin{minipage}{220pt}
\centering
\includegraphics[width=220pt, trim={0mm 0mm 0mm 0mm}, clip]{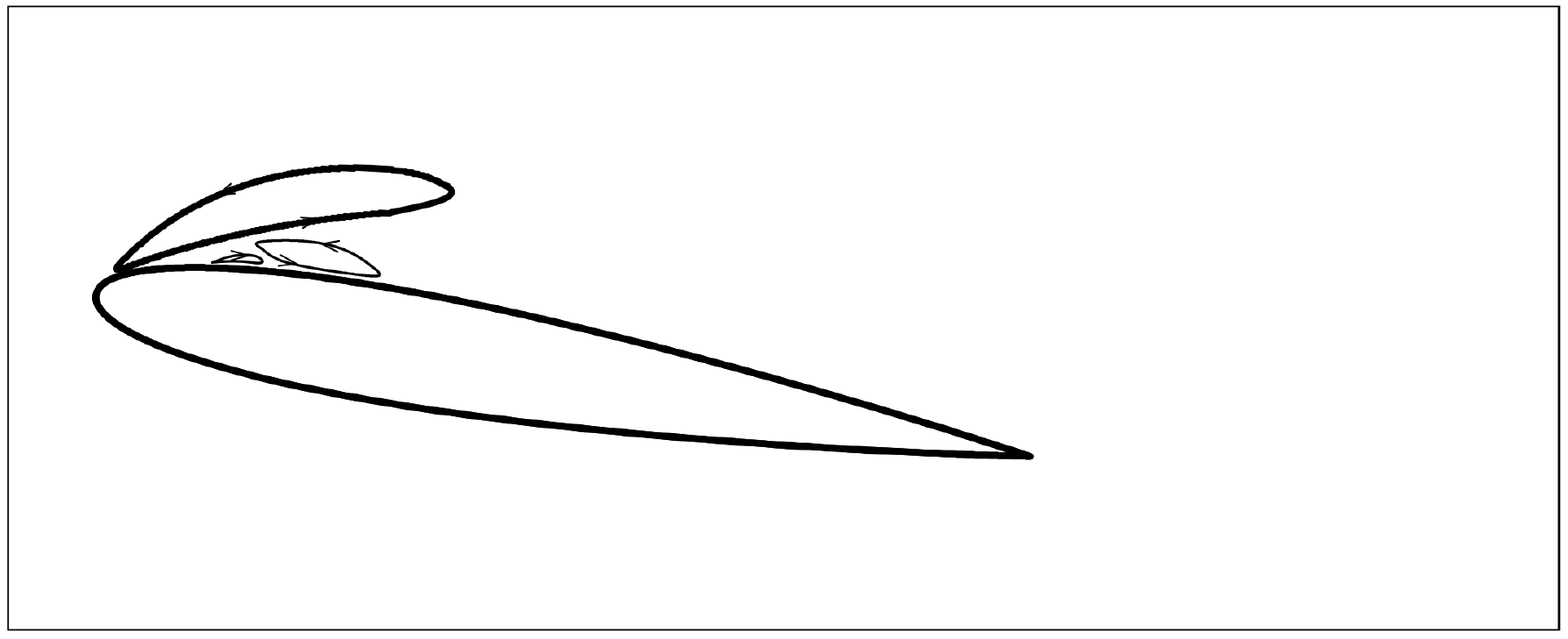}
\textit{$\Phi = 90^{\circ}$}
\end{minipage}
\medskip
\begin{minipage}{220pt}
\centering
\includegraphics[width=220pt, trim={0mm 0mm 0mm 0mm}, clip]{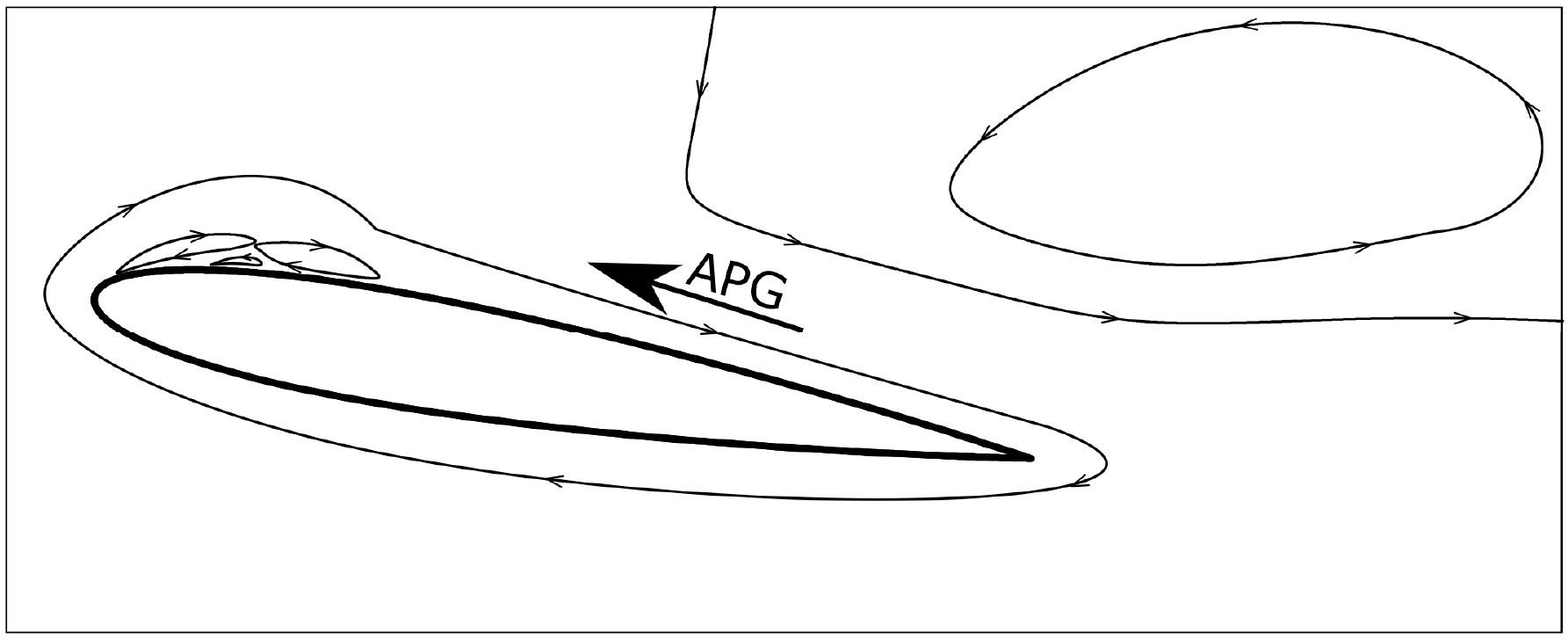}
\textit{$\Phi = 135^{\circ}$}
\end{minipage}
\medskip
\begin{minipage}{220pt}
\centering
\includegraphics[width=220pt, trim={0mm 0mm 0mm 0mm}, clip]{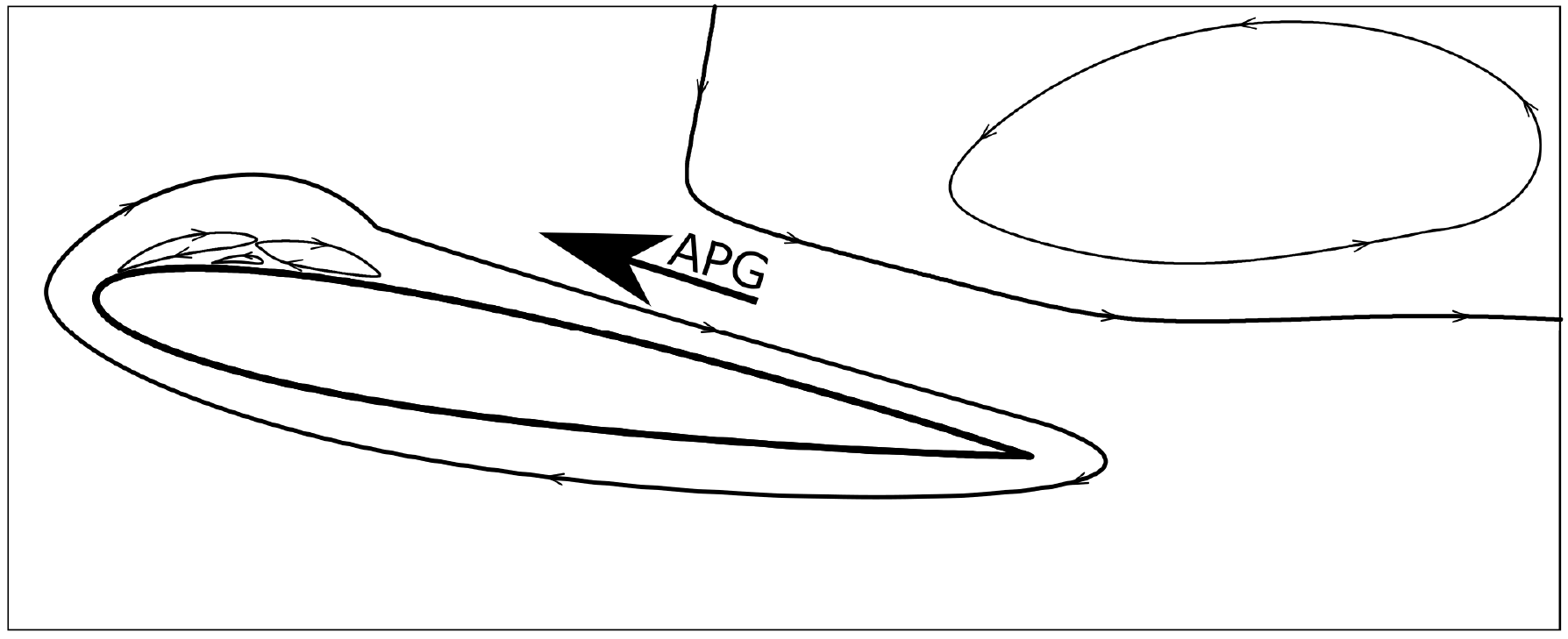}
\textit{$\Phi = 180^{\circ}$}
\end{minipage}
\medskip
\begin{minipage}{220pt}
\centering
\includegraphics[width=220pt, trim={0mm 0mm 0mm 0mm}, clip]{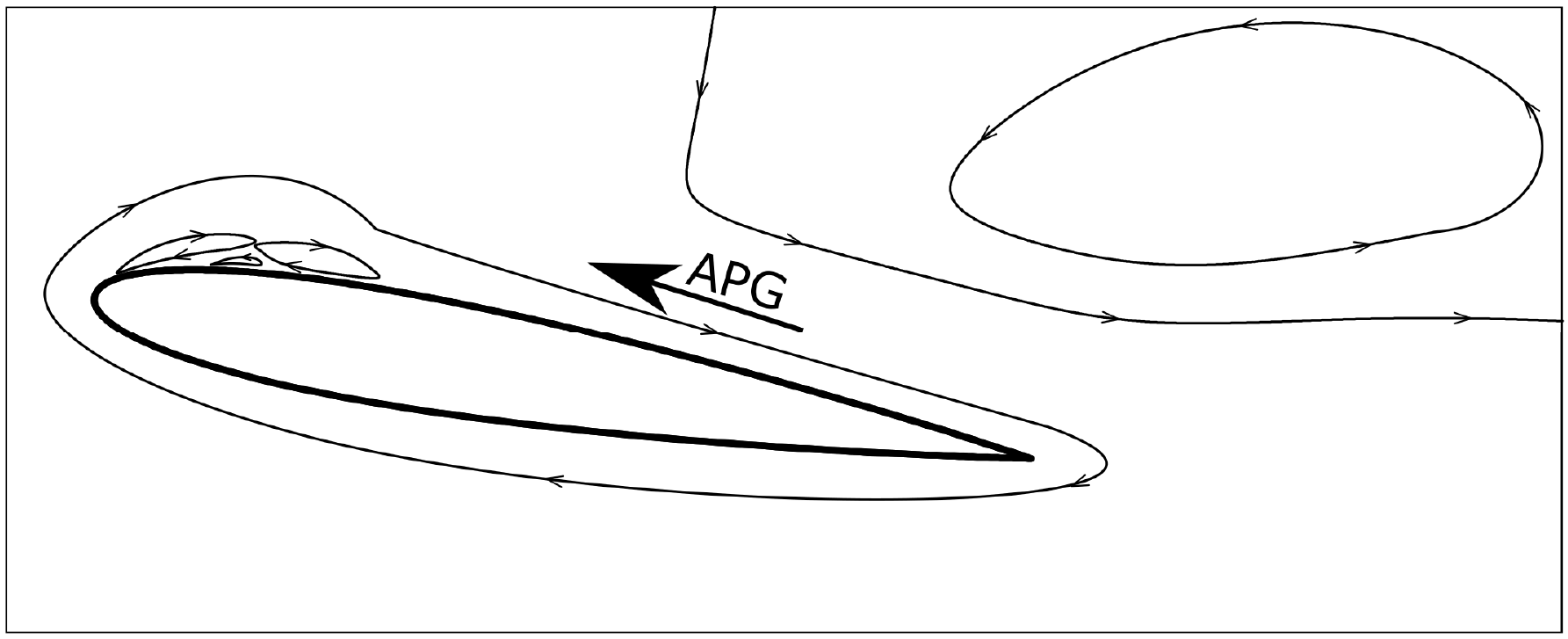}
\textit{$\Phi = 225^{\circ}$}
\end{minipage}
\medskip
\begin{minipage}{220pt}
\centering
\includegraphics[width=220pt, trim={0mm 0mm 0mm 0mm}, clip]{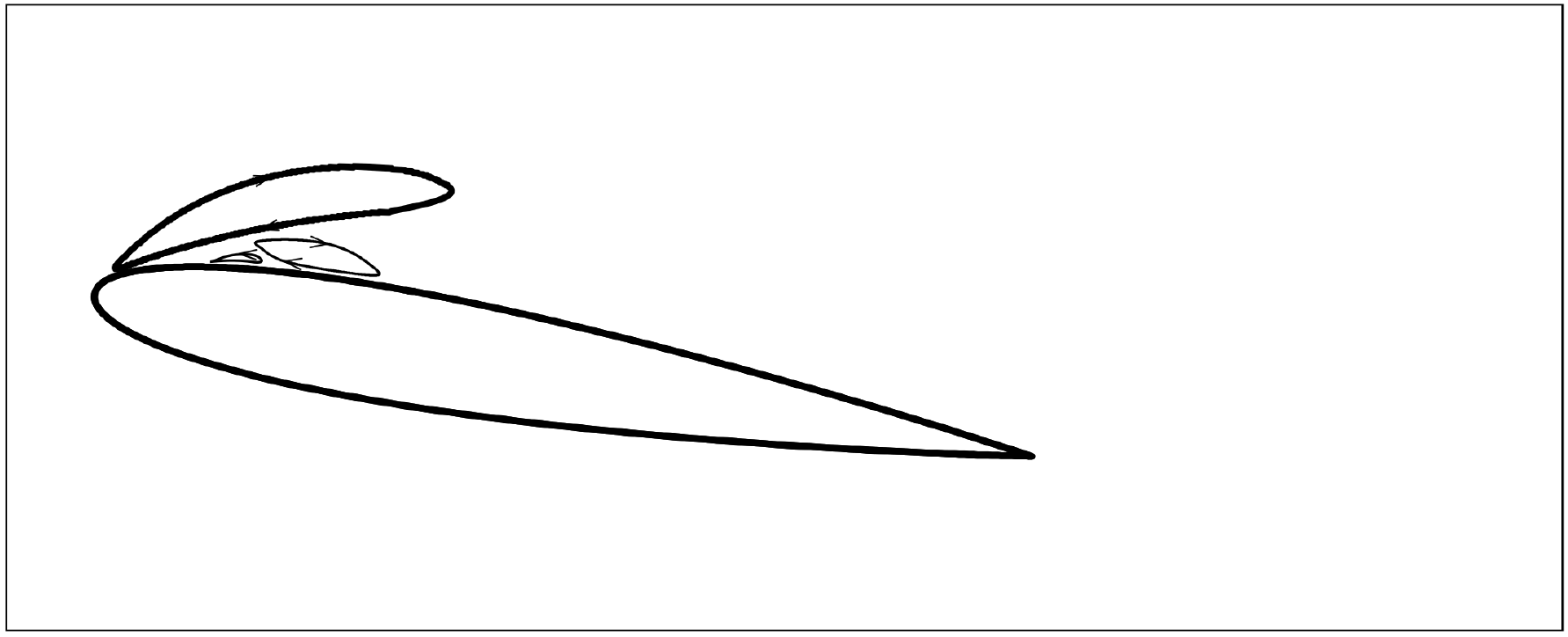}
\textit{$\Phi = 270^{\circ}$}
\end{minipage}
\begin{minipage}{220pt}
\centering
\includegraphics[width=220pt, trim={0mm 0mm 0mm 0mm}, clip]{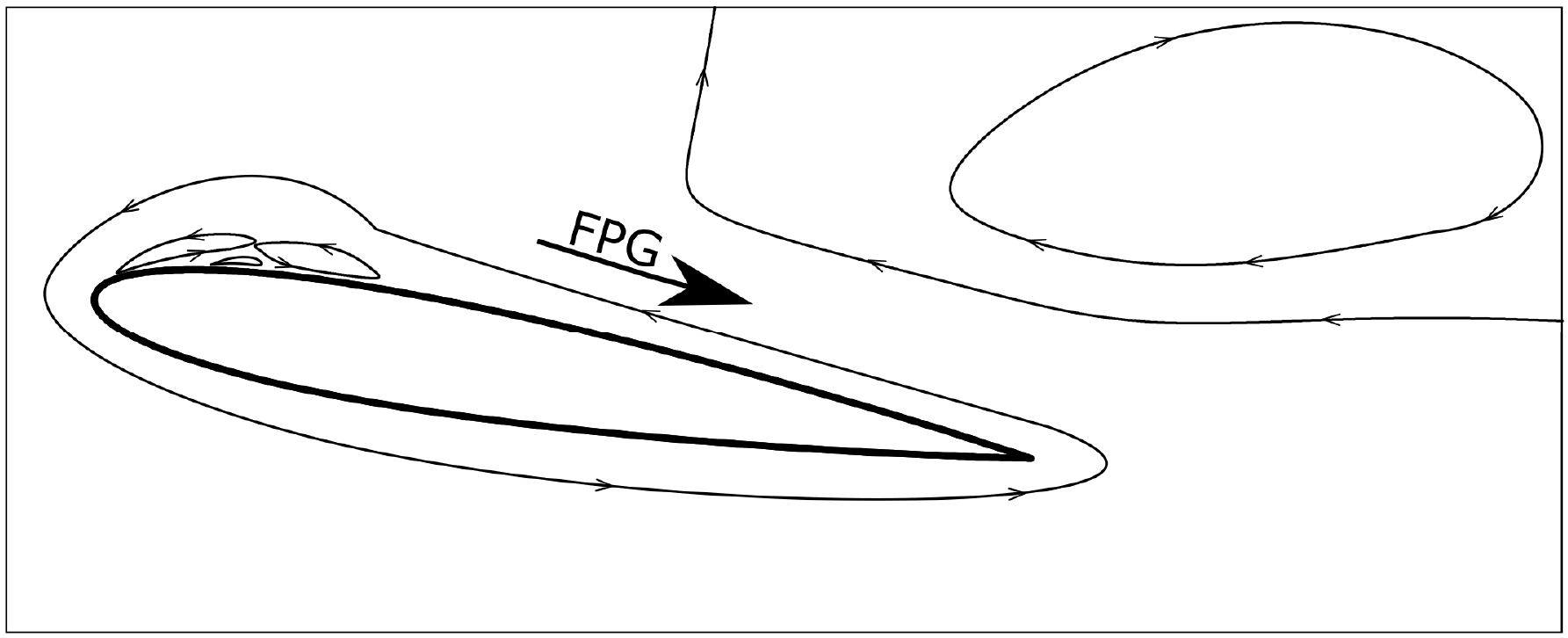}
\textit{$\Phi = 315^{\circ}$}
\end{minipage}
\caption{A model that describe the underlying mechanism behind the self-sustained quasi-periodic low-frequency f\/low oscillation featuring the bursting and reformation cycle of a laminar separation bubble. Line and arrow width displays the strength of the pressure gradient along the aerofoil chord, the oscillating-f\/low, and the UV of the TCV.}
\label{model}
\end{center}
\end{figure}

Most of the previous experimental and numerical work that captured the LFO reported that some disturbed cycles do not resemble other regular cycles,~\citet{tanaka2004flow} and~\citet{eljack2017high}. The underlying mechanism presented in $\S$~\ref{mechanism} is assumed to be perfectly periodic. If the LFO-control-parameters are perfectly synchronised in such a way that they oscillate in-phase, then they will oscillate in harmony and drive the f\/low into equilibrium periodically. Thus, the LFO process would repeat periodically. Whereas, when the LFO-control-parameters are not synchronised, for reasons that are not clear now, irregular disturbed cycles that do not resemble other regular cycles are occasionally produced. The reason behind such disturbed cycles is that the sequence of events as described in $\S$~\ref{mechanism} is interrupted, and the process is reversed before it is completed. For instance, in f\/igure~\ref{model}, at $\Phi = 45^{\circ}$ the f\/low-f\/ield is attaching and the lift coeff\/icient is increasing. If the LFO-control-parameters are synchronised, the sequence of events will continue as usual to that at $\Phi = 90^{\circ}$ and $135^{\circ}$. However, sometimes the process from $\Phi = 45^{\circ}$ goes back to $\Phi = 0^{\circ}$ instead of continuing to $\Phi = 90^{\circ}$ and the periodic cycle is interrupted.\newline
Another issue of major importance is the variation of the LFO cycle with the angle of attack and the route through which an aerofoil undergoes a full stall. As mentioned before, the oscillating-f\/low is present in all of the investigated angles of attack, including at zero angle of attack. The magnitude of the oscillation increases and the frequency of the oscillation decreases as the angle of attack is increased. As shown by~\citet{eljack2018bursting}, the mean f\/low-f\/ield is attached at the angle of attack of $9.25^{\circ}$ and fully separated at the angle of attack of $10.5^{\circ}$. At the angle of attack of $9.25^{\circ}$, the mean-f\/low is attached, the UV of the TCV occasionally expands; consequently, the LSB bursts and the f\/low-f\/ield separates intermittently. As the angle of attack increases, the UV expands more frequently. At the angle of attack of $9.8^{\circ}$, the LFO becomes fully developed and quasi-periodic. At higher angles of attack, the mean-f\/low is separated. Sometimes the UV expands, the f\/low attaches, and the conditions for the reformation of the LSB are met. Consequently, the f\/low-f\/ield remains separated with intermittent attachments. As the angle of attack is further increased, the UV expands less frequently, and the f\/low-f\/ield remains separated for longer time intervals. At a critical angle of attack, the UV never expands, the f\/low-f\/ield remains separated, and the aerofoil undergoes a full stall.

\subsection{Statistics of the conditionally phase-averaged oscillating-f\/low-f\/ield}
\begin{figure}
\begin{center}
\begin{minipage}{220pt}
\centering
\includegraphics[width=220pt, trim={0mm 0mm 0mm 0mm}, clip]{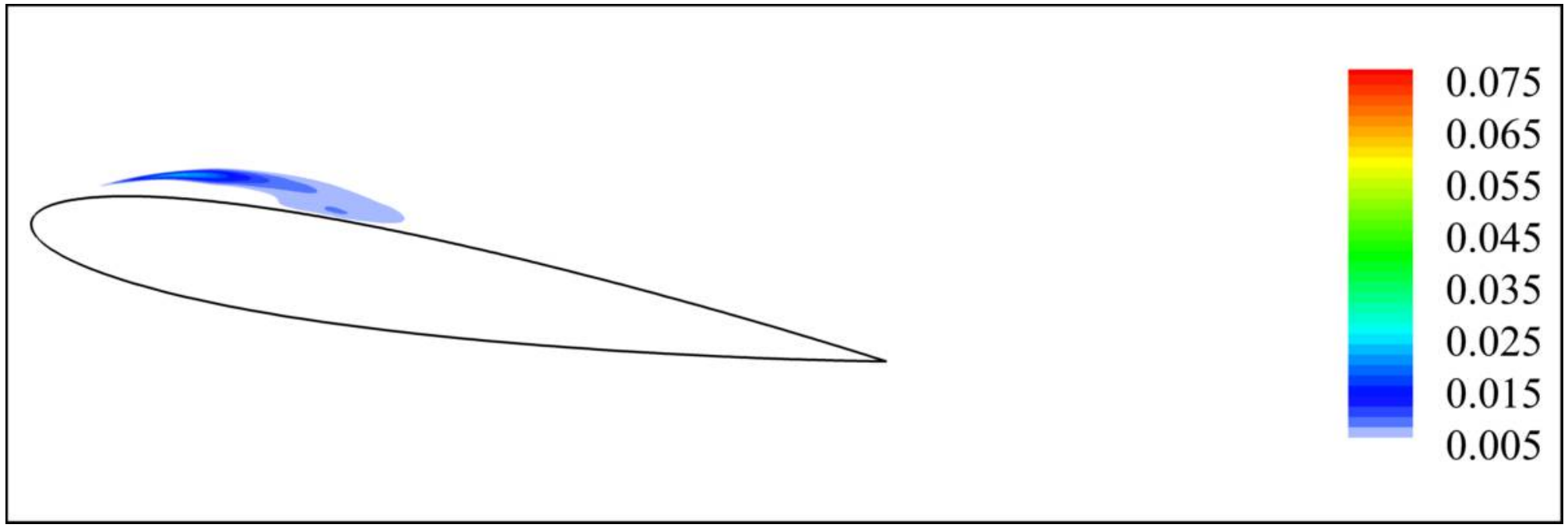}
\textit{$\alpha = 9.25^{\circ}$}
\end{minipage}
\medskip
\begin{minipage}{220pt}
\centering
\includegraphics[width=220pt, trim={0mm 0mm 0mm 0mm}, clip]{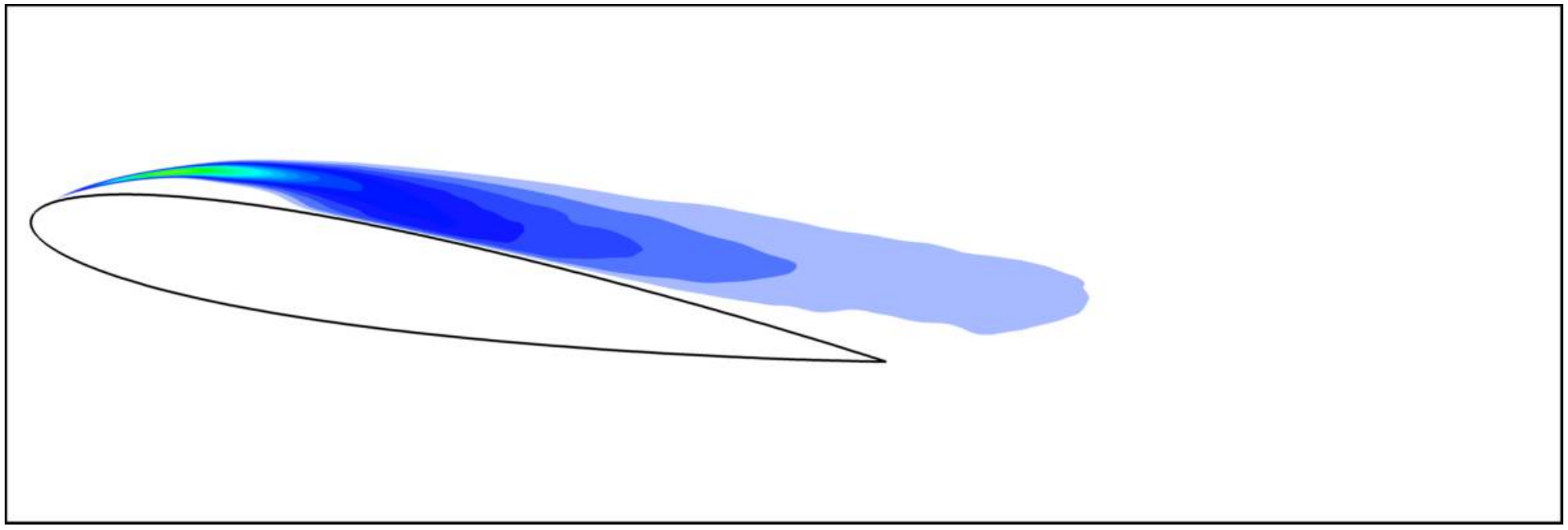}
\textit{$\alpha = 9.40^{\circ}$}
\end{minipage}
\medskip
\begin{minipage}{220pt}
\centering
\includegraphics[width=220pt, trim={0mm 0mm 0mm 0mm}, clip]{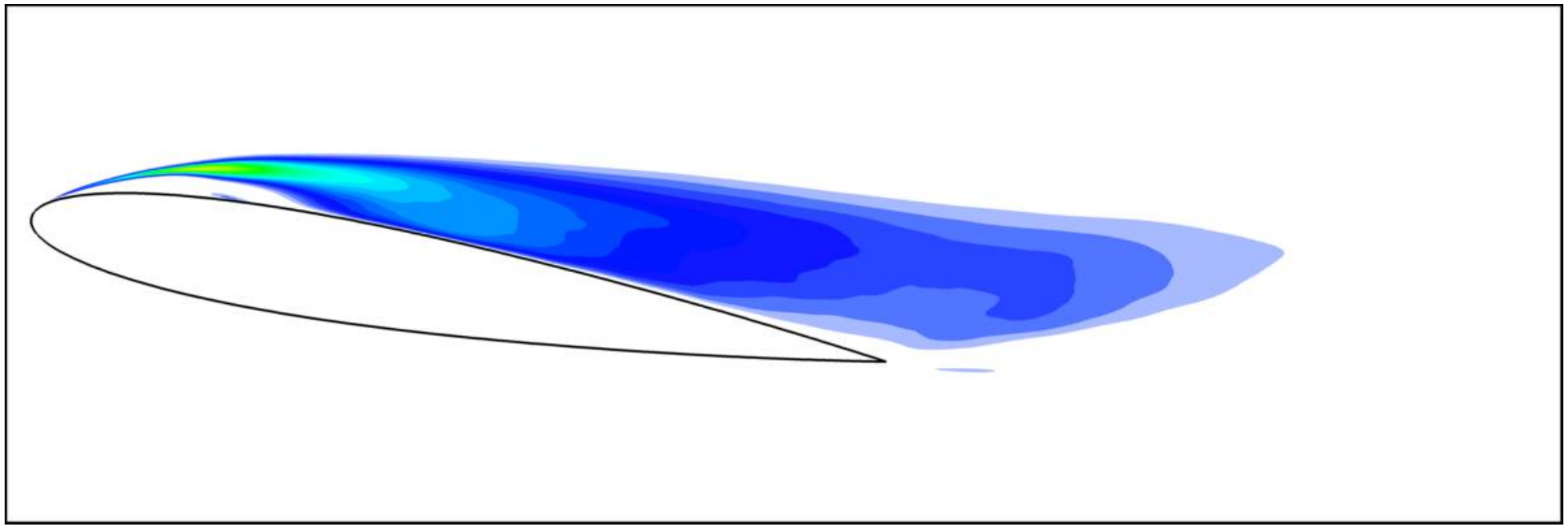}
\textit{$\alpha = 9.50^{\circ}$}
\end{minipage}
\medskip
\begin{minipage}{220pt}
\centering
\includegraphics[width=220pt, trim={0mm 0mm 0mm 0mm}, clip]{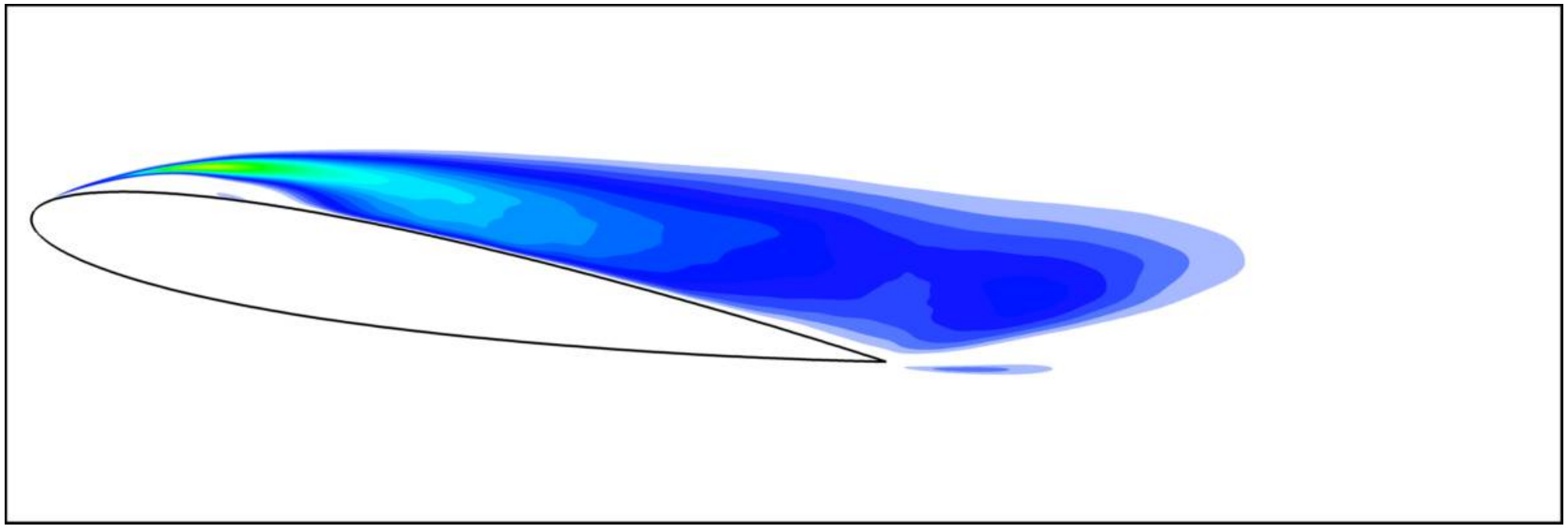}
\textit{$\alpha = 9.60^{\circ}$}
\end{minipage}
\medskip
\begin{minipage}{220pt}
\centering
\includegraphics[width=220pt, trim={0mm 0mm 0mm 0mm}, clip]{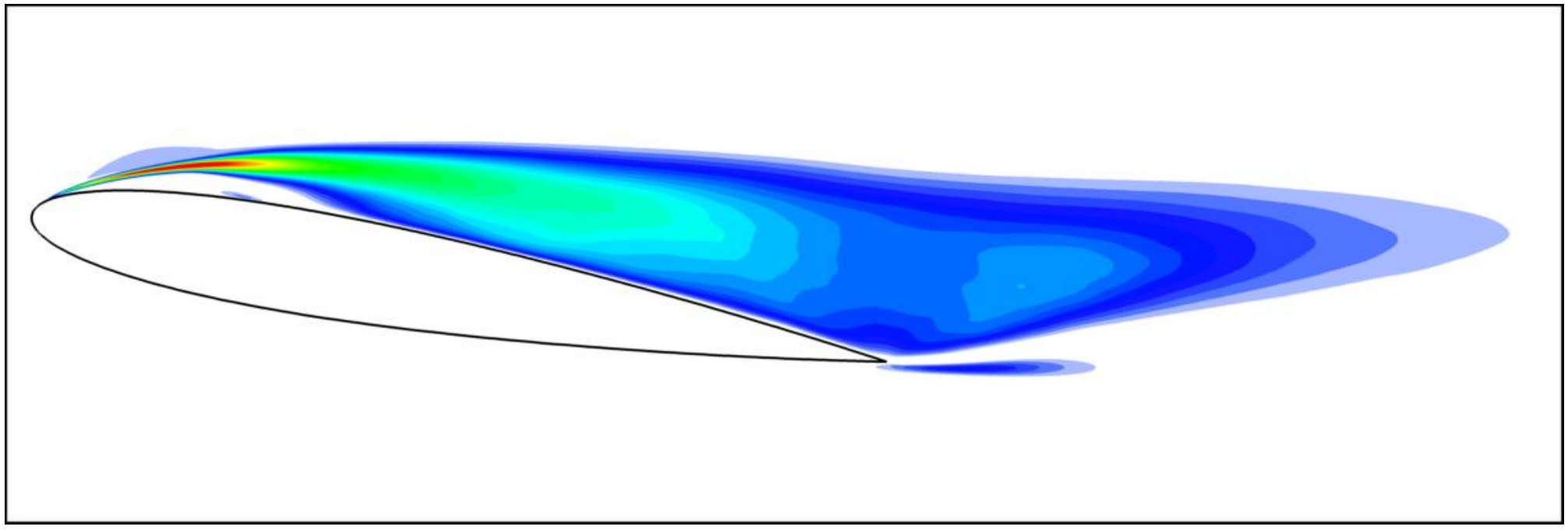}
\textit{$\alpha = 9.70^{\circ}$}
\end{minipage}
\medskip
\begin{minipage}{220pt}
\centering
\includegraphics[width=220pt, trim={0mm 0mm 0mm 0mm}, clip]{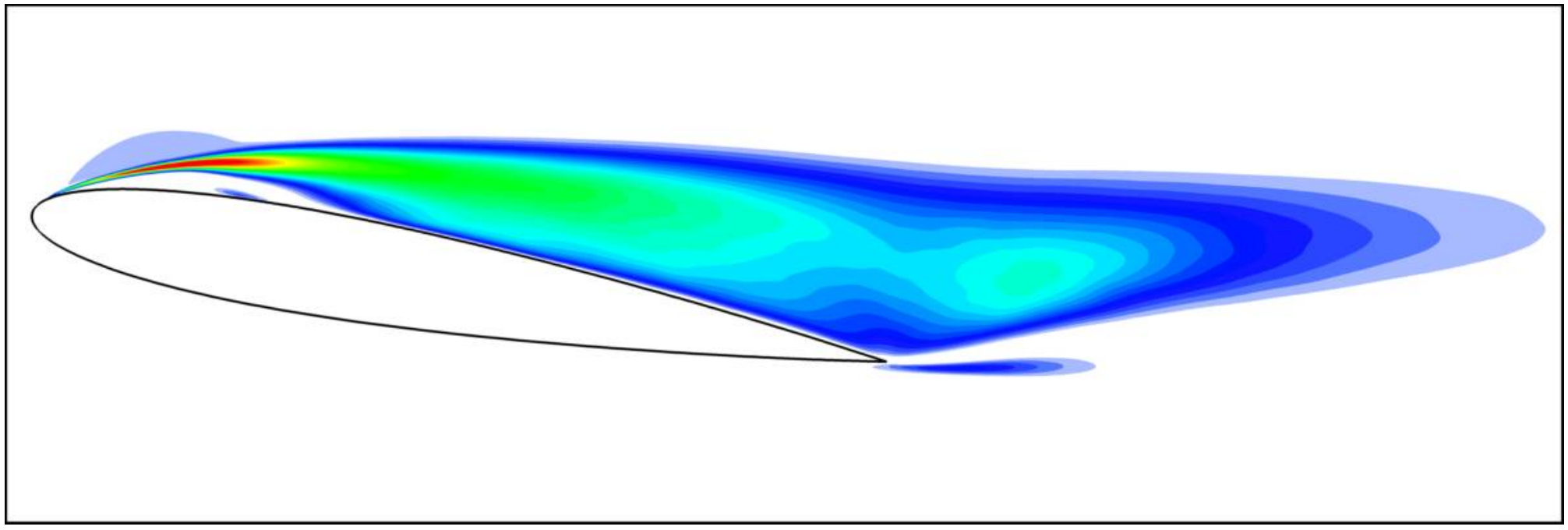}
\textit{$\alpha = 9.80^{\circ}$}
\end{minipage}
\medskip
\begin{minipage}{220pt}
\centering
\includegraphics[width=220pt, trim={0mm 0mm 0mm 0mm}, clip]{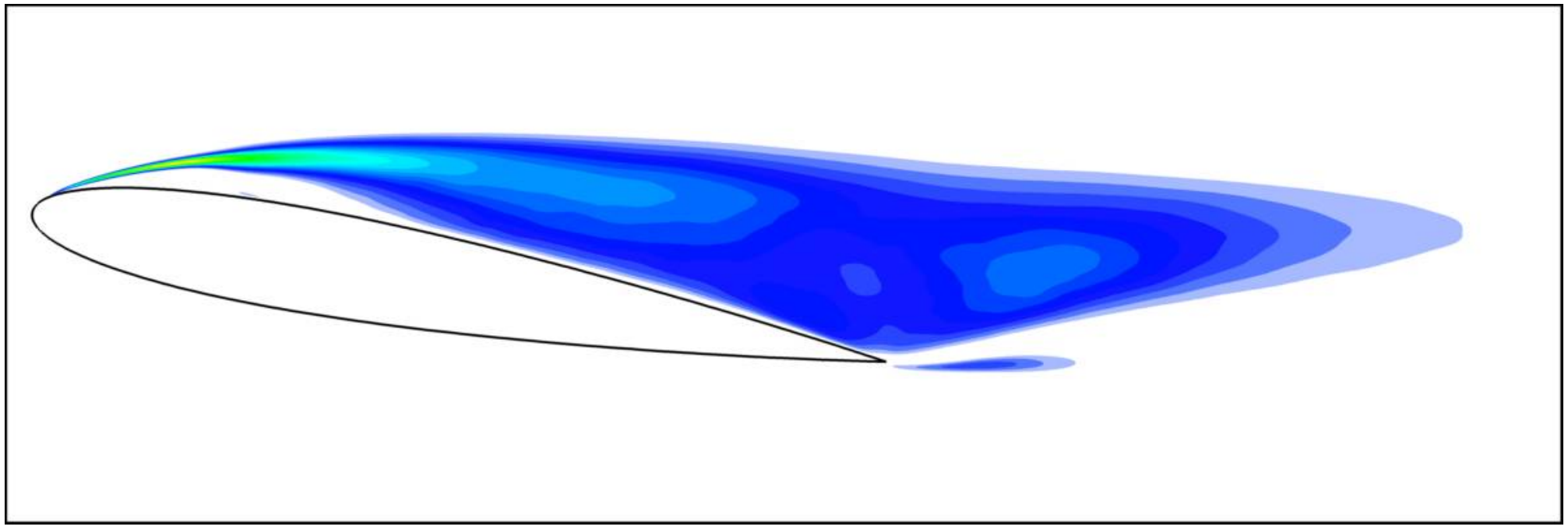}
\textit{$\alpha = 9.90^{\circ}$}
\end{minipage}
\medskip
\begin{minipage}{220pt}
\centering
\includegraphics[width=220pt, trim={0mm 0mm 0mm 0mm}, clip]{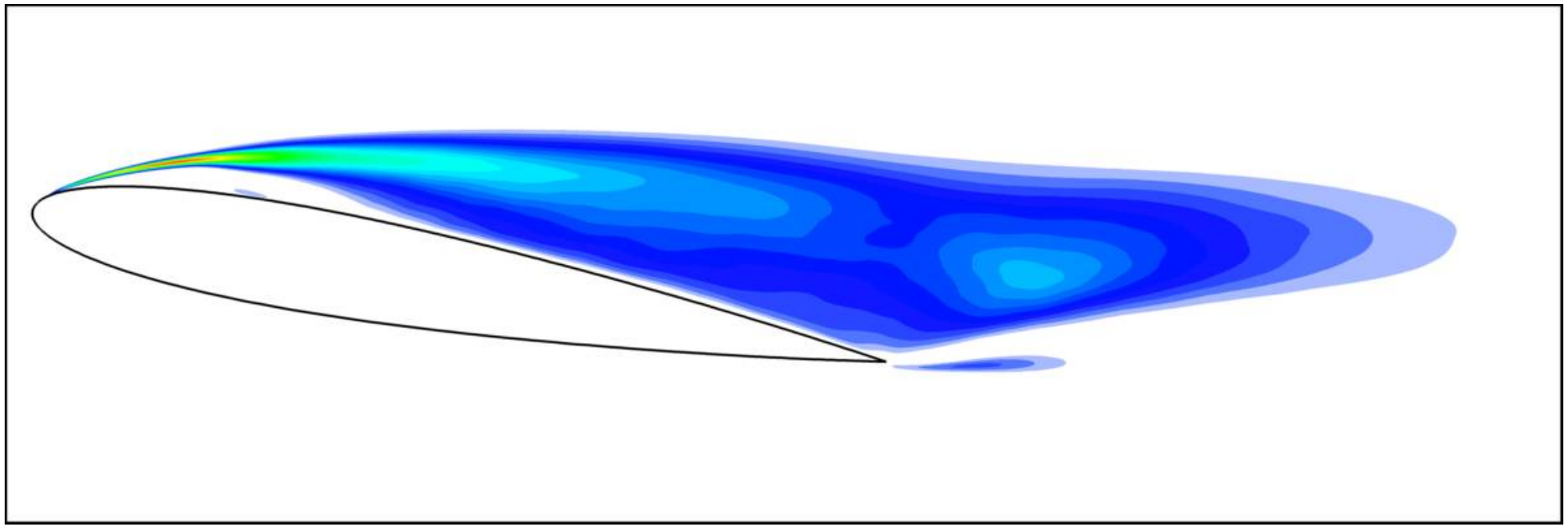}
\textit{$\alpha = 10.0^{\circ}$}
\end{minipage}
\medskip
\begin{minipage}{220pt}
\centering
\includegraphics[width=220pt, trim={0mm 0mm 0mm 0mm}, clip]{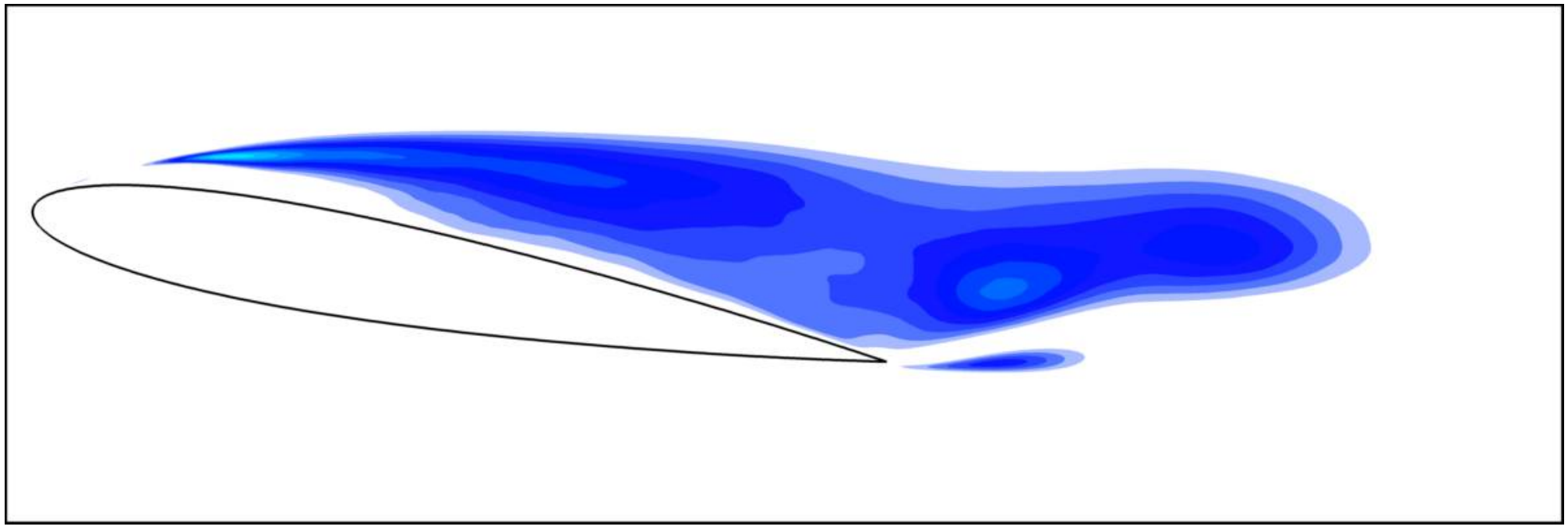}
\textit{$\alpha = 10.1^{\circ}$}
\end{minipage}
\begin{minipage}{220pt}
\centering
\includegraphics[width=220pt, trim={0mm 0mm 0mm 0mm}, clip]{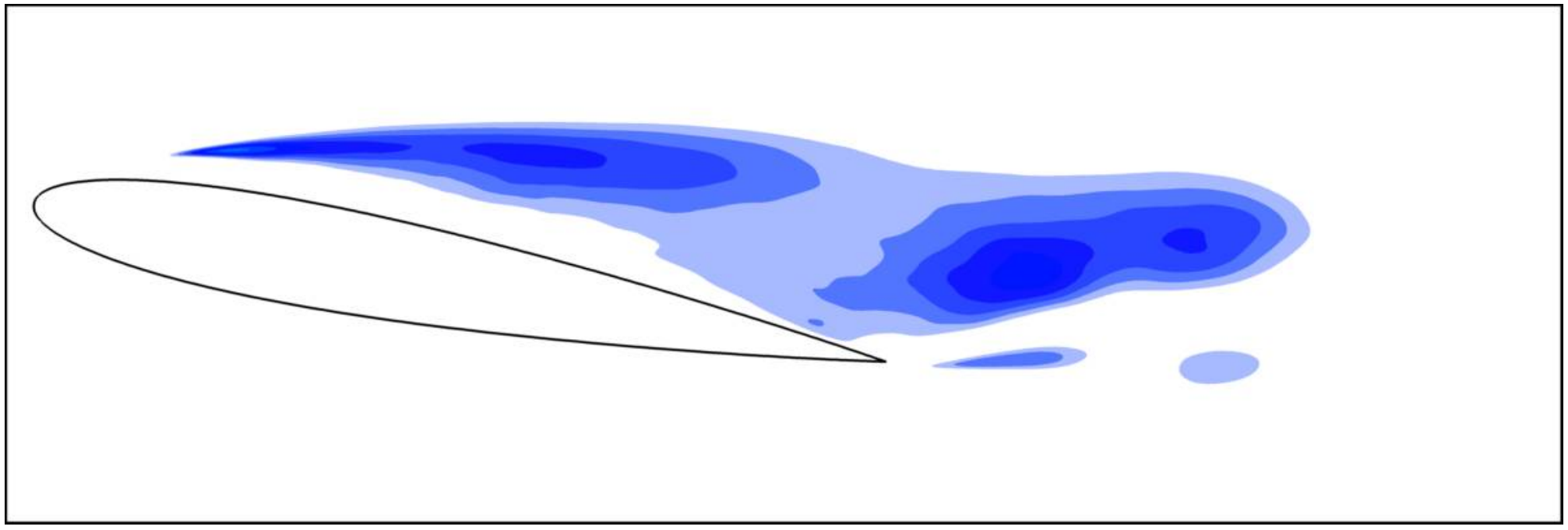}
\textit{$\alpha = 10.5^{\circ}$}
\end{minipage}
\caption{Colour maps of the variance of the phase-averaged streamwise velocity component, $\overline{\widetilde{u}^2}$, for the angles of attack $\alpha = 9.25^{\circ}$--$10.5^{\circ}$.}
\label{u2_phase}
\end{center}
\end{figure}

\begin{figure}
\begin{center}
\begin{minipage}{220pt}
\centering
\includegraphics[width=220pt, trim={0mm 0mm 0mm 0mm}, clip]{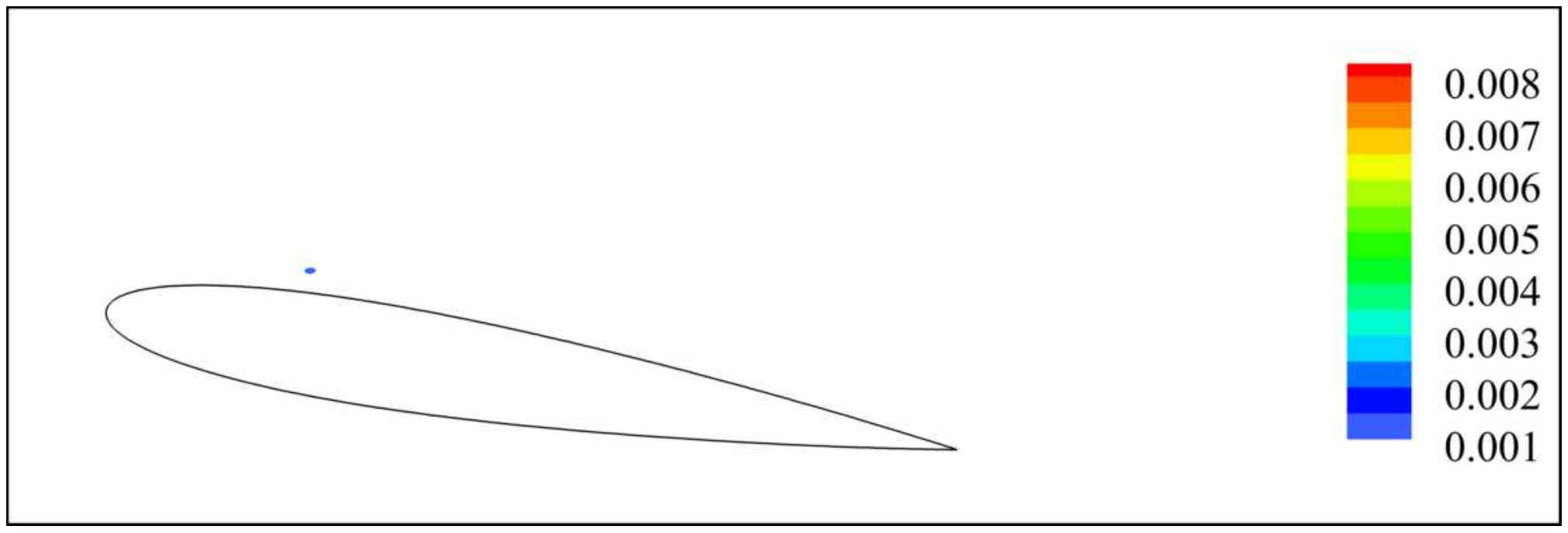}
\textit{$\alpha = 9.25^{\circ}$}
\end{minipage}
\medskip
\begin{minipage}{220pt}
\centering
\includegraphics[width=220pt, trim={0mm 0mm 0mm 0mm}, clip]{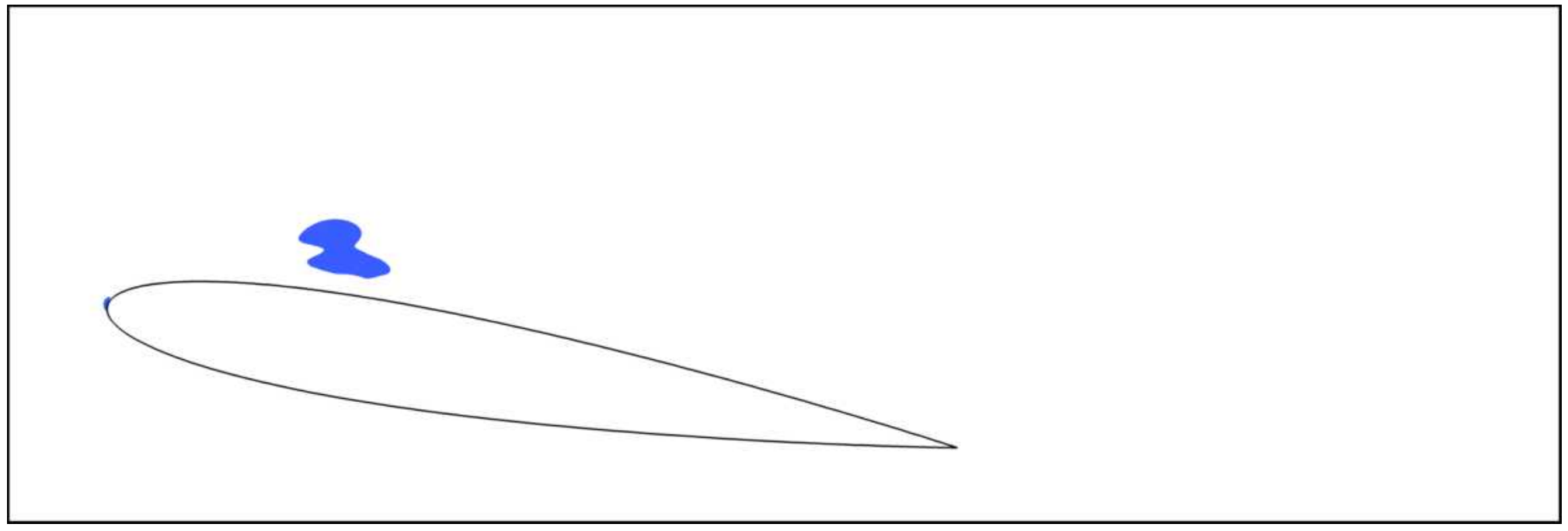}
\textit{$\alpha = 9.40^{\circ}$}
\end{minipage}
\medskip
\begin{minipage}{220pt}
\centering
\includegraphics[width=220pt, trim={0mm 0mm 0mm 0mm}, clip]{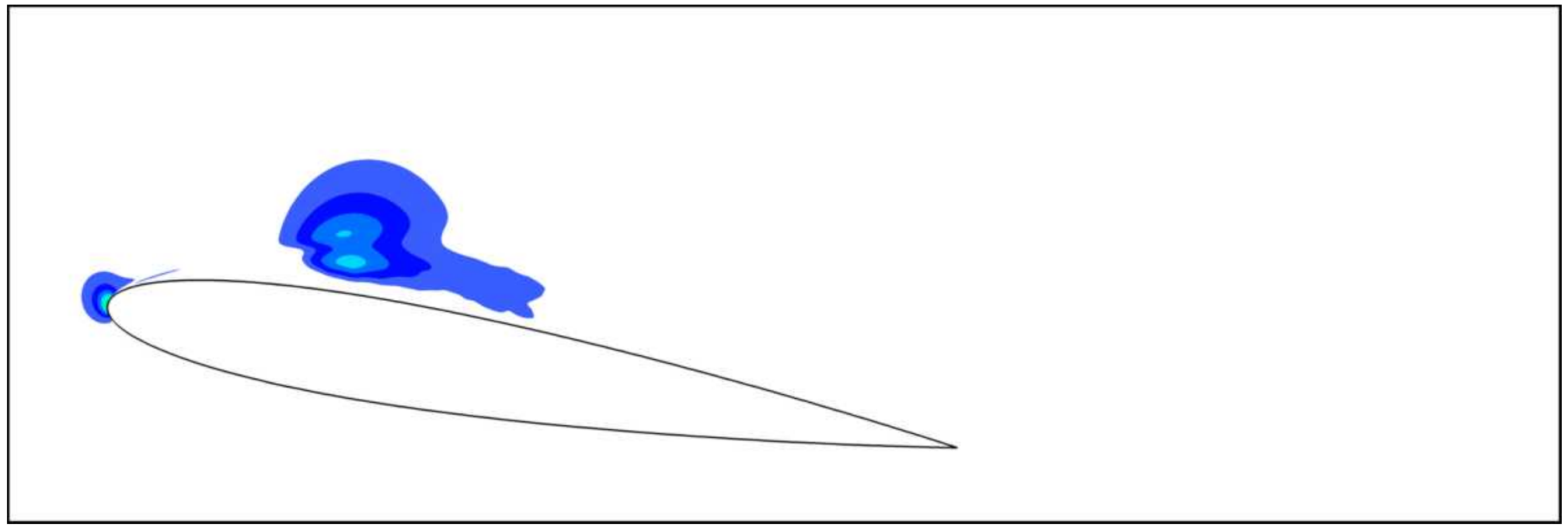}
\textit{$\alpha = 9.50^{\circ}$}
\end{minipage}
\medskip
\begin{minipage}{220pt}
\centering
\includegraphics[width=220pt, trim={0mm 0mm 0mm 0mm}, clip]{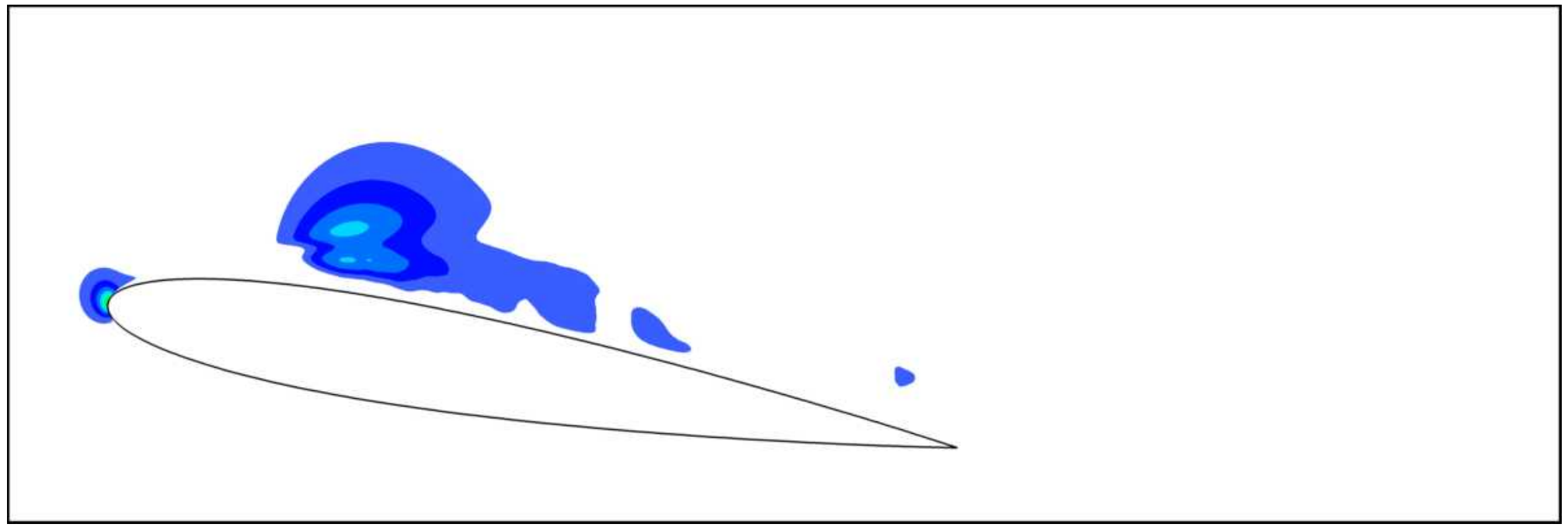}
\textit{$\alpha = 9.60^{\circ}$}
\end{minipage}
\medskip
\begin{minipage}{220pt}
\centering
\includegraphics[width=220pt, trim={0mm 0mm 0mm 0mm}, clip]{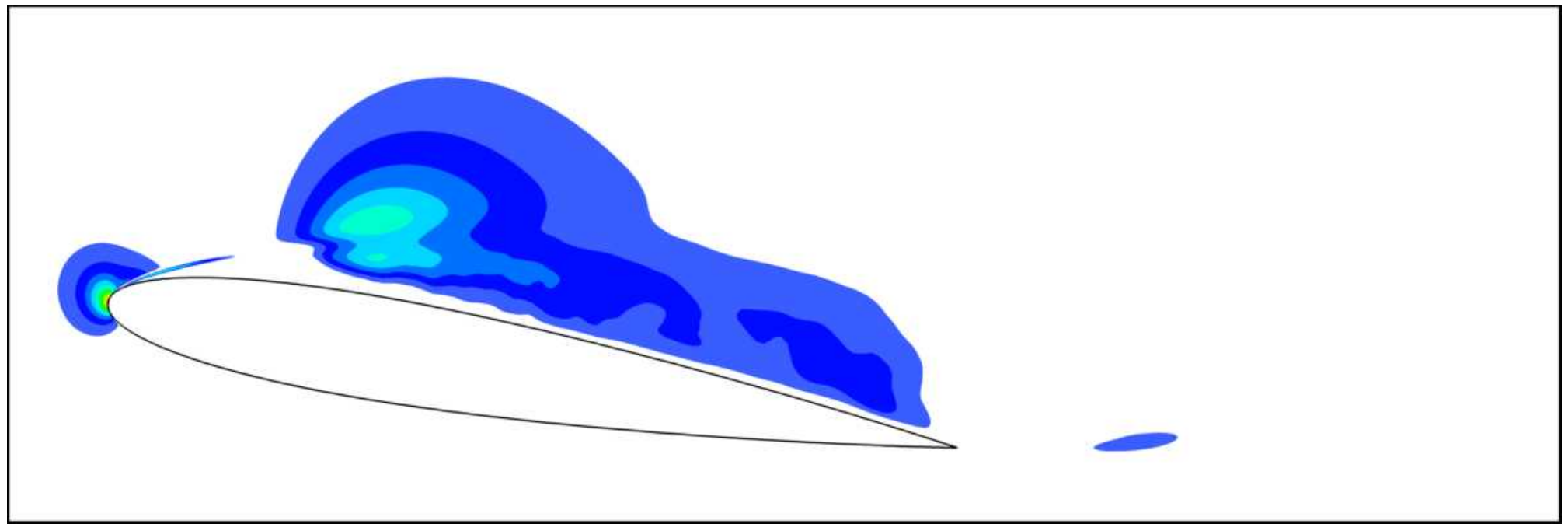}
\textit{$\alpha = 9.70^{\circ}$}
\end{minipage}
\medskip
\begin{minipage}{220pt}
\centering
\includegraphics[width=220pt, trim={0mm 0mm 0mm 0mm}, clip]{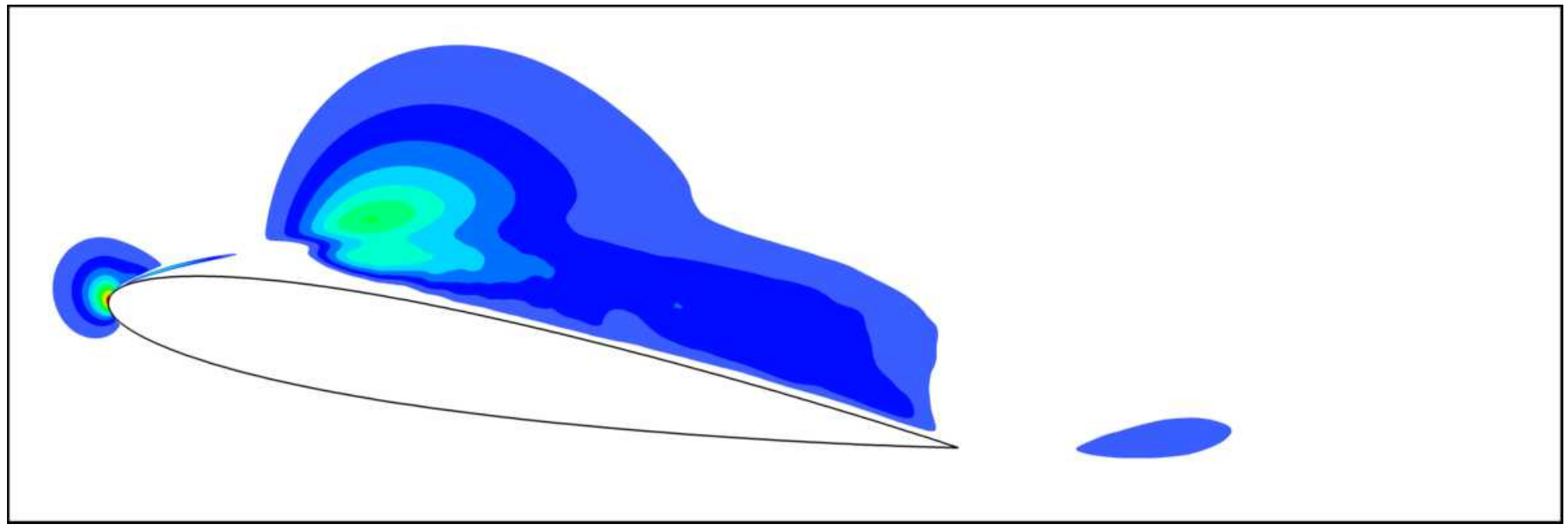}
\textit{$\alpha = 9.80^{\circ}$}
\end{minipage}
\medskip
\begin{minipage}{220pt}
\centering
\includegraphics[width=220pt, trim={0mm 0mm 0mm 0mm}, clip]{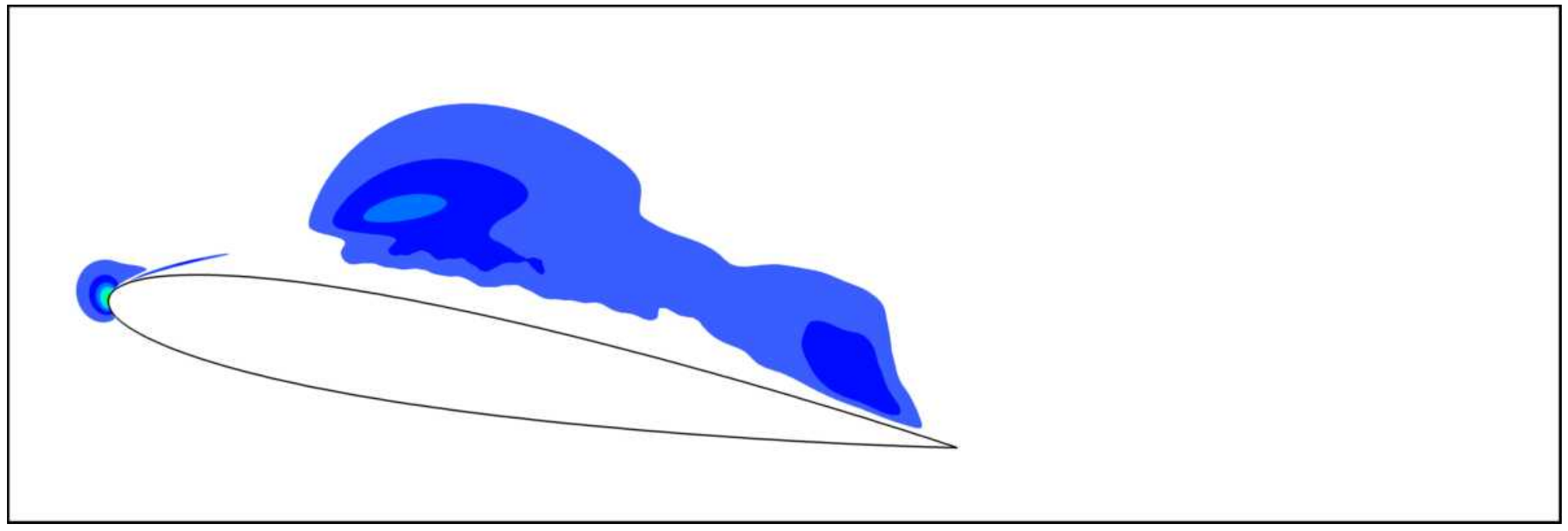}
\textit{$\alpha = 9.90^{\circ}$}
\end{minipage}
\medskip
\begin{minipage}{220pt}
\centering
\includegraphics[width=220pt, trim={0mm 0mm 0mm 0mm}, clip]{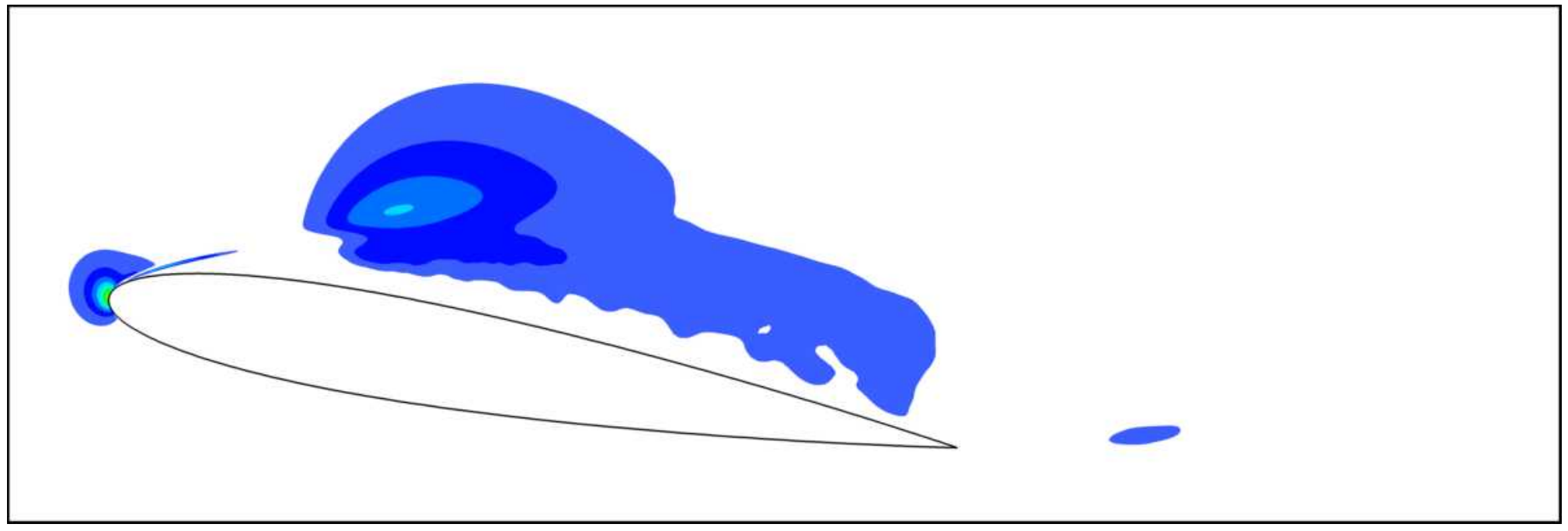}
\textit{$\alpha = 10.0^{\circ}$}
\end{minipage}
\medskip
\begin{minipage}{220pt}
\centering
\includegraphics[width=220pt, trim={0mm 0mm 0mm 0mm}, clip]{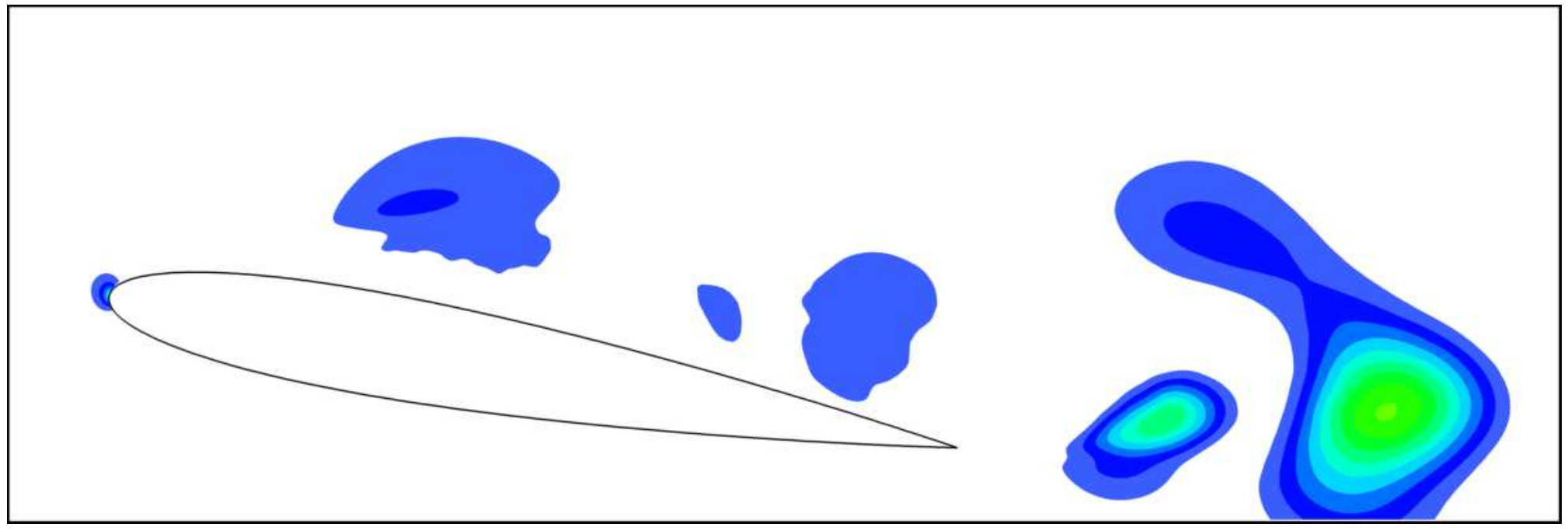}
\textit{$\alpha = 10.1^{\circ}$}
\end{minipage}
\begin{minipage}{220pt}
\centering
\includegraphics[width=220pt, trim={0mm 0mm 0mm 0mm}, clip]{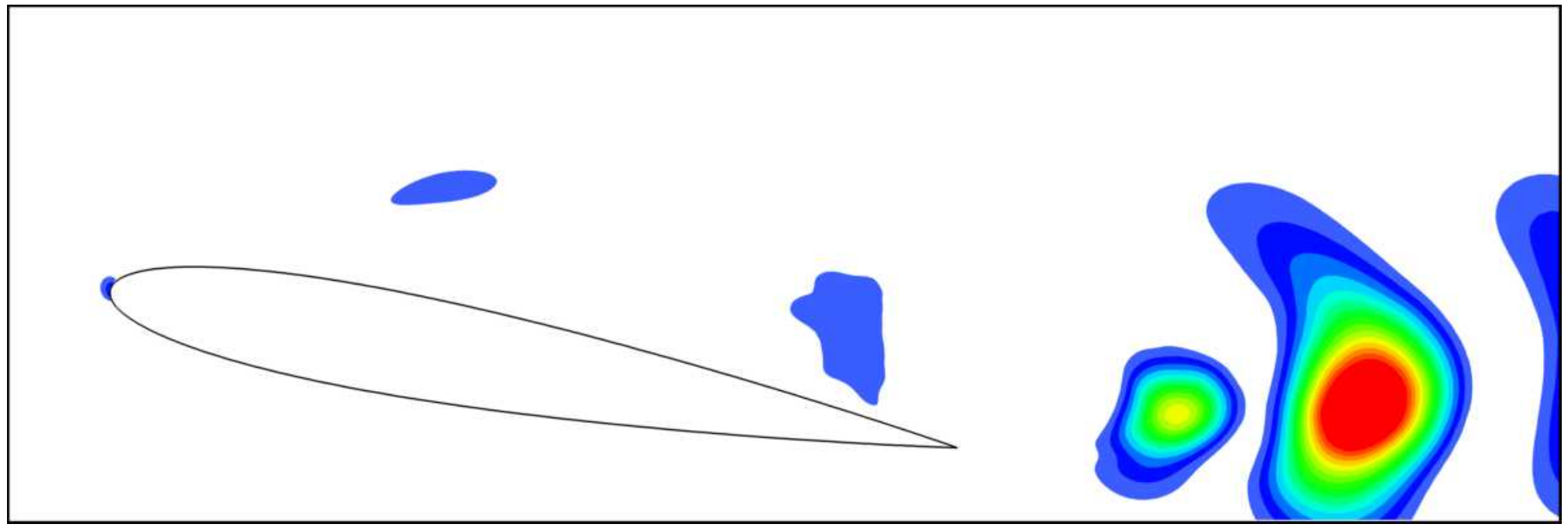}
\textit{$\alpha = 10.5^{\circ}$}
\end{minipage}
\caption{Colour maps of the variance of the phase-averaged wall-normal velocity component, $\overline{\widetilde{v}^2}$, for the angles of attack $\alpha = 9.25^{\circ}$--$10.5^{\circ}$.}
\label{v2_phase}
\end{center}
\end{figure}

\begin{figure}
\begin{center}
\begin{minipage}{220pt}
\centering
\includegraphics[width=220pt, trim={0mm 0mm 0mm 0mm}, clip]{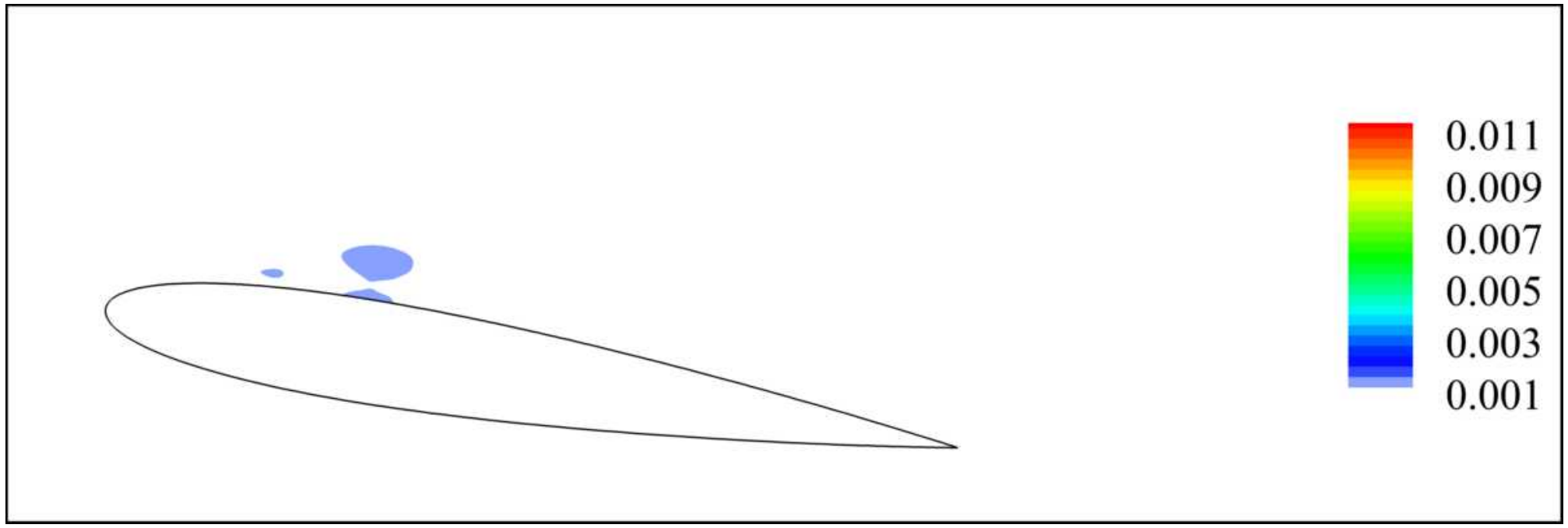}
\textit{$\alpha = 9.25^{\circ}$}
\end{minipage}
\medskip
\begin{minipage}{220pt}
\centering
\includegraphics[width=220pt, trim={0mm 0mm 0mm 0mm}, clip]{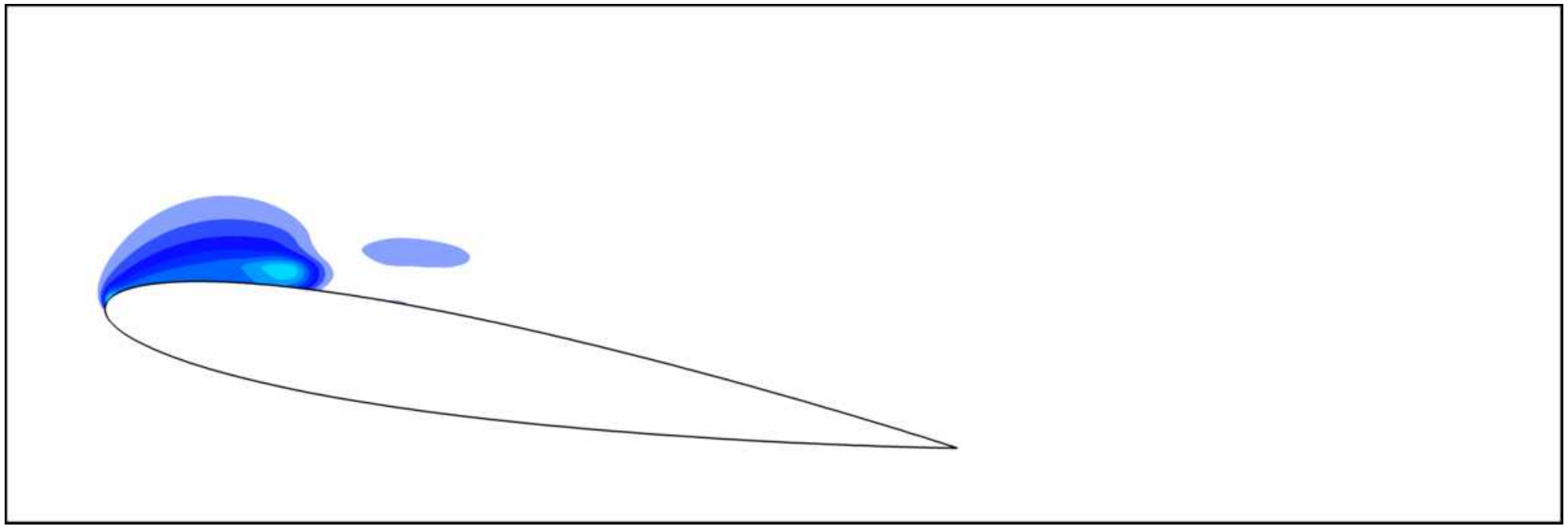}
\textit{$\alpha = 9.40^{\circ}$}
\end{minipage}
\medskip
\begin{minipage}{220pt}
\centering
\includegraphics[width=220pt, trim={0mm 0mm 0mm 0mm}, clip]{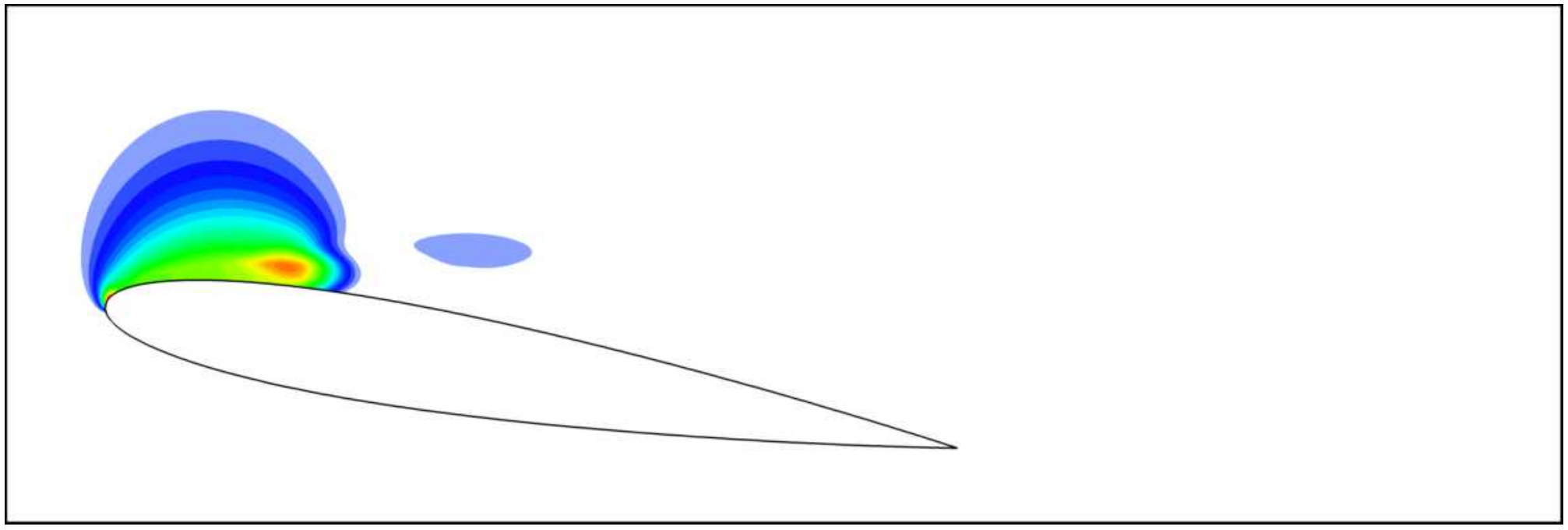}
\textit{$\alpha = 9.50^{\circ}$}
\end{minipage}
\medskip
\begin{minipage}{220pt}
\centering
\includegraphics[width=220pt, trim={0mm 0mm 0mm 0mm}, clip]{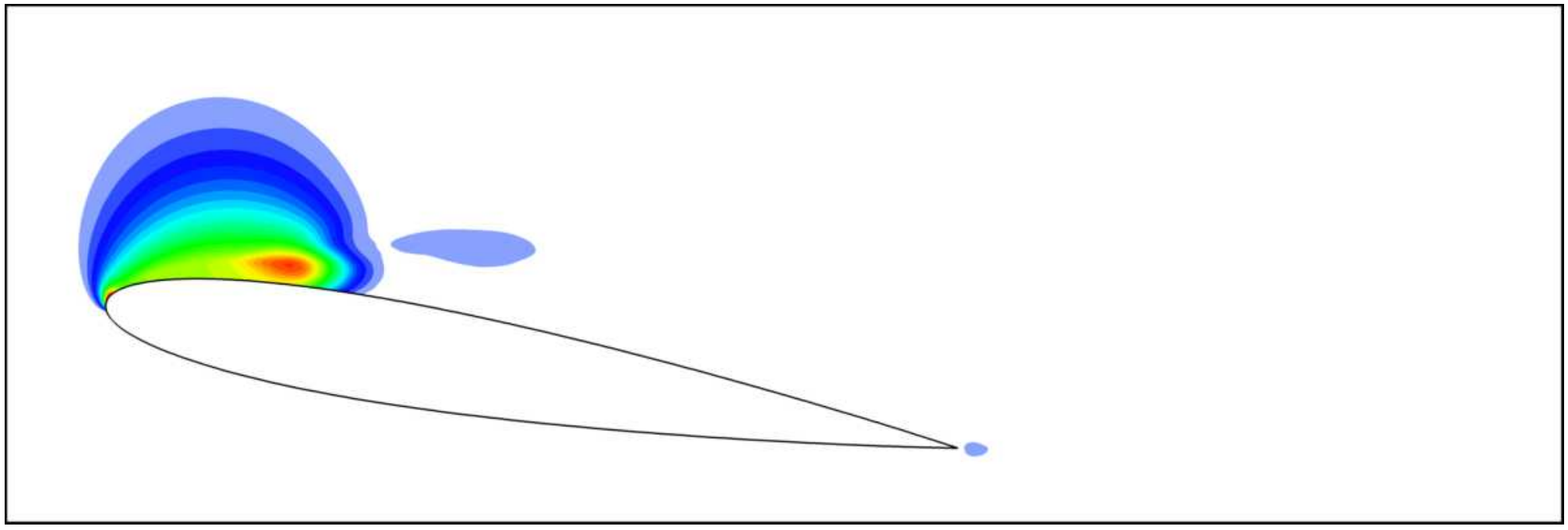}
\textit{$\alpha = 9.60^{\circ}$}
\end{minipage}
\medskip
\begin{minipage}{220pt}
\centering
\includegraphics[width=220pt, trim={0mm 0mm 0mm 0mm}, clip]{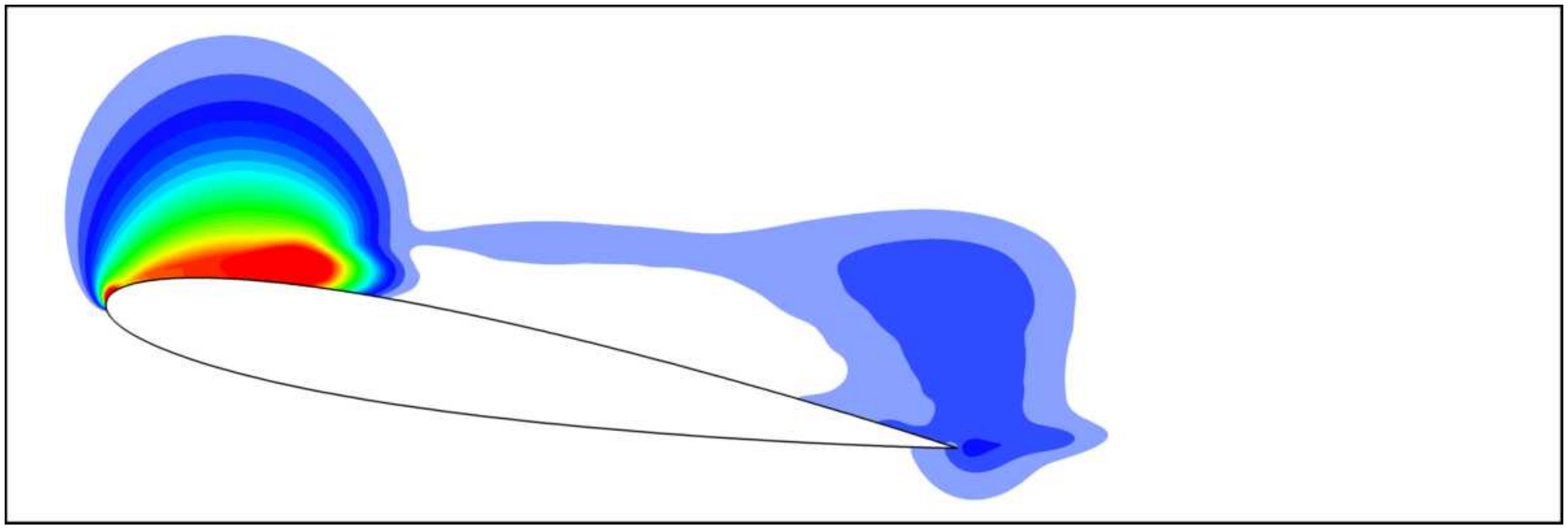}
\textit{$\alpha = 9.70^{\circ}$}
\end{minipage}
\medskip
\begin{minipage}{220pt}
\centering
\includegraphics[width=220pt, trim={0mm 0mm 0mm 0mm}, clip]{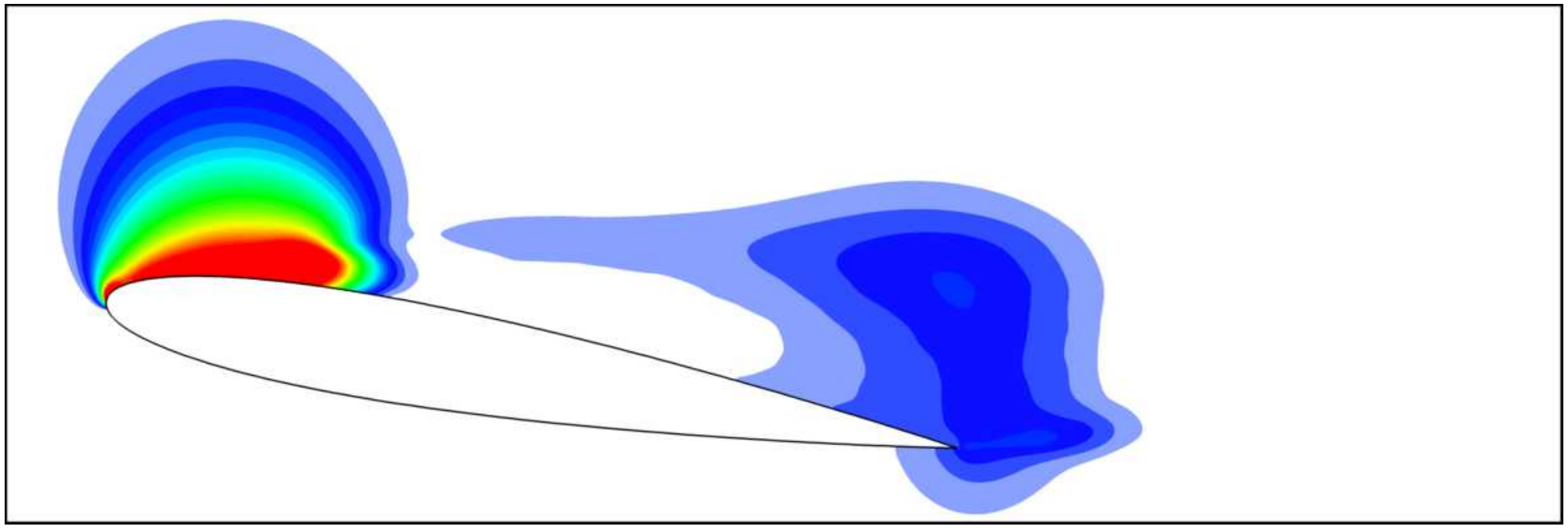}
\textit{$\alpha = 9.80^{\circ}$}
\end{minipage}
\medskip
\begin{minipage}{220pt}
\centering
\includegraphics[width=220pt, trim={0mm 0mm 0mm 0mm}, clip]{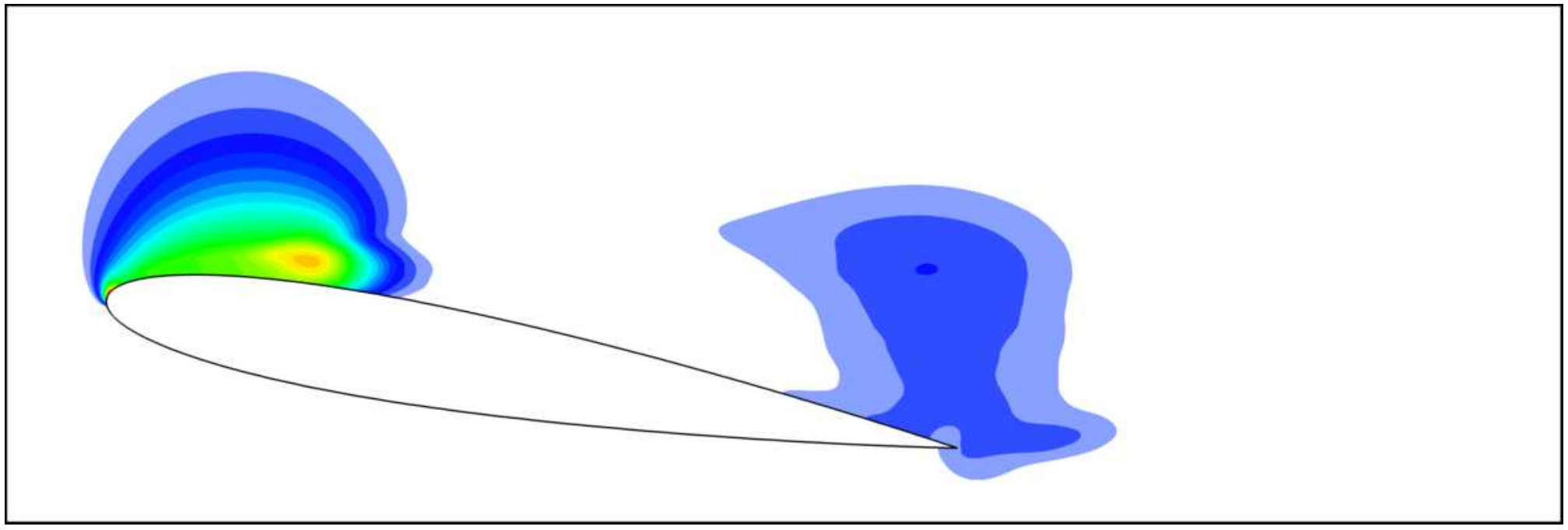}
\textit{$\alpha = 9.90^{\circ}$}
\end{minipage}
\medskip
\begin{minipage}{220pt}
\centering
\includegraphics[width=220pt, trim={0mm 0mm 0mm 0mm}, clip]{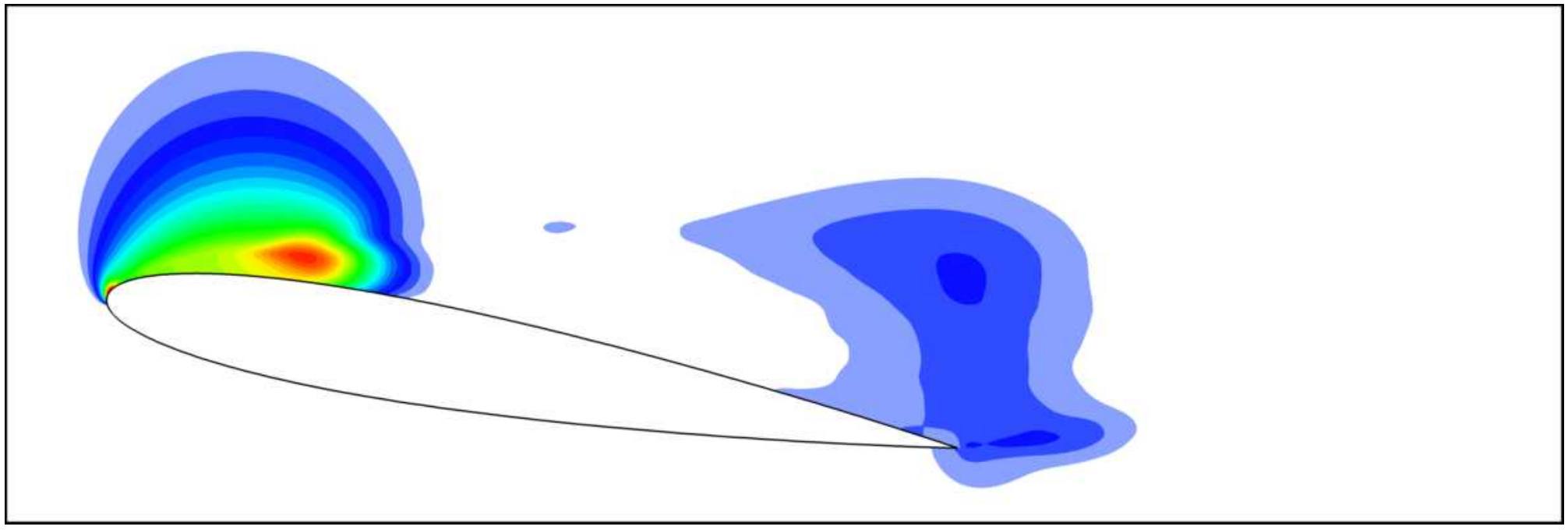}
\textit{$\alpha = 10.0^{\circ}$}
\end{minipage}
\medskip
\begin{minipage}{220pt}
\centering
\includegraphics[width=220pt, trim={0mm 0mm 0mm 0mm}, clip]{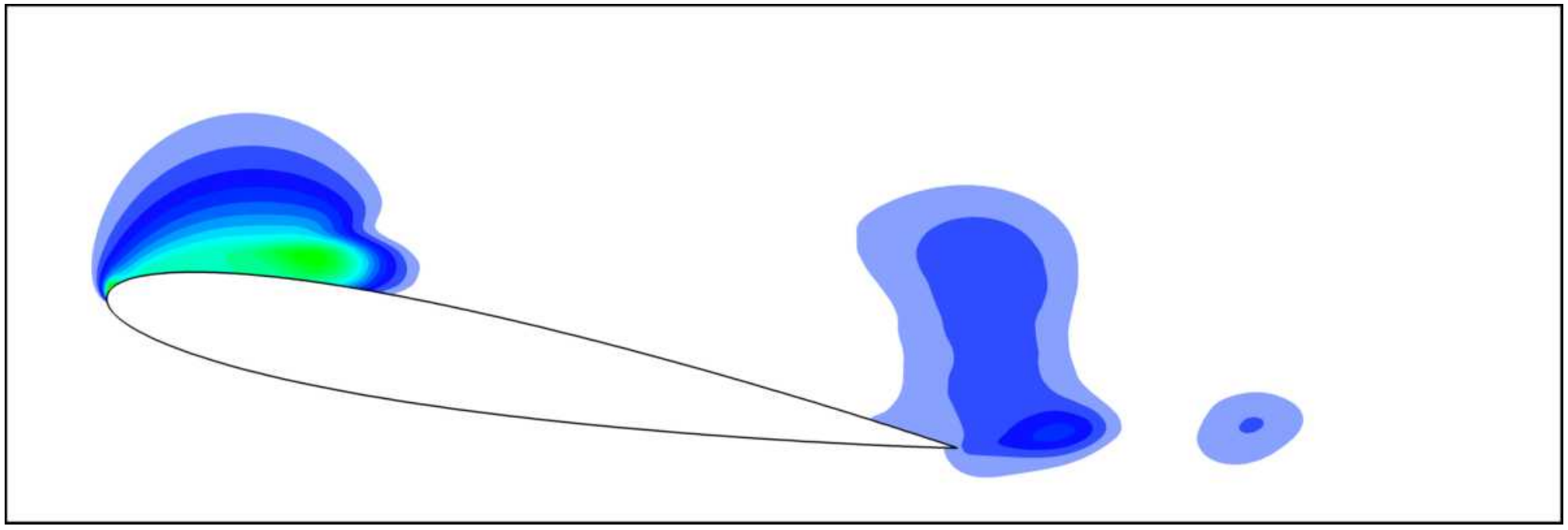}
\textit{$\alpha = 10.1^{\circ}$}
\end{minipage}
\begin{minipage}{220pt}
\centering
\includegraphics[width=220pt, trim={0mm 0mm 0mm 0mm}, clip]{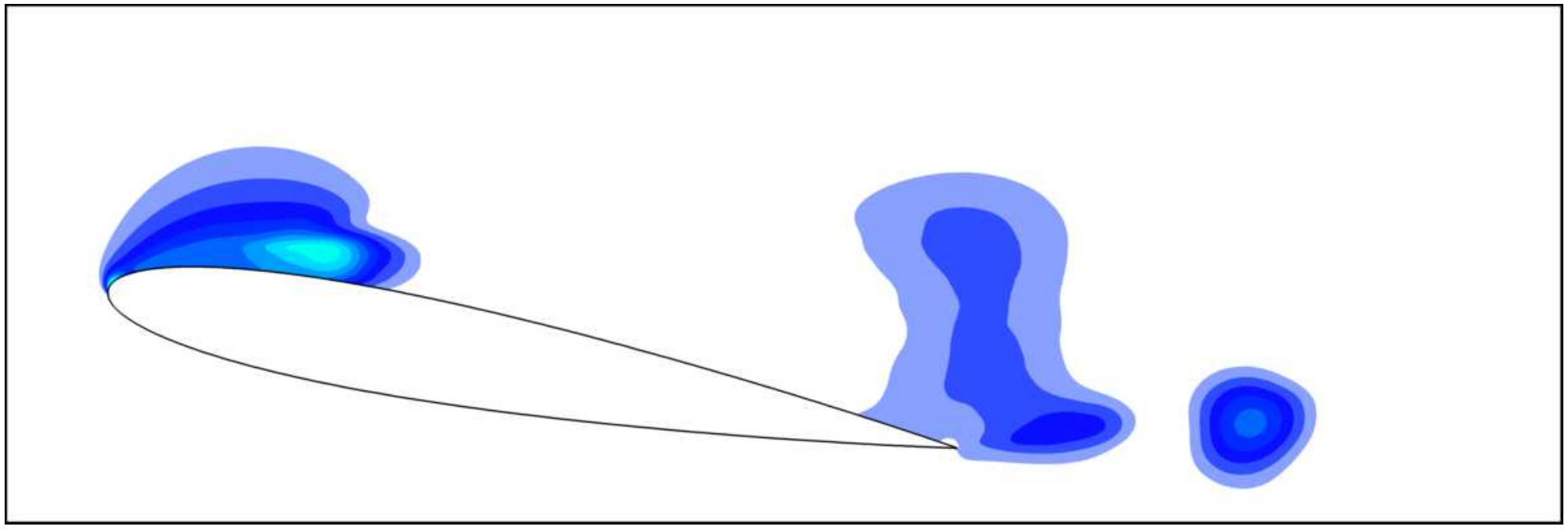}
\textit{$\alpha = 10.5^{\circ}$}
\end{minipage}
\caption{Colour maps of the variance of the phase-averaged pressure, $\overline{\widetilde{p}^2}$, for the angles of attack $\alpha = 9.25^{\circ}$--$10.5^{\circ}$.}
\label{p2_phase}
\end{center}
\end{figure}

A question was raised at this point of the analysis about the nature of the oscillating-f\/low over the LFO cycle. Does the oscillating-f\/low f\/lap up and down, does it pulsate in the streamwise direction or both? To shed some light on the behaviour of the oscillating-f\/low, the conditionally phase-averaged oscillating-f\/low was used to estimate the variance of the phase-averaged oscillating-streamwise-velocity, $\overline{\widetilde{u}^2}$, oscillating-wall-normal-velocity, $\overline{\widetilde{v}^2}$, and the oscillating-pressure, $\overline{\widetilde{p}^2}$.\newline
F\/igure~\ref{u2_phase} shows colour maps of the variance of the phase-averaged oscillating-streamwise-velocity, $\overline{\widetilde{u}^2}$, for the angles of attack $\alpha = 9.25^{\circ}$--$10.5^{\circ}$. At the angle of attack of $9.25^{\circ}$, the f\/igure displays no signif\/icant value of $\overline{\widetilde{u}^2}$ other than that along the shear layer. Thus, the f\/low is oscillating at this angle of attack with the minimum magnitude and mostly along the shear layer in the vicinity of the leading-edge. The f\/low oscillation extends along and around the shear layer at the angle of attack of $9.4^{\circ}$. The f\/low oscillation covers a wider region on the suction surface of the aerofoil, and has a more signif\/icant magnitude at higher angles of attack. The amplitude of oscillation continues to increase until it reaches a maximal value at the angle of attack of $9.8^{\circ}$. The magnitude of oscillation decreases as the angle of attack is further increased until it reaches an insignif\/icant amplitude at the angle of attack of $10.5^{\circ}$.\newline
Colour maps of the variance of the phase-averaged oscillating-wall-normal-velocity, $\overline{\widetilde{v}^2}$, for the angles of attack $\alpha = 9.25^{\circ}$--$10.5^{\circ}$ are shown in f\/igure~\ref{v2_phase}. As seen in the f\/igure, the wall-normal oscillations are mostly concentrated at the leading-edge and on the suction surface of the aerofoil. The signif\/icant oscillations of the wall-normal velocity component at the leading-edge are primarily due to the global oscillations in the f\/low-f\/ield. However, the maximum amplitude of oscillation at the angle of attack of $9.8^{\circ}$ is much smaller than that of the oscillating-streamwise-velocity component. Thus, the phase-averaged oscillating-f\/low is mostly pulsating in the streamwise direction rather than being f\/lapping up and down on the suction surface of the aerofoil. At the angles of attack of $10.1^{\circ}$ and $10.5^{\circ}$ the oscillations in the wall-normal velocity shifts to a wake like oscillating mode. Furthermore, the magnitude of $\overline{\widetilde{v}^2}$ is considerably reduced at the leading-edge. Thus, at these angles of attack the oscillations of the wall-normal velocity does not convert effectively into a global oscillations in the f\/low-f\/ield but rather converts into a stronger f\/lapping of the wake in the vicinity of the trailing-edge. Despite the relatively small magnitude of oscillation of the wall-normal velocity, it plays a profound role in initiating the global f\/low oscillation. Thus, it connects the tailing-edge and the leading-edge instabilities, and feeds the UV of the TCV.\newline
F\/igure~\ref{p2_phase} shows colour maps of the variance of the phase-averaged oscillating-pressure. The $\overline{\widetilde{p}^2}$ is distributed in accordance with the gradient of the oscillating-pressure along the aerofoil chord as seen in the f\/igure. The $\overline{\widetilde{p}^2}$ distribution highlights two important regions for the LFO and the stability of the LSB. The leading-edge where the LSB is formed and the triad of vortices are in action, and at the trailing-edge where the oscillating-pressure f\/luctuates signif\/icantly.

\subsection{Statistics of the ``turbulent-f\/luctuating'' f\/ield}
\begin{figure}
\begin{center}
\begin{minipage}{220pt}
\centering
\includegraphics[width=220pt, trim={0mm 0mm 0mm 0mm}, clip]{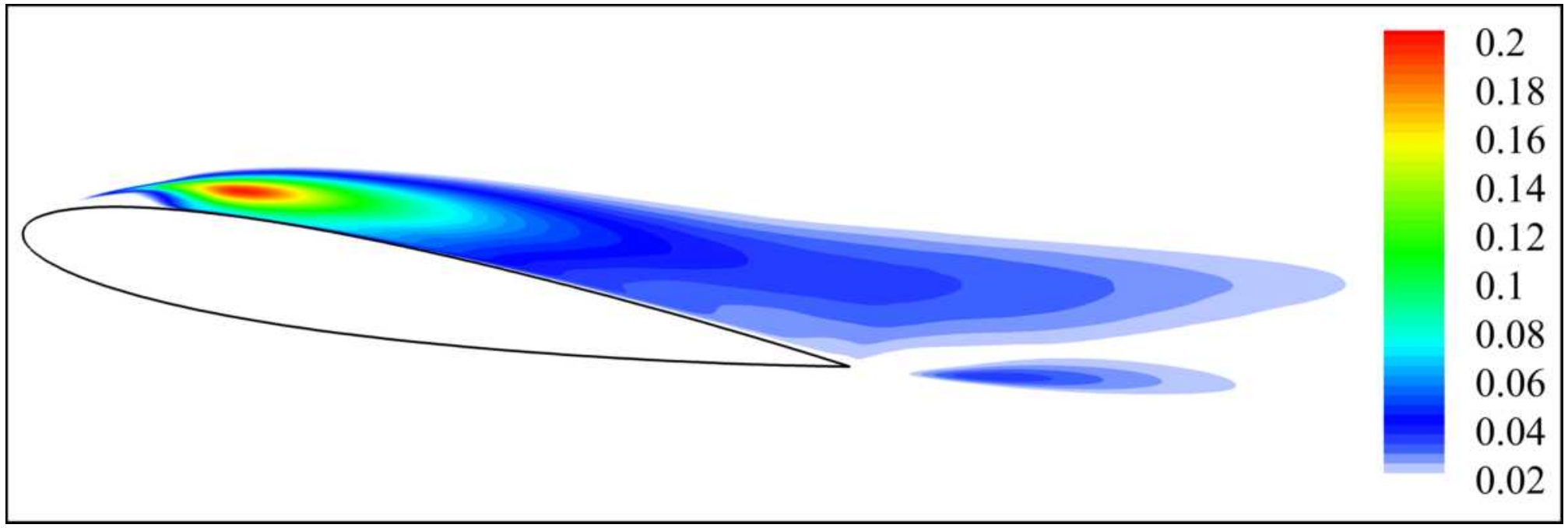}
\textit{$\alpha = 9.25^{\circ}$}
\end{minipage}
\medskip
\begin{minipage}{220pt}
\centering
\includegraphics[width=220pt, trim={0mm 0mm 0mm 0mm}, clip]{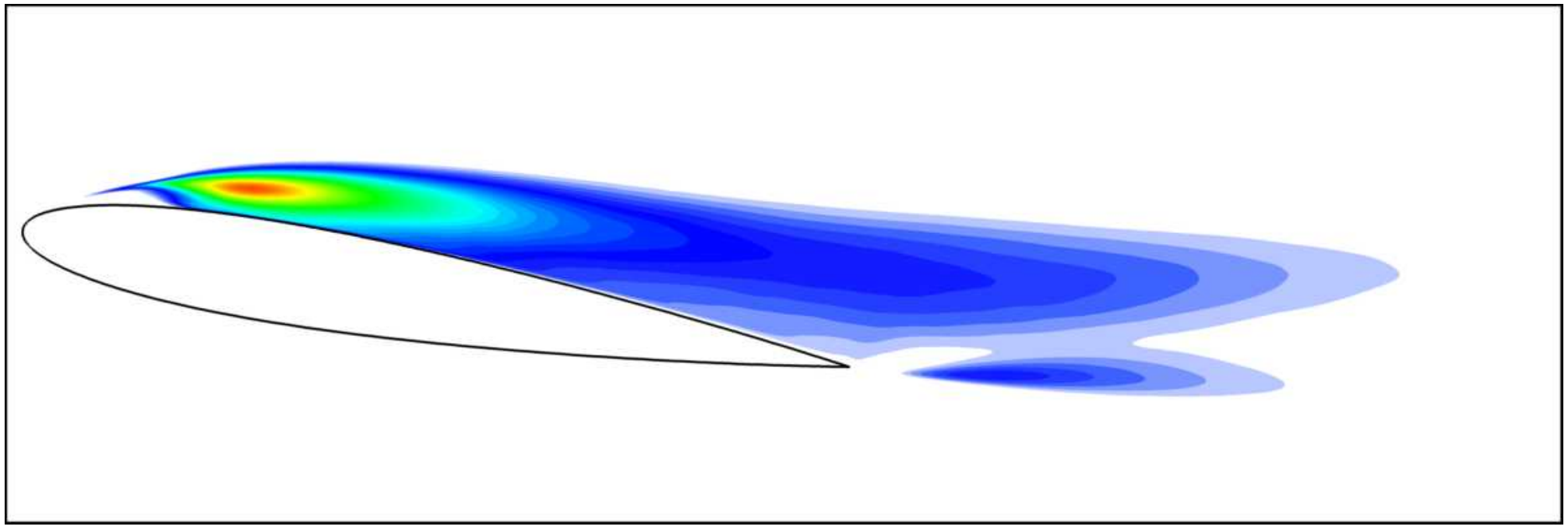}
\textit{$\alpha = 9.40^{\circ}$}
\end{minipage}
\medskip
\begin{minipage}{220pt}
\centering
\includegraphics[width=220pt, trim={0mm 0mm 0mm 0mm}, clip]{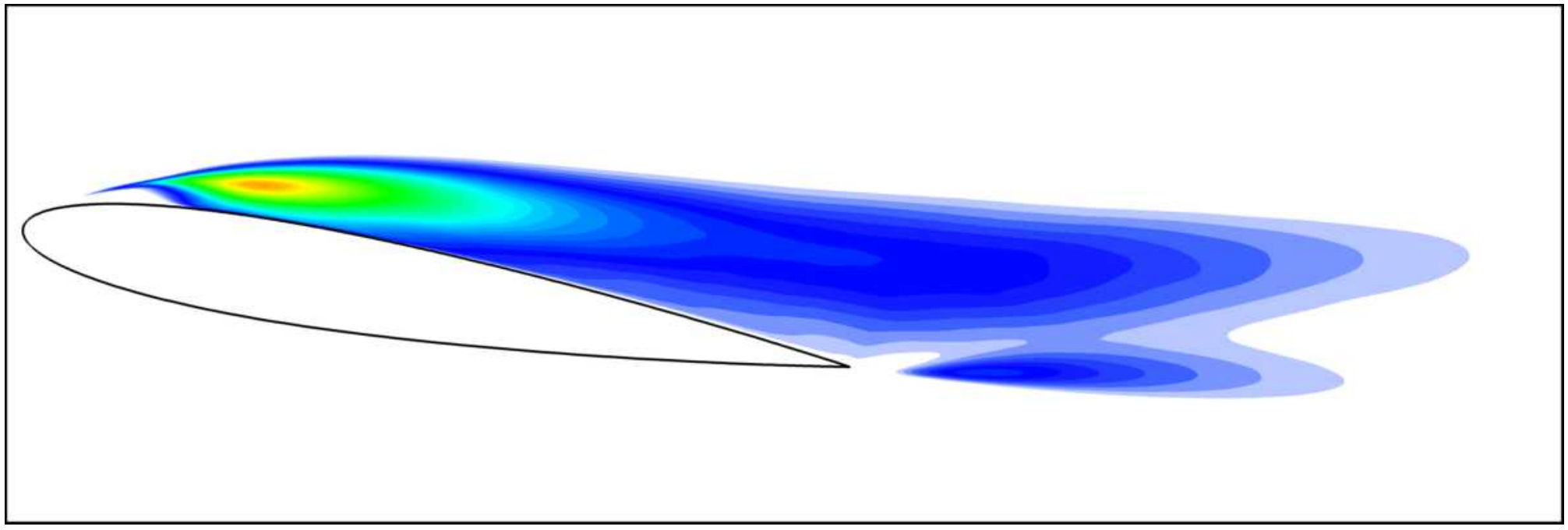}
\textit{$\alpha = 9.50^{\circ}$}
\end{minipage}
\medskip
\begin{minipage}{220pt}
\centering
\includegraphics[width=220pt, trim={0mm 0mm 0mm 0mm}, clip]{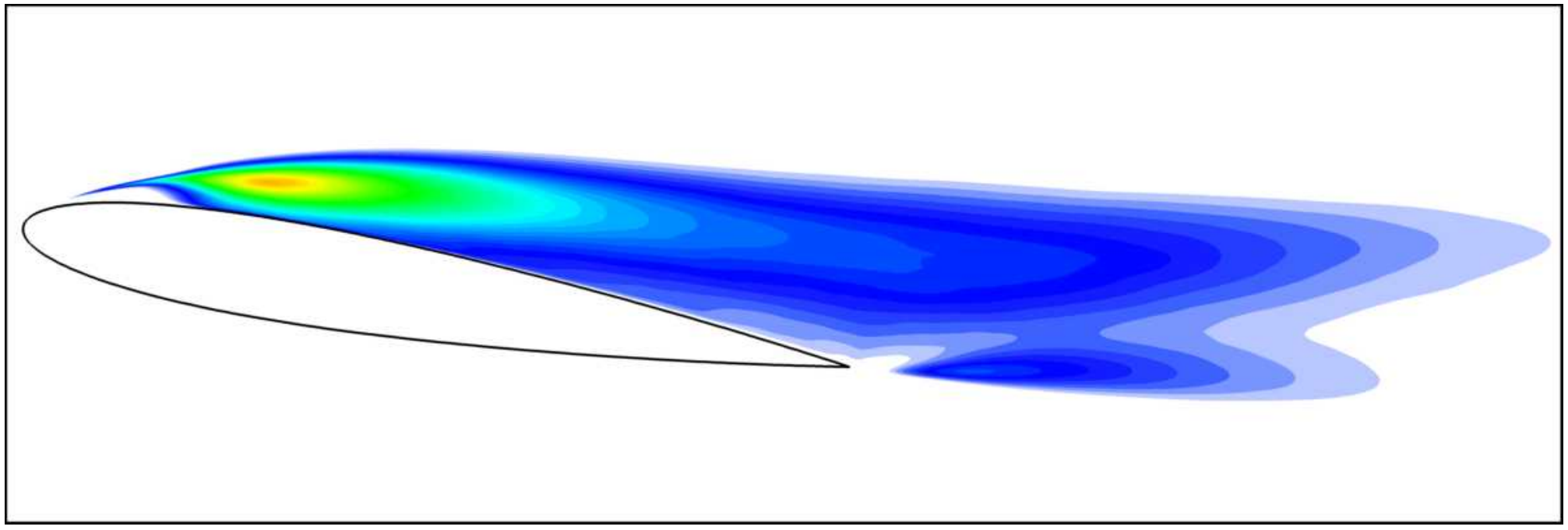}
\textit{$\alpha = 9.60^{\circ}$}
\end{minipage}
\medskip
\begin{minipage}{220pt}
\centering
\includegraphics[width=220pt, trim={0mm 0mm 0mm 0mm}, clip]{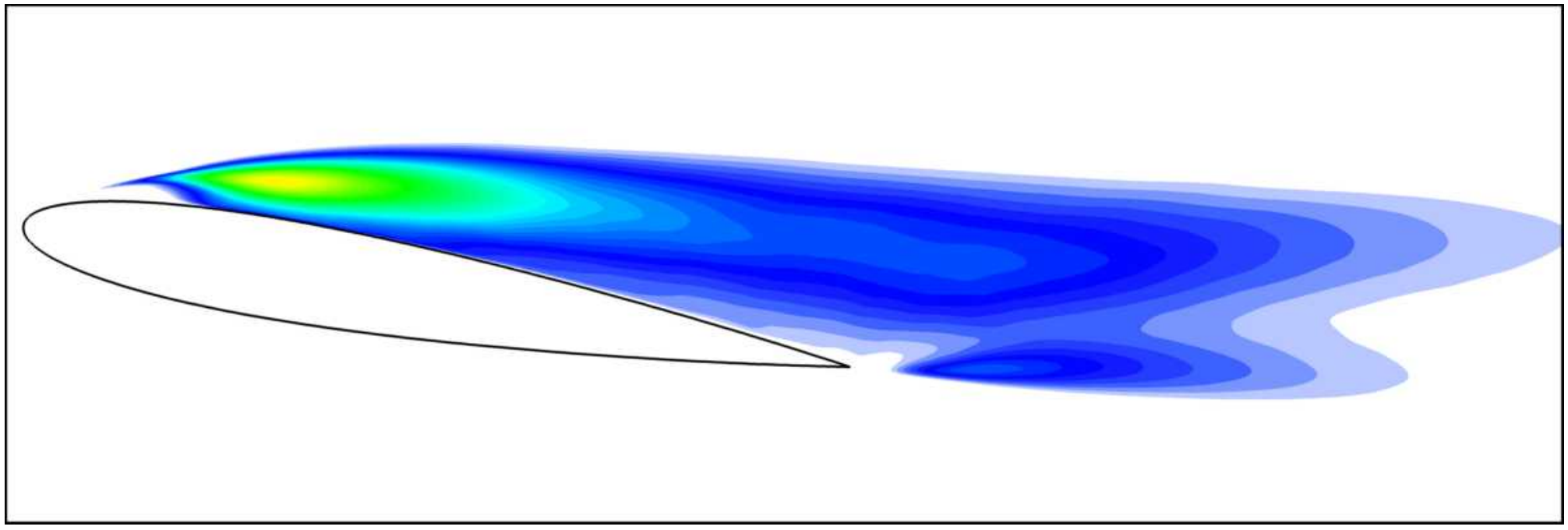}
\textit{$\alpha = 9.70^{\circ}$}
\end{minipage}
\medskip
\begin{minipage}{220pt}
\centering
\includegraphics[width=220pt, trim={0mm 0mm 0mm 0mm}, clip]{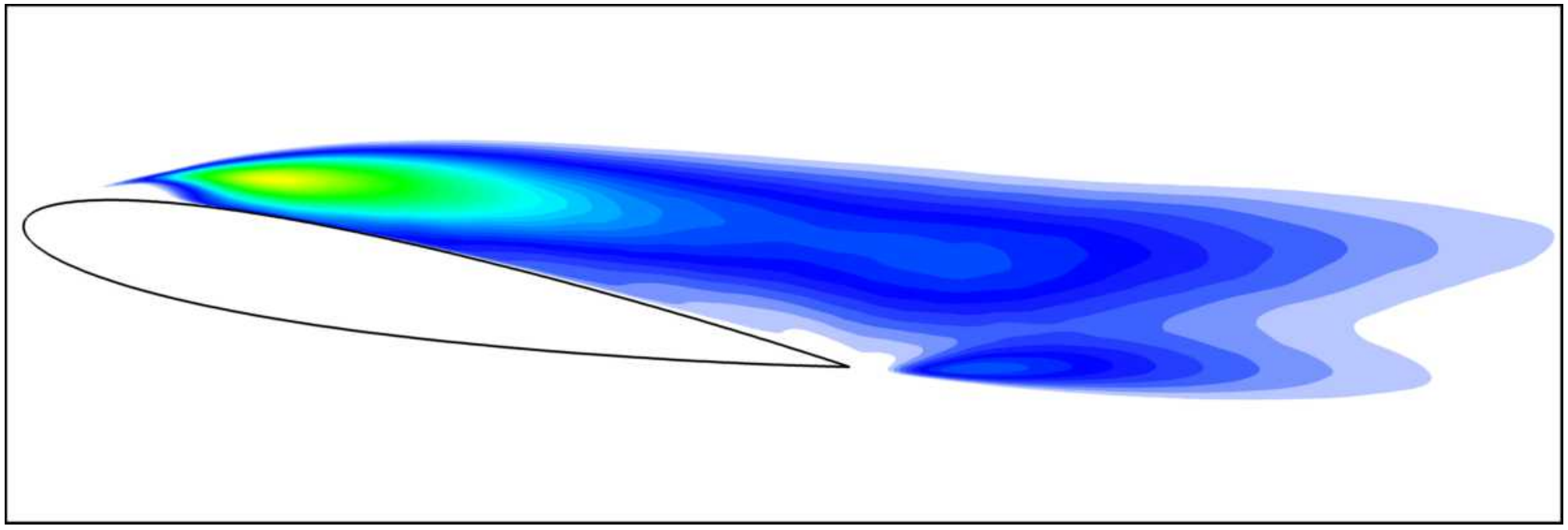}
\textit{$\alpha = 9.80^{\circ}$}
\end{minipage}
\medskip
\begin{minipage}{220pt}
\centering
\includegraphics[width=220pt, trim={0mm 0mm 0mm 0mm}, clip]{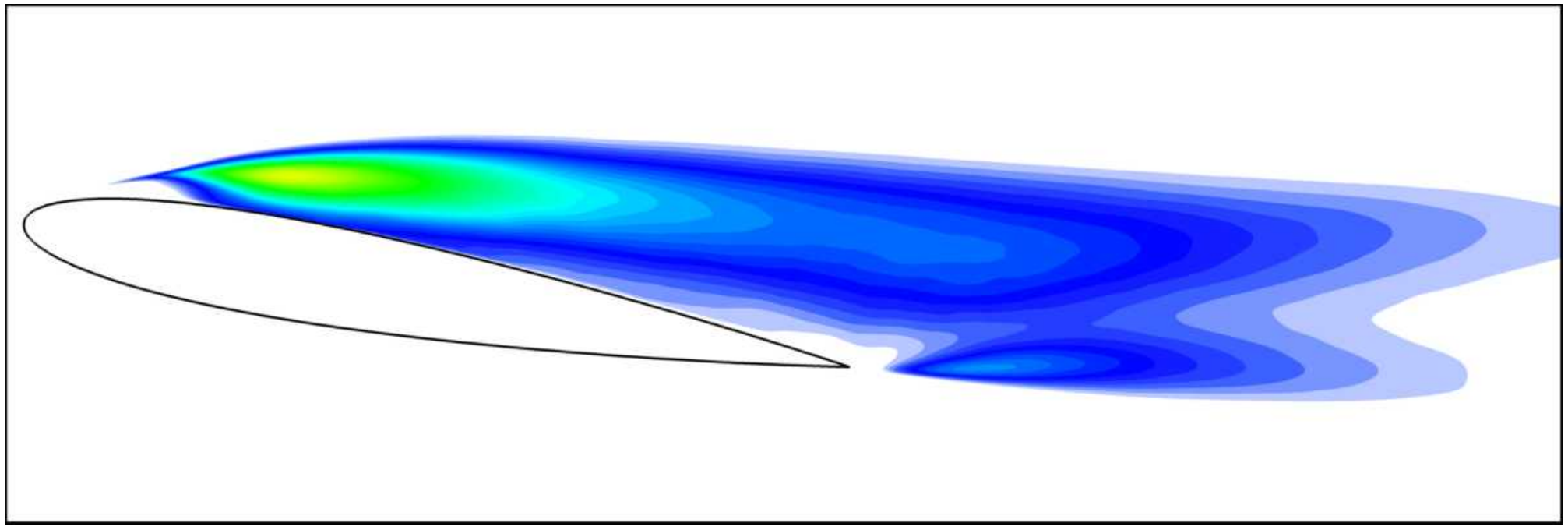}
\textit{$\alpha = 9.90^{\circ}$}
\end{minipage}
\medskip
\begin{minipage}{220pt}
\centering
\includegraphics[width=220pt, trim={0mm 0mm 0mm 0mm}, clip]{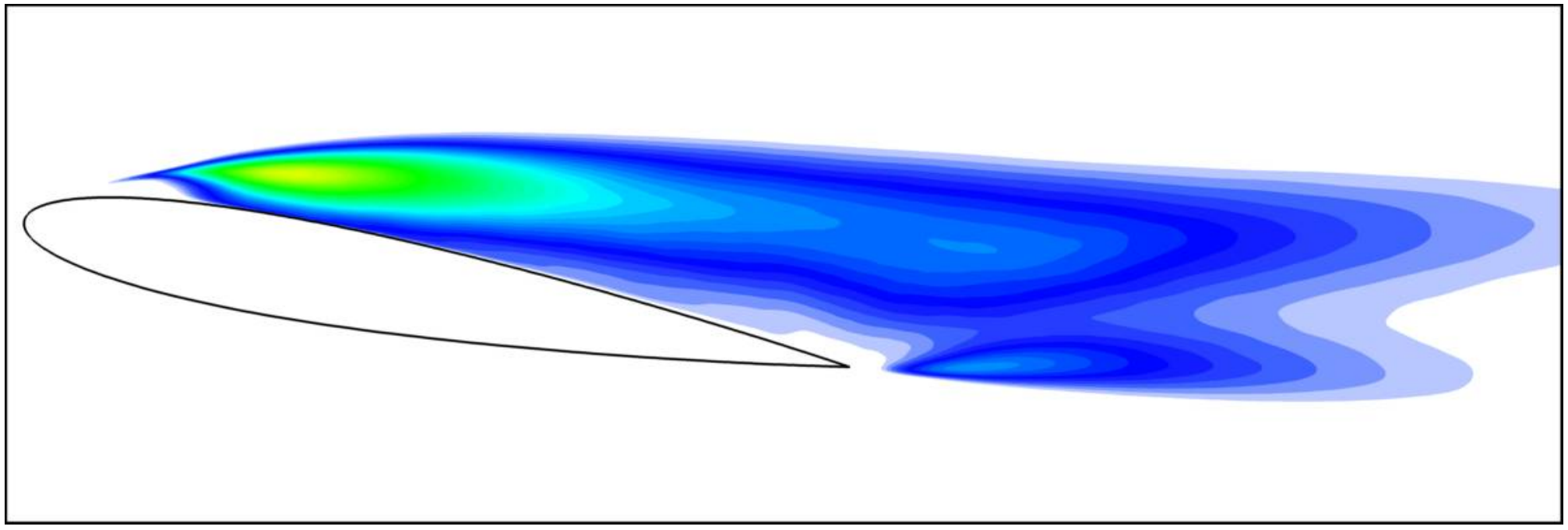}
\textit{$\alpha = 10.0^{\circ}$}
\end{minipage}
\medskip
\begin{minipage}{220pt}
\centering
\includegraphics[width=220pt, trim={0mm 0mm 0mm 0mm}, clip]{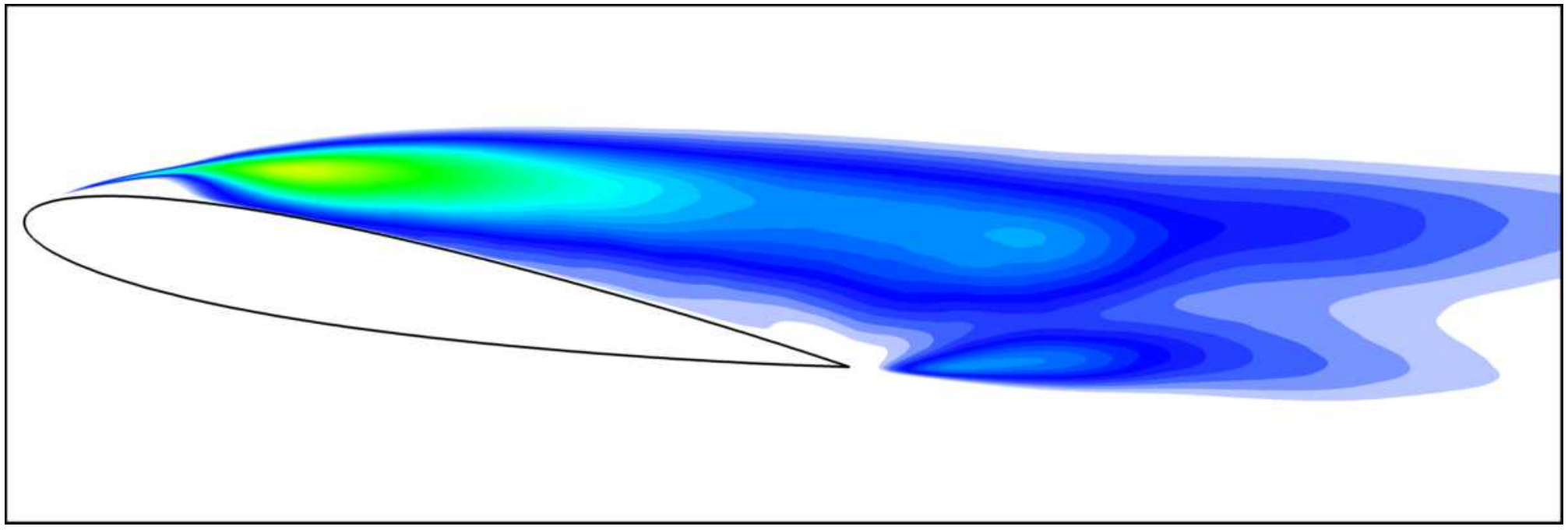}
\textit{$\alpha = 10.1^{\circ}$}
\end{minipage}
\begin{minipage}{220pt}
\centering
\includegraphics[width=220pt, trim={0mm 0mm 0mm 0mm}, clip]{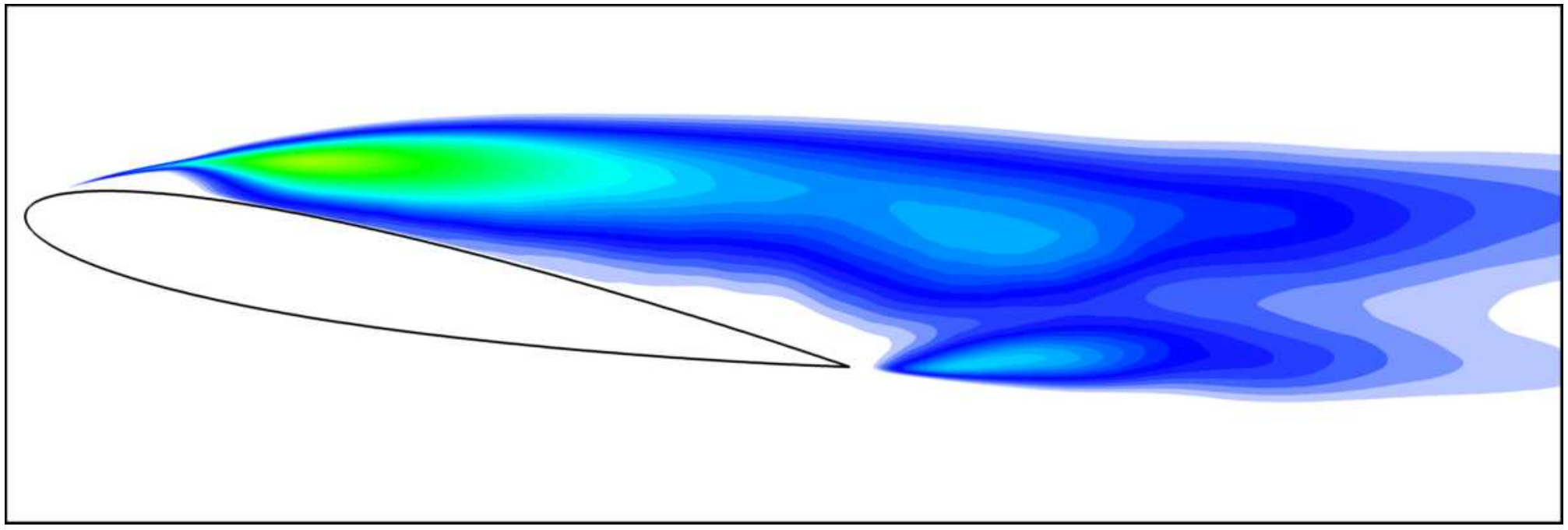}
\textit{$\alpha = 10.5^{\circ}$}
\end{minipage}
\caption{Colour maps of the variance of the ``turbulent f\/luctuations'' of the streamwise velocity component, $\overline{{u\myprime}^2}$, for the angles of attack $\alpha = 9.25^{\circ}$--$10.5^{\circ}$.}
\label{u2_turb}
\end{center}
\end{figure}

\begin{figure}
\begin{center}
\begin{minipage}{220pt}
\centering
\includegraphics[width=220pt, trim={0mm 0mm 0mm 0mm}, clip]{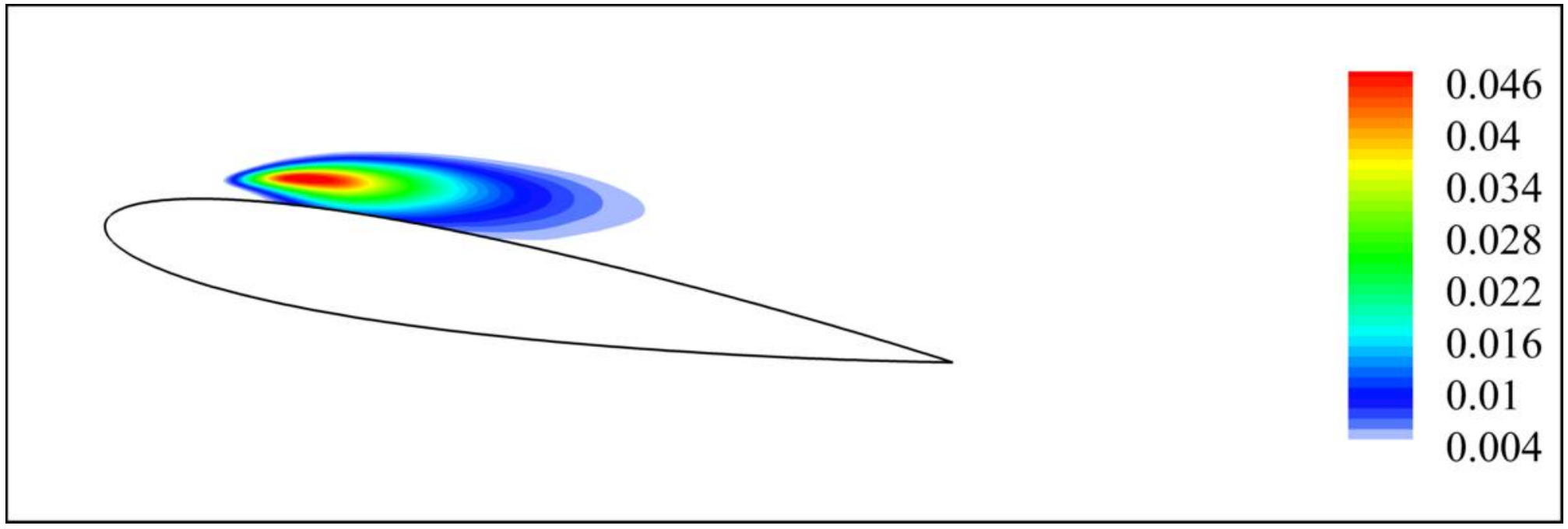}
\textit{$\alpha = 9.25^{\circ}$}
\end{minipage}
\medskip
\begin{minipage}{220pt}
\centering
\includegraphics[width=220pt, trim={0mm 0mm 0mm 0mm}, clip]{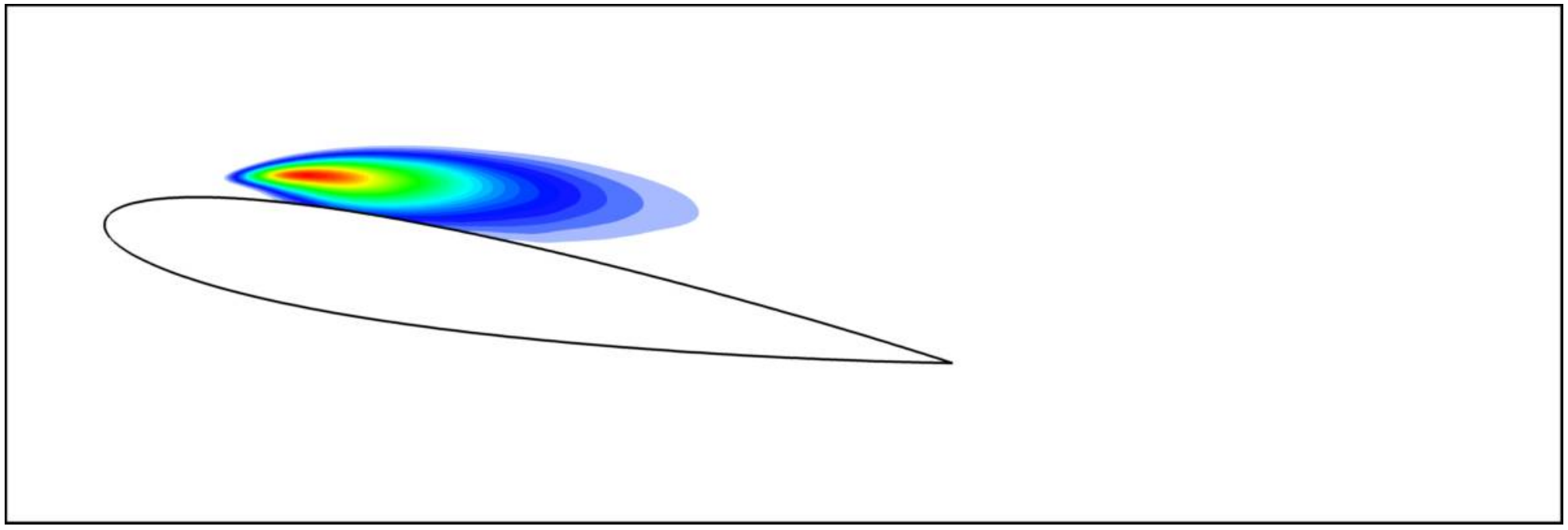}
\textit{$\alpha = 9.40^{\circ}$}
\end{minipage}
\medskip
\begin{minipage}{220pt}
\centering
\includegraphics[width=220pt, trim={0mm 0mm 0mm 0mm}, clip]{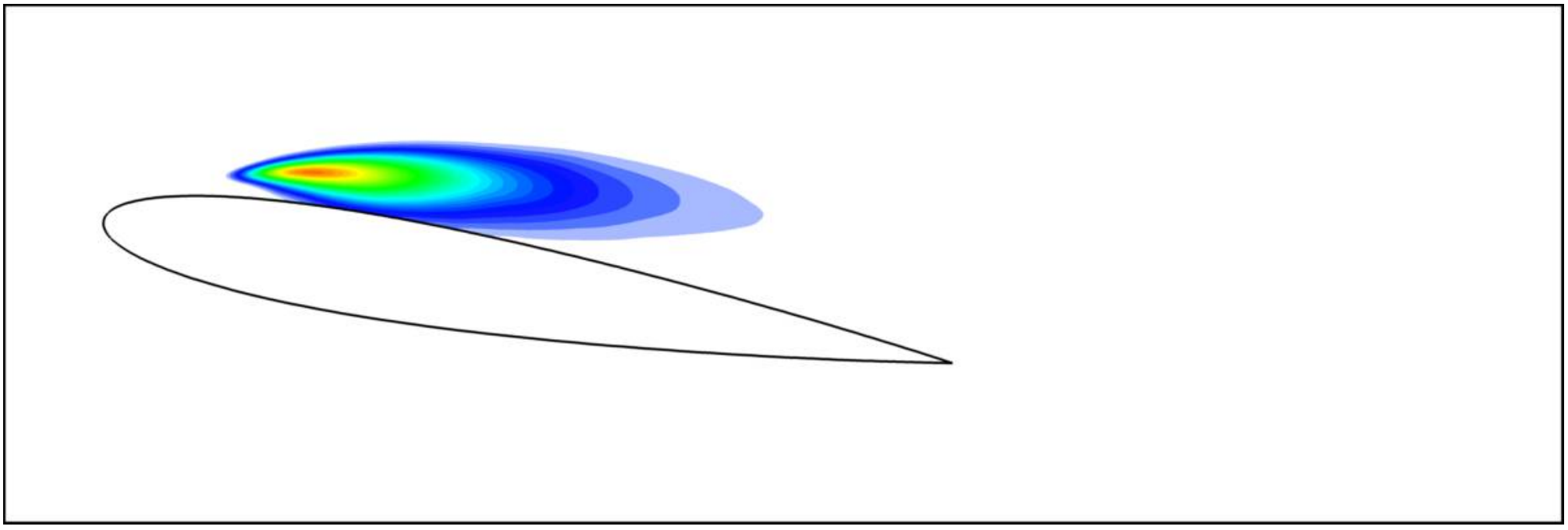}
\textit{$\alpha = 9.50^{\circ}$}
\end{minipage}
\medskip
\begin{minipage}{220pt}
\centering
\includegraphics[width=220pt, trim={0mm 0mm 0mm 0mm}, clip]{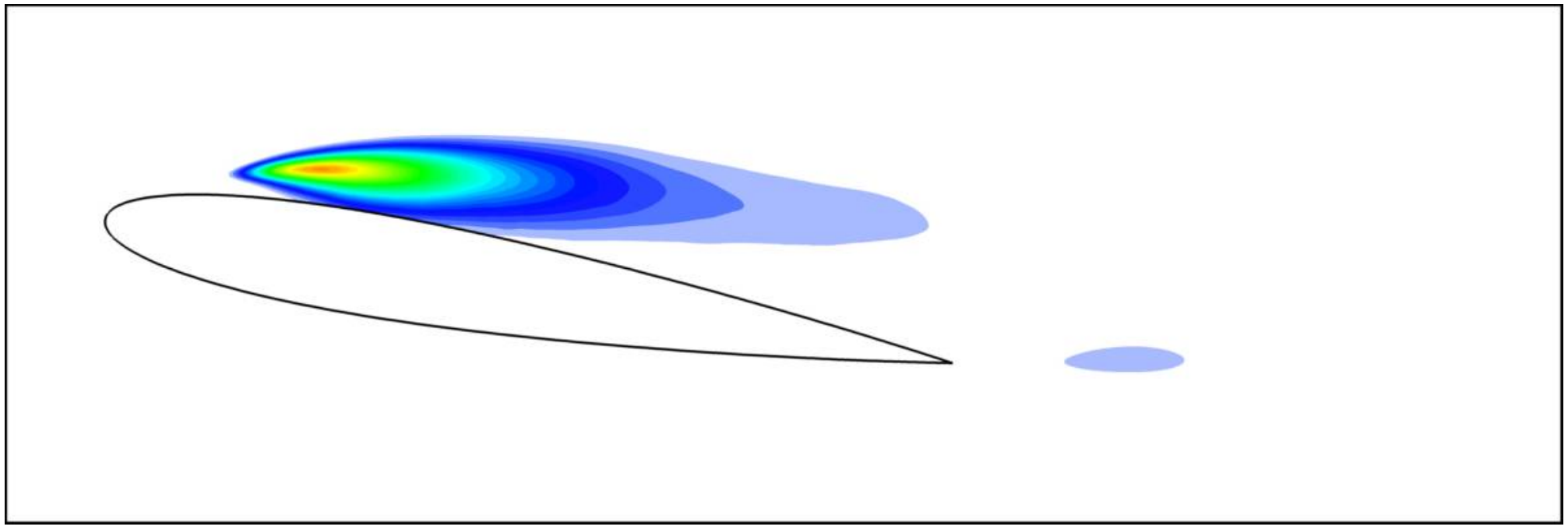}
\textit{$\alpha = 9.60^{\circ}$}
\end{minipage}
\medskip
\begin{minipage}{220pt}
\centering
\includegraphics[width=220pt, trim={0mm 0mm 0mm 0mm}, clip]{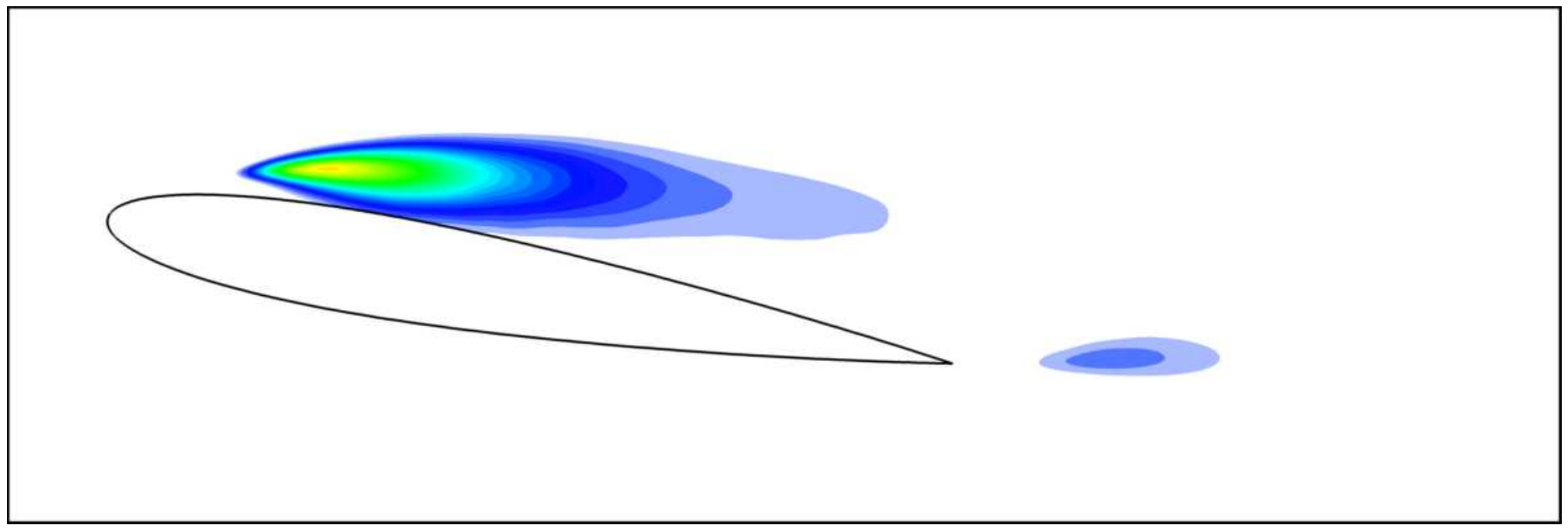}
\textit{$\alpha = 9.70^{\circ}$}
\end{minipage}
\medskip
\begin{minipage}{220pt}
\centering
\includegraphics[width=220pt, trim={0mm 0mm 0mm 0mm}, clip]{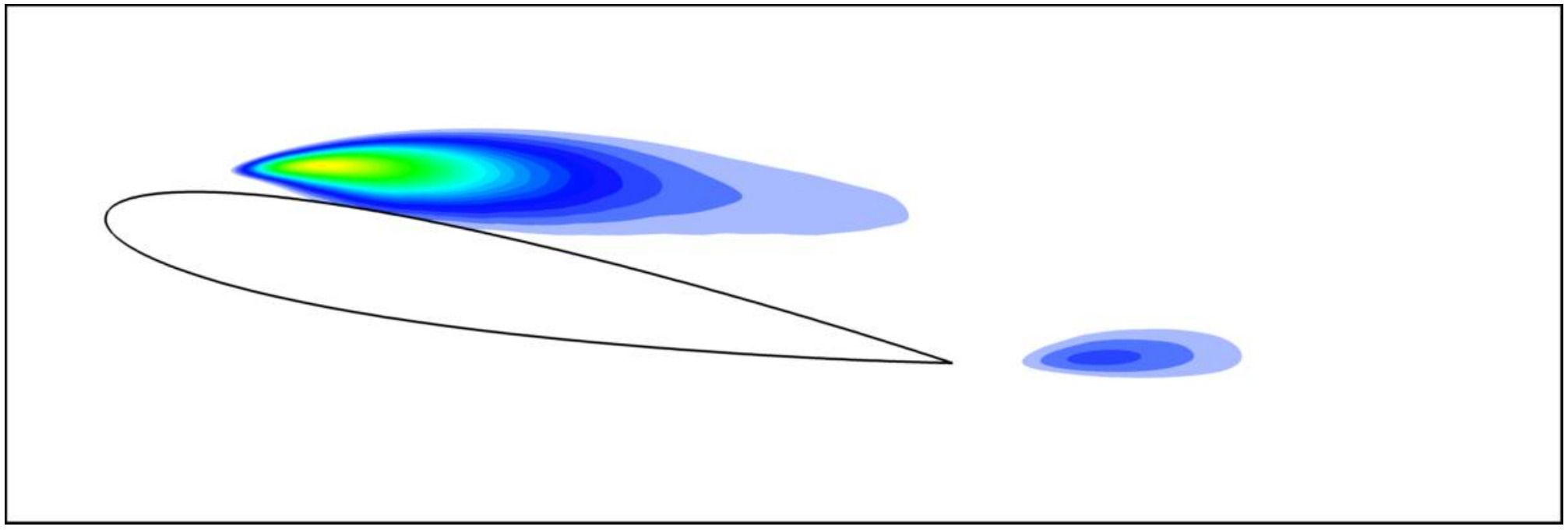}
\textit{$\alpha = 9.80^{\circ}$}
\end{minipage}
\medskip
\begin{minipage}{220pt}
\centering
\includegraphics[width=220pt, trim={0mm 0mm 0mm 0mm}, clip]{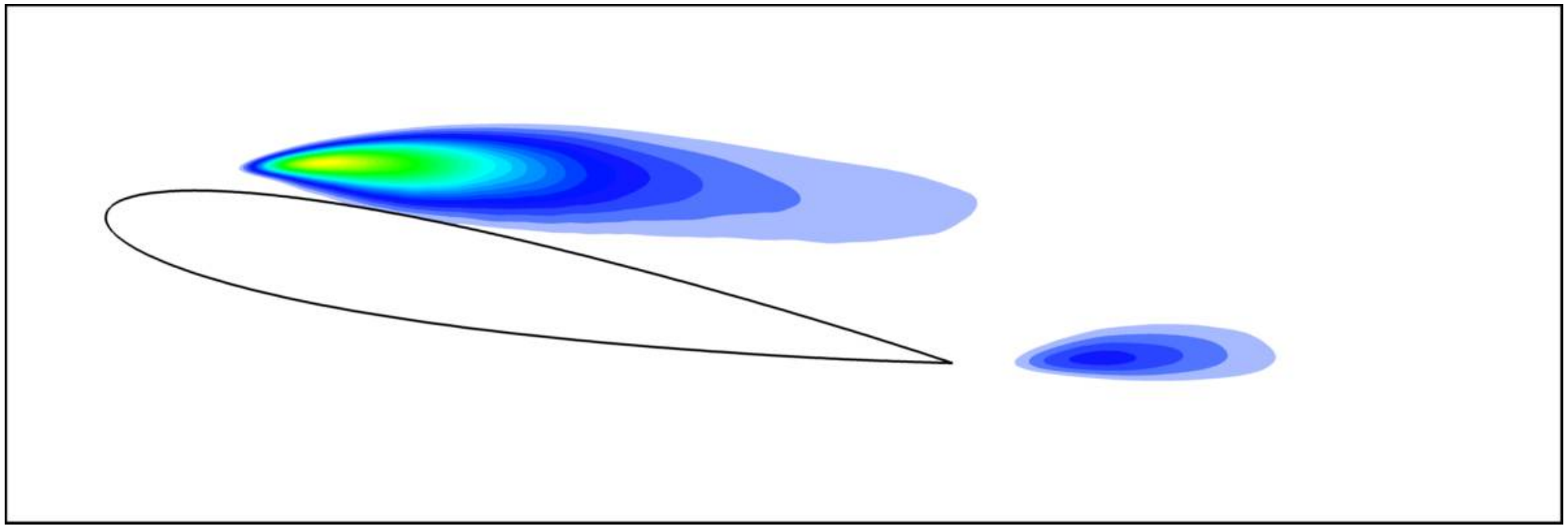}
\textit{$\alpha = 9.90^{\circ}$}
\end{minipage}
\medskip
\begin{minipage}{220pt}
\centering
\includegraphics[width=220pt, trim={0mm 0mm 0mm 0mm}, clip]{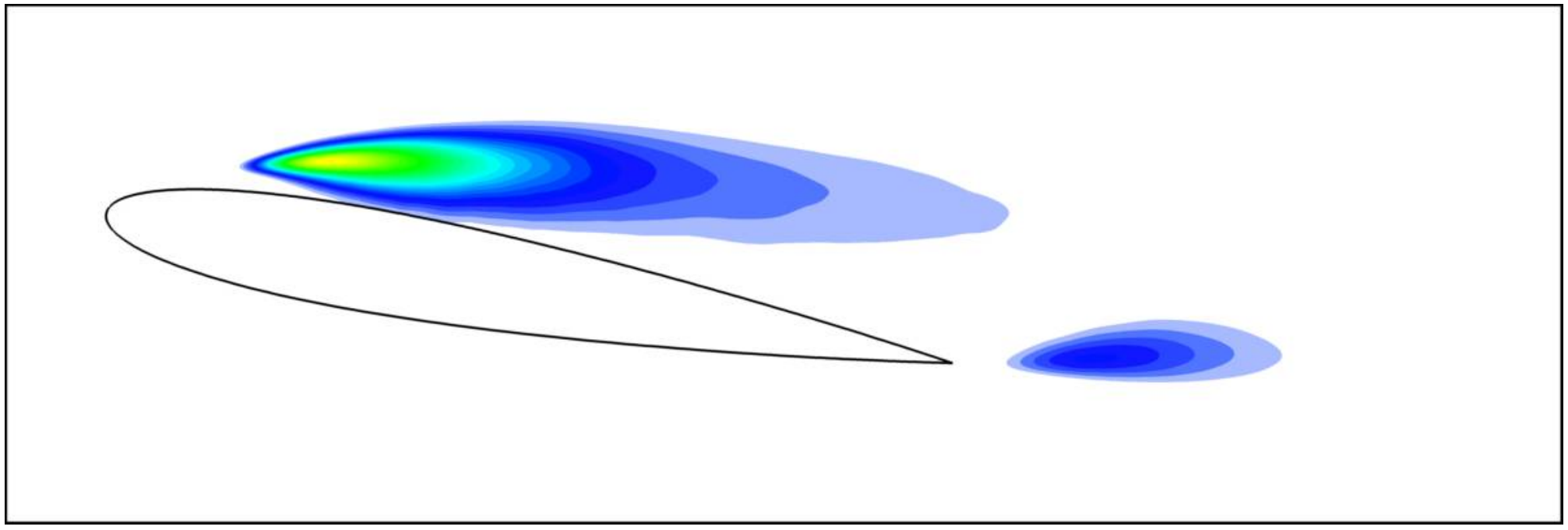}
\textit{$\alpha = 10.0^{\circ}$}
\end{minipage}
\medskip
\begin{minipage}{220pt}
\centering
\includegraphics[width=220pt, trim={0mm 0mm 0mm 0mm}, clip]{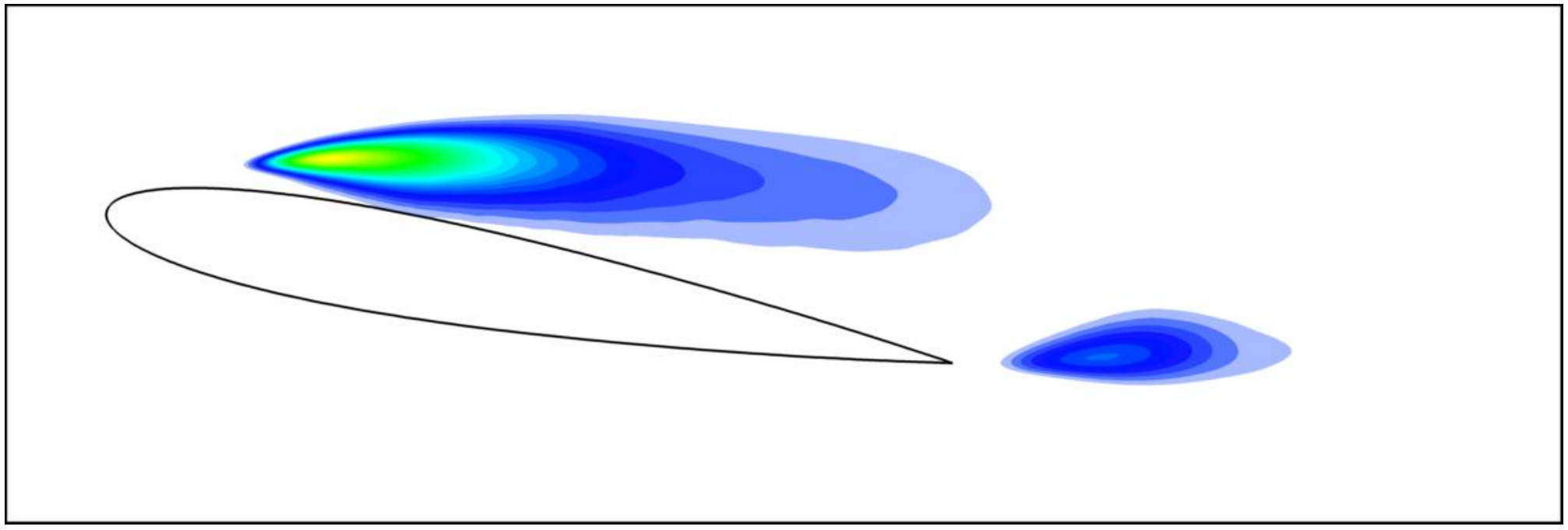}
\textit{$\alpha = 10.1^{\circ}$}
\end{minipage}
\begin{minipage}{220pt}
\centering
\includegraphics[width=220pt, trim={0mm 0mm 0mm 0mm}, clip]{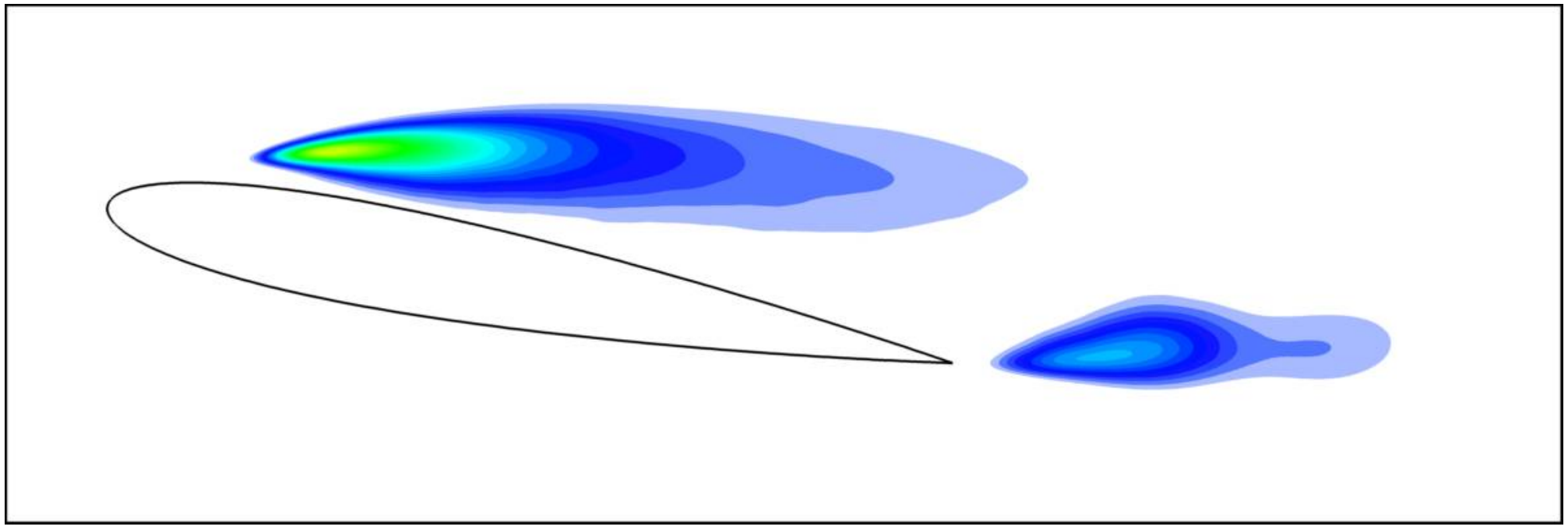}
\textit{$\alpha = 10.5^{\circ}$}
\end{minipage}
\caption{Colour maps of the variance of the ``turbulent f\/luctuations'' of the pressure, $\overline{{p\myprime}^2}$, for the angles of attack $\alpha = 9.25^{\circ}$--$10.5^{\circ}$.}
\label{p2_turb}
\end{center}
\end{figure}

The phase-averaged oscillations are characterised by large deterministic vortices that affects the f\/low-f\/ield globally. However, another type of oscillations is present in the f\/low-f\/ield. These are small random perturbations that grow in size and magnitude, breakup and merge, and very much characterise the turbulent f\/lows. The former is referred to as phase-averaged oscillations or the periodic component, and the latter is referred to as the ``turbulent f\/luctuations'' or the random component. The term is always mentioned in the text inside quotation mark to highlight the fact that the f\/low could be laminar, transitional, or turbulent; therefore, referring to the f\/luctuations as \textit{turbulent} does not imply that the f\/low underwent transition to turbulence.\newline
Equation~\ref{decomposition} shows how a f\/low variable $\psi$ is decomposed into mean, phase-averaged, and ``turbulent f\/luctuations''. The equation is used to estimate the variance of the ``turbulent f\/luctuations'' of the streamwise velocity component, $\overline{{u\myprime}^2}$, and the variance of the ``turbulent f\/luctuations'' of the pressure, $\overline{{p\myprime}^2}$. F\/igure~\ref{u2_turb} displays colour maps of the $\overline{{u\myprime}^2}$ for the angles of attack $\alpha = 9.25^{\circ}$--$10.5^{\circ}$. The streamwise f\/luctuations have signif\/icant magnitude along the separated shear layer and at the trailing-edge. It starts with a small amplitude at the laminar portion of the separated shear layer and increases along the shear layer. Hence, featuring the extraction of turbulent kinetic energy from the mean f\/low via the Kelvin--Helmholtz instability. At the angle of attack of $9.25^{\circ}$, the transition process takes place naturally by the Kelvin--Helmholtz instability, and the globally oscillating-f\/low does not affect the velocity gradient across the shear layer. Thus, the extraction of the turbulent kinetic energy from the mean and feeding it into the small-scales perturbations starts at the furthest upstream location along the shear layer, in the mean sense. As the angle of attack is increased, the velocity gradient across the separated shear layer is overwhelmed by the globally oscillating-f\/low. Consequently, the transition location oscillates signif\/icantly along the shear layer. Thus, the location of the maximum turbulent kinetic energy moves downstream, in the mean sense, as the angle of attack increases. At the angle of attack of $9.8^{\circ}$, the angle of maximum phase-averaged oscillations, $\overline{{u\myprime}^2}$ has a maximum value further downstream the leading-edge. The phase-averaged oscillations becomes insignif\/icant as the angle of attack is further increased. Consequently, the location where $\overline{{u\myprime}^2}$ has a maximum moves upstream as the angle of attack is increased above the angle of $9.8^{\circ}$.
The maximum value of $\overline{{u\myprime}^2}$ decreases as the angle of attack is increased as seen in the f\/igure. At the angle of attack of $9.8^{\circ}$, the f\/luctuating f\/low is overwhelmed by the LFO, and the ``turbulent f\/luctuations'' reduces signif\/icantly downstream the separated shear layer.\newline
There is no signif\/icant difference between the variance of the ``turbulent f\/luctuations'' of the wall-normal velocity component, $\overline{{v\myprime}^2}$, and the variance of the wall-normal velocity component, $\overline{{v\mydprime}^2}$. Obviously, this is because the phase-averaged component is much smaller compared to that of the ``turbulent f\/luctuations''. F\/igure~\ref{p2_turb} displays $\overline{{p\myprime}^2}$ for the angles of attack $\alpha = 9.25^{\circ}$ to $10.5^{\circ}$. As seen in the f\/igure, $\overline{{p\myprime}^2}$ has no signif\/icant value at the leading-edge and along the laminar portion of the separated shear layer. Again, this is the location where the inviscid absolute instability is at action and most of the pressure f\/luctuations are phase-averaged not ``turbulent''. Also, in the vicinity of the trailing-edge, the f\/luctuations in the pressure are due to large vortices and mostly phase-averaged.\newline
In summary, small-scale vortical structures that contribute to the ``turbulent f\/luctuations'' are primarily extracted and maintained by the Kelvin--Helmholtz instability along the shear layer. Thus, the variance of the ``turbulent f\/luctuations'' has signif\/icant magnitude along the separated shear layer in the vicinity of the leading-edge. Also, the instability at the trailing-edge contributes signif\/icantly to the small-scale perturbations. Thus, the variance of the ``turbulent f\/luctuations'' increases in the vicinity of the trailing-edge. On the contrary, large-scale vortices contribute to the variance of the phase-averaged oscillations. These vortices are generated and sustained by the inviscid instability across the laminar portion of the separated shear layer. Thus, the phase-averaged oscillations have signif\/icant amplitude on both sides of the laminar portion of the separated shear layer. Furthermore, these large vortices advect downstream and contribute signif\/icantly to the f\/low variations at the trailing-edge. Thus, the phase-averaged oscillations have a considerable magnitude in the vicinity of the trailing-edge. F\/inally, the aforementioned discussion conf\/irms the observations reported by~\citet{eljack2018bursting} that the perturbations in the streamwise velocity and the pressure are extracted by the global f\/low oscillation in addition to the Kelvin--Helmholtz instability. Hence, there is a signif\/icant difference between the phase-averaged and the f\/luctuating f\/ield. Whereas, f\/luctuations in the wall-normal velocity and the spanwise velocity are extracted from the mean f\/low exclusively by the Kelvin--Helmholtz instability. Thus, the phase-averaged component is very small compared to the ``turbulent f\/luctuations'' component.

\subsection{The f\/lat plate model}
As mentioned in $\S$~\ref{sec:intorudction},~\citet{gaster1967structure} wanted to eliminate the effect of the aerofoil geometry and generate data for different LSBs characteristics. He invented a brilliant model that allowed him to vary both the Reynolds number and the pressure distribution. The model was a f\/lat plate with an adjustable pressure distribution. However, does the physics of an LSB induced on a f\/lat plate and its associated LFO resemble that of an LSB formed naturally on the suction surface of an aerofoil? The answer to this question depends on whether a critical and profound difference is overlooked. For instance, the interaction of the instability at the trailing-edge of the aerofoil with the LSB and how it oscillates the f\/low-f\/ield globally cannot be modelled in the f\/lat plate model. The smart f\/lat plate and adjustable pressure distribution model has a major shortcoming that it can not account for the effect of the globally oscillating f\/low. In the light of the present discussion, the mechanism that controls the stability of the LSB, generates and sustains the LFO is dependent on the magnitude and direction of the oscillating-f\/low. The f\/low-f\/ield around the f\/lat plate can not rotate $360^{\circ}$. Thus, the LSB stability and the LFO phenomenon cannot be modelled using the f\/lat plate model. Unless somehow the globally oscillating f\/low around the aerofoil is taken into consideration. A simple trick to do so is to oscillate the free-stream at adjustable frequency and magnitude. However, the strength of the oscillating-f\/low is adjusted naturally by the dynamics of the f\/low in the case of the aerofoil. Whereas, it is imposed at a certain magnitude and frequency that do not change with the f\/low in the proposed f\/lat plate model.

\subsection{The linearized Navier--Stokes model}
Recently, solving linearised Navier--Stokes equations has been in vogue to model various f\/luid dynamics problems and come up with conditions of stability of the f\/low, and it proved to work very well for many problems. However, considering the aforementioned discussion, are such models relevant or applicable to this problem? The oscillating-pressure is much larger than $1\%$ of the mean free-stream pressure. The relevant oscillating-velocity at the laminar portion of the separated shear layer is also well above $1\%$ of the mean free-stream velocity. Thus, models where perturbations of the most amplif\/ied mode cannot be allowed to grow above $1\%$ of the base f\/low cannot be used to tackle this problem.

\subsection{The effect of compressibility on the LFO phenomenon}
Classically, this problem used to be investigated in compressible f\/low settings. Experiments are carried out in wind-tunnels rather than in water-tunnels and compressible Navier--Stokes equations are solved in numerical simulations rather than solving the incompressible set of equations. The vast majority of investigations being compressible is mainly due to the widely accepted hypothesis that there is an acoustic waves feed-back mechanism involved. It is assumed that the acoustic waves generated at the trailing-edge of the aerofoil travel upstream and interact with some receptivity mechanism in the vicinity of the leading-edge. Such feedback would force the shear layer to undergo early-transition and reattach the f\/low. The movement of the transition location along the shear layer, causing late-transition/early-transition, is synchronised with the f\/low separation and reattachment, respectively. However, it is shown in the present discussion that the movement of the transition location is interlinked with the oscillating-f\/low rather than the acoustic excitation scenario. Furthermore, it is shown that the oscillating-f\/low affects the velocity gradient across the separated shear layer; thus, affects the development of the Kelvin--Helmholtz instability. Hence, the underlying mechanism that generates and sustains the LFO phenomenon does not necessitate the f\/low to be compressible.\newline
The above discussion shows that the circumstances that govern the stability of the LSB and its associated LFO phenomenon can be met in both compressible and incompressible f\/lows. The question is whether the LFO can be captured in an incompressible f\/low. The answer is we do not know. Another question of major importance is does the compressibility of the f\/low and the acoustic feed-back affect the stability of the LSB and the LFO phenomenon? And the answer to this question is yes. For instance, f\/low forcing by an upstream source or due to a downstream source via acoustic waves feed-back could affect the evolution and breakdown of the Kelvin--Helmholtz instability. Hence, the forced shear layer becomes more energetic and has more momentum to reattach the f\/low compared to a naturally evolving shear layer in an incompressible f\/low with a clean free-stream. Furthermore, such energetic shear layer can attach the f\/low against more pronounced adverse pressure gradient. Consequently, the triad of vortices driven in such conditions will be more energetic compared to those driven in a naturally evolving f\/low. Thus, the f\/low forcing strengths the magnitude of the oscillations and widen the range of angles of attack at which the LFO phenomenon takes place. Therefore, despite the fact that the LFO could be generated and sustained, theoretically, in an incompressible f\/low; f\/low compressibility strengths the phenomenon and causes a more pronounced oscillations. On the contrary, it seems that f\/low forcing using selected set of frequencies could remove the low-frequency oscillation as reported by~\citet{zaman1987effect} and~\citep{eljack2018influence}. Such a conf\/licting role of the acoustic excitation was also noted by~\citet{zaman1989natural}.\newline
\citet{Rodriguez2013direct} and~\citet{Rodriguez2015flow} carried out several direct numerical simulations in which they solved the incompressible Navier--Stokes equations. They investigated the f\/low-f\/ield around a NACA-0012 aerofoil at Reynolds number of $5\times10^4$ and angles of attack of $5^{\circ}$, $8^{\circ}$, $9.25^{\circ}$, and $12^{\circ}$. The authors reported that they have started the simulations from an initially homogeneous f\/low-f\/ield that introduces some numerical disturbances. However, the authors did not state clearly whether they have captured a global low-frequency f\/low oscillation or not. They have not shown histories of the aerodynamic coeff\/icients; thus, the reader can not determine if the LFO is captured. The only low-frequency f\/low phenomenon reported by the authors is the shear layer f\/lapping on the suction surface of the aerofoil at Strouhal number of $0.021$. To the best of the authors knowledge, these are the only incompressible studies that investigated the low-frequency f\/low oscillation in the f\/low-f\/ield around an aerofoil at near stall conditions. However, further investigations are required before a conclusion on whether an LFO can be captured in an incompressible f\/low. Therefore, an experimental measurements in a water tunnel or solving the incompressible Navier--Stokes equations in a direct numerical simulation is merited. In the case of the numerical simulation, starting the simulation using clean f\/low might help in capturing the phenomenon. Whereas, in the water tunnel experiment, adjusting the free-stream turbulence to the right level is crucial for capturing the phenomenon.

\section*{Conclusions}
Large eddy simulations of the f\/low f\/ield around a NACA-0012 aerofoil were performed at a Reynolds number of $5\times10^4$ and Mach number of $0.4$ at various angles of attack near stall conditions to investigate the stability of the laminar separation bubble (LSB) and the underlying mechanism of the low-frequency f\/low oscillation (LFO). It is shown that a triad of three vortices, two co-rotating vortices (TCV) and a secondary vortex counter-rotating with them, is behind the quasi-periodic self-sustained bursting and reformation of the LSB and its associated LFO phenomenon. The gradient of the oscillating-velocity across the laminar portion of the separated shear layer drives the upstream vortex of the TCV (UV) and is faithfully aligned with it. The UV front tip forms a half-saddle on the aerofoil surface just upstream the separation point of the shear layer, and its rear end forms a full saddle. The downstream vortex of the TCV (DV) is submerged inside the region of the LSB and aligned such that its front tip forms a full saddle with the UV while its rear tip forms a half-saddle on the aerofoil surface further upstream the reattachment point. The secondary vortex lies beneath the TCV and counter-rotating with them. It acts as a roller support that facilitate the rotation and orientation of the TCV. A global oscillation in the f\/low-f\/ield around the aerofoil is observed in all of the investigated angles of attack, including at zero angle of attack. The magnitude of oscillation is increased as the angle of attack is increased. The strength and dynamics of the TCV determine the time-scale of the global f\/low oscillation. The f\/low switches between an attached-phase against an adverse pressure gradient and a separated-phase despite a favourable pressure gradient in a periodic manner with some disturbed cycles. When the direction of the oscillating-f\/low is clockwise, it adds momentum to the boundary layer and helps it to remain attached against the APG and vice versa. The point of maximum turbulent kinetic energy along the separated shear layer moves upstream when the oscillating f\/low is rotating in the clockwise direction causing an early-transition and vice versa. The best description of the mechanism is that of a whirligig. When the oscillating-f\/low rotates in the clockwise direction, the UV whirls in the clockwise direction and stores energy until it is saturated the process is then reversed. The strength and direction of rotation of the oscillating-f\/low; the transition location along the separated shear layer; the gradient of the oscillating-velocity across the laminar portion of the separated shear layer; the strength and dynamic of the triad of vortices; and the gradient of the oscillating-pressure along the aerofoil chord are interlinked, generate, and sustain the LFO. When these parameters oscillate in harmony, the LFO cycle repeats periodically. Whereas, when these parameters are not synchronised, a disturbed LFO cycle is produced, and the periodic LFO cycle is interrupted.\newline
The triad of vortices forms at moderate angles of attack, and before the LFO process begins. The magnitude of the oscillations increases and the frequency of the oscillations decreases as the angle of attack increases. At the beginning of the transition of the LFO at $\alpha=9.25^{\circ}$, the mean-f\/low is attached, the UV of the TCV occasionally expands; consequently, the LSB bursts and the f\/low-f\/ield separates intermittently. As the angle of attack increases, the UV expands more frequently and the f\/low separates. At the angle of attack of $9.8^{\circ}$, the LFO becomes fully developed and quasi-periodic. As the angle of attack further increases, the mean-f\/low becomes separated. Sometimes the UV expands, the f\/low attaches and the LSB reforms. Thus, the f\/low-f\/ield remains separated with intermittent attachments. The UV expands less frequently, and the f\/low-f\/ield remains separated for longer time intervals at higher angles of attack. At a critical angle of attack, the UV never expands, the f\/low-f\/ield remains separated, and the aerofoil undergoes a full stall. The presented mechanism also implies that the conditions for the LFO are met in both compressible and incompressible f\/lows. However, it is not clear yet whether the LFO phenomenon can be generated and sustained in an incompressible f\/low. Therefore, an experimental work in a water tunnel or solving the incompressible Navier--Stokes equations using direct numerical simulation is merited.

\section*{Acknowledgements}
All computations were performed on \verb"Aziz Supercomputer" at King Abdulaziz university's High Performance Computing Center (\url{http://hpc.kau.edu.sa/}). The authors would like to acknowledge the computer time and technical support provided by the center.

\bibliographystyle{jfm}
\bibliography{lfo2}

\newpage
\newpage
\newpage
\newpage
\newpage
\newpage
\newpage
\newpage
\newpage
\newpage
\newpage
\newpage
\newpage
\newpage
\newpage

\end{document}